\documentclass[floatfix,showpacs,aps,prc,nofootinbib,showkeys,superscriptaddress,reprint]{revtex4-1}

\usepackage[T1]{fontenc}
\usepackage{tgtermes}
\usepackage{graphicx}
\usepackage{xcolor}
\usepackage{dsfont}
\usepackage{bm}
\usepackage{slashed}
\usepackage{changepage} 
\usepackage{mathtools}
\usepackage{amsmath}
\usepackage{subfigure}
\usepackage{indentfirst}
\usepackage{hyperref}
\usepackage{booktabs, tabularx}
\usepackage{enumitem}
\usepackage[utf8]{inputenc}
\usepackage{lineno}

\hypersetup{
    colorlinks=true,
    linkcolor=blue,
    filecolor=magenta,      
    urlcolor=blue,
    citecolor=blue
}

\usepackage{amssymb}
\usepackage{multirow,array}
\usepackage{lipsum}
\usepackage{enumitem}

\newcommand{\be}{\begin{equation}}
\newcommand{\ee}{\end{equation}}
\newcommand{\bs}{\begin{subequations}}
\newcommand{\es}{\end{subequations}}
\newcommand{\beal}{\begin{align}}

\newcommand{\trento}{\texttt{T$_\mathrm{R}$ENTo}}
\newcommand{\URQMD}{\texttt{UrQMD}}
\newcommand{\SMASH}{\texttt{SMASH}}
\newcommand{\vishnew}{\texttt{VISHNew}}
\newcommand{\music}{\texttt{MUSIC}}

\newcommand{\sqrts}{\sqrt{s_\textrm{NN}}}
\newcommand{\Tsw}{T_{\text{sw}}}
\newcommand{\taufs}{\tau_{\text{fs}}}

\newcommand{\eq}[1]{Eq.~(\ref{#1})}

\newcommand{\fig}[1]{Fig.~\ref{#1}}
\newcommand{\figs}[1]{Figs.~\ref{#1}}

\newcommand{\Section}[1]{Section~\ref{#1}}
\newcommand{\Sections}[1]{Sections~\ref{#1}}

\newcommand{\Appendix}[1]{Appendix~\ref{#1}}

\newcommand{\Table}[1]{Table~\ref{#1}}

\DeclareMathOperator{\sign}{sign}

\begin{document}
\title{Multi-system Bayesian constraints on the transport coefficients of QCD matter}

\date{\today}


\author{D.~Everett}
\affiliation{Department of Physics, The Ohio State University, Columbus OH 43210.}

\author{W.~Ke}
\affiliation{Department of Physics, University of California, Berkeley CA 94270.}
\affiliation{Nuclear Science Division, Lawrence Berkeley National Laboratory, Berkeley CA 94270.}

\author{J.-F. Paquet}
\affiliation{Department of Physics, Duke University, Durham NC 27708.}

\author{G.~Vujanovic}
\affiliation{Department of Physics and Astronomy, Wayne State University, Detroit MI 48201.}

\author{S.~A.~Bass}
\affiliation{Department of Physics, Duke University, Durham NC 27708.}

\author{L.~Du}
\affiliation{Department of Physics, The Ohio State University, Columbus OH 43210.}

\author{C.~Gale}
\affiliation{Department of Physics, McGill University, Montr\'{e}al QC H3A\,2T8, Canada.}

\author{M.~Heffernan}
\affiliation{Department of Physics, McGill University, Montr\'{e}al QC H3A\,2T8, Canada.}

\author{U.~Heinz}
\affiliation{Department of Physics, The Ohio State University, Columbus OH 43210.}

\author{D.~Liyanage}
\affiliation{Department of Physics, The Ohio State University, Columbus OH 43210.}

\author{M.~Luzum}
\affiliation{Instituto  de  F\`{i}sica,  Universidade  de  S\~{a}o  Paulo,  C.P.  66318,  05315-970  S\~{a}o  Paulo,  SP,  Brazil. }

\author{A.~Majumder}
\affiliation{Department of Physics and Astronomy, Wayne State University, Detroit MI 48201.}

\author{M.~McNelis}
\affiliation{Department of Physics, The Ohio State University, Columbus OH 43210.}

\author{C.~Shen}
\affiliation{Department of Physics and Astronomy, Wayne State University, Detroit MI 48201.}
\affiliation{RIKEN BNL Research Center, Brookhaven National Laboratory, Upton NY 11973.}

\author{Y.~Xu}
\affiliation{Department of Physics, Duke University, Durham NC 27708.}


\author{A.~Angerami}
\affiliation{Lawrence Livermore National Laboratory, Livermore CA 94550.}

\author{S.~Cao}
\affiliation{Department of Physics and Astronomy, Wayne State University, Detroit MI 48201.}

\author{Y.~Chen}
\affiliation{Laboratory for Nuclear Science, Massachusetts Institute of Technology, Cambridge MA 02139.}
\affiliation{Department of Physics, Massachusetts Institute of Technology, Cambridge MA 02139.}

\author{J.~Coleman}
\affiliation{Department of Statistical Science, Duke University, Durham NC 27708.}

\author{L.~Cunqueiro}
\affiliation{Department of Physics and Astronomy, University of Tennessee, Knoxville TN 37996.}
\affiliation{Physics Division, Oak Ridge National Laboratory, Oak Ridge TN 37830.}

\author{T.~Dai}
\affiliation{Department of Physics, Duke University, Durham NC 27708.}

\author{R.~Ehlers}
\affiliation{Department of Physics and Astronomy, University of Tennessee, Knoxville TN 37996.}
\affiliation{Physics Division, Oak Ridge National Laboratory, Oak Ridge TN 37830.}

\author{H.~Elfner}
\affiliation{GSI Helmholtzzentrum f\"{u}r Schwerionenforschung, 64291 Darmstadt, Germany.}
\affiliation{Institute for Theoretical Physics, Goethe University, 60438 Frankfurt am Main, Germany.}
\affiliation{Frankfurt Institute for Advanced Studies, 60438 Frankfurt am Main, Germany.}

\author{W.~Fan}
\affiliation{Department of Physics, Duke University, Durham NC 27708.}

\author{R.~J.~Fries}
\affiliation{Cyclotron Institute, Texas A\&M University, College Station TX 77843.}
\affiliation{Department of Physics and Astronomy, Texas A\&M University, College Station TX 77843.}

\author{F.~Garza}
\affiliation{Cyclotron Institute, Texas A\&M University, College Station TX 77843.}
\affiliation{Department of Physics and Astronomy, Texas A\&M University, College Station TX 77843.}

\author{Y.~He}
\affiliation{Key Laboratory of Quark and Lepton Physics (MOE) and Institute of Particle Physics, Central China Normal University, Wuhan 430079, China.}

\author{B.~V.~Jacak}
\affiliation{Department of Physics, University of California, Berkeley CA 94270.}
\affiliation{Nuclear Science Division, Lawrence Berkeley National Laboratory, Berkeley CA 94270.}

\author{P.~M.~Jacobs}
\affiliation{Department of Physics, University of California, Berkeley CA 94270.}
\affiliation{Nuclear Science Division, Lawrence Berkeley National Laboratory, Berkeley CA 94270.}

\author{S.~Jeon}
\affiliation{Department of Physics, McGill University, Montr\'{e}al QC H3A\,2T8, Canada.}



\author{B.~Kim}
\affiliation{Cyclotron Institute, Texas A\&M University, College Station TX 77843.}
\affiliation{Department of Physics and Astronomy, Texas A\&M University, College Station TX 77843.}

\author{M.~Kordell~II}
\affiliation{Cyclotron Institute, Texas A\&M University, College Station TX 77843.}
\affiliation{Department of Physics and Astronomy, Texas A\&M University, College Station TX 77843.}

\author{A.~Kumar}
\affiliation{Department of Physics and Astronomy, Wayne State University, Detroit MI 48201.}


\author{S.~Mak}
\affiliation{Department of Statistical Science, Duke University, Durham NC 27708.}

\author{J.~Mulligan}
\affiliation{Department of Physics, University of California, Berkeley CA 94270.}
\affiliation{Nuclear Science Division, Lawrence Berkeley National Laboratory, Berkeley CA 94270.}

\author{C.~Nattrass}
\affiliation{Department of Physics and Astronomy, University of Tennessee, Knoxville TN 37996.}

\author{D.~Oliinychenko}
\affiliation{Nuclear Science Division, Lawrence Berkeley National Laboratory, Berkeley CA 94270.}



\author{C. Park}
\affiliation{Department of Physics, McGill University, Montr\'{e}al QC H3A\,2T8, Canada.}

\author{J.~H.~Putschke}
\affiliation{Department of Physics and Astronomy, Wayne State University, Detroit MI 48201.}

\author{G.~Roland}
\affiliation{Laboratory for Nuclear Science, Massachusetts Institute of Technology, Cambridge MA 02139.}
\affiliation{Department of Physics, Massachusetts Institute of Technology, Cambridge MA 02139.}

\author{B.~Schenke}
\affiliation{Physics Department, Brookhaven National Laboratory, Upton NY 11973.}

\author{L.~Schwiebert}
\affiliation{Department of Computer Science, Wayne State University, Detroit MI 48202.}

\author{A.~Silva}
\affiliation{Department of Physics and Astronomy, University of Tennessee, Knoxville TN 37996.}

\author{C.~Sirimanna}
\affiliation{Department of Physics and Astronomy, Wayne State University, Detroit MI 48201.}

\author{R.~A.~Soltz}
\affiliation{Department of Physics and Astronomy, Wayne State University, Detroit MI 48201.}
\affiliation{Lawrence Livermore National Laboratory, Livermore CA 94550.}

\author{Y.~Tachibana}
\affiliation{Department of Physics and Astronomy, Wayne State University, Detroit MI 48201.}

\author{X.-N.~Wang}
\affiliation{Key Laboratory of Quark and Lepton Physics (MOE) and Institute of Particle Physics, Central China Normal University, Wuhan 430079, China.}
\affiliation{Department of Physics, University of California, Berkeley CA 94270.}
\affiliation{Nuclear Science Division, Lawrence Berkeley National Laboratory, Berkeley CA 94270.}

\author{R.~L.~Wolpert}
\affiliation{Department of Statistical Science, Duke University, Durham NC 27708.}


\collaboration{The JETSCAPE Collaboration}

\begin{abstract}
We study the properties of the strongly-coupled quark-gluon plasma with a multistage model of heavy ion collisions that combines the \trento{} initial condition ansatz, free-streaming, viscous relativistic hydrodynamics, and a relativistic hadronic transport. A model-to-data comparison with Bayesian inference is performed, revisiting assumptions made in previous studies. The role of parameter priors is studied in light of their importance towards the interpretation of results. We emphasize the use of closure tests to perform extensive validation of the analysis workflow before comparison with observations. Our study combines measurements from the Large Hadron Collider and the Relativistic Heavy Ion Collider, achieving a good simultaneous description of a wide range of hadronic observables from both colliders. The selected experimental data provide reasonable constraints on the shear and the bulk viscosities of the quark-gluon plasma at $T\sim$ 150--250 MeV, but their constraining power degrades at higher temperatures $T \gtrsim 250$~MeV. Furthermore, these viscosity constraints are found to depend significantly on how viscous corrections are handled in the transition from hydrodynamics to the hadronic transport. Several other model parameters, including the free-streaming time, show similar model sensitivity, while the initial condition parameters associated with the \trento{} ansatz are quite robust against variations of the particlization prescription. We also report on the sensitivity of individual observables to the various model parameters. Finally, Bayesian model selection is used to quantitatively compare the agreement with measurements for different sets of model assumptions, including different particlization models and different choices for which parameters are allowed to vary between RHIC and LHC energies. 
\end{abstract}

\maketitle

\tableofcontents


\section{Introduction}

One of the primary goals of the heavy ion program pursued at the Relativistic Heavy Ion Collider (RHIC) and the Large Hadron Collider (LHC) is a quantitative understanding of the many-body properties of nuclear matter under extreme conditions as described by Quantum Chromodynamics (QCD). Lattice QCD calculations \cite{Borsanyi:2013bia,Bazavov:2014pvz} predict that hot and dense nuclear matter undergoes a cross-over transition near a temperature of 150 MeV at vanishing net baryon density, changing from hadronic degrees of freedom to a deconfined phase of nuclear matter --- the quark-gluon plasma (QGP). While lattice QCD calculations nowadays provide very precise first-principles information on the equilibrium properties of QCD matter across this transition from hadrons to color-deconfined QGP \cite{Borsanyi:2013bia, Bazavov:2014pvz}, the transport properties of QCD matter, such as its shear and bulk viscosities (which are of critical importance for its collective dynamical behavior in heavy-ion collisions), so far remain under the purview of phenomenology.\footnote{%
    See Ref.~\cite{Shen:2020gef} for a recent overview.} 
Calculating the viscosities of QCD from first principles remains challenging, in particular in the range of temperatures reached in heavy-ion collisions ($T\sim$ 150--500 MeV) where non-perturbative effects are important.\footnote{%
    See for example Ref.~\cite{Bazavov:2019lgz} and references therein for recent efforts at evaluating the QGP shear viscosity on the lattice, and Refs. \cite{Lu:2011df, Rose:2017bjz, Rose:2020lfc, Karsch:2007jc, NoronhaHostler:2008ju, Arnold:2006fz, Ghiglieri:2018dib} for other approaches to calculating these viscosities.}

The distributions and correlations of hadrons produced in heavy-ion collisions provide a complementary approach to quantify these transport properties of a QCD medium in the laboratory. Multistage dynamical models of heavy-ion collisions based on relativistic viscous hydrodynamics have been successful in describing a wide range of soft ($p_{\text{T}} \lesssim 2$ GeV) hadronic observables from RHIC and LHC \cite{Heinz:2013th, Gale:2013da, deSouza:2015ena, Schenke:2020mbo}. It was realized early on that hadronic observables measured in heavy-ion collisions are sensitive to the shear \cite{Romatschke:2007mq, Song:2007fn} and bulk \cite{Song:2008hj, Denicol:2009am} viscosities of QCD matter. Early works in constraining these transport coefficients with hydrodynamic simulations of heavy ion collisions generally focused on the shear viscosity ({\it c.f.} the reviews \cite{Gale:2013da, Heinz:2013th}) which was approximated as having a constant ratio to the entropy density $\eta/s$ (i.e. a constant {\it specific} shear viscosity). Contemporary efforts attempt to constrain the temperature and chemical potential dependence of the specific shear and bulk viscosities, namely $(\eta/s)(T, \mu_B)$ and $(\zeta/s)(T, \mu_B)$ \cite{hotQCDwhitepaper, Akiba:2015jwa}. Large-scale model-to-data comparisons are necessary to achieve this goal, given the considerable challenge of constraining the full functional form of $(\eta/s)(T, \mu_B)$ and $(\zeta/s)(T, \mu_B)$ in the relevant temperature and chemical potential ranges.

The primary goal of this work is to explore phenomenological constraints on the temperature dependence of both the specific shear and bulk viscosities of the nuclear matter produced in relativistic heavy ion collisions at high energy, where net baryon density is negligible at midrapidity. We use a modern multistage model to simulate heavy ion collisions and perform a Bayesian parameter estimation  on all the associated model parameters, including shear and bulk viscosities. The current analysis focuses on soft hadron measurements in the central rapidity region, as the available data are precise and they are arguably the observables whose theoretical uncertainties are best under control.

This work builds upon recent Bayesian analyses of soft hadronic observables \cite{Petersen:2010zt, Novak:2013bqa, Sangaline:2015isa, Bernhard:2015hxa, Bernhard:2016tnd}, in particular Ref.~\cite{Bernhard:2019bmu}. One difference of the current analysis with these previous works is our use of a more flexible parametrization of the model; our more general functional forms for $(\eta/s)(T)$ and $(\zeta/s)(T)$ and the wider range of viscosity values allowed (``the priors'') in the model-to-data comparison plays an important role in this new analysis. The role of the emulator is also discussed in greater depth, in particular the importance of closure tests as a validation of the Bayesian approach. For the first time, we quantify the uncertainties of $(\eta/s)(T)$ and $(\zeta/s)(T)$ stemming from known theoretical uncertainties in modeling viscous corrections to particlization; in the accompanying Letter \cite{Everett:2020xug} we use Bayesian model averaging to arrive at improved estimates by combining the constraints from different particlization models. The effects of the shear relaxation time on our constraints on $(\eta/s)(T)$ and $(\zeta/s)(T)$ are also quantified. We further explore constraining $(\eta/s)(T)$ and $(\zeta/s)(T)$ by combining experimental measurements from both RHIC and the LHC, and explore the subtleties of including measurements from different experiments.

Whereas one of the features of our work is the exploration of theoretical uncertainties associated with the viscous corrections to particlization, and attempts to minimize the latter by calibrating the model with $p_T$-integrated flow observables \cite{Song:2010mg}, an orthogonal approach is pursued in the recent works \cite{Nijs:2020ors, Nijs:2020roc} which restrict their analysis to a single particlization model and instead explore additional gains in precision by including additional $p_T$-differential flow observables.

Our work is performed within the greater scope of the JETSCAPE Collaboration \cite{Putschke:2019yrg}. The physics objectives for the Collaboration are (i) to obtain a complete and reliable space-time description of the quark-gluon plasma by performing a state-of-the-art calibration of its key properties using the large body of available experimental data, and (ii) to use this calibrated dynamical evolution model for performing precision studies of penetrating and hard probes of the evolving medium. Our focus here is on the first objective.

Unless noted otherwise, all equations in this manuscript use natural units ($\hbar = c = k_{B} = 1$). We use the ``mostly-minus'' metric signature: $g^{\mu\nu} = \text{diag}(1, -1, -1, -1)$ in Cartesian coordinates. In general, position and momentum four-vectors are denoted with capital letters, and their components with Greek indices, e.g. the space-time position four-vector $X$ has components $X^{\mu}$; three-vectors are denoted with boldface, and their components with Latin indices, e.g. the spatial position three-vector $\mathbf{X}$ has components $X^{i}$. The present work focuses on midrapidity observables from high-energy heavy-ion collisions where longitudinal boost-invariance is a good approximation. Such systems are most efficiently implemented using Milne coordinates $X^\mu=(\tau,\bm{x}_\perp,\eta_s)$, with $\tau=\sqrt{t^2{-}z^2}$ and $\eta_s=\frac{1}{2}\ln[(t{+}z)/(t{-}z)]$.

\section{Inference using Bayes' Theorem}
\label{sec:applying_bayes}

During the evolution of heavy ion collisions, the systems probe a wide range of many-body regimes of Quantum Chromodynamics. Because of the microscopic system size and its ultra-fast dynamics, this many-body physics must be inferred from the only information experimentally accessible in heavy ion collisions: the energy-momentum spectra and correlations of final-state particles that hit the detectors. These particles include stable (under the strong interaction) hadrons such as pions, kaons, protons, unstable hadronic resonances, and also electroweak bosons.

Given the complicated and non-linear correlations between parameters of the many-body dynamics and the observables, it is rare that a given physical effect --- for example the quark-gluon plasma having a certain temperature-dependent viscosity --- can be associated with a single observable constructed from final state particles. In general, quite different physical scenarios for the various evolution stages of the collisions may lead to quantitatively similar predictions for any single final state observable. Disentangling different physical scenarios requires simultaneous analysis of large sets of complementary observables, each having been measured with finite uncertainties. This is a formidable challenge, known as the ``inverse problem'' of complex models. It makes the field of heavy ion collisions a natural candidate for the application of advanced statistical methods to study many-body QCD.

From an abstract point of view, systematically performing statistical inference with a complex model requires a formalism or set of axioms that can form the basis for ``plausible reasoning''. The rules of Bayesian probability offer a natural formalism to systematically tackle such inverse problems. A Bayesian definition of the conditional probability of some proposition $A$ given known information $B$, denoted $\mathcal{P}(A|B)$, is a quantification of our degree of belief about $A$, or its ``plausibility''. A simple but powerful identity, which follows from the product rule of probability, is Bayes' theorem:
\begin{equation}
\label{eq1}
    \mathcal{P}(A|B) = \frac{\mathcal{P}(B|A)\,\mathcal{P}(A)}  {\mathcal{P}(B)}.
\end{equation}
Here the probability $\mathcal{P}(A)$ represents our prior knowledge or belief about the proposition $A$, $\mathcal{P}(B|A)$ is the likelihood for $B$ to be true if the proposition $A$ holds, and the normalization $\mathcal{P}(B) = \int dA\, \mathcal{P}(B|A)\,\mathcal{P}(A)$ is called the overall ``Bayes evidence'' for the information $B$ to be true. The left hand side of Eq.~(\ref{eq1}) is known as the posterior probability distribution (``posterior'' in short) for the proposition $A$ given the information $B$. Theorem (\ref{eq1}) allows us to invert the order of conditioning when we want to perform plausible reasoning or inference about some proposition $A$ with knowledge $B$ in hand. A more thorough introduction to Bayesian inference can be found in Ref. \cite{Sivia2006}. 

Broadly, we could quantify many different propositions. For example, one may want to quantify the likely values of transport coefficients such as shear and bulk viscosity, discussed in more detail later, given measured values for the soft ($p_T\lesssim 2$ GeV) hadronic observables such as multiplicities or mean transverse momenta. Another example of immediate interest is quantifying the likely values of transport coefficients controlling the energy-momentum exchange of hard partons and the quark-gluon plasma, given hard ($p_T\gtrsim 8$ GeV) observables such as the nuclear modification factor of inclusive hadron production. Any proposition regarding the physics of heavy ion collisions can in principle be quantified using Bayesian inference.

When performing inference for heavy-ion collisions, we only have the experimentally measured observables, complemented by first-principles physics considerations, to guide us towards what could plausibly be the dynamics of the collision. To make quantitative statements about the quark-gluon plasma requires a dynamical model for the entire heavy ion collision evolution. Broadly, the model is the relation between quantities of interest (such as the viscosities) and the observables: a ``map'' from the model parameters to the observables.

Many widely-used models for heavy ion collisions include a stage of evolution that is described by hydrodynamics, where dynamical properties are specified by a set of transport coefficients, such as the shear and bulk viscosities. Thus, a major goal of the phenomenology of heavy ion physics has been to infer the viscosities of the QGP given hydrodynamic models for its expansion. However, only the embedding of these macroscopic hydrodynamic models within a more sophisticated class of ``hybrid models'' \cite{Bass:2000ib, Nonaka:2006yn, Hirano:2007ei, Petersen:2008dd, Song:2010aq, Heinz:2011kt, Song:2013qma, Zhu:2015dfa, Ryu:2017qzn}, which add modules for describing on a more microscopic level the early pre-hydrodynamic and late hadronic rescattering and freeze-out stages, has enabled a fully quantitative modeling of heavy-ion collisions. The quantitative predictive power and precision of these modern dynamical approaches have now opened the door for serious efforts to tackle the ``inverse problem'':  instead of providing a range of model predictions based on a limited scan of the model parameter space using subjective selection criteria, the community is now beginning to exploit the full set of available experimental observations to provide quantitative estimates for the model parameters, with quantified uncertainties.

The method of inferring the likely model parameters given the observed data with use of Bayes' theorem is called Bayesian parameter estimation. This is the natural tool for estimating the likely values of the viscosities of the QGP, given a hydrodynamic model. This will occupy the largest portion of this work, beginning with a description of the hybrid model in \Section{sec:model_overview} and a description of parameter estimation in \Sections{sec:priors} through \ref{sec:param_est_syst_uncert}. In \Section{sec:model_sensitivity}, we take a brief aside to examine a measure of our model's sensitivity to changes in the parameters. This helps us understand which of the observables carry information about parameters, thereby also allowing us to interpret the effects of different observables on the posterior. 

Evidently, the likely values of any parameters depend not only on the observed data, but also on the model at hand. Some parameters that have meaning in a given model may not be relevant for a different model. Therefore, in general, it does not suffice to take a single model that works reasonably well, perform parameter estimation, and conclude from the estimations that the system is quantitatively understood. It is likely that there exists a ``superior'' model. In this context ``superiority'' should not only be measured by how well a model can fit a set of data. With sufficient complexity, any model can begin to overfit the data from a given experiment. Such overfitting leads to a model having a less universal applicability: when the model is applied to make a prediction in a previously unexplored region, it can perform very poorly. The best model to explain a given set of data should balance ``simplicity'' with accuracy in reproducing the experimental data.

A statistical tool for judging the relative merit of two models is the Bayes factor (the ratio of their Bayesian evidences as defined in Eq.~(\ref{eq1}) which weighs both their ability to fit measurements and their simplicity. Choosing between competing models using Bayes factors is called Bayesian model selection. We will use the Bayes factor to compare various choices of hydrodynamic hybrid models for heavy ion collisions in \Section{sec:model_selection}. It can happen that, after performing model selection for competing models that share certain common parameters, none of the models studied is strongly favored over all others by the available experimental data. In this situation, one might want to estimate the posterior for the shared parameters by averaging over the considered models \cite{Vardanyan_2011}. This procedure of Bayesian Model Averaging is explored in \cite{Everett:2020xug}, where it is used to obtain estimates for the shear and bulk viscosity that include the known theoretical model uncertainty regarding particlization, i.e. the process of converting hydrodynamic output into momentum spectra for finally observed particles.

\section{Model Overview}
\label{sec:model_overview}

Measurements of soft ($p_T{\,\lesssim\,}2$ GeV) particles from heavy ion collisions can be described well by multistage dynamical simulations whose core is relativistic viscous hydrodynamics. Features of many-body QCD enter hydrodynamics through medium properties such as the equation of state and transport coefficients (e.g. the shear and bulk viscosities). The collision dynamics preceding the applicability of hydrodynamics are described separately with a ``pre-hydrodynamic stage'' model. Following the hydrodynamic evolution, the last stage of the collision proceeds microscopically via hadronic transport.

The multistage model used in this work combines the following ingredients:
\begin{itemize}
    \item initial energy deposition from the colliding nuclei, given by the \trento\ ansatz \cite{Moreland:2014oya, trento_code}, followed by free-stream\-ing expansion \cite{Liu:2015nwa, Broniowski:2008qk,fs_code}; 
    \item relativistic viscous hydrodynamic evolution \cite{Schenke:2010nt, Schenke:2010rr, Paquet:2015lta, 2000JCoPh.160..241K} employing a lattice QCD based equation of state \cite{Bazavov:2014pvz, Bernhard:2018hnz, eos_code} and flexible temperature dependent parametrizations of the first-order transport coefficients;
    \item conversion of the nuclear fluid into particles employing the Cooper-Frye approach, where several different models are used for mapping the fluid's energy-momentum tensor to hadronic momentum distributions \cite{McNelis:2019auj,is3d_code};
    \item hadronic rescatterings in the hadronic transport model \SMASH\ \cite{Weil:2016zrk, smash_code} until kinetic freeze-out.
\end{itemize}%

In the following subsections we present details on each of these ingredients.

\subsection{Initial stage model}
\label{sec:model:init}

First-principles descriptions of the pre-hydrodynamic stage of a heavy ion collision remain challenging. However, many microscopic details of this early stage of the collisions are thought to be irrelevant \cite{Florkowski:2017olj} for the initialization of hydrodynamics at a time $\taufs$ of order 0.1--1 fm/$c$ (which, in kinetic theory language, requires only the first and second momentum moments of the microscopic distribution function). We therefore parametrize the initial conditions of hydrodynamics at $\taufs$, assuming that the system got to this time from $\tau=0^+$ by free-streaming. Evidence from other microscopic theories of the early dynamics in heavy-ion collisions \cite{Vredevoogd:2008id, Keegan:2016cpi, Kurkela:2018vqr, Kurkela:2018wud, Schlichting:2019abc} suggests that (at least for systems that are sufficiently weakly coupled to admit a kinetic theory description of their microscopic dynamics \cite{Kurkela:2019set}) coarse-grained (i.e. long wavelength) features of the energy-momentum tensor of a conformal system follow a relatively simple evolution similar to the free-streaming approach employed here. The energy deposition at $\tau=0^+$ is parametrized with the \trento{} ansatz. This approach should be judged by the flexibility of the combination of \trento{} with free-streaming, rather than by each component individually. Both components are now described in more detail.

\subsubsection{Energy deposition at $\tau=0^+$: \rm{T$_\mathrm{R}$ENTO}}

We use Woods-Saxon profiles to describe the distribution of nucleons in each colliding heavy nucleus.\footnote{%
    The Woods-Saxon parameters used for Au are radius $R=6.38$\,fm and surface thickness $a=0.535$\,fm while those for Pb are $R=6.62$\,fm and $a=0.546$\,fm.}
In the plane transverse to the collision axis the energy\footnote{%
    While initial studies used \trento\ to parametrize entropy density \cite{Moreland:2014oya, Bernhard:2016tnd, Ke:2016jrd} we follow later works \cite{Moreland:2018gsh, Bernhard:2019bmu} which use an energy density parametrization before free-streaming.}
deposition immediately following the impact between the two nuclei is parametrized with the \trento\ ansatz. This model parametrizes the transverse energy deposition via a reduced thickness function $T_R$,
\begin{eqnarray}
    \bar\epsilon(\mathbf{x}_\perp) 
    &\equiv& \frac{dE}{d\eta d^2\mathbf{x}_\perp} 
    = N T_R(\mathbf{x}_\perp; p),  
\label{eq:TrentoEd}
\\
\label{eq:TR}
    T_R(\mathbf{x}_\perp; p) &=& \left(\frac{T^p_A(\mathbf{x}_\perp) + T^p_B(\mathbf{x}_\perp)}{2}\right)^{1/p},
\end{eqnarray}
where $N$ is a normalization parameter estimated by comparison with data, and $T_A$ and $T_B$ represent the participant nucleon areal densities of the two nuclei.\footnote{%
    $T_{A/B}$ should not be confused with the similar nuclear thickness function that often appears in the optical Glauber model.}
In Eq.~(\ref{eq:TrentoEd}) we assume longitudinal boost invariance for the collision system, and $\bar\epsilon(\mathbf{x}_\perp)\equiv\lim_{\tau\to0+}\tau\epsilon(\tau,\mathbf{x}_\perp,\eta{=}0)$. The continuous parameter $p$ defines a family of mappings from $T_{A}$ and $T_B$ to the energy deposition. Specific values of $p$ are known to reproduce certain features of other initial condition models \cite{Moreland:2014oya}. When $p\to 0$, $T_R \to \sqrt{T_A T_B}$ and the model shares similar relations between eccentricities ($\epsilon_2, \epsilon_3$) and centrality \cite{Moreland:2014oya} as those predicted by the IP-Glasma initial condition model \cite{Schenke:2012wb}. This $\sqrt{T_A T_B}$-scaling is also expected from imposing the conservation of longitudinal momentum with a flux-tube profile ansatz for energy density along space-time rapidity \cite{Shen:2020jwv}. For $p=1$, the energy deposition is equivalent to the wounded nucleon Glauber model \cite{Miller:2007ri}. Nucleon positions are sampled from the Woods-Saxon profiles of the heavy nuclei; a parameter $d_{\min}$ controls the minimum distance between any pair of nucleons to mimic the short range repulsion of the nucleon potential.\footnote{%
    By default \trento\ sets a fixed value for $d_{\min}$ but we treat the corresponding volume $d_{\min}^3$ as variable when performing our parameter estimation. In particular, we will assign a uniform prior for $d_{\min}^3$.}
Nuclear collisions are generated by performing binary-nucleon inelastic collisions in a Monte-Carlo procedure.\footnote{%
    We define minimum bias events in \trento\ as events with at least a single binary collision. Events that produce no particles because they have no switching hypersurface into the hydrodynamic phase are included with zero multiplicity in the centrality selection.} 
The collision probability for two nucleons separated by an impact parameter $b$ is
\begin{equation}
\frac{dP}{d^2b} = 1 - \exp \left[ -\sigma_{gg} \int d^2 \mathbf{x}_\perp \rho\left(\mathbf{x}_\perp{+}\frac{\mathbf{b}}{2}\right) \rho\left(\mathbf{x}_\perp{-}\frac{\mathbf{b}}{2}\right) \right].
\end{equation}
$\sigma_{gg}\left(\sqrt{s}\right)$ is an effective nucleon opacity fixed by the proton-proton inelastic cross-section at a given beam energy,
\begin{equation}
    \sigma_{pp}^{\textrm{in}}\left(\sqrt{s}\right) = \int d^2b\, \frac{dP}{d^2b}(\mathbf{b}, \sigma_{gg}),
\end{equation}
and $\rho$ is the longitudinally-integrated  nucleon density per unit area in the transverse plane,
\begin{eqnarray}
    \rho(\mathbf{x}_\perp) &=& \int_{-\infty}^{\infty} \frac{dz}{\left(2\pi w^2\right)^{3/2}}\exp\left(-\frac{\mathbf{x}_\perp^2+z^2}{2w^2}\right),
\end{eqnarray}
i.e. the nucleon thickness function. Here, we have modeled the density distribution of a nucleon as a three-dimensional Gaussian function with a width parameter $w$. Finally, each participant nucleon contributes to the participant density function,
\begin{equation}
    T_{A}(\mathbf{x}_\perp) = \sum_{i\in A} \gamma_i \rho(\mathbf{x}_\perp{-}\mathbf{x}_{i,\perp}),
\end{equation}
and similarly for nucleus $B$, where $\mathbf{x}_{i,\perp}$ is the position of participant nucleon in the transverse plane, and each nucleon's contribution is individually modulated by a stochastic factor $\gamma_{i}$ sampled from a $\Gamma$-distribution with unit mean and standard deviation parameter $\sigma_k$.\footnote{%
    By default \trento\ sets a fixed value for $k = 1/\sigma^2_k$, but we treat $\sigma_k$ as variable when performing our parameter estimation.}
This additional source of fluctuations is introduced to model the large multiplicity fluctuations observed in minimum-bias proton-proton collisions.

\subsubsection{Free-streaming}
\label{sec:model:fs}

In this study, a non-trivial initial energy-momentum tensor is generated via a free-streaming model. It takes the initial energy density profile by \trento\ and dynamically generates non-zero components for the entire energy-momentum tensor. This section provides details pertaining to how this is achieved.

The energy density (\ref{eq:TrentoEd}) generated with \trento\ at $\tau_0=0$ is assumed to represent massless degrees of freedom with a locally isotropic momentum distribution $f(X; P)$ centered at zero transverse momentum whose effective temperature (or mean $P_T$) varies with position $X$. The initial phase-space distribution is evolved by free-streaming, i.e. it is assumed to solve the collisionless Boltzmann equation
\be
    P^{\mu} \partial_{\mu} f(X;P) = 0,
\ee
which has the solution 
\be
    f(t,\boldsymbol{x};P) = f\bigl(t_0, \boldsymbol{x}{-}\boldsymbol{v}(t{-}t_0); P\bigr).
\label{free-streaming-soln}
\ee
For massless particles $P^0 = |\boldsymbol{P}|$ and $\hat{p}^{\mu} \equiv P^{\mu} / P^0 = (1, \boldsymbol{v})$ where $|\boldsymbol{v}|=c$, i.e. all particles move with the speed of light. It is convenient to define a moment $F(X; \Omega_p)$ of the distribution function by
\begin{equation}
    F(X;\Omega_p) \equiv \frac{g}{(2\pi)^3} \int P_0^3\, dP_0\, f(X;P) 
\end{equation}
where $g$ is a degeneracy factor. Then the stress tensor at any time $t > t_0$ is given by
\begin{eqnarray}
\label{Tmn_fs}
    T^{\mu\nu}(t,\boldsymbol{x}) &=&  \int d\Omega_p\, \hat{p}^{\mu} \hat{p}^{\nu}\, F(t, \boldsymbol{x}; \Omega_p) \nonumber \\
    &=& \int d\Omega_p\, \hat{p}^{\mu} \hat{p}^{\nu}\, F\bigl(t_0, \boldsymbol{x}{-}\boldsymbol{v}(t{-}t_0\bigr) ; \Omega_p),
\end{eqnarray}
where $\Omega_p$ is the solid angle in momentum space. The second equality was obtained by inserting the free-streaming solution (\ref{free-streaming-soln}). 

Since we assume that the initial distribution function is locally isotropic in momentum space, the moment $F(t_0)$ can be related to the initial energy density by normalization:
\begin{equation}
    T^{tt}(t_0, \boldsymbol{x}) =  \mathcal{N} F(t_0, \boldsymbol{x}).
\label{eq:FS_ed_mapping}
\end{equation}
In three spatial dimensions $\mathcal{N} = 4\pi$. Here we assume longitudinal boost invariance, rendering the momentum distribution essentially two-dimensional, and therefore use $\mathcal{N} = 2\pi$ in the mapping (\ref{eq:FS_ed_mapping}). A more detailed description of the boost invariant case can be found in \cite{Liu:2015nwa, Broniowski:2008qk,fs_code}. 

Free-streaming is stopped at a longitudinal proper time $\tau_{\text{fs}}$, and the free-streamed energy-momentum tensor (\ref{Tmn_fs}) is matched to that of viscous hydrodynamics through Landau matching conditions. The energy density $\epsilon$ in the local rest frame (LRF) and the flow velocity $u^{\mu}$ are the eigenvalue and timelike eigenvector of $T^{\mu\nu}$:
\be
    u_{\mu}T^{\mu}_{\nu} = \epsilon u_{\nu}.
\ee
The shear stress tensor is also matched exactly and given by the traceless and transverse projection of the stress tensor:
\be
    \pi^{\mu\nu} = \Delta^{\mu\nu}_{\alpha\beta}T^{\alpha\beta}.
\ee
where the transverse-traceless projector is defined by 
\begin{equation}
    \Delta^{\mu\nu}_{\alpha\beta} \equiv \frac{1}{2}(\Delta^{\mu}_{\alpha} \Delta^{\nu}_{\beta} + \Delta^{\nu}_{\alpha} \Delta^{\mu}_{\beta}) - \frac{1}{3}\Delta^{\mu\nu}\Delta_{\alpha\beta}
\label{eq:delta_munu_alphabeta}
\end{equation}
and 
\be
    \Delta^{\mu\nu} \equiv g^{\mu\nu} - u^{\mu}u^{\nu}
\ee
projects onto the spatial directions in the local rest frame. The LRF is defined by $u^{\mu}_{\rm LRF} = (1, \bf{0})$. Because the free-streaming dynamics continuously drive the system out of equilibrium, the size of the initial shear stress tensor grows with the free-streaming time $\tau_{\text{fs}}$. We have checked in our numerical calculations that the second-order viscous hydrodynamics can reliably evolve the free-streaming $\pi^{\mu\nu}$ tensor to near local equilibrium for $\tau_{\text{fs}} \lesssim 1.5$\,fm/$c$.

The combination of thermal and bulk pressure, $p+\Pi$, is given by
\begin{equation}
    p(\epsilon)+\Pi=\frac{\epsilon-T^{\mu}_{\phantom{\mu}\mu}}{3}.
\end{equation}
Since we assume massless degrees of freedom (i.e. conformal symmetry) during the free-streaming stage, the energy-momentum tensor (\ref{Tmn_fs}) is traceless, $T^{\mu}_{\phantom{\mu}\mu}{\,=\,}0$.
The QGP fluid described by viscous hydrodynamics, on the other hand, is characterized by an equation of state that breaks conformal symmetry by interactions and non-zero quark masses: $p_{_{\text{QCD}}}(\epsilon) < \epsilon/3$. Matching of the free-streamed energy-momentum tensor to the hydrodynamic one thus entails a non-zero, positive initial value for the fluid's bulk viscous pressure at $\taufs$: 
\begin{equation}
\label{Pi_init}
    \Pi=\frac{\epsilon}{3}-p_{_{\text{QCD}}}(\epsilon) \ge 0.
\end{equation}
Note that for an expanding system a negative bulk viscous pressure is expected in the Navier-Stokes limit; the hydrodynamic evolution therefore quickly erases the positive initial value (\ref{Pi_init}) and switches its sign. Persistence effects from the initially positive bulk viscous pressure depend on the value of the bulk relaxation time, which is studied in \Appendix{app_bulk_relax}. Different pre-hydrodynamic evolution models in which interaction effects break the conformal symmetry may lead to different initial conditions for the bulk viscous pressure, even in sign; this might be worth investigating in future studies.

In the present model, the initialization time for hydrodynamics is the free-streaming time $\tau_{\text{fs}}$. It is common for model calculations to assume that this hydrodynamic initialization time is the same for all centralities and/or for different collision systems.\footnote{%
    See, e.g., Ref.~\cite{Oliinychenko:2015lva} for examples and exceptions.}
There are, however, reasons to believe that systems with higher energy densities ``hydrodynamize'' faster \cite{Basar:2013hea}. To capture this effect we parametrize the free-streaming time $\taufs$ to include a dependence on the initially deposited transverse energy density $\bar\epsilon$ from Eq.~(\ref{eq:TrentoEd}):
\begin{equation}
   \taufs = \tau_R \left(\frac{\langle \bar\epsilon \rangle}{\bar\epsilon_R}\right)^{\alpha}.
   \label{eq:taufs}
\end{equation}
Here $\tau_R$ is a normalization factor for the duration of the free-streaming stage, and the parameter $\alpha$ controls its dependence on the magnitude of the ``average'' initial energy density in the transverse plane, defined by
\begin{equation}
\label{eps_ave}
    \langle \bar\epsilon \rangle \equiv \frac{\int d^2x_\perp\, {\bar\epsilon}^2(\mathbf{x}_\perp)}{\int d^2x_\perp\, \bar\epsilon(\mathbf{x}_\perp)}.
\end{equation}
We choose $\bar\epsilon_R = 4.0$\,GeV/fm$^2$ as an arbitrary reference scale; the resulting posteriors for $\alpha$ and $\tau_R$ depend on this choice. 

\subsection{Relativistic viscous hydrodynamics}
\label{sec:hydro}

After Landau matching, the energy-momentum tensor of the system is evolved with second-order relativistic viscous hydrodynamics \cite{Denicol:2012es, Schenke:2010nt, Schenke:2010rr, Shen:2014vra, Paquet:2015lta}. In this work we use the dissipative fluid dynamics code MUSIC \cite{hydro_code} and focus on the midrapidity region where we can approximate the dynamics as effectively (2+1)-dimensional with longitudinal boost-invariance. Conservation of energy and momentum
\begin{eqnarray}
    \partial_{\mu}T^{\mu\nu} &=& 0,
\label{cons}
\\
    T^{\mu\nu} &=& \epsilon u^\mu u^\nu - (p+\Pi)\Delta^{\mu\nu} + \pi^{\mu\nu}
\label{eq:T_munu}
\end{eqnarray}
provides evolution equations for the energy density and flow for given viscous flows. The latter (i.e. the shear stress $\pi^{\mu\nu}$ and bulk viscous pressure $\Pi$ in Eq.~(\ref{eq:T_munu})) follow their own relaxation equations:
\begin{align}
    \tau_{\Pi }\dot{\Pi}+\Pi &= -\zeta \theta -\delta _{\Pi \Pi }\Pi \theta
    + \lambda _{\Pi \pi }\pi ^{\mu \nu }\sigma_{\mu \nu }\;,
\label{relax_eqn_PI}
\\
    \tau_{\pi }\dot{\pi}^{\left\langle \mu \nu \right\rangle }+\pi ^{\mu \nu }
    &= 2\eta \sigma ^{\mu \nu }-\delta _{\pi \pi }\pi ^{\mu \nu }\theta
    +\varphi_{7}\pi _{\alpha }^{\left\langle \mu \right. }\pi ^{\left. \nu \right\rangle \alpha } \notag 
    \\
    &\ -\tau _{\pi \pi }\pi _{\alpha }^{\left\langle \mu \right. }\sigma^{\left. \nu \right\rangle \alpha }+\lambda _{\pi \Pi }\Pi \sigma ^{\mu\nu}.
\label{relax_eqn_pi}
\end{align}
Here $\dot{\Pi} = u^\lambda \partial_\lambda \Pi$, $\dot{\pi}^{\langle\mu\nu\rangle} = \Delta^{\mu\nu}_{\alpha\beta} u^\lambda \partial_\lambda \pi^{\alpha\beta}$, $\theta = \partial_\lambda u^\lambda$, and $\sigma^{\mu\nu} = \Delta^{\mu\nu}_{\alpha\beta}\partial^\alpha u^\beta$, with $\Delta_{\alpha\beta}^{\mu\nu}$ from Eq.~(\ref{eq:delta_munu_alphabeta}).

The equilibrium properties of QCD matter are encoded in the equation of state which enters Eq.~(\ref{eq:T_munu}) through the pressure $p{\,=\,}p(\epsilon)$. The near-equilibrium dynamics of QCD matter is controlled by first and second-order transport coefficients that enter in Eqs.~(\ref{relax_eqn_PI},\ref{relax_eqn_pi}). The first-order transport coefficients are the shear and bulk viscosities, $\eta$ and $\zeta$. Second-order transport coefficients entering into our hydrodynamic equations are $\delta _{\Pi \Pi }$, $\lambda _{\Pi \pi }$, $\delta _{\pi \pi }$, $\varphi _{7}$, $\tau _{\pi \pi }$, and $\lambda_{\pi\Pi }$, as well as the shear and bulk relaxation times $\tau _{\pi }$ and $\tau _{\Pi }$.

For the equilibrium properties we use an equation of state matched to (i) a lattice calculation \cite{Bazavov:2014pvz} at high temperatures and (ii) a hadron resonance gas at lower temperatures (see Refs.~\cite{Bernhard:2018hnz, eos_code} for details). The hadron content of the resonance gas is chosen to be consistent with that of the hadronic afterburner \SMASH \cite{Weil:2016zrk} used in this work. While this consistency is important, the matching procedure does carry some uncertainties (see. e.g., Ref.~\cite{Auvinen:2020mpc}) which are not explored in this work.

The shear and bulk viscosities, $\eta$ and $\zeta$, are parametrized as functions of temperature, and measurements are used to constrain them.\footnote{\label{fn10}%
    In general, if conserved charges are taken into account (which is not done here), the transport coefficients also depend on chemical potentials.}
They are discussed in more detail below.

With even less quantitative theoretical guidance available than for the first-order transport coefficients, the second-order ones should similarly be parametrized and constrained from measurements. Some studies\footnote{%
    The effects of second-order transport coefficients on the hydrodynamic evolution were numerically checked in \cite{Liu-thesis, Schenke:2019pmk} and found to be small. The hydrodynamic equations of motion used here are based on the Grad approach and constructed such that the second-order transport effects are assumed to be small compared to the first-order ones \cite{Denicol:2014loa}.} 
suggest, however, that the second-order transport coefficients have a smaller effect than first-order ones on the evolution of the plasma. In this work we therefore apply the same strategy as for the first-order transport coefficients $\eta$ and $\zeta$ only to the shear relaxation time $\tau _{\pi}$, whereas all other second-order transport coefficients are related to first-order ones by using parameter-free relations derived in a moment expansion of the Boltzmann equation \cite{Denicol:2014vaa}.

\begin{figure}[!htb]
\includegraphics[width=9cm]{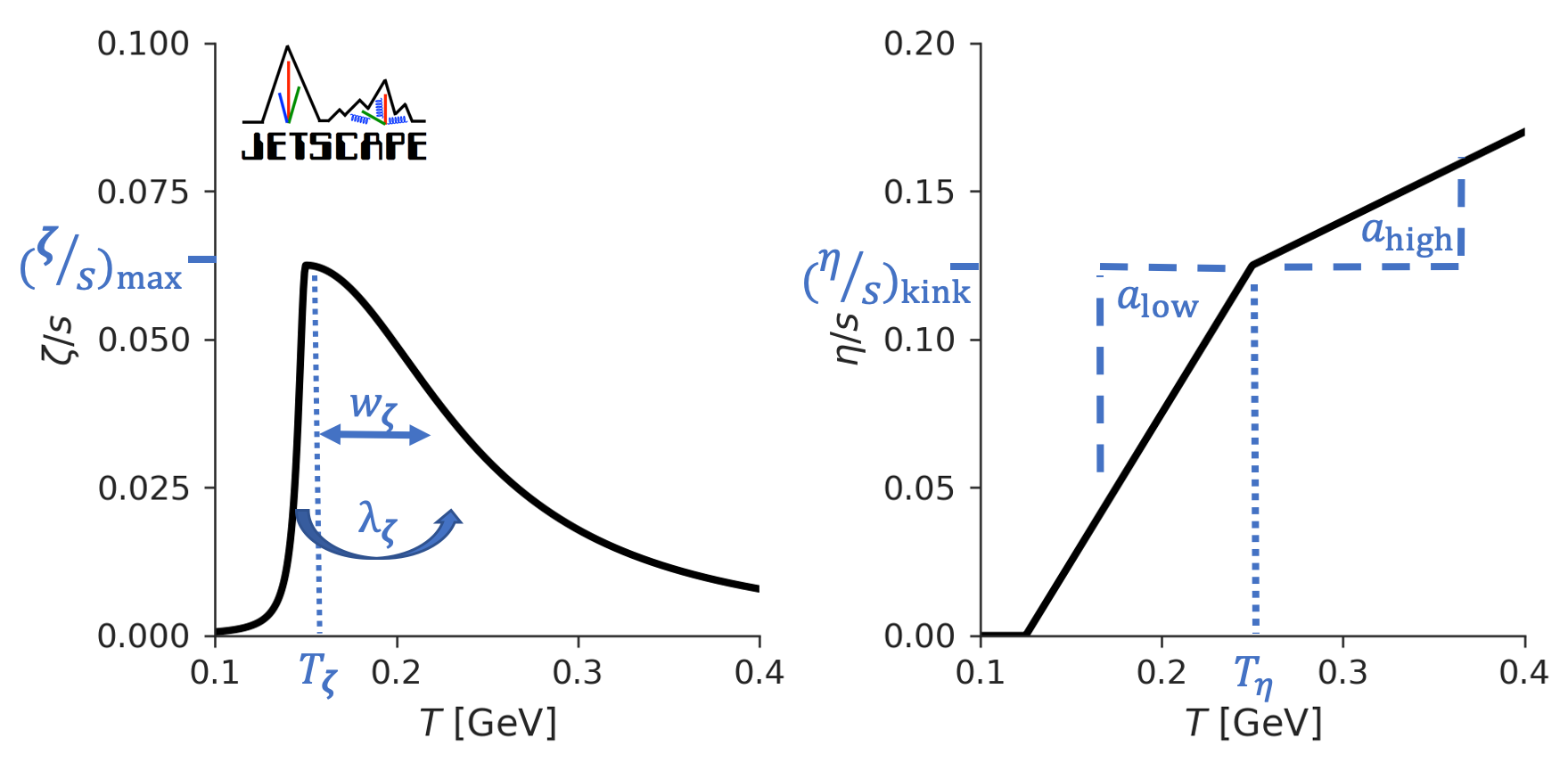}
    \caption{Depictions of the parametrizations of specific bulk (left) and shear (right) viscosity as functions of temperature. The specific bulk viscosity has the form of a skewed Cauchy distribution, while the specific shear viscosity is piecewise-linear, with in general two different slopes. Both shear and bulk viscosities are required to be positive-definite to satisfy the second law of thermodynamics. The example for $(\eta/s)$ shown here has a positive low-temperature and high-temperature slope ($a_{\rm low},a_{\rm high}$>0).}
\label{fig:eta_zeta_pzations}
\end{figure}

All transport coefficients depend on the equilibrium properties of the system, which we characterize with the temperature $T$.$^{\ref{fn10}}$ 
We parametrize the ratios of shear and bulk viscosity to entropy density --- the unitless specific viscosities --- instead of parametrizing the viscosities themselves. A depiction of the parametrizations for the specific bulk and shear viscosities is shown in Fig.~\ref{fig:eta_zeta_pzations}. 

For the specific shear viscosity, $\eta/s$, we assume that it has a single inflection point at or above the deconfinement transition \cite{Niemi:2015qia}. The position of this inflection point in temperature, $T_{\eta}$, is a parameter, as is the value of $\eta/s$ at this point, $(\eta/s)_{\rm kink}$. A linear dependence of $\eta/s$ on temperature is assumed, with slopes $a_{\rm low}$ below and $a_{\rm high}$ above the inflection point, with both positive and negative slopes allowed. Negative values for $\eta/s$ are not allowed. The formula for this parametrization is
\begin{equation}
\label{positivity}
    \frac{\eta}{s}(T) = \max\left[\left.\frac{\eta}{s}\right\vert_{\rm lin}\!\!\!(T),0\right],
\end{equation}
with
\begin{eqnarray}
    \left.\frac{\eta}{s}\right\vert_{\rm lin}\!\!\!(T) &=& a_{\rm low}\, (T{-}T_{\eta})\, \Theta(T_{\eta}{-}T)+ (\eta/s)_{\rm kink}
\nonumber\\
    && +\, a_{\rm high}\, (T{-}T_{\eta})\, \Theta(T{-}T_{\eta}).
\end{eqnarray}
Theoretically expected is a negative slope at temperatures below $T_{\eta}$, i.e. $a_{\rm low}{\,<\,}0$, and a positive slope at temperature above $T_{\eta}$, i.e. $a_{\rm high}{\,>\,}0$ \cite{Csernai:2006zz}. Nevertheless, in this work, we will allow both slopes to take negative and positive values: the aim is to ascertain whether phenomenological constraints are consistent with the theoretical expectations.

For the specific bulk viscosity, we assume that it peaks near the deconfinement temperature and that this single peak can be captured with a skewed Cauchy distribution:

\begin{eqnarray}
    \frac{\zeta}{s}(T) &=& \frac{(\zeta/s )_{\max}\Lambda^2}{\Lambda^2+ \left( T-T_\zeta\right)^2},\\
    \Lambda &=& w_{\zeta} \left[1 + \lambda_{\zeta} \sign \left(T{-}T_\zeta\right) \right]\nonumber.
\end{eqnarray}
Here $T_\zeta$ is the position and $(\zeta/s)_{\max}$ the value of the peak; $w_\zeta$ and $\lambda_{\zeta}$ control the width and skewness of the Cauchy distribution, respectively. Allowing for a non-vanishing skewness is a generalization compared to Ref.~\cite{Bernhard:2019bmu}.

Previous studies \cite{Kharzeev:2007wb, Karsch:2007jc, NoronhaHostler:2008ju, Rose:2020lfc, Arnold:2006fz} suggest that $\zeta/s$ for QCD peaks near the deconfinement transition. The functional form of its temperature-dependence is still not well understood. Below the transition ($T\lesssim 150$\,MeV), the bulk viscosity is understood to be non-zero. We emphasize that we do not attempt to describe the dependence of the bulk viscosity below the particlization temperature of our model (discussed in the next section) which is never smaller than 135\,MeV. The fact that our parametrization of $(\zeta/s)(T)$ rapidly approaches zero at low temperature should therefore not be read as a physical feature: this low temperature range is never described by the hydrodynamic code, but rather microscopically by a hadronic transport model. While we thus cannot make any statements about the bulk viscosity of QCD matter at these low temperatures it has recently been estimated in the \SMASH{} transport model \cite{Rose:2020lfc}.

Previous theoretical work \cite{Baier:2007ix, Bhattacharyya:2008jc, Denicol:2014vaa,Florkowski:2015lra, Czajka:2017wdo, Ghiglieri:2018dgf} suggests that, in the absence of conserved charges, the shear relaxation time can be well captured by following temperature dependence:
\begin{equation}
    T \tau_\pi(T)= b_{\pi}\frac{\eta}{s}(T)
\end{equation}
where $b_\pi$ is a constant that we consider unknown. The linearized causality bound \cite{Pu:2009fj} requires $b_\pi{\,\ge\,}(4/3)/(1{-}c_s^2){\,\ge\,}2$. Refs.~\cite{Baier:2007ix, Bhattacharyya:2008jc, Denicol:2014vaa, Florkowski:2015lra, Czajka:2017wdo} showed for a variety of weakly and strongly coupled theories other than QCD that this causality bound is respected, with $b_\pi$ varying between ${\sim}2$ and ${\sim}6$; we use these values to limit the prior range explored for $b_\pi$ in our parameter estimation. 

Previous investigations of the effects of the shear relaxation time and other second-order transport coefficients on soft hadronic observables have found them to be of modest phenomenological importance \cite{Song:2008si, Liu-thesis, Bernhard:2015hxa, Schenke:2019pmk}, consistent with general theoretical expectations (see e.g. Ref.~\cite{Schaefer:2014awa}). Nevertheless varying the shear relaxation time in this work provides additional quantitative insights into the typical magnitude of effects from a second-order coefficient on the Bayesian constraints for the first-order transport coefficients.

\subsection{Particlization}
\label{sec:model:particlization}

Particlization is not a physical process but a change of language from a description in terms of macroscopic fluid dynamical degrees of freedom to a microscopic kinetic description in terms of particles with positions and momenta. We here implement it on a surface of constant ``switching'' or ``particlization'' temperature $\Tsw$. In principle, this translation requires simultaneous applicability of both approaches. Since hydrodynamics rapidly breaks down below the confinement transition because the mean-free path increases as a consequence of color neutralization, while the strongly-coupled nature of the color confinement process itself makes kinetic theory inapplicable during the phase transition, this condition puts rather tight theoretical constraints on the temperature range for the particlization procedure. We here impose these constraints through a prior range within which we sample the particlization temperature. As we will see below, experimental data on hadronic yields provide strong constraints on $\Tsw$, given the assumption that particlization happens at $\mu_i=b_i \mu_B + s_i \mu_S + q_i \mu_Q$ with $\mu_B{\,=\,}\mu_S{\,=\,}\mu_Q{\,=\,}0$ (Eq.~\ref{chemeq}). 

The Cooper-Frye \cite{Cooper:1974mv, Cooper:1974qi} prescription for particlization \cite{Huovinen:2012is} is used to convert all the energy and momentum of the fluid into hadrons on the switching hypersurface $\Sigma$. The formula for the Lorentz-invariant particle momentum spectrum of particles of species $i$ with degeneracy $g_i$ in terms of their kinetic phase-space distribution $f_i(X;P)$ is given by
\begin{equation} 
\label{CFEqn}
    P^0 \frac{dN_i}{d^3P} = \frac{g_i}{(2\pi)^3} \int_{\Sigma} d^3\sigma_{\mu} P^{\mu} f_i(X;P).
\end{equation}
The integral goes over the switching hypersurface $\Sigma$ with normal vector $\sigma_{\mu}(X)$. The distribution function $f_i(X;P)$ must be chosen such that it reproduces the hydrodynamic energy-momentum tensor of the fluid on the particlization surface:
\be
\label{Tmunu_kinetic}
    T^{\mu\nu}(X) = \sum_i g_i \int \frac{d^3P}{(2\pi)^3P^0} P^{\mu}P^{\nu} f_i(X;P). 
\ee
Without any hydrodynamic information on all the infinitely many other momentum moments of the distributions functions, and no hydrodynamic information on how to split the fluid $T^{\mu\nu}$ into contributions from different hadron species $i$ as written in Eq.~(\ref{Tmunu_kinetic}), this leaves infinitely many choices for the distribution functions $f_i(X,P)$. If the QGP fluid were an ideal fluid in perfect local kinetic and chemical equilibrium, the choice for $f_i(X,P)$ would be unambiguous: it would be of local equilibrium form \cite{Cooper:1974mv, Cooper:1974qi}, with the local rest frame velocity provided by hydrodynamics and the temperature and chemical potentials fixed by the local-rest-frame energy density and the chemically equilibrated hadronic particle densities. In this case hadronic chemical potentials would be constrained by the equilibrium relations 
\begin{equation}
\label{chemeq}
    \mu_i = b_i \mu_B + s_i \mu_S + q_i \mu_Q
\end{equation}
where $\mu_B$, $\mu_S$ and $\mu_Q$ are the chemical potentials associated with the conserved baryon number, strangeness and electric charge, and $(b_i,s_i,q_i)$ are the baryon, strangeness and electric charges carried by hadron species $i$. All these chemical potentials are zero in this work, $\mu_B{\,=\,}\mu_S{\,=\,}\mu_Q{\,=\,}0$, reflecting the approximately zero net baryon, strangeness and electric charge near midrapidity at top RHIC and LHC collision energies. As it is, the QGP is a dissipative fluid with nonzero dissipative flows contributing to $T^{\mu\nu}$ on the particlization surface, and since hydrodynamics does not provide any microscopic information on how the system evolved to this surface we are left with a large and irreducible ambiguity as to the choice of local momentum distributions for the different hadron species \cite{Dusling:2009df, Molnar:2014fva, Damodaran:2017ior, Damodaran:2020qxx}.

Working off the hypothesis that the color-confining hadronization process is very strongly coupled and involves a multitude of different possible color-neutralizing microscopic channels \cite{Heinz:1999kb}, and with strong support from experimental observations \cite{Andronic:2005yp, Andronic:2017pug} we will assume that the chemical potentials for the different hadronic species satisfy the above chemical equilibrium relations (\ref{chemeq}) at particlization as long as we perform it soon after completion of hadronization. However, the dissipative flows in $T^{\mu\nu}$ reflect {\it deviations of the hadrons'\ momentum distributions and yields from local thermodynamic equilibrium}, arising from dissipative corrections. To specify these deviations one might want to argue that the distribution functions $f_i(X;P)$ should solve a set of coupled Boltzmann equations, but this begs the question what should be assumed for the form of the collision terms and the initial conditions (both of which must be expected to be strongly affected by the proximity in space and time of the preceding hadronization process about whose microscopic dynamics we know very little). 

In such a situation of irreducible theoretical ambiguity it makes sense to ask what constraints the experimental data may provide. To pose this question we here consider four different models of viscous corrections to the local equilibrium distribution function:
\begin{enumerate}[topsep=1pt,itemsep=0ex,partopsep=1ex,parsep=1ex]
    \item Grad's method \cite{Grad} (a.k.a. 14-moments method in the relativistic context \cite{Israel:1976tn, Israel:1979wp, Teaney:2003kp, Dusling:2009df, Monnai:2009ad, Dusling:2011fd, Denicol:2012cn});
    \item the first-order Chapman-Enskog (CE) expansion in the Relaxation Time Approximation \cite{chapman1990mathematical, 
    ANDERSON1974466, Jaiswal:2014isa};
    \item the Pratt-Torrieri-Bernhard (PTB) modified equilibrium distribution \cite{Pratt:2010jt, Bernhard:2018hnz};
    \item the Pratt-Torrieri-McNelis (PTM) modified equilibrium distribution \cite{Pratt:2010jt, McNelis:2019auj}.
\end{enumerate}

Given the values of the 10 components of the energy-momentum tensor, these models are used to determine how energy and momentum are distributed among hadronic species and across momentum. By performing Bayesian parameter estimation, combined with Bayesian Model Selection techniques using these four models, we aim to estimate the theoretical uncertainty in the extraction of the transport coefficients resulting from the viscous corrections at particlization. 

We briefly describe the four models individually. For a more in-depth review and comparison of these models we refer the reader to Ref.~\cite{McNelis:2019auj}.

\subsubsection{Linear viscous corrections: Grad \& Chapman-Enskog}

The Grad and Chapman-Enskog methods have both been used extensively in the literature to study particle production in heavy ion collisions. They give corrections which are linear in the dissipative currents $\pi^{\mu\nu}$ and $\Pi$. Both ans\"atze should only be valid when these corrections to the thermal equilibrium distribution are small. In practice this approximation is often pushed to the limit or even beyond. In the following we describe the Grad and Chapman-Enskog methods in turn. We then discuss regularization that is applied similarly to both approaches when large viscous corrections are encountered.

\paragraph{Grad (or 14-moments) approximation:}
What we refer to as ``Grad's method'' assumes that the correction to the local equilibrium distribution function can be expanded in powers of hadronic momentum. Including only the terms relevant for a system without a conserved charge yields 
\begin{equation}
    \delta f_i =  f_{\text{eq}, i}  \bar{f}_{\text{eq}, i} c_{\mu\nu}P^{\mu}P^{\nu}
\end{equation}
where $\bar{f}_{\text{eq}, i} \equiv 1{-}\Theta f_{\text{eq}, i}$, and $\Theta$ is 1 for fermions and ${-}1$ for bosons. Assuming that the coefficients $c_{\mu\nu}$ are species-independent and requiring the Landau-matching conditions yields the following expression for the viscous correction in terms of the dissipative currents: 
\begin{eqnarray}
    \delta f^{\text{Grad}}_i = f_{\text{eq}, i} \bar{f}_{\text{eq}, i} \bigl[ \Pi \left( A_{T}m_i^2{+}A_E(u{\cdot}P)^2  \right)\ && 
\nonumber\\
    + A_{\pi}\pi^{\mu\nu}P_{\langle \mu}P_{\nu \rangle}  \bigr]&&. 
\end{eqnarray}
Here $A_{T}$, $A_{E}$, and $A_{\pi}$ are combinations of thermodynamic moments of the equilibrium distribution described in Ref.~\cite{McNelis:2019auj}, $m_i$ is the mass of the hadron species $i$, $P_{\langle \mu}P_{\nu \rangle}=\Delta_{\mu\nu}^{\alpha\beta}\, P_\alpha P_\beta$, and $\Delta_{\mu\nu}^{\alpha\beta}$ is defined in Eq.~(\ref{eq:delta_munu_alphabeta}). 

\paragraph{Linearized Chapman-Enskog expansion in the relaxation time approximation (CE RTA):}
The Chapman-Enskog (CE) expansion is a method to solve the Boltzmann equation by expanding in Knudsen number, which is the dimensionless ratio of microscopic to macroscopic length scales in the system. Although this series can be written down for a more general collisional kernel we here use the relaxation-time
approximation (RTA) \cite{Bhatnagar:1954zz, ANDERSON1974466},
\begin{equation}
    P^{\mu}\partial_{\mu}f = -\frac{u \cdot P}{\tau_\mathrm{rel}} (f - f_{\text{eq}}),
\end{equation}
where $f_{\text{eq}}$ is the local equilibrium distribution function, and the relaxation time $\tau_\mathrm{rel}$ is assumed to be species- and momentum-independent. Expanding the distribution function in a series around the local equilibrium distribution, called the Chapman-Enskog series, and keeping only the first-order correction one finds
\begin{equation}
    f = f_{\text{eq}} - \frac{\tau_\mathrm{rel}}{u \cdot P} P^{\mu}\partial_{\mu}f_{\text{eq}} + \mathcal{O}\left(\partial^2 \right).
\end{equation}
Using the zeroth order conservation laws to rewrite derivatives of the temperature and flow velocity, as well as the Navier-Stokes relations $\Pi = -\zeta \theta $ and $\pi^{\mu\nu} = 2 \eta \sigma^{\mu\nu}$ we finally obtain
\begin{eqnarray}
    \delta f^{\text{CE RTA}}_i = f_{\text{eq}, i} \bar{f}_{\text{eq}, i}
    \left[  \frac{\Pi}{\beta_{\Pi}}\left( \frac{(u{\,\cdot\,}P) \mathcal{F}}{T^2} - \frac{P{\,\cdot\,}\Delta{\,\cdot\,}P}
    {3(u{\,\cdot\,}P)T}\right) \right.\ && 
\nonumber\\ 
    \left. + \frac{ \pi_{\mu\nu}P^{\langle \mu}P^{\nu \rangle}}{2\beta_{\pi}(u \cdot P)T} \right].&&\quad
\end{eqnarray}
Again we refer to \cite{McNelis:2019auj} for the definitions of $\mathcal{F}$, $\beta_\pi$ and $\beta_\Pi$.

\paragraph{Handling large viscous corrections:}
The Grad and Chapman-Enskog momentum distributions discussed above assume $|\delta f| \ll f_{\rm eq}$. The viscous correction $\delta f$ scales linearly with the shear stress $\pi^{\mu\nu}$ and the bulk viscous pressure $\Pi$. It also scales either quadratically or linearly with the hadron momentum $P$. There are thus values of $\pi^{\mu\nu}$ and $\Pi$ for which $|\delta f| > f_{\rm eq}$ even for moderate (thermal) momenta. Moreover, even for small values of $\pi^{\mu\nu}$ and  $\Pi$, $|\delta f| > f_{\rm eq}$ at sufficiently large momenta.

In hydrodynamic simulations of heavy-ion collisions it is thus not uncommon to encounter $|\delta f| > f_{\rm eq}$ in certain phase-space regions. Even though these regions are usually small enough to not contribute significantly to experimental observables, from a practical point of view one does need to specify a hadronic momentum distribution even when $|\delta f| \sim f_{\rm eq}$.
This is commonly achieved by regulating the Grad or Chapman-Enskog viscous corrections to prevent $|\delta f| > f_{\rm eq}$. In this work this is achieved locally by setting 
\begin{equation}
    \delta f \rightarrow \sign(\delta f) \min(f_{\rm eq}, |\delta f|)
\end{equation}
in every cell. 

The need for regulation of the linearized viscous corrections has motivated models that attempt to resum the viscous corrections to all orders. We now discuss two such prescriptions. 

\subsubsection{Exponentiated viscous corrections: Pratt-Torrieri-McNelis and Pratt-Torrieri-Bernhard}

The approaches described in this subsection rely on the development of positive definite ``modified equilibrium'' distributions \cite{Pratt:2010jt, Bernhard:2018hnz, McNelis:2019auj} that include the effects of the viscous pressures in the argument of the exponential function that characterizes the equilibrium distribution. 

\paragraph{Pratt-Torrieri-McNelis (PTM):}
The Pratt-Torrieri-McNelis (PTM) distribution \cite{Pratt:2010jt, McNelis:2019auj} is defined as follows:
\begin{equation}
    f_{\text{PTM}} = \mathcal{Z} \left[ \exp\left(\frac{ \sqrt{|\mathbf{P'}|^2 + m^2} }{T + \beta_{\Pi}^{-1} \Pi \mathcal{F}}\right) + \Theta \right]^{-1}.
\end{equation}
Here the spatial momentum components have been transformed as $P_i = A_{ij}P'_j$ where
\begin{equation}
    A_{ij} \equiv \left(1+ \frac{\Pi}{3\beta_{\Pi}}\right)\delta_{ij} + \frac{\pi_{ij}}{2\beta_{\pi}}.
\end{equation}
The PTM ansatz has the feature that expanding to first order in the dissipative currents yields the usual linear Chapman-Enskog viscous correction discussed above. The yield of each hadron is corrected from its equilibrium yield by a scaling factor $\mathcal{Z}$ which depends on the bulk viscous pressure as well as the hadron mass as described in Ref.~\cite{McNelis:2019auj}.

\paragraph{Pratt-Torrieri-Bernhard (PTB):}
The Pratt-Torrieri-Bernhard (PTB) distribution \cite{Pratt:2010jt, Bernhard:2018hnz} is defined by
\begin{equation}
    f_{\text{PTB}} = \frac{\mathcal{Z}_{\Pi}}{\text{det} \Lambda} \left[\exp\left(\frac{ \sqrt{|\mathbf{P'}|^2 + m^2} }{T}\right) + \Theta \right ]^{-1},
\end{equation}
where $\mathcal{Z}_{\Pi}$ is a scaling factor described in Ref.~\cite{Bernhard:2018hnz, McNelis:2019auj} which again depends on the bulk viscous pressure but is species-independent. $\Lambda$ is a momentum-transformation matrix operating on the spatial momentum components as $P_i = \Lambda_{ij}P'_j$ with
\begin{equation}
    \Lambda_{ij} \equiv \left(1+ \lambda_{\Pi}\right)\delta_{ij} + \frac{\pi_{ij}}{2\beta_{\pi}}.
\end{equation}
In particular, $\lambda_{\Pi} \neq \Pi/(3\beta_{\Pi})$; instead, it is adjusted such that the total pressure of the system is matched. This method parametrizes the effect of the bulk viscous pressure on the particle yields and momentum spectra, and it does not reduce to the linear Chapman-Enskog correction in the limit of small $\Pi$. 

The PTB distribution was used in the recent Bayesian analysis of Ref.~\cite{Bernhard:2019bmu}. It should be noted that, in contrast to the (unregulated) linearized Grad and Chapman-Enskog distributions, for both PTB and PTM distributions the matching constraint  (\ref{Tmunu_kinetic}) is not satisfied exactly when the viscous stresses are large \cite{McNelis:2019auj}. The slight matching inconsistencies introduced by the different regulation schemes discussed above were quantitatively studied in \cite{McNelis:2019auj} and found to be acceptable in practice. For other approaches to regulate the viscous corrections to the distribution functions during particlization we refer the interested reader to Refs.~\cite{Romatschke:2003ms, Romatschke:2004jh, Martinez:2012tu, Florkowski:2013lya, Florkowski:2014bba, Tinti:2015xwa, Molnar:2016vvu, Tinti:2018nrp, Tinti:2018qfb}.

We remind the reader that in the presence of bulk viscous stresses the bulk viscous corrections to the distribution function $f_i$ shift the chemical equilibrium yields of the hadron species $i$ \cite{Dusling:2011fd} and thereby have the potential to affect the particlization temperature $\Tsw$ at which the hadron yields are consistent with the equilibrium relations (\ref{chemeq}) (see discussion in the following subsection).

\subsection{Hadronic transport}
\label{sec3D}

In our hybrid model we transition to microscopic hadronic Boltzmann dynamics, simulated with the kinetic evolution code \SMASH{} \cite{Weil:2016zrk,smash_code}, by imposing particlization at the switching temperature $\Tsw$ as described above. After particlization of the fluid, the resulting hadrons are allowed to scatter, form resonances, and decay. \SMASH{} solves a tower of coupled Boltzmann equations for a system of hadronic resonances:
\be
    P^{\mu} \partial_{\mu} f_i(x;P) = C[f_i],
\ee
where $f_i$ is the distribution function for hadronic species $i$ and $C[f_i]$ is the collision term describing all scattering, resonance formation, and decays involving particle species $i$.

Past phenomenological studies \cite{Nonaka:2006yn, Hirano:2007ei, Petersen:2008dd, Song:2010aq, Heinz:2011kt, Song:2013qma, Zhu:2015dfa, Ryu:2017qzn} have found that including a hadronic afterburner improves the ability of a hydrodynamic model to describe the spectra of heavier hadronic states, such as protons. This transport approach allows different species to reach chemical and kinetic freezeout dynamically. This contrasts with other approaches where chemical and kinetic freezeout are enforced at specific temperatures.\footnote{%
    For example, the partial chemical equilibrium approach \cite{Hirano:2002ds} enforces chemical freezeout at a given temperature in ideal hydrodynamics, by introducing chemical potentials to conserve all hadronic multiplicities to a chosen chemical freeze-out values. This was a popular procedure before the widespread availability of hybrid codes (see e.g. \cite{Song:2010aq, Heinz:2011kt} for comparisons of these two approaches).}
At particlization, the momentum distributions and particle yields already deviate from their equilibrium relations at that temperature due to shear and bulk viscous effects. After switching to the afterburner, they continue to evolve until yields and momentum distributions cease changing.  Most hadronic yields vary by less than 20\% as a consequence of inelastic collisions in the afterburner phase, and the particlization temperature $\Tsw$ is therefore sometimes associated with a chemical freeze-out temperature \cite{Song:2010aq}. However, baryon and anti-baryon yields may change more significantly, due to the large annihilation cross section \cite{Bass:2000ib,Steinheimer:2012rd}. As a result, this is not a precise relationship and we indeed find somewhat lower values for $\Tsw$ than the canonical chemical freeze-out temperatures extracted from static thermal model fits such as those in Refs.~\cite{Andronic:2005yp, Andronic:2017pug}.

We note that none of the parameters in the \SMASH{} afterburner are varied in this work. We did validate, however, that the afterburner used in this work (\SMASH{}) agrees well with the popular \URQMD{} implementation that has been used for decades. This comparison is discussed in Appendix \ref{app:smash}. 

\paragraph*{Treatment of the $\sigma$ meson:}

At particlization the hydrodynamic energy-momentum tensor is converted into hadrons assuming the system has the thermodynamic properties of a hadron resonance gas. Though the $\sigma$ meson can be formed as a resonance in the $\pi+\pi$ scattering channel, it has been shown in Ref.~\cite{Broniowski:2015oha} that the contribution to the partition function from $\sigma$ meson exchange (an isoscalar-scalar channel) is almost perfectly canceled by the repulsive isotensor-scalar channel in $\pi+\pi$ scattering. Based on this observation, it is generally agreed that the $\sigma$ meson should be omitted from isospin-averaged hadron resonance gas models \cite{Broniowski:2015oha}. This is the approach we use in this work: the $\sigma$ meson is \emph{not} sampled at particlization, and correspondingly it is also omitted in the construction of the equation of state in the hadronic phase.\footnote{%
    More details about the construction of the equation of state are provided in Appendix~\ref{appendix:eos}. The physical effects on observables from excluding the $\sigma$ meson from the hadron gas are studied in Appendix~\ref{app:sigma_effect}.}
In the hadronic afterburner, we still allow \SMASH{} to dynamically form and decay $\sigma$ resonances because they are an essential ingredient in explaining the $\pi+\pi$ cross section in \SMASH{}.  We note for reference that the Bayesian analysis in Ref.~\cite{Bernhard:2019bmu} did include the $\sigma$ meson in both the sampling at particlization and the construction of the hadronic equation of state, making this one of its differences with the current analysis.

This concludes the discussion of the dynamical evolution model used in this work. We now proceed to a discussion of the statistical tools used in its calibration with experimental data.

\section{Specifying prior knowledge}
\label{sec:priors}

\begin{table*}[!hbt]
\noindent\makebox[\textwidth]
{%
\footnotesize
\begin{tabular}{||p{0.18\linewidth}|p{0.11\linewidth}|p{0.15\linewidth}||p{0.18\linewidth}|p{0.08\linewidth}|p{0.15\linewidth}||}
Norm. Pb-Pb 2.76 TeV & $N$[2.76 TeV] & {[}10, 20{]} & temperature of $(\eta/s)$ kink & $T_{\eta}$ & {[}0.13, 0.3{]} GeV \\
Norm. Au-Au 200 GeV & $N$[0.2 TeV] & {[}3, 10{]} & $(\eta/s)$ at kink & $(\eta/s)_{\rm kink}$ & {[}0.01, 0.2{]} \\
generalized mean & $p$ & {[}--0.7, 0.7{]} & low temp. slope of $(\eta/s)$ & $a_{\text{low}}$ & {[}--2, 1{]} GeV$^{-1}$ \\
nucleon width & $w$ & {[}0.5, 1.5{]} fm & high temp. slope of $(\eta/s)$ & $a_{\text{high}}$ & {[}--1, 2{]} GeV$^{-1}$ \\
min. dist. btw. nucleons & $d_{\text{min}}^3$ & {[}0, 1.7$^3${]} fm$^3$ & shear relaxation time factor & $b_{\pi}$ & {[}2, 8{]} \\
multiplicity fluctuation & $\sigma_k$ & {[}0.3, 2.0{]} & maximum of $(\zeta/s)$ & $(\zeta/s)_{\text{max}}$ & {[}0.01, 0.25{]} \\
free-streaming time scale & $\tau_R$ & {[}0.3, 2.0{]} fm/$c$ & temperature of $(\zeta/s)$ peak & $T_{\zeta}$ & {[}0.12, 0.3{]} GeV \\
free-streaming energy dep. & $\alpha$ & {[}--0.3, 0.3{]} &  width of $(\zeta/s)$ peak & $w_{\zeta}$ & {[}0.025, 0.15{]} GeV \\
particlization temperature & $\Tsw$ & {[}0.135, 0.165{]} GeV & asymmetry of $(\zeta/s)$ peak & $\lambda_{\zeta}$ & {[}--0.8, 0.8{]}
\end{tabular} 
}
\caption{A list of all priors used (see Sec.~\ref{sec:model_overview} for the definitions of the model parameters). All prior distributions are assumed to be uniform and nonzero within the range quoted, and zero outside. The Table does not exhibit the step functions that ensure non-negativity of the shear viscosity at all temperatures (see Eq.~(\ref{positivity})).}
\label{prior_table}
\end{table*}

Before using a set of measurements to perform Bayesian inference, one must quantify the current state of knowledge. If we want to infer the likely values of model parameters, given some observed data, then we must quantify our belief about the model parameters before we see the data. If we want to compare models, given some observed data, we must quantify the likely values of each model's parameters, as well as our belief in each model, before seeing the data. Broadly, before we use Bayesian inference to address a question in light of some observed data, our ``prior'' encodes our current state of knowledge before we have seen the selfsame data. 

It is essential that the construction of the prior distribution should not be informed by the same data that will be used in performing parameter estimation. In particular, the posterior of earlier analyses that used the same data sets should not in any way be used as a prior for a new analysis: it would be an attempt to use the same measurements twice, as well as being likely inconsistent given differences in the models. 

When selecting a prior different factors must be considered. Theoretical constraints are important, including self-consistency concerns for the model and/or conservation laws or symmetries that must be respected, all of which are problem-specific. Within these constraints, a range of different priors is possible. There are methods aimed at reducing subjectivity in the choice of priors; examples include maximum-entropy priors \cite{Jaynes:1957zza}. In this work we focus on a careful selection of the range of the priors, rather than on the exact form of the prior probability distribution for each parameter. In the following two paragraphs we illustrate this selection process for a subset of the parameters.

\paragraph{Initial conditions:} There are physical constraints on the prior for the width parameter $w$ in \trento{}: when choosing a reasonable range of values one must keep in mind that the electric charge radius of the proton is about $0.9$\,fm. The width parameter $w$ in \trento{} should likely not be allowed to be much smaller or larger than this value.
\paragraph{Transport coefficients:} The ranges of allowed shear and bulk viscosities are important priors. First, both viscosities should be non-negative, to ensure the second law of thermodynamics, i.e. a positive entropy production rate. At the opposite end of the allowed range, the applicability of hydrodynamics becomes debatable when the viscous part of the energy-momentum tensor is large compared to the ideal part. This can happen when the shear and bulk viscosities are large. For self-consistency it is thus desirable that the prior ranges of $\eta/s$ and $\zeta/s$ exclude unreasonably large values. Though exploring large values of the viscosities may be physically interesting, it would push the hydrodynamic component of our model outside its regime of validity. If the experimental data require larger viscosities than included in our prior, this should be visible in our posterior distributions for these transport coefficients, as well as in a low value for the model evidence, i.e. a bad fit of the data for any viscosity within the allowed prior range. The shear relaxation time also has physical constraints. As discussed in Section \ref{sec:hydro}, there is a minimum value for $b_\pi=T \tau_\pi/(\eta/s)$ set by causality of the linearized equations, $b_\pi \gtrsim 2$, and theoretical evaluations of $b_\pi$ within a number of microscopic theories ranging from weakly to strongly coupled provide some theoretical guidance for the most likely range of this parameter. 

\begin{figure}[!htb]
\includegraphics[trim=0 0 0 15, clip, width=8cm]{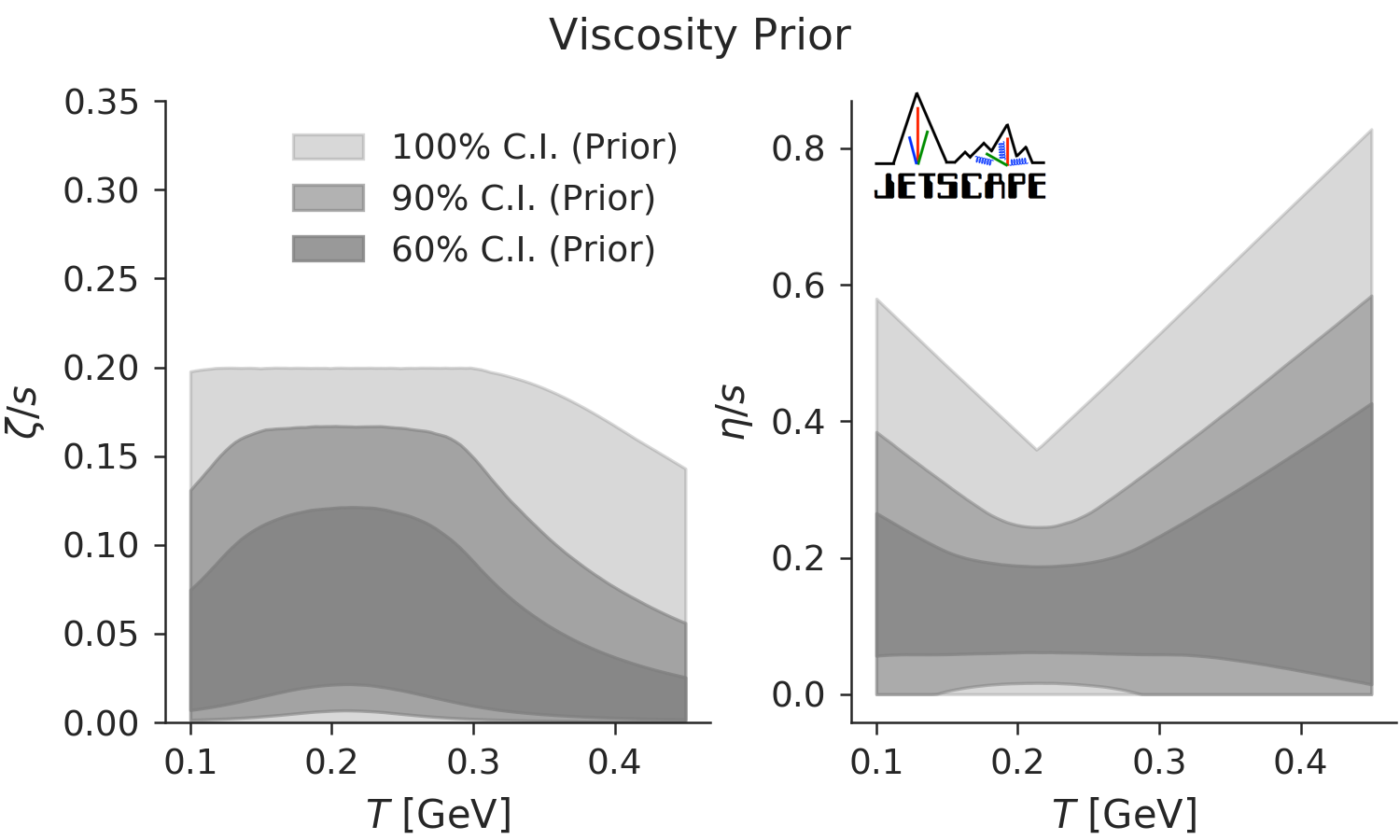}
\caption{Credible intervals of the prior probability density for the specific bulk (left) and shear (right) viscosities that we use when performing Bayesian parameter estimation. The 60\%, 90\% and 100\% credible intervals (C.I.) are shown.}
\label{viscous_prior}
\end{figure}

In the present analysis, for simplicity all of the parameters (denoted by the vector $\boldsymbol{x}$) are assigned a uniform prior probability density $\mathcal{P}(\boldsymbol{x})$ on a finite range. These ranges are listed in Table~\ref{prior_table}; as discussed above, they have been chosen with certain theoretical biases. The priors for different parameters are assumed to be independent, so that the joint prior is simply given by their product,
\begin{eqnarray}
   \mathcal{P}(\boldsymbol{x}) \propto \prod_{i} \Theta(x_i-x_{i,\min})\Theta(x_{i,\max}-x_i),
\end{eqnarray}
where $i$ runs over all the model parameters in $\boldsymbol{x}$. Note that uniform priors are not uninformative priors. Moreover, the choice of priors in principle affects the results of the Bayesian parameter estimation, especially in situations where the data do not have sufficient information to correct prior prejudice. For instance, in this work, we require $(\eta/s) (T)$ and $(\zeta/s) (T)$ to be given by specific parametrizations, with each of the parameters sampled from a uniform prior. The resulting prior for $(\eta/s)(T)$ is, however, not uniform as a function of temperature; thus, our choice of parametrization informs our prior. A plot showing credible intervals \emph{for the prior} for the shear and bulk viscosities is shown in Fig.~\ref{viscous_prior}.
%
%
We see that this prior encapsulates our belief that the bulk viscosity should have a peak somewhere near the deconfinement transition temperature, and that the specific shear viscosity reaches a minimum in that region. 

Nevertheless, we used a broad prior for $\eta/s$, allowing it to take either a maximum or a minimum in the deconfinement region. By doing so we tried to limit the theoretical bias of our prior for $(\eta/s)(T)$. When selecting the priors for the remaining model parameters we followed similar considerations, with the goal of ensuring that our posterior parameter constraints will be guided as much as possible by the heavy-ion data and not by prior prejudice. 

It is important, however, to understand that in practice theoretical bias can never be fully avoided. In many cases it can be helpful or even needed: if highly constraining data are lacking, exploring the reaction of the posterior distribution to different prior theoretical assumptions can yield useful insights into the variability and reliability of model predictions and suggest strategies for decreasing this variability with targeted theoretical efforts. The Bayesian methodology accepts the reality of theoretical bias, but at the same time accounts for it quantitatively in the posterior probability distribution for the model parameters $\boldsymbol{x}$. Sensitivity to our prior assumptions is further explored in Sec.~\ref{sec:priorsensitivity}. 

\section{Bayesian Parameter Estimation with a Statistical Emulator}
\label{sec:bayes_param_est_w_emu}

In this section, we describe the statistical methodology used to estimate the likely model parameters by comparison with the experimental data. This non-trivial problem is tackled by the application of Bayesian parameter estimation with a physical model surrogate or ``emulator''. The use of a model emulator is necessary when the physical simulation is computationally intensive. Performing a Bayesian parameter estimation requires evaluating the  model's prediction on arbitrary points in the relevant region of the parameter space. In theory, this involves evaluating the model for millions of different values for initial conditions, viscosities and other model parameters. Given a model that requires nearly $\mathcal{O}(10^3)$ CPU-hours to run a single set of input parameters, this would quickly become intractable as a method of exploring the posterior. The model emulator is designed precisely to solve this problem. The emulator can be understood as a computationally fast interpolation of the physical model simulation with an estimate of the interpolation uncertainty. The model is evaluated on a sample set of points in the parameter space, and the model's predictions at these points are used to infer the predictions at other points in parameter space through the use of an emulator. Such an emulator dramatically reduces the numerical cost of estimating the posterior; the trade-offs are that it assumes a certain smoothness in the behavior of the model outputs, as well as introduces an additional source of uncertainty in the analysis: emulator uncertainty. In addition to describing Bayesian parameter estimation in general, we also discuss specifically the role of the emulator. Our discussion in this section builds upon Refs.~\cite{Petersen:2010zt, Novak:2013bqa, Sangaline:2015isa, Bernhard:2015hxa, Bernhard:2016tnd, Moreland:2018gsh, Bernhard:2019bmu} where many of these techniques were previously applied to Bayesian parameter estimation in relativistic heavy-ion physics. Additional information on Bayesian inference can be found in e.g. Ref.~\cite{Sivia2006}.

\subsection{Overview of Bayesian Parameter Estimation}
\label{overview_param_est}

Bayesian parameter estimation is a systematic approach to infer the probability distribution of model parameters ($\boldsymbol{x}$) by comparing theoretical calculations ($\mathbf{y}_{\boldsymbol{x}}$) to experimental data ($\mathbf{y}_{\exp}$). The starting point is the prior distribution $\mathcal{P}(\boldsymbol{x})$ that encodes the current state of knowledge regarding the model parameters $\boldsymbol{x}$ before making comparison with data (Section~\ref{sec:priors}). 
The posterior distribution of model parameters after model-to-data comparison, $\mathcal{P}(\boldsymbol{x}|\mathbf{y}_{\exp})$, is given by Bayes' theorem,
\begin{equation}
    \mathcal{P}(\boldsymbol{x}|\mathbf{y}_{\exp}) =  \frac{ \mathcal{P}(\mathbf{y}_{\exp}|\boldsymbol{x}) \mathcal{P}(\boldsymbol{x}) }{ \mathcal{P}(\mathbf{y}_{\exp}) },
    \label{eq:likelihood}
\end{equation}
where $\mathcal{P}(\mathbf{y}_{\exp}|\boldsymbol{x})$ is the ``likelihood'' that the model agrees with experimental measurement, given the parameters $\boldsymbol{x}$, and the normalization $\mathcal{P}(\mathbf{y}_{\exp})$ is called the ``Bayesian evidence''. The exact form of the likelihood is often unknown, as it depends on the probability distribution of the experimental and theoretical uncertainties. In this work, we follow the common assumption that the likelihood can be taken to be a multivariate normal distribution. This choice is justified when uncertainties are normally distributed. With this choice of likelihood function, the logarithm of $\mathcal{P}(\mathbf{y}_{\exp}|\boldsymbol{x})$ contains the quadratic form of the difference between the measurement and the prediction $\Delta \mathbf{y}_{\boldsymbol{x}} =\mathbf{y}_{\boldsymbol{x}} - \mathbf{y}_{\exp}$,
\begin{equation}
    \label{eq:log_likelihood}
    \log \left[\mathcal{P}(\mathbf{y}_{\exp}|\boldsymbol{x})\right] = -\frac{1}{2}\ln  \left[(2\pi)^{n} \det\Sigma\right]-\frac{1}{2}\Delta \mathbf{y}_{\boldsymbol{x}}^T\Sigma^{-1}\Delta \mathbf{y}_{\boldsymbol{x}}.
\nonumber
\end{equation}
Here, $n$ is the number of observation points (i.e. the length of the vector $\mathbf{y}_{\exp}$), and $\Sigma$ is a covariance matrix that encodes both experimental and model uncertainties, as well as correlations among uncertainties. These correlations are generally not readily available experimentally. As such the treatment of uncertainties can become a relatively complex question. We discuss the treatment of uncertainties and the covariance matrix separately in Section~\ref{section:bayes:uncertainties} below.

In principle, in order to calculate the posterior, one is faced with the task of calculating the evidence $\mathcal{P}(\mathbf{y}_{\exp})$. For many problems of interest the required high dimensional integration can be numerically challenging or even intractable. Fortunately, when performing Bayesian parameter estimation, knowledge of the relative probability of different points in parameter space is sufficiently interesting in itself. That is, as the evidence $\mathcal{P}(\mathbf{y}_{\exp})$ does not depend on the parameters $\boldsymbol{x}$, it is sufficient to consider the proportionality
$$\mathcal{P}(\boldsymbol{x}|\mathbf{y}_{\exp}) \propto   \mathcal{P}(\mathbf{y}_{\exp}|\boldsymbol{x}) \mathcal{P}(\boldsymbol{x}).$$ 
Methods for estimating the posterior which take advantage of this include Markov Chain Monte Carlo. Therefore, when we discuss or plot the posterior of parameter estimates throughout this section, we implicitly mean the unnormalized posterior. Hence, we are interested in the relative probability density of each parameter set, and {\it not} the absolute probability. 

Because the plotted posterior for the model parameters in general does not contain information about this normalization, it is imperative to check the level of agreement between the posterior prediction of observables to assess quantitatively how well the model can describe the experimental data. As was explained in \Section{sec:applying_bayes}, it is meaningless to ponder on the posterior estimates of parameters for a model which poorly explains the observed data.
Thus, in \Section{sec:calib_combined}, we will also explore how well the model observables sampled from the posterior fit the experimental data. An estimation of the evidence $\mathcal{P}(\mathbf{y}_{\exp})$ becomes necessary if we want to compare models in a Bayesian framework and this will be discussed in \Section{sec:model_selection}. 

\paragraph*{Simultaneous constraints from multiple collision systems:}
%
When combining constraints from different experiments, RHIC and LHC for example, the joint likelihood function is assumed to be the product of the individual likelihoods for each system:
\begin{equation}
    \mathcal{P}(\mathbf{y}^{\rm LHC}_{\exp}, \mathbf{y}^{\rm RHIC}_{\exp}  |\boldsymbol{x}) = \mathcal{P}(\mathbf{y}^{\rm LHC}_{\exp} |\boldsymbol{x}) \mathcal{P}(\mathbf{y}^{\rm RHIC}_{\exp} |\boldsymbol{x}).
\end{equation}
The parameter values that maximize the joint likelihood strike a compromise between maximizing the individual likelihoods.

Importantly, one must decide which parameters are shared for the different collision systems.
Naively, one could expect that all parameters should be shared; in reality this depends partly on how the model parameters were defined.

We highlight that comparisons with measurements can always help determine if model assumptions need to be relaxed. If RHIC and LHC measurements could be described independently by the model but not simultaneously, it could be an indication that the $\sqrts{}$ dependence of certain parameters needs to be revisited, i.e., that it may not be correct to enforce the same value of certain parameters at RHIC and the LHC. We will compare more complex models which relax some of these assumptions by estimating Bayes factors in \Section{sec:bayes_factor_tension}. 

Inclusion of data at two very different collision energies raises the question where and how we make allowance for $\sqrts{}$ dependence of the model parameters for which we  interrogate the data for constraints. Answers are provided in the following paragraphs:

\paragraph{Initial stage model:} Because \trento{} is a parametric initial condition model, not a dynamical one, many of its parameters should, in principle, be beam-energy dependent.\footnote{%
    For example, in the color glass condensate effective theory for QCD at very high energies, the only relevant scale is the saturation scale $Q_s$, which controls correlations in the transverse direction and which runs with the energy of the collision system \cite{Gelis:2014qga}. This suggests that the nucleon width in \trento{} should perhaps have a similar $\sqrts{}$ dependence.}
Generically, we assume that at high collision energies the parameters that we try to extract from experiment evolve sufficiently slowly with $\sqrts{}$ that their change from RHIC to LHC can be ignored. As an exception we retain the $\sqrts{}$ dependence of the normalization $N$ of the energy density in \trento{}, because it is directly responsible in our model for the large increase of mid-rapidity particle and energy production from RHIC to LHC. Rather than parametrizing its $\sqrts{}$ dependence, we simply use two independent normalizations at $\sqrts{}=200$ and 2760\ GeV, labeled by $N$[0.2\,TeV] and $N$[2.76\,TeV], respectively. We also point out that in \Section{sec:bayes_factor_tension} we use Bayesian Model Selection to explore whether experimental data would prefer a dependence of the nucleon width $w$ in \trento{} on $\sqrts{}$. The free-streaming time (\ref{eq:taufs}) is allowed to depend on $\sqrts{}$ implicitly, through the deposited energy density.

\paragraph{Transport coefficients:} The specific shear and bulk viscosities, as well as the second-order transport coefficients in our hydrodynamic approach, are medium properties that (for systems without conserved charges) depend only on the temperature of the plasma. Their parametrizations as functions of temperature, $(\eta/s)(T)$ and $(\zeta/s)(T)$, are therefore independent of $\sqrts{}$.

\paragraph{Particlization:} We use the same particlization temperature $\Tsw$ at RHIC and at the LHC. As discussed in Secs.~\ref{sec:model:particlization} and \ref{sec3D}, particlization is assumed to happen at $\Tsw$ with chemical potentials (\ref{chemeq}) that, in a static environment, reflect chemical equilibrium relations between the hadronic yields at $\Tsw$. It is known that, when initialized in this way at $\Tsw$, hadronic transport changes these yield ratios much more slowly than they would in a chemically equilibrated fluid \cite{Bass:1999tu, Bass:2000ib, Heinz:1999kb, Song:2010aq, Ryu:2017qzn}. The finally observed hadronic chemical composition is therefore largely set at particlization \cite{Bass:1999tu}, linking the switching temperature $\Tsw$ to the chemical decoupling temperature.\footnote{%
    Let us emphasize again that, for the same values of $\Tsw$ and $\mu_i$ (\ref{chemeq}), bulk viscous pressure corrections (which are different at RHIC and LHC, due to different expansion rates) will lead to different hadron yields $dN_i/dy$ in collisions at RHIC and LHC energies. To the best of our knowledge this dissipative correction to thermal equilibrium model fits of hadron yield ratios has not been previously studied, but it is automatically taken into account in our Bayesian inference analysis.} 
At both collision energies matter with approximately zero baryon chemical potential is produced which passes through the hadronization phase transition at the same temperature. Based on strong experimental and theoretical evidence \cite{Bass:1999tu, Heinz:1999kb, Andronic:2005yp, Andronic:2017pug} we expect the chemical equilibrium relations (\ref{chemeq}) between the chemical potentials of the different hadronic species to be broken soon after hadronization is complete, which should occur at the same value for $\Tsw$ at both collision energies. We consider this a combined theoretical and empirical prior for $\Tsw$.

\subsection{Physical model emulator} 
\label{section:bayes:GP}

Throughout this study, we define an emulator as a map from a point in the multidimensional parameter space to the mean vector and covariance matrix of the distribution of all the predicted model observables of interest. This map provides a non-parametric estimation of the physical model calculations at arbitrary points in the region of the parameter space of interest. It is ``non-parametric'' because predictions at novel parameter points are not made by constructing explicit functional relations between model predictions and parameter inputs, but rather are obtained by modeling the way that predictions at any point are correlated with known calculations at other parameter points. This sample of points in parameter space where we know the physical model calculations are called the design points ($\boldsymbol{x}_i; i=1,\dots, m$). 

The parameter design samples are chosen carefully using the Latin hypercube sampling technique, which randomly fills the volume of parameter space yielding a uniform distribution for each parameter while maximizing the distance between adjacent points. For models with sufficient smoothness, the number of design points necessary to achieve a certain level of prediction accuracy scales linearly with the dimension of the parameter space\footnote{%
    This scaling of interpolation uncertainty with design size is explored in \cite{Nijs:2020roc} for a different set of observables.}
\cite{Loeppky}. In this work we have taken a Latin hypercube design of $500$ points. At each design point, the full model is run $2500$ times for each collision system (Au+Au at RHIC or Pb+Pb at LHC), each time with a randomly fluctuating initial condition. The $2500$ events are then ordered according to the yield of charged hadrons in each event $dN_{\rm ch}/d\eta$ to define centrality classes, and observables are averaged over the fluctuating events in centrality bins which match those given by the experimental measurements.

With the training data available, the following steps define the construction and training of the emulator. 

\paragraph{Dimensionality reduction via Principal Component Analysis:}
When comparing the model output with experimental data, we are faced with the large dimensionality of the output. Many of the model observables carry correlated information. As a simple example, an increased normalization of initial energy density increases pion multiplicity in all centrality bins. Therefore, the predicted value of multiplicity at different centralities effectively equals a single degree-of-freedom in response to the change of the normalization parameter. To put it another way, a small subspace of the full model output carries nearly all of the information about the model parameters. Therefore, we apply principal component analysis as a dimensionality reduction method. Suppose an array of observations $y_i$ ($i=1,\dots,n$) are calculated at each of the $m=500$ design points $j$. They are organized as an $n\times m$ matrix $Y$ with elements $y_{ij}$. First, for each of the observables $y_i$, we compute its mean $\mu_i$  and standard deviation $\sigma_i$ over the sample of $m$ design points. Then, each of the $n$ observables is standardized by subtracting the mean and dividing by the standard deviation, yielding an $n\times m$ matrix $\tilde Y$ with elements $\tilde{y}_{ij} = (y_{ij}{-}\mu_j)/\sigma_j$ for $j=1,\dots, m$. Secondly, we define a new set of ``observables'' $z_{i}$ which are linear combinations of the standardized observables: $z_{i} = O_{ik}\tilde{y}_{k}$. One seeks an optimized set of $z_i$ such that the linear correlations between different $z$-observables vanish:
\begin{eqnarray}
    \langle\delta z_i \delta z_j\rangle &=&\frac{1}{m}\sum_{k=1}^m (O\tilde{Y})_{ik} (O\tilde{Y})_{jk} = \frac{1}{m} \bigl(O (\tilde{Y}\tilde{Y}^T) O^T\bigr)_{ij}
\nonumber\\
    &=& \lambda_i \delta_{ij} \equiv \textrm{diag}\{\lambda_1, \cdots, \lambda_n\}, 
\end{eqnarray}
where $\delta z_i$ denotes the deviation of the $z_i$ from their mean. Therefore, the coefficients $O_{ij}$ that define $z_i$ are simply the elements of the orthogonal matrix that diagonalizes the covariance matrix of $\tilde{y}_i$. This optimized set of $z_i$ are the so-called principal components. The rows of $O$ are organized such that the eigenvalues $\lambda_i$, which are the variances of the $z_i$, have a descending order. In this way, each successive principal component explains less variance in the standardized observables. This allows us to reduce the standardized observable space to a much smaller subspace, which captures most of the information about the parameters. One should remember that the principal component analysis can only remove linear correlations among observables, so it is important to check that there are no significant non-linear correlations. This is demonstrated in \Appendix{pca_valid}.

In our experience, a very small fraction of the total number of principal components is generally sufficient to capture most of the model observables' dependence on the parameters. This follows from the strong linear correlations present in many pairs of observables. Pairs of observables with stronger linear correlations carry less mutual information about the parameters; knowledge of one observable is nearly sufficient to know the value of the other. Gaussian processes are only trained on this subset of dominant principal components.

From a practical point of view, we also observed that it is important not to include too many principal components: this limits the risk that the emulator overfits the noise present in the simulation data.

\paragraph{Interpolating each Principal Component by a Gaussian Process:}
Each dominant principal component is interpolated with a unique Gaussian process. The spirit of a Gaussian process regressor is to infer the outputs of the target (scalar) function $y=M(x)$\footnote{In this context, the output of the target function is one of the dominant principal components.} by a distribution of functions denoted by $\mathcal{GP}$: $f(x) \sim \mathcal{GP}(\textrm{mean}(x), \textrm{cov}(x, x'))$. This distribution is a multivariate normal distribution specified by a mean $\mu(x)$ and a covariance $\textrm{cov}(x,x')$, so that the expectation value of the output at a given $x$ is
\begin{eqnarray}
    \langle  f(x) \rangle = \textrm{mean}(x),
\end{eqnarray}
and the correlation of the output between two independent inputs $x, x'$ is
\begin{eqnarray}
    \langle \delta f(x) \delta f(x') \rangle = \textrm{cov}(x, x'),
\end{eqnarray}
where $\delta f(x) = f(x) - \textrm{mean}(x)$.

To find the desired distribution of functions that emulates $M(x)$, one starts with a distribution that is completely agnostic to the target function $M(x)$. In this study this distribution, referred to as the unconditioned Gaussian process, is assumed to have mean $\mu(x) = 0$\footnote{%
    It can happen that near the boundaries of parameter space the model prediction for some principal component is nonzero. In this case it may be beneficial to include a non-zero mean function in the Gaussian Process. We do not explore this in this work.}
and a covariance function $k(x, x')$ (the so-called kernel function). A Gaussian process makes a prediction at $m_{\star}$ novel inputs $X_{\star}$ according to the correlations with known values that have been calculated at the $m$ training inputs $X$. Consistency requires that the joint distribution of outputs at both training and novel inputs is also multivariate normal with zero mean,
\begin{gather}
 \begin{bmatrix} 
 \mathbf{f}(X) \\ 
 \mathbf{f}(X_{\star} ) 
 \end{bmatrix}
 \sim
 \mathcal{N} 
 \left( 
 \mathbf{0},
  \begin{bmatrix}
   K( X, X ) & K( X, X_{\star} ) \\
   K( X_{\star}, X ) & K( X_{\star}, X_{\star} ),
   \end{bmatrix}
  \right)
\end{gather}
where $K( X, X_{\star} )$ is the $m \times m_{\star}$ matrix whose elements are composed of the pointwise covariances $k(\mathbf{x}_p, \mathbf{x}_q)$ between pairs of training points $\mathbf{x}_p$ and prediction points $\mathbf{x}_q$.
Then, one conditions the random vector $\mathbf{f}(X)$ on the training outputs $M(X)$ to obtain the probability distribution of $\mathbf{f}(X_*)$ given training data. 
The mean and covariance can be obtained by the properties of the multivariate normal distribution,
\begin{eqnarray}
\label{eq:conditioned-GP}
    \mathbf{f}(X_*) &\sim& \mathcal{GP}\left(\textrm{mean}(X_*), \textrm{cov}(X_*, X_*)\right)
\\
\label{eq:conditioned-mean}
    \textrm{mean}(X_*) &=& K(X_*,X)\left[K(X,X)\right]^{-1}M(X),
\\\nonumber
    \textrm{cov}(X_*, X_*) &=& K(X_*,X_*)\\
\label{eq:conditioned-cov}
    &-& K(X_*,X)\left[K(X,X)\right]^{-1}K(X,X_*).
\end{eqnarray}
Focusing on a single novel input, the prediction with uncertainty quantification of the target function is $M(x_*)\approx \textrm{mean}(x_*) \pm \sqrt{\textrm{cov}(x_*, x_*)}$. Eqs.~(\ref{eq:conditioned-mean}), (\ref{eq:conditioned-cov}) imply that if $x_*$ coincides with one of the training inputs then the mean agrees with the training output with vanishing uncertainty.

In addition to the training data, choosing the kernel function $k(x, x')$ is another key step in Gaussian process regression. An independent kernel function $k(\mathbf{x}_p, \mathbf{x}_q)$ is assigned to each dominant principal component, and is given by the sum of a squared-exponential kernel $k_{\rm exp}(\mathbf{x}_p, \mathbf{x}_q)$ and white-noise kernel $k_{\rm noise}(\mathbf{x}_p, \mathbf{x}_q)$,
\begin{equation}
    k(\mathbf{x}_p, \mathbf{x}_q) = k_{\rm exp}(\mathbf{x}_p, \mathbf{x}_q) + k_{\rm noise}(\mathbf{x}_p, \mathbf{x}_q).
\end{equation}
The squared-exponential kernel is given by 
\begin{equation}
    k_{\rm exp}(\mathbf{x}_p, \mathbf{x}_q) = C^2 \exp \left( -\frac{1}{2} \sum_{i=1}^{s} \frac{|x_{p, i} - x_{q, i}|^2}{l_i^2} \right)
\end{equation}
where $C^2$ is the unknown auto-correlation hyperparameter. The index $i$ runs over all $s$ parameters, and each parameter is assigned an uncertain hyperparameter $l_i$. This length-scale $l_i$ controls the smoothness of the response of the principal component output to a change in the $i^{\rm th}$ parameter. 
The white-noise kernel is given by 
\begin{equation}
    k_{\rm noise}(\mathbf{x}_p, \mathbf{x}_q) = \sigma_{\rm noise}^2 \delta_{p, q}
\end{equation}
where $\delta_{p, q}$ is the Kronecker delta, while $\sigma_{\rm noise}$ is an uncertain hyperparameter controlling the amount of statistical spread present in the principal component. The $\sigma_{\rm noise}$ is present because our model calculations average over a finite number of initial conditions and a finite number of particles. 

All of the hyperparameters $C, l_i$ and $\sigma_{\rm noise}$ are assigned a possible window, and then simultaneously optimized inside this window such that they maximize the likelihood of fit of the Gaussian process to the training calculations. This likelihood includes a complexity penalty, to reduce the potential for overfitting. This procedure is automated, and performing emulator validation is necessary to check that each kernel function has hyperparameters which are not underfit or overfit ~\cite{scikit-learn}.

\paragraph{Reconstructing the observables:}
The predictions for principal components are then grouped and transformed back into the observables via the inverse PCA transformation. Variances of those non-dominant principal components on which we did not train Gaussian processes are included as prediction uncertainty. These neglected principal components in fact behave similarly to noise terms. We use this feature to actually replace them by white noise (variance which is uncorrelated point-to-point in parameter space) terms as an estimation of their contributed uncertainty. 

A more detailed description of the above procedure can be found in~\cite{Bernhard:2018hnz}. We note that our use of transverse-momentum-integrated observables, principal component analysis, and Gaussian process model emulator for performing Bayesian parameter estimation for heavy-ion collisions is very similar to those put forward in the seminal study \cite{Novak:2013bqa}. 

\subsection{Treatment of uncertainties} \label{section:bayes:uncertainties}

We divide our uncertainties into three different sources: experimental uncertainties, interpolation and statistical model uncertainties, as well as systematic model discrepancies. 

\paragraph{Experimental uncertainties:}
%
In general, experimental collaborations do not report the error covariance matrix between different observables, or for different momentum bins of the same observable. As such, we only have access to the systematic uncertainties of individual observables or bins, with limited or no information on possible correlations. Assuming no correlations among the errors associated with the $n$ observables results in a diagonal covariance matrix for the experimental systematic covariance:
\begin{equation}
    \Sigma_{\textrm{sys}} = \textrm{diag}(\sigma_{\textrm{sys},1}^2, \cdots, \sigma_{\textrm{sys},n}^2).
\end{equation}

In principle, the systematic uncertainties have nonzero correlations. Although it is possible to use our model to estimate correlations between the mean values of different observables, we do not have information about other sources of systematic uncertainties. Without knowledge of the experimental covariance matrix we can only make assumptions regarding the form and magnitude of the correlations. We have tested the effect of this approach on the parameter posteriors in \Appendix{app_exp_cov}. However, we did not use this approach in general in the body of this work.

\paragraph{Interpolation and statistical model uncertainties:}
%
The statistical uncertainty which is present in our model calculations results primarily from averaging over a finite number of fluctuating initial conditions, and to a lesser extent sampling a finite number of particles during particlization. These result in a statistical spread in each of the principal components (recall from \Section{section:bayes:GP} that it is the principal components that are interpolated, not the individual observables). 

The total interpolation uncertainty is
\begin{eqnarray}
   \Sigma_{\text{interp}} = \Sigma_{\text{trunc.}} + \Sigma_{\text{GP}}.
\end{eqnarray}
The covariance $\Sigma_{\text{GP}}$ contains the total covariance of all the Gaussian Processes (one for each dominant principal component), including both interpolation and statistical uncertainties. The covariance $\Sigma_{\text{trunc.}}$ contains the total covariance of all the remaining principal components to which Gaussian processes were not fit and which were replaced by noise terms. 

\paragraph{Additional systematic model discrepancy:}
%
Our model of heavy ion collisions is unavoidably imperfect. Therefore there exist additional sources of systematic discrepancy in our model when we use it to describe physical observations. Quantifying and interpreting the associated uncertainties presents a challenging problem \cite{Brynjarsdottir_2014}. 

Some of these uncertainties could in principle be quantified, although for practical reasons we will not attempt to do so in this work. For example, it would be meaningful to vary all second-order transport coefficients of the hydrodynamic model. In this work we took only a small step in this direction by varying the shear relaxation time.

Other model uncertainties/biases are more difficult to study. For example, our ansatz of initial conditions for hydrodynamics may not be sufficiently flexible to capture all relevant features of the early pre-hydrodynamic evolution in heavy ion collisions. Accounting for these model limitations systematically is still challenging.

In Ref.~\cite{Bernhard:2019bmu} a parametrized systematic model discrepancy was included; this single percentage uncertain parameter was meant as a proxy for all systematic model discrepancies. This parameter was added in quadrature to the covariance matrix of the Gaussian process for each principal component, in the form of a diagonal matrix parametrized by the single parameter $\sigma_m$. That is, to every principal component was added the same systematic uncertainty in percentage. This results in a complicated distribution of the uncertainty across observables, depending on the linear transformation from principal components to observables.

While the above approach may be a step in the right direction, we worry about the crudeness of this distribution of systematic uncertainty across the observables. In consequence, we did not include this type of uncertainty in our current analysis.

\subsection{Sampling of the posterior} \label{section:bayes:sampling}

In our case the posterior is a 18-dimensional probability distribution.\footnote{There are 16 shared parameters and one additional parameter per collision system (the \trento{} normalization).} Estimation of the posterior is accomplished via Markov Chain Monte Carlo algorithms \cite{Hogg:2017akh}. Typical algorithms, including the Metropolis-Hastings algorithm, are able to estimate the shape of the posterior without knowledge of its normalization. 

Efficient and accurate Markov Chain Monte Carlo algorithms are now readily available, thanks to their widespread use in other fields (e.g. in cosmology). This includes nested sampling, Hamiltonian methods, and parallel tempering ~\cite{brooks2011handbook}. In this work, we used an implementation of parallel tempering \cite{Vousden_2015}; the algorithm showed good convergence in sampling our posterior and at the same time made it possible to estimate the Bayesian evidence. The latter is discussed in \Section{sec:model_selection}. While this algorithm does provide an estimate for the evidence we will use this information only for performing model selection; for parameter estimation the unnormalized posteriors were used and plotted.

\subsection{Maximizing the posterior} \label{section:bayes:MAP}
Although the primary result of parameter estimation is the posterior distribution, it is also useful to calculate the point in parameter space which maximizes the posterior. This is referred to as the Maximum a Posteriori (MAP) set of parameters. Because throughout this work we use priors which are uniform distributions, the MAP parameters are those which maximize the likelihood function; that is, the parameters which optimize the fit to the experimental data.

\section{Closure Tests}
\label{sec:closure_tests}

Closure tests are required to ensure that, in a situation with known model parameters, the Bayesian inference framework correctly reproduces them from a set of ``mock data''. These data are obtained by running the model to generate predictions instead of real data for the observables which one intends to use for the model calibration. Ideally, the posterior of a closure test should approach a delta-function around the true value of the model parameters, $\mathcal{P}(\boldsymbol{x}|\mathbf{y}_{\hat{\boldsymbol{x}}}) \rightarrow \delta(\boldsymbol{x}{-}\hat{\boldsymbol{x}})$. In practice, this delta function is always smeared by the uncertainties present in the Bayesian parameter estimation. 

A first source of uncertainties is in the model calculations used instead of experimental data: since for heavy-ion collisions the initial conditions fluctuate stochastically and running the model is expensive, statistical uncertainties of the ``mock data'' are often larger than those from real experiments, and these will propagate non-trivially and contribute to the width for the parameter posterior. Additional uncertainties are contributed by the emulator: (i) statistical uncertainties from the calculations used to train the emulator; (ii) interpolation uncertainty from the limited number of parameter samples used to train the emulator; and (iii) the limited number of principal components that are interpolated via Gaussian processes. Finally, there might be partial degeneracies in the model which makes it difficult to match a unique set of observables to a unique set of model parameters. Even if a sufficiently large set of observables is used to avoid exact degeneracies, approximate degeneracies can persist until all the uncertainties decrease below a certain threshold. 

Closure tests provide a way to identify such potential issues and, for a chosen set of observables, quantify the effect of these types of uncertainties on the parameter estimation before any comparison with measurements is performed. Closure tests can also help clarify the level of constraint on the model parameters that can be expected given the emulator uncertainties. These two aspects of closure tests are not independent; however they are sufficiently different objectives that they benefit being discussed separately.

\subsection{Validating Bayesian inference with closure tests}

In what follows we show a sample set of closure tests. They employ the same emulator as will later be used for comparison with experimental data.

\begin{figure*}[!htb]
\noindent\makebox[\textwidth]{%
  \centering
  \begin{minipage}{0.47\textwidth}
    \includegraphics[trim=0 0 0 20, clip, width=\textwidth]{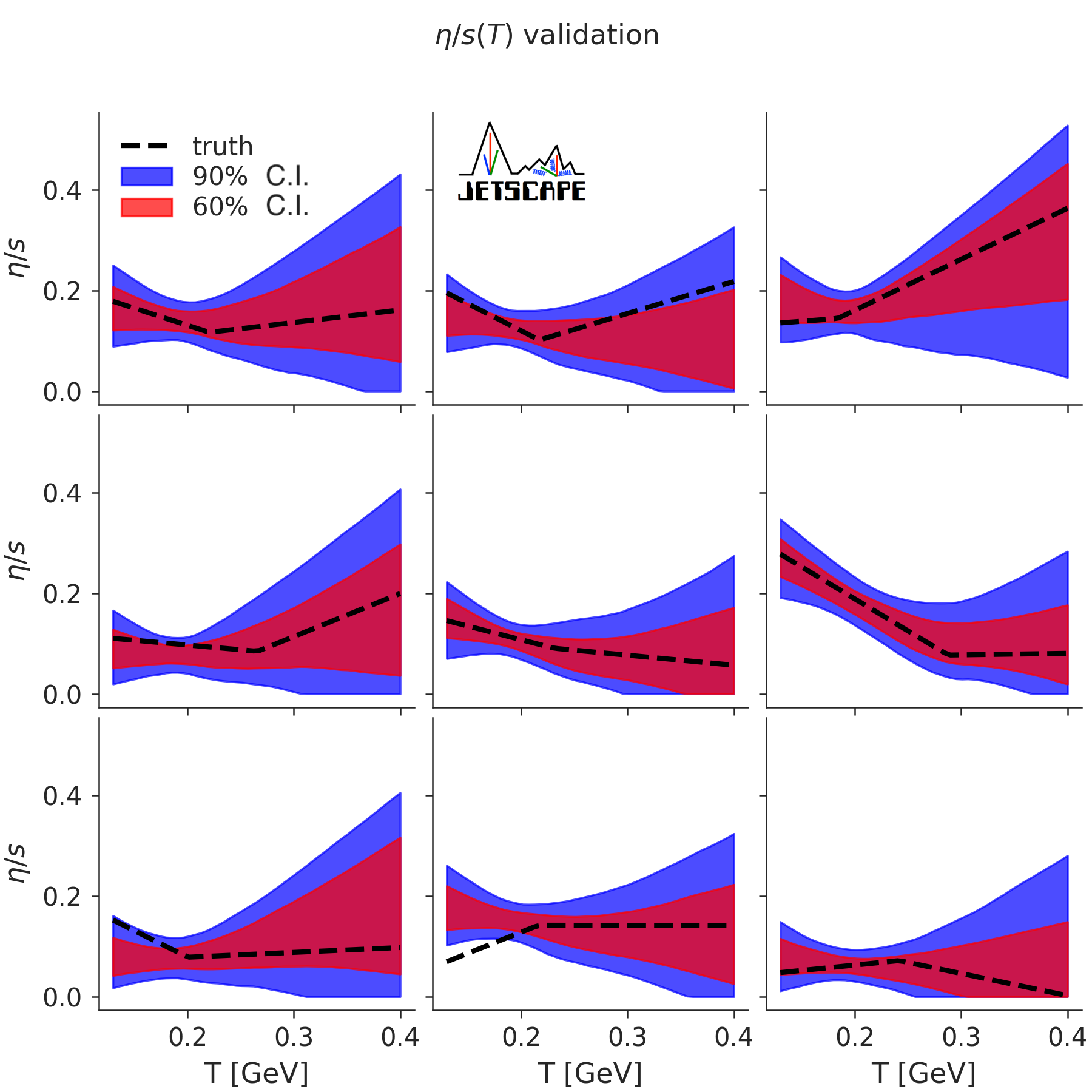}
  \end{minipage}
  \begin{minipage}{0.47\textwidth}
    \includegraphics[trim=0 0 0 20, clip, width=\textwidth]{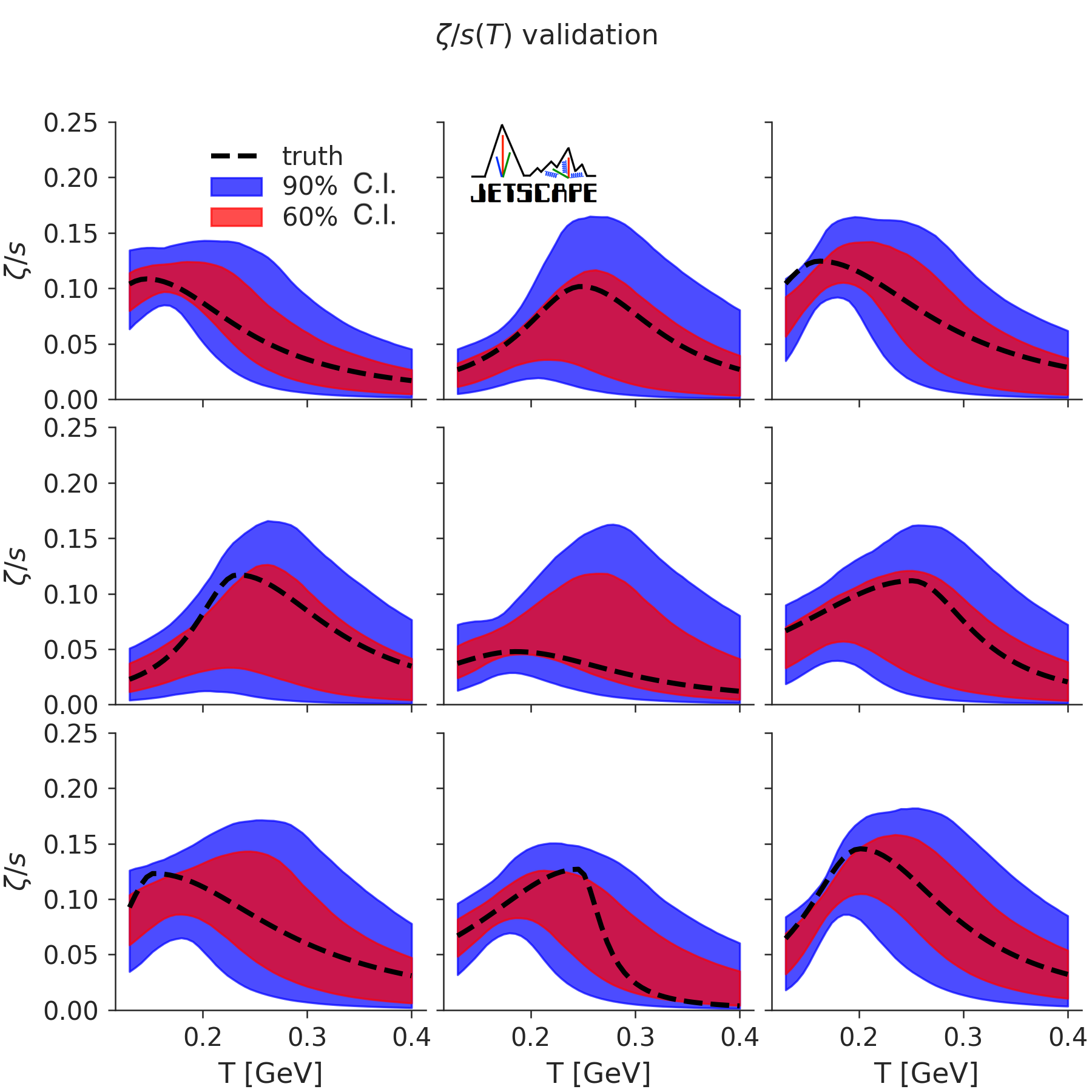}
  \end{minipage}
  }
  \caption{
  Closure tests of the specific shear (left) and bulk (right) viscosities using 9 sets of validation points in the parameter space. Performed with the emulator for Pb-Pb $\sqrts{} = 2.76$ TeV collisions. Shown in blue are the $90$\% credible intervals and in red the $60$\% credible intervals. }
  \label{closure_plots}
\end{figure*}

We proceed as follows:
\begin{enumerate}[topsep=1pt,itemsep=0ex,partopsep=1ex,parsep=1ex]
    \item We generate a set of design points ($\boldsymbol{x}_{i}; i=1,\dots, m_v$) for training the model emulator, and a separate set of design points for validation ($\hat{\boldsymbol{x}}_{i}; i=1,\dots, \hat{m}_v$).
    \item We perform full model calculations at both the training and validation design points and compute final state observables.
    \item We perform principal component analysis on the \emph{training} calculations, and fit a Gaussian process to each retained principal component.
    \item For each point $i$ in the \emph{validation} set, we use the trained emulator to perform parameter estimation using the calculated model observables at validation point $\hat{\boldsymbol{x}}_{i}$ as the ``data''. 
    \item We compare the posterior $\mathcal{P}(\boldsymbol{x}|\mathbf{y}_{\hat{\boldsymbol{x}}_{i}})$ to the known true values $\hat{\boldsymbol{x}}_{i}$.
\end{enumerate}

Our emulator uses 500 design points. At each design point we use the full model to compute predicted values for all observables that will also be used in the calibration with real data (see Sec.~\ref{sec:post_param}). As discussed previously, our model includes statistical fluctuations, which arise from averaging over a finite number of initial conditions (2500 hydrodynamic events per design point), as well as Cooper-Frye sampling each particlization hypersurface a finite number of times (at least $10^5$ particles sampled per hydrodynamic event). We use $10$ principal components, which explain approximately $98\%$ of the model variance for Pb-Pb data at $\sqrts{}=2.76$\ TeV. These uncertainties, combined with the emulator uncertainty discussed above, lead to a finite spread of our posterior $\mathcal{P}(\boldsymbol{x}|\mathbf{y}_{\hat{\boldsymbol{x}}_{i}})$. What we can verify is how often the true parameters lie within given regions of credibility.

Figure~\ref{closure_plots} shows the result of our closure tests for $9$ sets of validation points. We focus on the specific shear and bulk viscosities of the QGP, $\eta/s$ and $\zeta/s$. The parametrization of these physical quantities involves non-linearly correlated parameters. These parameters are no more than a few degrees of freedom we put into the functional form of $(\eta/s)(T), (\zeta/s)(T)$; therefore, it is of less physical importance to focus on the posterior of these parameters individually. Instead, what we show in Fig.~\ref{closure_plots} are their resultant posterior for $\eta/s$ and $\zeta/s$ as functions of $T$, compared to the assumed ``true values'' shown as dashed black lines). Red and blue bands show the 60\% and 90\% credible intervals of the estimation; at different temperatures these credible intervals are determined independently. The results demonstrate that the functional shapes of the ``true'' viscosity-to-entropy ratios are well enclosed by the inferred 60\% and 90\% confidence regions. 

\subsection{Guiding analyses with closure tests}

Figure~\ref{closure_plots} provides convincing evidence that the emulators and the Bayesian analysis are performing well. Importantly, it also provides a wealth of information on the eventual results of the Bayesian parameter estimation when performed with real data.

Recall that the result of the Bayesian analysis depends on a variety of factors, including (i) the choice of observables, (ii) the uncertainty on these observables, and (iii) the uncertainty of the emulator. In an ideal world, the emulator uncertainty would be much smaller than the uncertainties on the observables; in such a scenario, the result of the Bayesian analysis would be essentially independent of the emulator. This is always the goal one should strive for; however, in practice this is difficult to achieve, and most 
Bayesian parameter estimations are performed under much less ideal conditions.

For the case shown in Fig.~\ref{closure_plots}, emulator uncertainties are not negligible. However, given the emulator we use in this work, a comparison of the closure test in Fig.~\ref{closure_plots} with the prior from Fig.~\ref{viscous_prior} demonstrates that the current Bayesian parameter estimation method, as well as the selection of observables, have the best constraining power for $\eta/s$ and $\zeta/s$ at low temperatures. This is expected since these temperatures are closer to the switching temperature between hydrodynamics and the hadronic transport model, and much of the space-time volume explored by the expanding medium is characterized by such moderate temperatures \cite{Shen:2013vja}. The closure tests also indicate that the uncertainty on the viscosities is large at higher temperatures. We believe this could be a consequence of the smaller amount of time spent by the systems at high temperatures \cite{Shen:2013vja}, decreasing the sensitivity of the observables' response to the transport parameters in this temperature region. 

Additional observables or collision energies may help improve these constraints on the viscosities of QCD. For example, emission of electromagnetic radiation puts somewhat stronger weight on the earlier and shorter-lived hot fireball regions than hadrons do \cite{Shen:2013vja}. On the other hand, electromagnetic observables are plagued by larger statistical and systematic uncertainties. Closure tests can be used exactly for the purpose of assessing the value of adding such additional measurements even before such data are available: they allow for quantifying the contribution of different observables towards constraining the properties of the quark-gluon plasma. In the future this could be an important tool to guide the priorities of experimental campaigns. Observables contribute differently to constraining different model parameters: by quantifying the effect of adding a new observable, or reducing the uncertainty on an existing one, one can provide meaningful feedback which measurements should be prioritized. These methods are closely related to those employed in "Bayesian Experimental Design" ~\cite{chaloner1995}. 

One caveat to be kept in mind in this context is that closure tests evidently rely on the correctness of the underlying physics model. When we compare to actually experimentally observed data, we cannot assume that our model provides an exact description of the observables even when given the right choice of parameters \cite{Brynjarsdottir_2014}. The systematic model discrepancy must not be forgotten. Hence, the result of a closure test should not be taken as the final word: the importance of a given observable in constraining model parameters may need to be revisited when physics tested by this observable is modified in the model. 

In spite of these unavoidable limitations, closure tests can provide important guidance to experimental collaborations to help determine which observables can best constrain physical parameters.

\section{Bayesian parameter estimation\\ using RHIC and LHC measurements}
\label{sec:post_param}

In this Section we perform a Bayesian parameter estimation with RHIC and LHC measurements. We focus on constraints for the shear and bulk viscosities provided by transverse-momentum-integrated data from LHC and RHIC. We perform these first analyses for a specific model of viscous corrections at particlization, the Grad model (see Section~\ref{sec:model:particlization}). The effect of using different viscous corrections as well as other systematic uncertainties of the model are quantified in the next section.

\subsection{Constraints on $\eta/s$ and $\zeta/s$\\ from Pb-Pb measurements at $\sqrts{}=2.76$\ TeV}

We first study the parameter estimates including only the data from Pb-Pb collisions at $\sqrts{} = 2.76$\ TeV. We use the following measurements from the ALICE collaboration:
\begin{itemize}
    \item the charged particle multiplicity $dN_{\text{ch}}/d\eta$~\cite{Aamodt:2010cz} for bins in $0{-}70$\% centrality;
    \item the transverse energy $dE_T/d\eta$ \cite{Adam:2016thv} for bins in $0{-}70$\% centrality;
    \item the multiplicity $dN/dy$ and mean transverse momenta $\langle p_T \rangle $ of pions, kaons and protons \cite{Abelev:2013vea} for bins in $0{-}70$\% centrality;
    \item the two-particle cumulant harmonic flows $v_n\{2\}$ for $n=2,3,4$, for bins in $0{-}70$\% centrality for $n{\,=\,}2$, and for bins in $0{-}50$\% centrality for $n=3$ and $4$ \cite{ALICE:2011ab};
    \item the fluctuation in the mean transverse momentum $\delta p_T / p_T$ \cite{Abelev:2014ckr} for bins $0{-}70$\% centrality.
\end{itemize}

Before being reduced by principal component analysis this data set represents 123 ``independent observables'', given that measurements at different centralities are treated as separate observables. We found that 10 principal components (linear combinations of observables) are sufficient to capture most of the sensitivity of these observables to the full set of parameters: they capture more than $98$\% of the variance. This number of dominant principal components represents 8\% of the total number of observables. Thus there is a significant amount of redundant information in the observables with respect to our model parameters. We highlight that we tested the effect on our analysis of reducing the number of principal components: we determined that our results are robust with respect to the number of principal components used. The results of this test are presented in \Appendix{pca_valid}. 

We note that all observables used in this analysis are $p_T$-integrated. Observables which are differential in transverse momentum undeniably carry additional information about the medium \cite{Nijs:2020ors, Nijs:2020roc}. There is reasonable evidence that low-$p_T$ ($p_T\lesssim 1.5$~GeV) information is generally included in $p_T$-integrated observables \cite{Novak:2013bqa}. The higher-$p_T$ range ($p_T\gtrsim 1.5$\,GeV) tends to have larger modeling uncertainties, if only from viscous corrections at particlization which can be very significant at higher transverse momenta. At sufficiently high $p_T$, hadron production is beyond the realm of hydrodynamics altogether; this threshold is not known precisely, but even a breakdown at $p_T\gtrsim 2-3$ GeV would not be wholly surprising. Because of these limitations, there is a risk that inferences using observables in the higher-$p_T$ range  ($p_T\gtrsim 1.5$\,GeV) lead to more precise but unreliable constraints on the parameters. While both avenues are worth exploring, in the present analysis we opt for the more conservative approach of using $p_T$-integrated observables that introduce less model bias, while also studying in greater detail model uncertainties.

\begin{figure}[!htb]
\centering
\includegraphics[trim=0 0 0 15, clip, width=8cm]{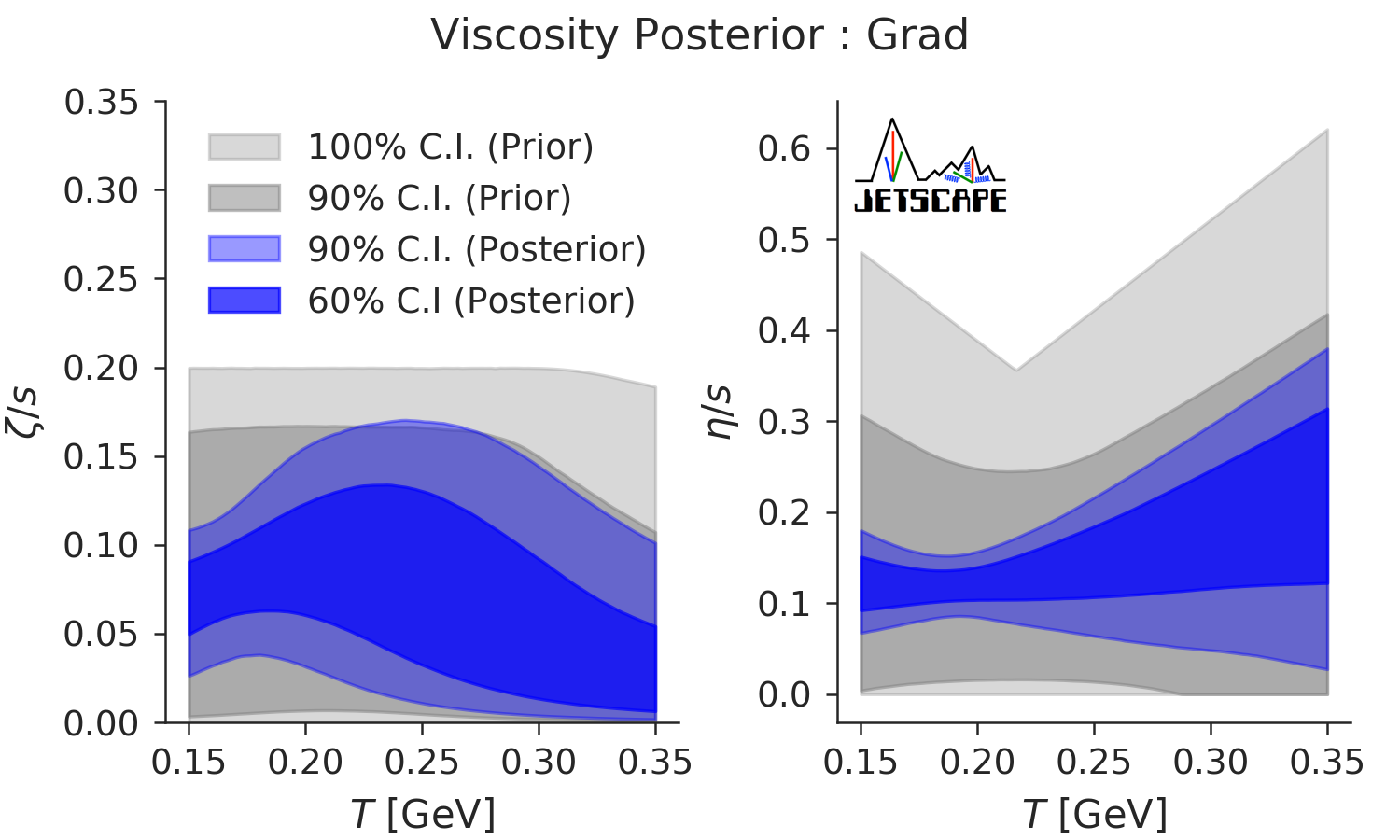}
\caption{The posterior for specific bulk (left) and shear (right) viscosities resulting from a Grad viscous correction model parameter estimation using ALICE data for Pb-Pb collisions at $\sqrts{} = 2.76$ TeV}
\label{grad_lhc_posterior}
\end{figure}

The posteriors for the shear and bulk viscosities are shown in Fig.~\ref{grad_lhc_posterior}. Recall that this result is for a {\it single} viscous correction model, the Grad viscous correction.

We first note a general feature which will remain when we examine other viscous corrections and include more systems: the constraint on the shear and bulk viscosities is best near the switching temperature $\Tsw$. This was already observed in the closure tests performed in Sec.~\ref{sec:closure_tests}. The viscous corrections in the particlization procedure depend on the magnitude of shear stress $\pi^{\mu\nu}$ and bulk pressure $\Pi$ on the switching surface, making the model predictions sensitive to the viscosities near these temperatures. As we have discussed in the closure test, the uncertainties in $\zeta/s$ and $\eta/s$ are larger in the high temperature region. We see that for the bulk viscosity in particular, our 90\% posterior credible interval is only slightly smaller than our prior above 250 MeV. 

\subsection{Constraints on $\eta/s$ and $\zeta/s$\\
from Au-Au measurements at $\sqrts{}=0.2$~TeV}

We also examine the constraints on the viscosities provided by the existing data for Au-Au collisions at $\sqrts{} = 200$\,GeV. Heavy-ion collisions at RHIC provide complimentary information, having smaller temperatures and a shorter lifetime than collisions at the LHC. We use the following experimental measurements from the STAR Collaboration:
\begin{itemize}
    \item the yields $dN/dy$ and mean transverse momenta $\langle p_T \rangle $ of pions and kaons for bins in 0--50\% centrality \cite{Abelev:2008ab}; 
    \item the two-particle cumulant harmonic flows $v_n\{2\}$ for $n=2,3$ for bins in 0--50\% centrality \cite{Adams:2004bi, Adamczyk:2013waa}.
\end{itemize}
We remark that because of the tension between STAR and PHENIX measured proton yields at mid-rapidity in Au-Au collisions at $\sqrts{} = 200$\,GeV \cite{Abelev:2008ab,Adler:2003cb}, we have deliberately left the proton yield and mean transverse momentum out of the current comparison.\footnote{%
    Moreover, both measurements \cite{Abelev:2008ab,Adler:2003cb} show notable excess of proton production over anti-proton production, suggesting the importance of including a non-zero baryon chemical potential ($\mu_B$) in our calculation. The current study assumes $\mu_B=0$ in both initial condition and dynamical evolution, and improvement should be considered in future studies.}

The above includes 29 observables, again counting centralities as separate observables. After performing principal component analysis, we kept 6 principal components (equivalent to 21\% of the total number of observables), which explain more than 98\% of the variance of the observables across the parameter space.

\begin{figure}[!htb] 
\centering
\includegraphics[trim=0 0 0 15, clip, width=8cm]{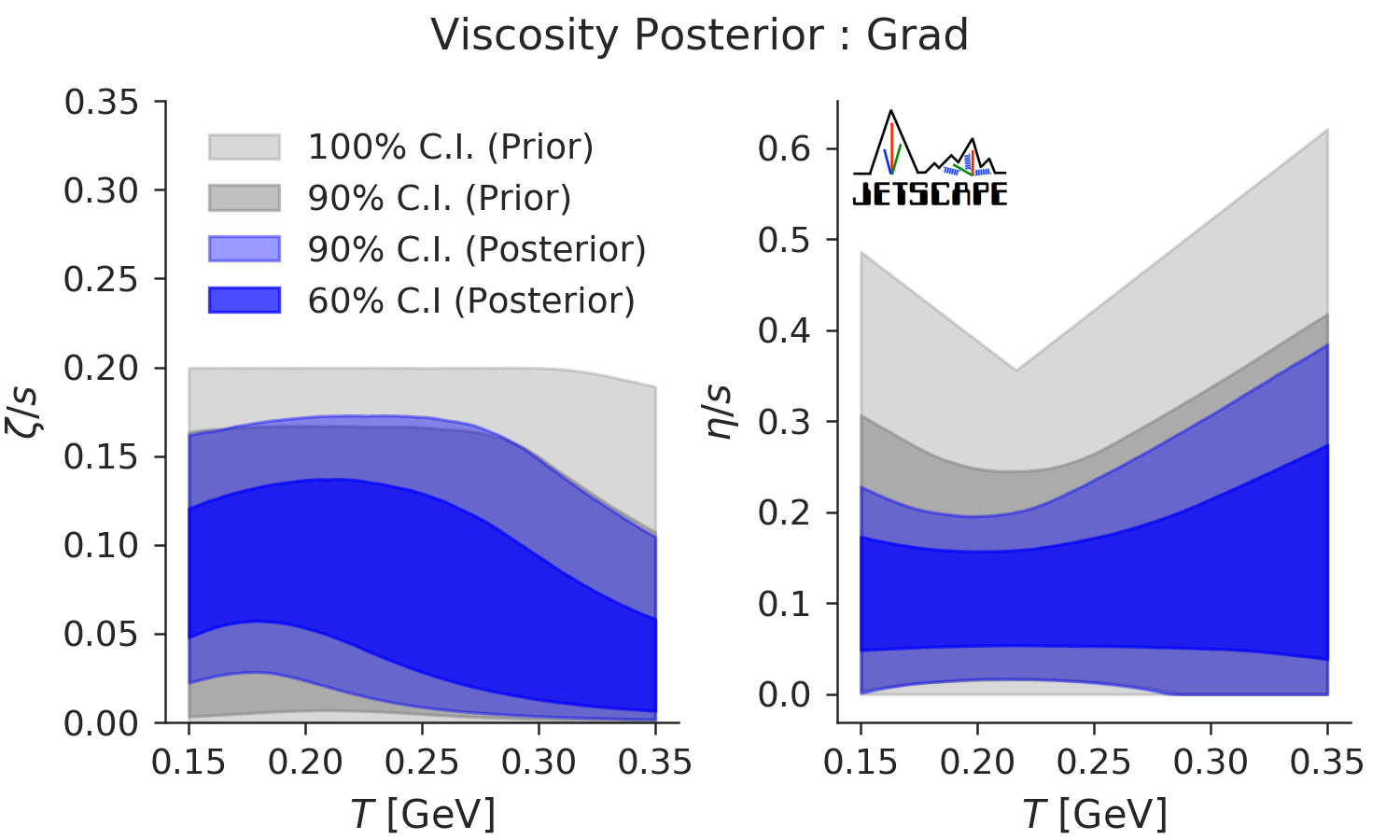}
\caption{The posterior for specific bulk (left) and shear (right) viscosities resulting from a model parameter estimation using STAR data for Au-Au collisions at $\sqrts{} = 200$\,GeV.}
\label{grad_rhic_posterior}
\end{figure}

The estimated viscosities using only these measurements from RHIC, again for the Grad viscous correction, are shown in Fig.~\ref{grad_rhic_posterior}. 
%
%
The posteriors for specific bulk and shear viscosity when calibrating against only RHIC data have in general different features than those given by the LHC data. For instance, we see that a large specific bulk viscosity is allowed near the switching temperature. Also, the 90\% credible interval for the specific shear viscosity extends to lower values for these data than the LHC data; only using these RHIC observables, a specific shear viscosity which is nearly zero ($\eta/s < 0.03$) is consistent with the data.

It is important to note that not only the specific bulk and shear viscosity parameters have different posteriors, but in general the entire parameter posterior will be different when we use RHIC observables rather than LHC observables. The two are compared for a different subset of model parameters in \Appendix{app:post_LHC_RHIC_separate}. 

\subsection{Viscosity estimation and model accuracy for combined RHIC \& LHC data}
\label{sec:calib_combined}

\begin{figure}[!htb] 
\centering
\includegraphics[trim=0 0 0 17, clip, width=8cm]{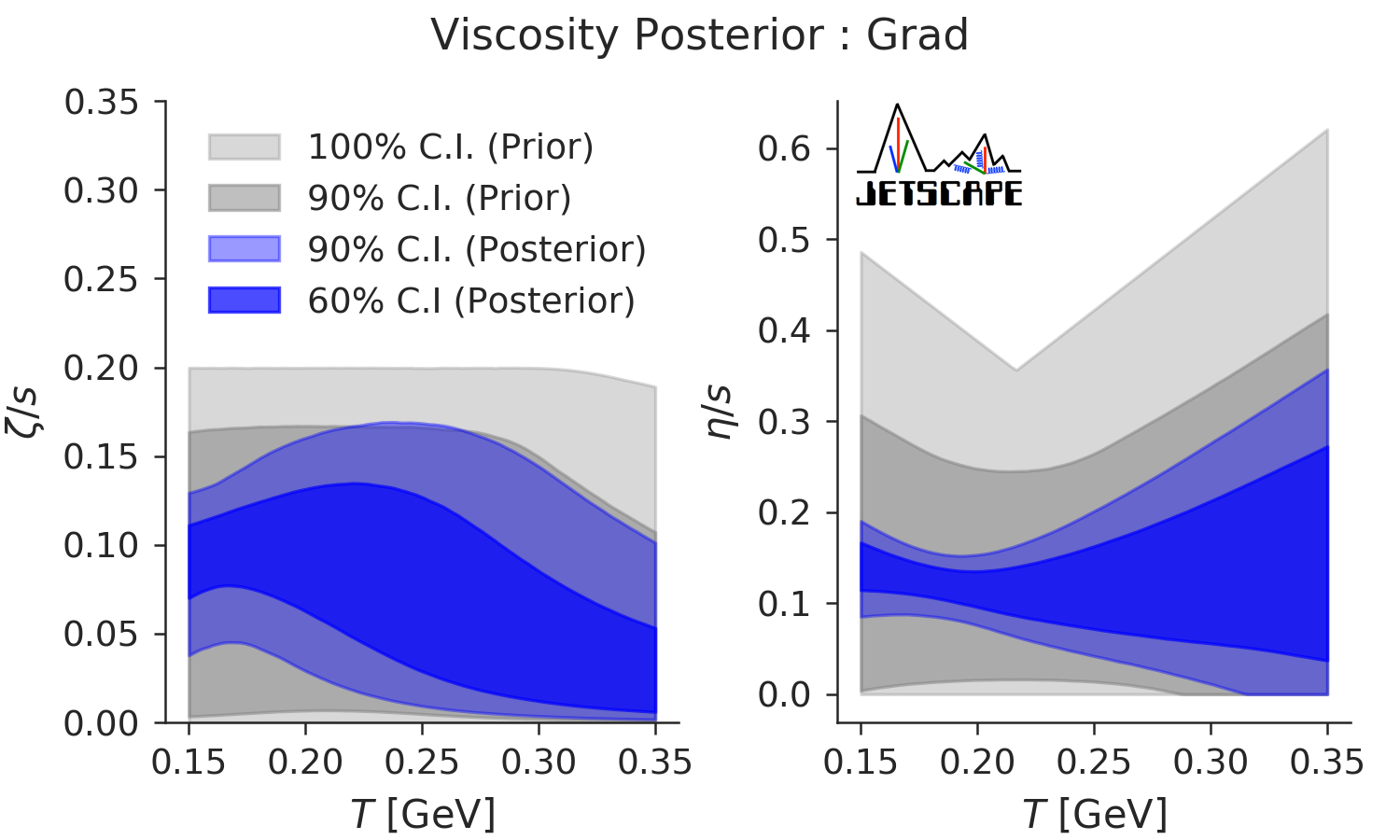}
\caption{The posterior for specific bulk (left) and shear (right) viscosities resulting from a model parameter estimation using combined data for Au-Au collisions at $\sqrts{} = 200$ GeV and Pb-Pb collisions at $\sqrts{} = 2.76$ TeV. }
\label{grad_lhc_rhic_posterior}
\end{figure}

Reviewing Figs.~\ref{grad_lhc_posterior} and \ref{grad_rhic_posterior} we find that the observables at the LHC give stronger constraints on the slope of the specific shear viscosity at large temperature. It is the general expectation that higher $\sqrts{}$ collisions at the LHC are more sensitive to the transport coefficient at high temperature. This conclusion was verified quantitatively in previous Bayesian parameter estimation \cite{Pratt:2015zsa,Sangaline:2015isa}. For the present analysis, we do caution that we currently use a different number of observables at RHIC and the LHC; consequently, we are not in a position to compare systematically the constraining power of the two collision energies at the moment. We do expect RHIC and LHC data to be complementary, and we proceed to a combined Bayesian parameter estimation for Pb-Pb at $\sqrts{} = 2.76$\,TeV and Au-Au at $\sqrts{} = 200$\,GeV collisions. For this combined analysis, the viscosity posterior for the Grad viscous correction is shown in Fig.~\ref{grad_lhc_rhic_posterior}. 

As discussed in Section~\ref{overview_param_est}, all parameters are held the same for the two systems except for their overall normalizations of the initial conditions --- $N$[2.76\,TeV] and $N$[0.2\,TeV]. Recall that model parameters being kept constant does not imply that the effective physical quantities are the same at RHIC and the LHC. For example, the transport coefficients are temperature dependent, and the free-streaming time depends on $\sqrts{}$ and centrality through the total energy of the event.

The information gained by fitting both systems slightly reduces the width of the credible intervals for the specific shear and bulk viscosities at temperatures above 250 MeV; the 90\% confidence band in the posterior for specific shear and bulk viscosity is slightly smaller than the credible intervals given by calibrating against either one of these two systems alone. This illustrates the added constraining power accessed by combining the two data sets. 

\begin{figure}[!htb] 
\centering
\includegraphics[trim=0 0 0 25, clip, width=8cm]{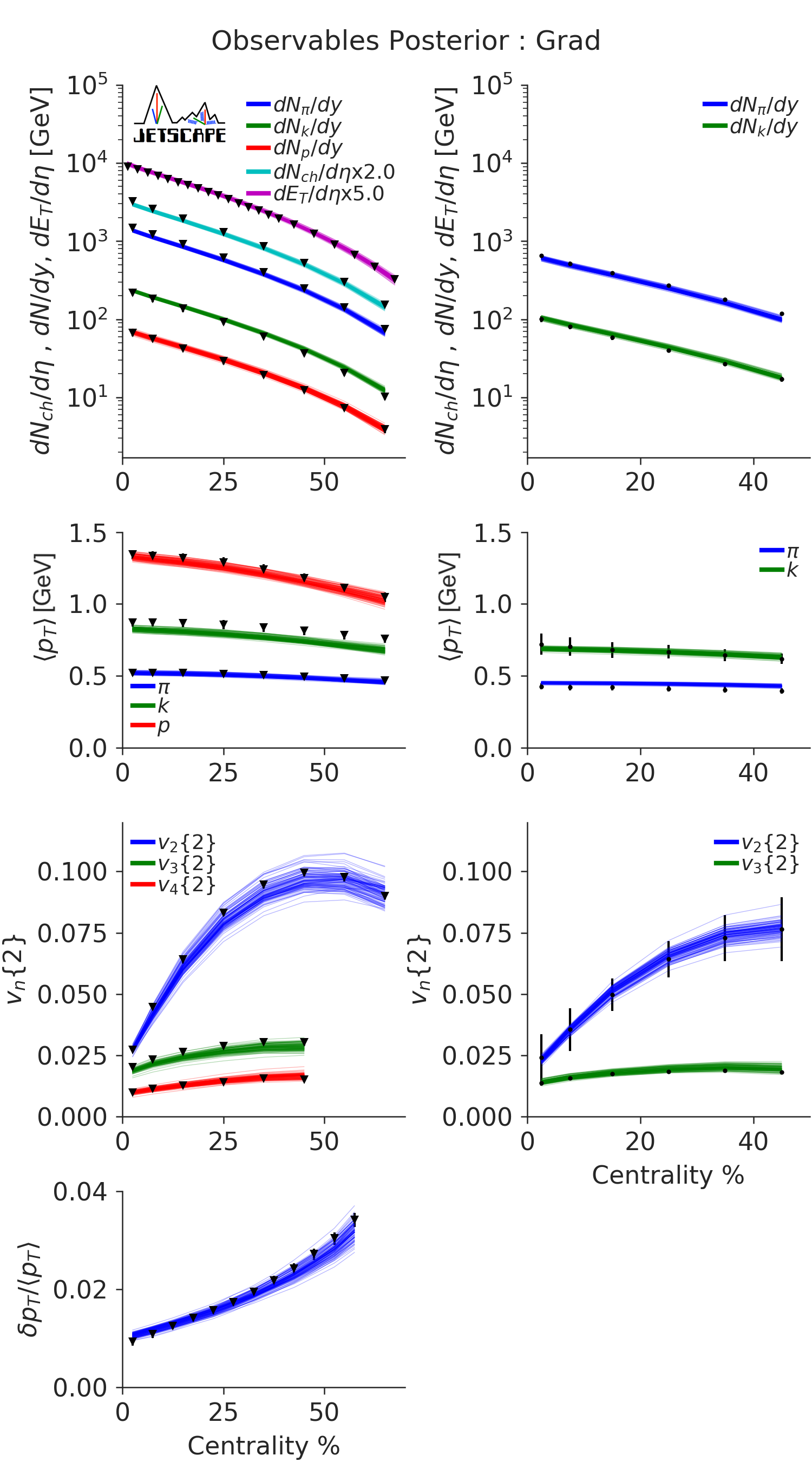}
\caption{The observables predicted by the Grad viscous correction emulator, drawn from the posterior resulting from the combined fit of ALICE data (left) for Pb-Pb collisions at $\sqrts{} = 2.76$ TeV and STAR data (right) for Au-Au collisions at $\sqrts{} = 200$ GeV. The simultaneous fit yields model observables which agree within ${\sim}20\%$ of experimental measurements.}
\label{grad_lhc_rhic_expt_fit}
\end{figure}

The simultaneous fit to experimental observables is shown in Fig.~\ref{grad_lhc_rhic_expt_fit}, where we have plotted the emulator prediction for the observables at one hundred parameter samples drawn randomly from the posterior. 
%
%
Note that, in spite of some undeniable tension in the simultaneous fit of ALICE and STAR data (for example in the mean transverse momenta of kaons), our hybrid model can describe simultaneously all of the observables we considered for the two systems to within 20\% of the experimental results. As discussed earlier, this is important: our confidence in the significance of this section's parameter estimates rests on a good description of the experimental data when sampling model parameters according to their posterior probability distribution. 

\begin{figure}[!htb] 
\centering
\includegraphics[trim=0 0 0 25, clip, width=8cm]{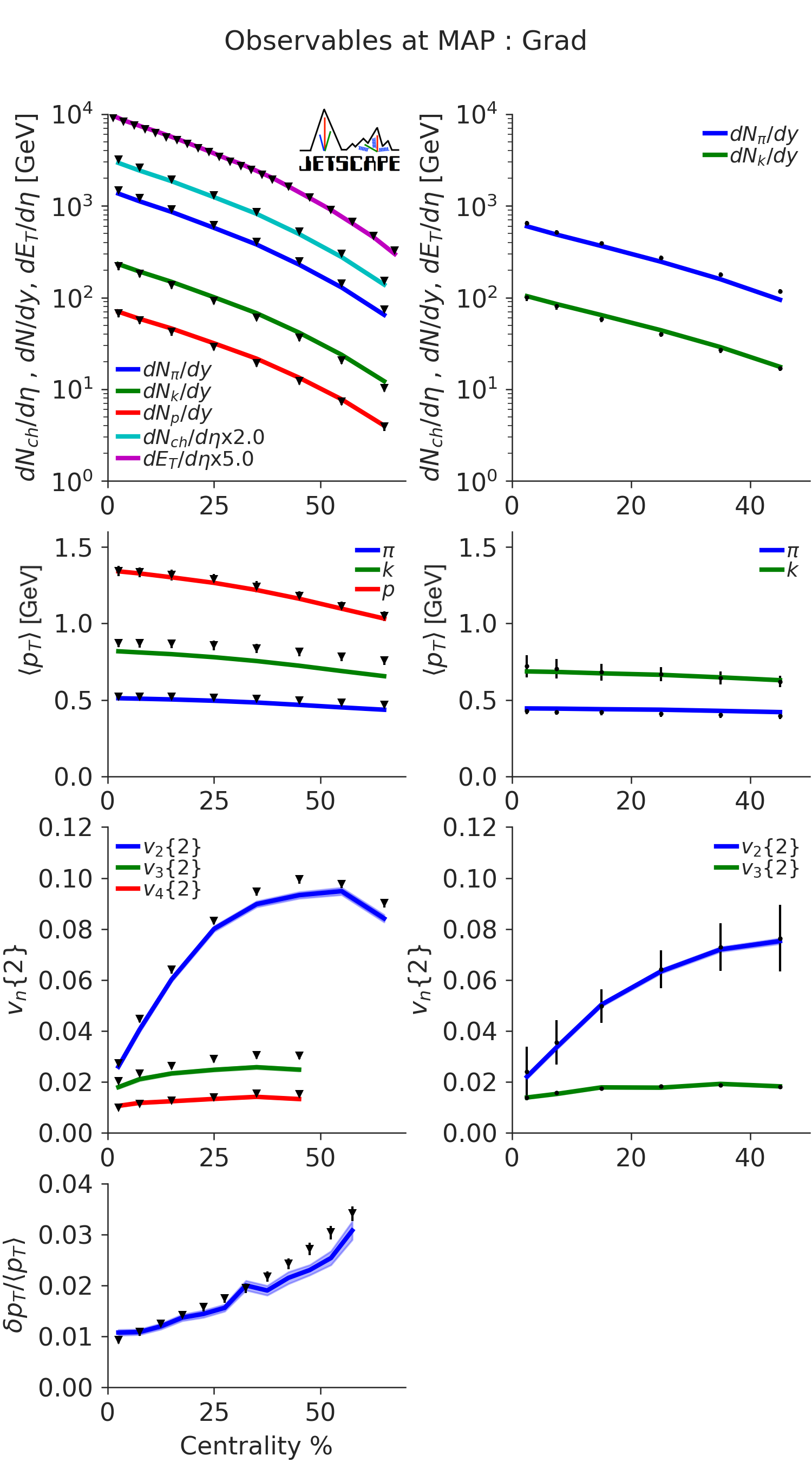}
\caption{The observables resulting from averaging over five thousand fluctuating events for each system, run with the MAP parameters of the combined calibration of ALICE data for Pb-Pb collisions at $\sqrts{} = 2.76$ TeV and STAR data for Au-Au collisions at $\sqrts{} = 200$ GeV. Results are shown for the Grad viscous correction. Shaded bands around model predictions reflect the variance arising from initial state fluctuations combined with statistical fluctuations from the particlization sampler. Pb-Pb $\sqrts{}=2.76$ TeV events are shown at left and Au-Au $\sqrts{}=0.2$ TeV events at right.}
\label{grad_lhc_rhic_MAP_observables}
\end{figure}

As a final emulator validation, we have calculated the Maximum A Posteriori (MAP) parameters of the Grad viscous correction model. Using these parameters, we simulated 5,000 fluctuating events and performed centrality averaging. The comparison between the hybrid model prediction at the MAP parameters and the experimental data are shown in Fig.~\ref{grad_lhc_rhic_MAP_observables}, and MAP parameters for the Grad, Chapman-Enskog and Pratt-Torrieri-Bernhard models are listed in Table~\ref{table_MAP_grad}.\footnote{%
    For reasons explained in Sec.~\ref{sec:uncerntainty:deltaf}, the Pratt-Torrieri-McNelis model (PTM) is left out from Table~\ref{table_MAP_grad}.}

\begin{table}[!htb]
\centering
\begin{tabular}{|l|l|l|l|}
\hline
Parameter & Grad & CE & PTB \\ \hline
$N$[2.76\,TeV] & 14.2 & 15.6 & 13.2\\ \hline
$N$[0.2\,TeV] &  5.73 & 6.24 & 5.31\\ \hline
$p$ &  0.063 & 0.063 & 0.139\\ \hline
$\sigma_k$ & 1.05 & 1.00 & 0.98\\ \hline
$w$\,[fm] &  1.12 & 1.19 & 0.81\\ \hline
$d_{\text{min}}^3$\,[fm$^3$] &  2.97 & 2.60 & 3.11\\ \hline
$\tau_R$\,[fm/$c$] &  1.46 & 1.04 & 1.46\\ \hline
$\alpha$ &  0.031 & 0.024 & 0.017\\ \hline
$T_\eta$\,[GeV] & 0.223 & 0.268 & 0.194\\ \hline
$a_\text{low}$\,[GeV$^{-1}$] & --0.776 & --0.729 & --0.467\\ \hline
$a_\text{high}$\,[GeV$^{-1}$] &  0.37 & 0.38 & 1.62\\ \hline
$(\eta/s)_{\text{kink}}$ & 0.096 & 0.042 & 0.105 \\ \hline
$(\zeta/s)_{\text{max}}$ & 0.133 & 0.127 & 0.165\\ \hline
$T_\zeta$\,[GeV] &  0.12 & 0.12 & 0.194\\ \hline
$w_{\zeta}$\,[GeV] &  0.072 & 0.025 & 0.026\\ \hline
$\lambda_{\zeta}$ & --0.122 & 0.095 & --0.072\\ \hline
$b_{\pi}$ &  4.65 & 5.62 & 5.54\\ \hline
$\Tsw{}$\,[GeV] & 0.136 & 0.146 & 0.147\\ \hline
\end{tabular}
\caption{Table of MAP parameters of the Grad, Chapman-Enskog (CE) and Pratt-Torrieri-Bernhard (PTB) viscous correction models, from combined RHIC and LHC data. }
\label{table_MAP_grad}
\end{table}

Because our prior for each of these parameters was uniform on a finite range, the parameters which maximize the posterior also maximize the likelihood function; this means that they also optimize the fit to the experimental data (i.e. minimize $\chi^2$).

\section{Parameter estimation and systematic model uncertainties}
\label{sec:param_est_syst_uncert}

In this section, we continue our exploration of the estimated parameter posterior for the combined LHC Pb-Pb $\sqrts = 2.76$ TeV and RHIC Au-Au $\sqrts = 0.2$ TeV data. We identify and discuss some of the largest sources of theoretical uncertainty in the physical model, and the effect these uncertainties have on constraining the viscosities of QCD. 

The first source of uncertainty that we investigate in \Section{sec:uncerntainty:deltaf} originates from mapping the hydrodynamic fields to hadronic momentum distributions, the ``viscous corrections'' at particlization, discussed in \Section{sec:model:particlization}. Recall that the results from the previous section were for a specific choice of viscous corrections, the Grad model.

The viscous corrections correspond to one source of uncertainty in the transition from hydrodynamics to particles. A second source is the determination of the particlization hypersurface, which in this work is defined at a fixed switching temperature $\Tsw{}$. We discuss the dependence of our results on this switching temperature in \Section{sec:uncerntainty:Tsw}. We discuss at the same time the transition between the early stage of the model and hydrodynamics, which we find exhibits clear correlation with the switching temperature.

Finally, we discuss the effect of second-order transport coefficients, as quantified with the shear relaxation time in \Section{sec:uncerntainty:taupi}.

\subsection{Mapping hydrodynamic fields to hadronic momentum distributions}
\label{sec:uncerntainty:deltaf}

\begin{figure}[tb] 
\centering
\includegraphics[trim=0 0 0 17, clip, width=8cm]{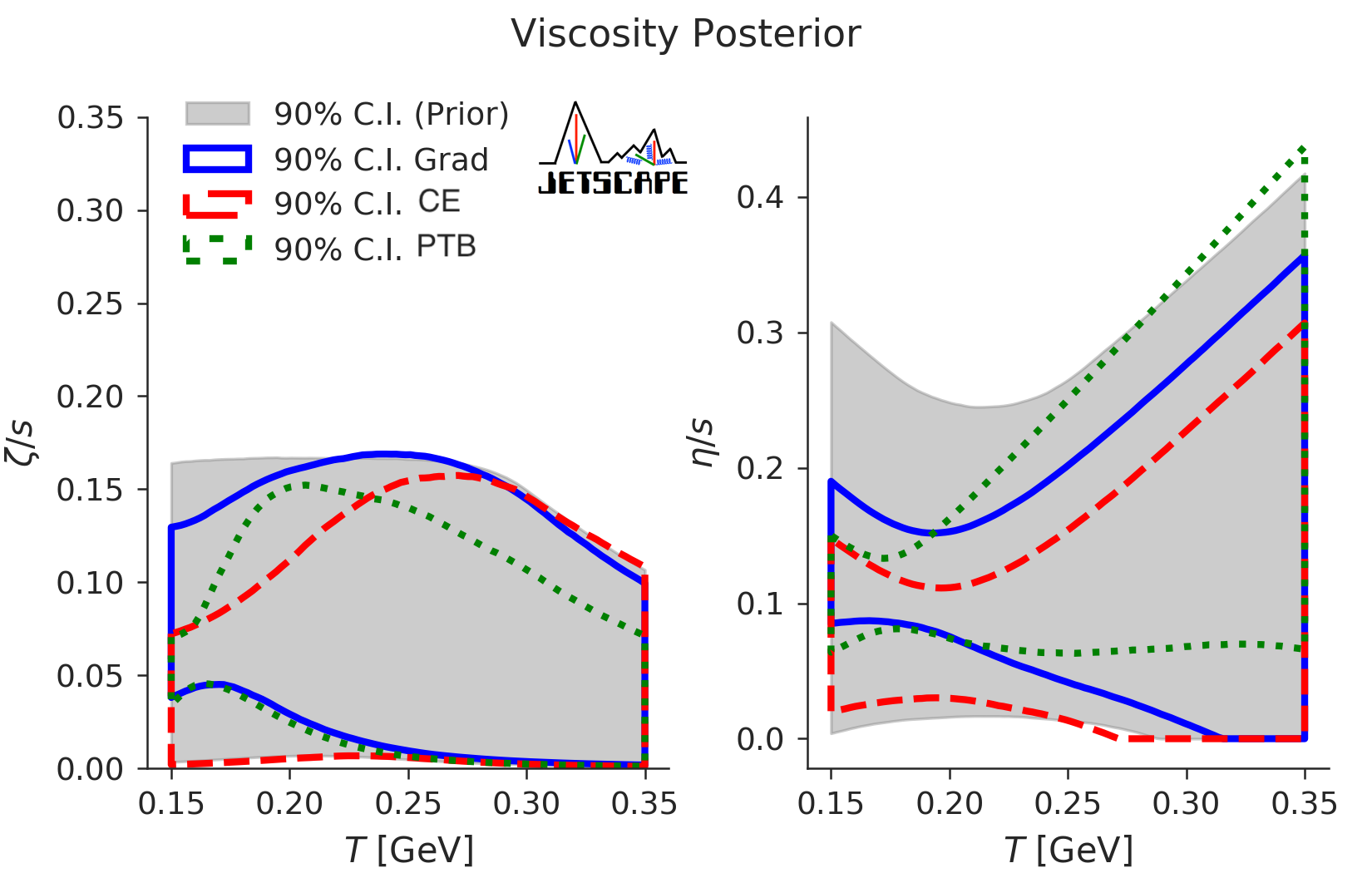}
\caption{The 90 \% credibility intervals for the prior (gray shaded area) and for the posteriors (colored outlines) of the specific bulk (left) and shear (right) viscosities, for three viscous correction models: Grad (blue), Chapman-Enskog (CE, red) and Pratt-Torrieri-Bernhard (PTB, green). The Pratt-Torrieri-McNelis (PTM) posterior is not shown, but is nearly identical with the Chapman-Enskog result. 
}
\label{compare_visc_posteriors}
\end{figure}

As discussed in \Section{sec:model:particlization}, there are still significant uncertainties in matching the energy-momentum tensor from hydrodynamics to a unique hadronic momentum distribution. Inevitably this affects the results of the phenomenological constraints we obtain on the shear and bulk viscosity of QCD: the posterior for every model parameter depends on the choice of viscous correction at particlization. Recall that in this work, we chose to study four different models of viscous corrections (see \Section{sec:model:particlization}): (i) Grad (``14-moments''); (ii) Chapman-Enskog in Relaxation Time Approximation; (iii) an exponentiated version of the Chapman-Enskog model referred to as ``Pratt-Torrieri-McNelis''; and (iv) an additional exponentiated model of viscous corrections referred to as ``Pratt-Torrieri-Bernhard''. In our tests, we found that the posteriors for the exponentiated Chapman-Enskog ansatz called ``Pratt-Torrieri-McNelis'' were always very similar to the results for the linearized Chapman-Enskog ansatz. To avoid clutter we therefore decided not to show the posteriors for the Pratt-Torrieri-McNelis model, neither in this Section nor anywhere else in this work. We begin with the marginalized posteriors for the QGP viscosities, shown in \fig{compare_visc_posteriors}.
%
%
The Figure exhibits clear differences in the experimentally preferred shear and bulk viscosities for the different viscous correction models. Remember that it is important to read \fig{compare_visc_posteriors} with respect to the parameter prior whose 90\% credibility region is indicated by the gray shaded area. A posterior that covers the same area as the prior should be interpreted as indication for weak or even absence of experimental constraints. On the other hand, a posterior that systematically excludes certain regions of the prior provides good evidence that parameter values in these excluded regions are disfavored by data.

For the bulk viscosity (left), we can see in \fig{compare_visc_posteriors} that each of the different viscous correction models excludes only relatively small regions of the prior. For all four particlization models the constraints on $\zeta/s$ are tighter at lower temperatures than at higher ones. However, the $\zeta/s$ regions favored by each model at low temperature differ from each other: the Grad viscous correction model favors a larger $\zeta/s$ where the Chapman-Enskog model favors lower values, with the Pratt-Torrieri-Bernhard model lying in between. We note in particular that the Pratt-Torrieri-Bernhard posterior is very narrow at low temperature. We understand this to be a consequence of mean transverse momenta and harmonic flows being very sensitive to the specific bulk viscosity near the switching temperature for the Pratt-Torrieri-Bernhard viscous correction model. We quantify and revisit this difference in sensitivity of the viscous correction models in Section~\ref{sec:model_sensitivity}.

Overall, only large values of $\zeta/s$ at low temperature are excluded by all three viscous correction models. As such, our constraints on the bulk viscosity are limited, especially after accounting for the model uncertainty introduced by the viscous corrections.

For the shear viscosity shown in the right panel of \fig{compare_visc_posteriors} we encounter a similar situation: limited constraints on $\eta/s$ at higher temperatures, and exclusion of large values of $\eta/s$ at low temperature by all viscous correction models. Overall, shear viscosity is best constrained at temperatures around 200\,MeV.

From the results of this section, we see that viscous corrections represent a considerable uncertainty in constraining the QGP shear and bulk viscosities. It is important to remember that all viscous correction models studied in this work are based on relatively simple assumptions. The capacity of any of these models to describe correctly the momenta and chemistry of a realistic out-of-equilibrium system of hadrons is still under investigation (see Ref.~\cite{Damodaran:2017ior, Damodaran:2020qxx} and references therein for a recent overview). For instance, all of these particlization models assume that the hydrodynamic shear stress is shared ``democratically'' among the hadronic species. This approximation greatly simplifies the models, but microscopic transport theory suggests that it may not be suitable for heavier hadrons such as protons \cite{Molnar:2014fva}. Additional theoretical efforts (see e.g. Refs.~\cite{Dusling:2009df, Molnar:2014fva, Wolff:2016vcm, Damodaran:2017ior, Damodaran:2020qxx}) may be able to shed more light on this question and provide additional insights that can be used for tightening our prior assumptions in future Bayesian analyses. Until this happens the particlization model uncertainty must be considered as ``irreducible'' and is best accounted for by Bayesian Model Averaging as reported in \cite{Everett:2020xug}. 

\subsection{Transition to and from hydrodynamics: initial state and switching temperature}
\label{sec:uncerntainty:Tsw}

The previous section focused on the uncertainty originating from transitioning from hydrodynamics to a particle description of the system. This transition occurs on a hypersurface defined by a temperature $\Tsw{}$. Recall that this switching temperature is also a model parameter, allowed to vary between $135$ and $165$~MeV.

The other transition point to hydrodynamics is the time at which hydrodynamics is initialized with the energy-momentum tensor from the preceding free-streaming evolution (\Section{sec:model:init}). This hypersurface is defined at a constant proper time $\taufs{}$, the value of which depends on two parameters as defined in \eq{eq:taufs}. The hydrodynamic initial conditions on this hypersurface further depend on the initial condition parameters of the \trento{} ansatz. In this section, we discuss the posterior of $\Tsw{}$, $\taufs{}$ and the \trento{} parameters, how they are correlated, and how they are affected by the viscous correction models discussed in the previous section.

\begin{figure*}[!htb] 
\centering
\includegraphics[width=18cm]{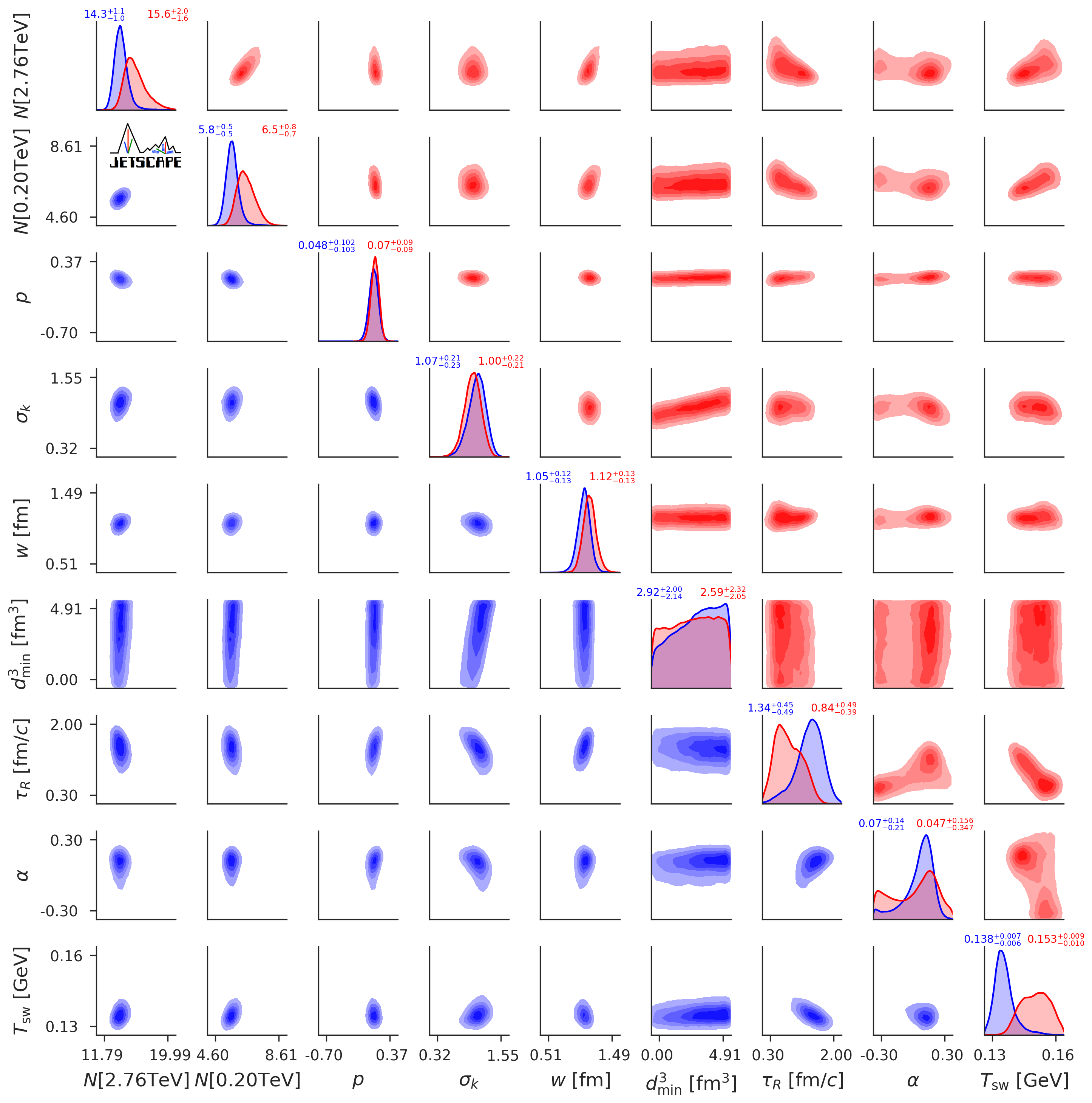}
\vspace*{-3mm}
\caption{
The posterior for Grad (blue) and Chapman-Enskog (red) viscous corrections for select parameters related to the initial state, pre-hydrodynamic evolution and switching temperature. The histograms on the diagonal are the marginal distributions for each parameter, with appended numbers denoting the median and the left and right limits of the $90$\% credible interval. Off-diagonal histograms display the joint-posterior of each pair of parameters, marginalized over all others.}
\label{initial_state_posterior}
\end{figure*}

\fig{initial_state_posterior} provides a dimensionally reduced representation of the joint posterior probability distribution for all model parameters except those related to shear and bulk stress, for two viscous correction models, Grad (blue) and Chapman-Enskog (red). Histograms on the diagonals are the marginalized one-dimensional posteriors for each parameter. The off-diagonal histograms are the joint posteriors for each pair of parameters, marginalized over all others. 

One observes that, within the chosen prior range, the normalizations of the initial energy density for the two systems $N[2.76$\,TeV$]$ and $N[0.2$\,TeV$]$ are well constrained by the observables, for both viscous correction models, but with slightly shifted peak values. Note that, since the final multiplicities are fixed by experiments, lower normalization factors for the initial energy (and hence entropy) density reflect larger viscous heating effects during the subsequent dynamical evolution. The amount of viscous heating is also affected by the particlization temperature $\Tsw$, with lower values of $\Tsw$ corresponding to longer lifetimes of the hydrodynamic stage. 

We further find that the estimation of the generalized mean parameter $p$ is the same for the two viscous correction models, close to $p=0$. We verified that $p$ is also close to zero ($p\approx 0.1$) with the Pratt-Torrieri-Bernhard viscous model. These $p$-values are consistent with previous studies which also used the Pratt-Torrieri-Bernhard viscous correction model but differed in some other model details \cite{Bernhard:2018hnz}. The result $p\approx 0$ seems to be remarkably robust across all existing Bayesian inference analyses of high-energy heavy-ion collision data \cite{Bernhard:2016tnd, Bernhard:2018hnz, Moreland:2018gsh, Bernhard:2019bmu}. We note that \trento{} with $p=0$ shares important aspects of fluctuating collision geometry with phenomenologically successful initial condition models based on saturation physics. For example, $p=0$ predicts that the energy deposition is proportional to $\sqrt{T_A T_B}$ as discussed in Sec.~\ref{sec:model:init}. This unique feature of the initial local energy density being solely a function of the product $T_A(x_\perp)T_B(x_\perp)$ is also found in the pQCD+saturation based EKRT initial condition model \cite{Niemi:2015qia},\footnote{%
    Though the EKRT model used a different parametrization for the relation between energy density and $T_A T_B$, its functional form agrees very well with the $\sqrt{T_A T_B}$ relation for typical nuclear thickness functions obtained for lead nuclei.}
and in models with approximate longitudinal boost-invariance the $\sqrt{T_A T_B}$ dependence can be motivated by simple arguments based on conservation of energy and momentum during the initial energy deposition process \cite{Shen:2020jwv}. Earlier studies \cite{Moreland:2014oya, Bernhard:2016tnd} further noted that for $p\approx 0$ \trento{} can reproduce the centrality dependent 2-particle cumulant eccentricity $\epsilon_2$ and triangularity $\epsilon_3$ of the IP-Glasma initial condition model \cite{Schenke:2012wb}. However, one should keep in mind that the two models have very different participant scaling of local energy deposition. According to Eqs.~(\ref{eq:TrentoEd},\ref{eq:TR}), \trento{} for $p=0$ sets the initial local energy density proportional to $\sqrt{T_A T_B}$, but the IP-Glasma model predicts a $T_A T_B$ scaling immediately after the collision \cite{Lappi:2006hq}. The two models also have different levels of granularity and fluctuation in the energy deposition \cite{Schenke:2012wb}. Moreover, studies \cite{Gale:2012rq, McDonald:2016vlt} that used the IP-Glasma model to initialize the hydrodynamics defined centrality differently from the present and earlier studies using the \trento{} model  \cite{Bernhard:2016tnd,Moreland:2018gsh}. All these differences convoluted into the comparison of centrality dependent $\epsilon_2$ and $\epsilon_3$ between the two models. Therefore, \trento{} ($p=0$) should not be considered a substitute for these theories based on saturation physics but rather taken as an efficient parametrization of general geometric features shared by these initial state models that happen to be preferred by the experimental data.

The nucleon width $w$, which controls the transverse length scale of energy fluctuations in the initial state, is also well constrained by the data and found to be about $1.1$\,fm, nearly independent of the two viscous correction models. A smaller value for this nucleon width ($w \approx 0.9$~fm) was found, however, with the Pratt-Torrieri-Bernhard viscous correction.

In general, our conclusions for the \trento{} parameters is that they do not appear to be highly sensitive to the choice of viscous correction model at particlization. However, the particlization uncertainty should not be entirely ignored as it can be larger than the width of the posteriors for each of these parameters.

\begin{figure}[tb] 
\centering
\includegraphics[trim=0 0 0 17, clip, width=7cm]{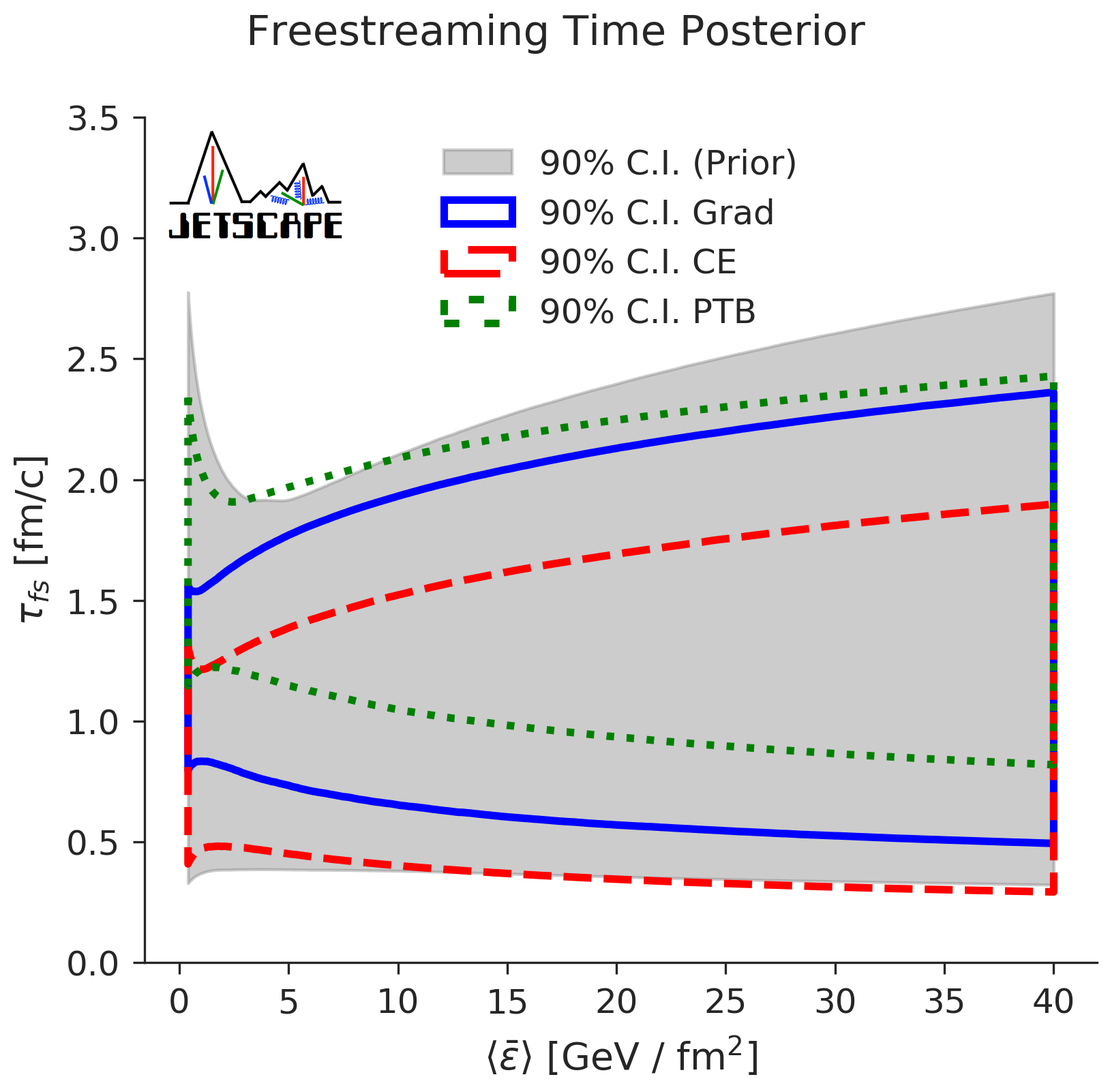}
\caption{
The 90\% posterior credible intervals for the free-streaming time, as a function of the initial average transverse energy density defined in Eq.~(\ref{eps_ave}), resulting from parameter estimation using combined data for Au-Au collisions at $\sqrts{} = 200$\,GeV and Pb-Pb collisions at $\sqrts{} = 2.76$\,TeV. The Grad model is shown with solid blue, Chapman-Enskog with dashed red, and Pratt-Torrieri-Bernhard with dotted green lines.}
\label{tau_fs_posterior}
\end{figure}

The posteriors for the free-streaming time scale $\tau_R$ and the associated energy dependence parameter $\alpha$ are not easily interpreted; they are correlated by our parametrization (\eq{eq:taufs}) of the effective free-streaming time $\taufs$. We can point out, however, that the posteriors for the Grad and Chapman-Enskog models are quite different for these two parameters; in the case of the Chapman-Enskog model, the posterior for $\alpha$ is bimodal. It is not clear whether the peak in $\alpha$ near $-0.3$ is a local maximum or if there exists a global maximum in the posterior for values of $\alpha$ less than $-0.3$. We cannot currently differentiate between these two scenarios.

We can take a closer look at the posterior of the free-streaming time by plotting it as a function of the physical scale $e_0$, which is the magnitude of the average initial energy density in the transverse plane.
This is shown in \fig{tau_fs_posterior}. We see that the 90\% credible interval for the energy dependence of the free-streaming time is not well constrained, and that it is consistent with having no energy dependence. What is constrained is the overall magnitude of the free-streaming time. The Chapman-Enskog model has a posterior which prefers smaller free-streaming times, while the Pratt-Torrieri-Bernhard model prefers the largest free-streaming time of all particlization models studied. 

It is generally expected that collisions with higher energy density will hydrodynamize more rapidly \cite{Basar:2013hea}. In our model this would correspond to $\alpha<0$. The peak at $\alpha>0$ in the posterior for $\alpha$ is at variance with this expectation. One should remember, however, that our pre-hydrodynamic model does not naturally lead to hydrodynamization which is instead enforced by hand at $\taufs$. As such, it is conceptually problematic to associate our free-streaming time $\taufs$ with a hydrodynamization time. As discussed in connection with \eq{Pi_init}, hard-matching an energy-momentum tensor from a conformally invariant pre-hydrodynamic evolution model without thermalization to dissipative hydrodynamics with a non-conformal EoS leads to a (possibly large) {\it positive} (i.e. unphysical) initial value for the bulk viscous pressure whose subsequent decay can have counter-intuitive effects on the subsequent hydrodynamic flow and its dependence on $\taufs$. Recent studies demonstrate that this problem persists when the free-streaming module is replaced by a thermalizing but conformal effective kinetic theory, and that the magnitude of the mismatch depends on centrality \cite{NunesdaSilva:2020bfs}. Although we have not been able to fully dissect the mechanisms leading to positive preferred values for $\alpha$ in our analysis, we strongly suspect that these issues play a role. 

Turning to the later stages of the collision, we now look at the posterior for the switching temperature in \fig{initial_state_posterior}. Its marginalized posterior turns out to be quite different for the two particlization models. We find that for the selected experimental observables the effects of increasing the magnitude of the bulk viscous pressure or increasing $\Tsw{}$ are qualitatively similar. We verified that if we hold all other parameters fixed while increasing the switching temperature from 135 MeV to 165 MeV, the mean transverse momenta of pions and protons is reduced and the number of protons is increased. On the other hand, holding $\Tsw{}$ fixed and increasing $(\zeta/s)_{\text{max}}$ has the same effect. Because the Grad model prefers a large specific bulk viscosity near switching, it also prefers a lower switching temperature. 

Although some of the parameters which define the \trento{} model are well constrained, their interpretation is not always straightforward. To what extent $p\approx0$ in the \trento{} model provides support for saturation physics or is mostly a consequence of energy-momentum conservation combined with approximate boost invariance at high collision energies deserves further study. Similarly, we found our posterior for the energy density dependence of the free-streaming time difficult to understand. Further theoretical understanding of these issues is likely to find its way into improved models. For example, there is considerable room for improving the description of the pre-hydrodynamic evolution stage. It is encouraging that, using the likely values for the \trento{} model parameters, we are able to describe our experimental observables with good accuracy. Still, there is obvious value in seeking models which can not only fit the experimental data, but at the same time offer a coherently and consistently interpretable physical picture.

\subsection{Second-order transport coefficients: shear relaxation time}\label{sec:uncerntainty:taupi}

As discussed in \Section{sec:hydro}, second-order transport coefficients are treated differently in this work than first-order ones: the shear and bulk viscosities are parametrized, while the second-order transport coefficients are related to the first-order ones through relations derived in kinetic theory. The one exception is the shear relaxation time $\tau_\pi$, whose normalization is allowed to vary in a wide range. This allows for a precise quantification of the effect of a second-order transport coefficient on phenomenological constraints on $\eta/s$ and $\zeta/s$.

\begin{figure}[tb] 
\centering
\includegraphics[trim=0 0 0 17, clip, width=8cm]{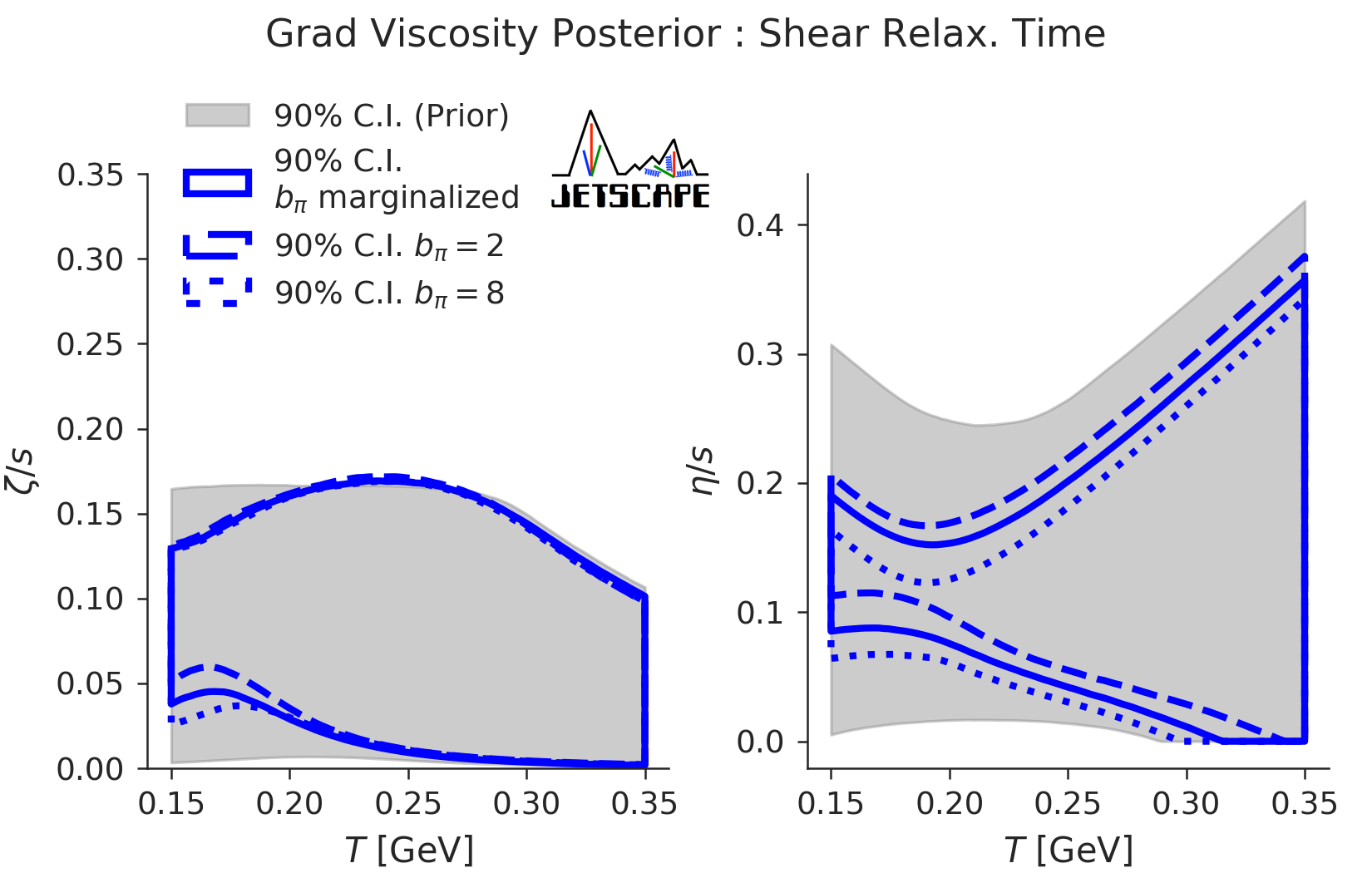}
\caption{The posterior of specific bulk(left) and shear(right) viscosities, depending on whether one marginalizes over the shear relaxation time factor $b_{\pi}$ (solid blue) or fixes it (dashed or dotted blue). The shear relaxation time and magnitude of $\eta/s$ are seen to be inversely related when fitting the LHC and RHIC data. }
\label{visc_post_shear_relax}
\end{figure}

Figure~\ref{visc_post_shear_relax} examines the extent to which the shear relaxation time normalization factor $b_{\pi}$ affects the posterior of first-order transport coefficients. 
%
%
When trying to fit the experimental data, we see that in general smaller values of the shear relaxation time lead to larger shear viscosities, and vice versa. This is because increasing either $b_{\pi}$ or $\eta/s$ tends to reduce the harmonic flows, for instance. 

Some sensitivity to the shear relaxation time factor $b_{\pi}$ may be caused by the use of free-streaming as a pre-hydrodynamic model. Indeed, free-streaming can generate large values of $\pi^{\mu\nu}$. The subsequent size of $\pi^{\mu\nu}$ in hydrodynamics, following free-streaming, is governed by the overall size of $\tau_{\pi}$ which controls the relaxation towards the Navier-Stokes limit $\pi^{\mu\nu}_{\rm NS} = 2 \eta \sigma^{\mu\nu}$. Previous viscous hydrodynamics studies, using different initial conditions (not including free-streaming), found a smaller sensitivity to $b_\pi$~\cite{Luzum:2008cw, Song:2009gc}. Note however that these studies did not include higher harmonic flows ($v_3$, $v_4$, \ldots) which we find to have a stronger sensitivity to $b_{\pi}$ than $v_2$. This will be discussed in \Section{sec:model_sensitivity}.

Since there is significant uncertainty induced by the shear-relaxation time on $\eta/s$ (and to a lesser extent on $\zeta/s$), future efforts should consider varying other second-order transport coefficients. Among those, the most important is likely the bulk relaxation time.
Performing a systematic analysis of all second-order transport coefficients would be a significant future undertaking, in part because their parametric dependence must be specified, and a prior needs to be fixed before performing parameter estimation.\footnote{%
    Existing microscopic calculations such as Ref.~\cite{Denicol:2014vaa} can help constrain the parametric dependence and the priors.  For example, Ref.~\cite{Denicol:2014vaa} finds $\delta_{\pi \pi} = \frac{4}{3} \tau_{\pi} + \mathcal{O}((m/T)^2)$. One could assume $\delta_{\pi \pi} \propto \tau_{\pi}$ with a parameter being a proportionality factor of order 1.}
Non-conformal hydrodynamic has a large number of second-order transport coefficients \cite{Denicol:2012cn}; relatively little is known for many of them. There is value in simultaneously studying these transport coefficients theoretically and phenomenologically. Even if these coefficients cannot be constrained from measurements, their influence on other model parameters should be studied.

\section{Sensitivity to prior knowledge and assumptions}
\label{sec:priorsensitivity}

Each model parameter in this study is assumed to have a uniform prior probability density over a finite range. This range represents crucial prior knowledge or assumptions. Our wider, less subjective prior (Table \ref{restricted_prior_table}, middle column) almost completely\footnote{%
    We do not include a curvature parameter for the shear viscosity at high temperatures as was done in Ref.~\cite{Bernhard:2019bmu} since there it was found, within the given prior limits, to be rather poorly constrained.} 
encloses as a subspace the narrower, more subjective prior range postulated in Ref.~\cite{Bernhard:2019bmu} (Table \ref{restricted_prior_table}, right column).

\begin{table}
\begin{tabular}{|l|l|l|}
\hline
parameter & full prior range & restricted prior \\   & & range or value \\ \hline
$\alpha$ & $[-0.3, 0.3]$ & $0.0$       \\ \hline
$T_{\eta}$\ [GeV] & $[0.13, 0.3]$ &$0.154$        \\ \hline
$a_{\mathrm{low}}$ [GeV$^{-1}$] & $[-2, 1]$ & $0.0$  \\ \hline
$\lambda_\zeta$ & $[-0.8, 0.8]$ & $0$       \\ \hline
$b_{\pi}$ & $[2, 8]$ & $5$      \\ \hline
$p$ & $[-0.7, 0.7]$ & $[-0.5, 0.5]$     \\ \hline
$w$ [fm] & $[0.5, 1.5]$ & $[0.5, 1.0]$     \\ \hline
$\tau_R$ [fm/$c$] & $[0.3, 2]$ & $[0.3, 1.5]$      \\ \hline
$(\zeta/s)_{\mathrm{max}}$ & $[0, 0.25]$ & $[0.01, 0.1]$       \\ \hline
$T_{\zeta}$ [GeV] & $[0.12, 0.3]$ & $[0.15, 0.2]$       \\ \hline
$w_{\zeta}$ [GeV] & $[0.025, 0.15]$ & $[0.025, 0.1]$ \\ \hline
\end{tabular}
\caption{
Table of full (left) and restricted (right) parameter ranges. The restricted prior is similar to the prior employed in Ref.~\cite{Bernhard:2019bmu}.}
\label{restricted_prior_table}
\end{table}

\begin{figure}[!htb]
  \centering
   \includegraphics[trim=0 0 0 17, clip, width=8cm]{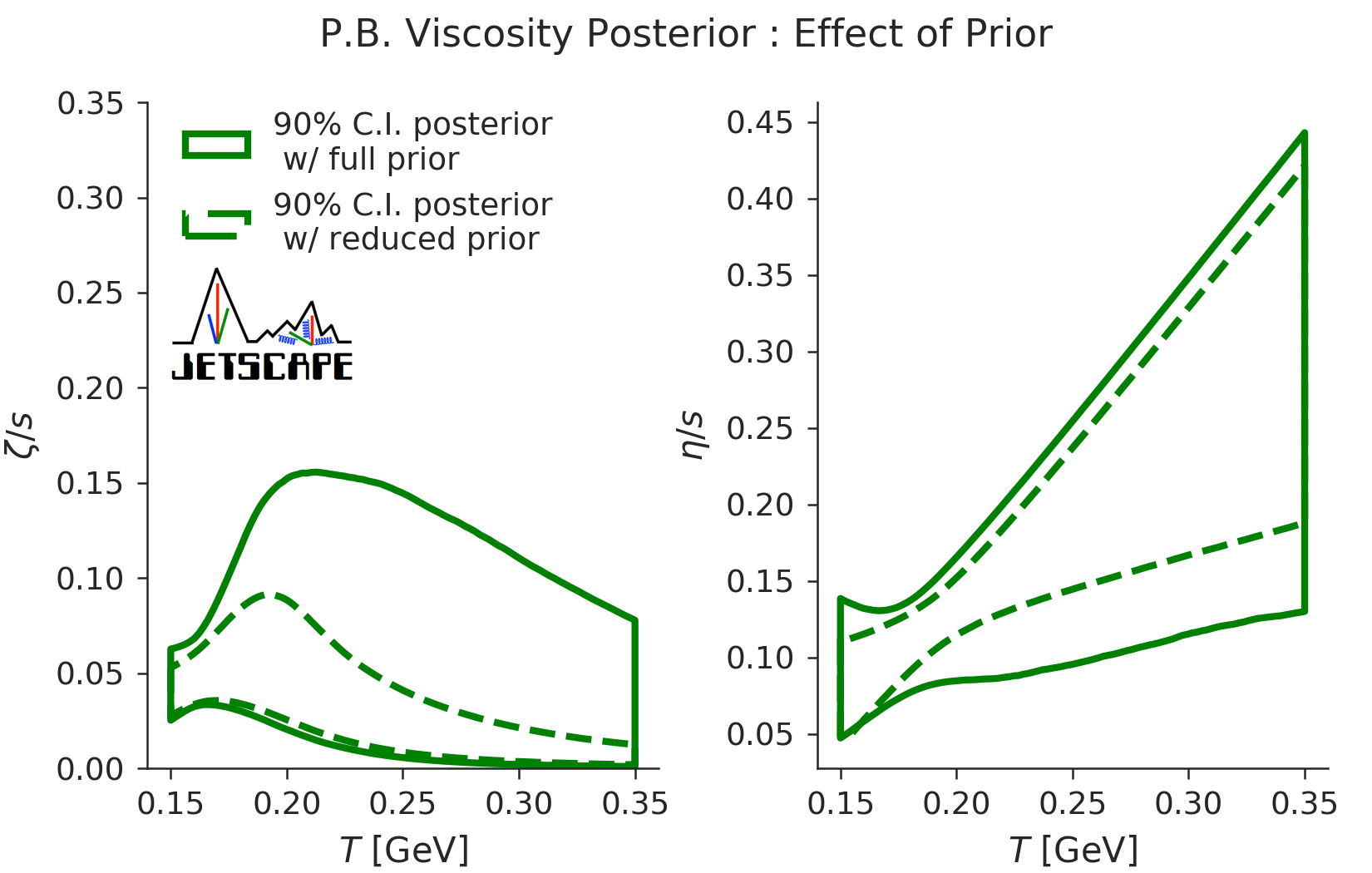}
    \caption{
    The 90\% posterior credible intervals of the specific bulk (left) and shear (right) viscosities for the Pratt-Torrieri-Bernhard viscous correction model, including only observables from LHC Pb-Pb collisions at $\sqrts=2.76$\,TeV, depending on whether one uses a more informed or less informed prior.}
    \label{fig:visc_post_red_prior}
\end{figure}

By comparing the posteriors for these different prior parameter ranges we assess the sensitivity of our inference to prior knowledge. This is illustrated in Fig.~\ref{fig:visc_post_red_prior} for one of the particlization models studied in this work (the Pratt-Torrieri-Bernhard model \cite{Pratt:2010jt, Bernhard:2018hnz}). We compare the posteriors for the specific shear and bulk viscosities using either the more or less subjective priors described above. These posteriors were obtained via Bayesian parameter estimation using only the ALICE Pb-Pb measurements at $\sqrts = 2.76$ TeV. 

%

Clearly the more subjective prior drastically reduces the width of the credible intervals of the posterior for both the shear and bulk viscosities. Table \ref{restricted_prior_table} shows that, in addition to narrower prior ranges for the shear and bulk viscosities, the more restrictive prior assumed additional information about the initial conditions, and the shear relaxation time (which is a second-order transport coefficient). 

Since the posterior of any Bayesian inference is proportional to the product of the prior and likelihood function, a tightening of the prior automatically also causes the posterior to tighten. Insofar the results shown in Fig.~\ref{fig:visc_post_red_prior} are in principle expected. However, the observed large sensitivity of the posterior (in particular for the bulk viscosity) to the prior suggests that the constraining power of the experimental data is still limited, and that for a fully uninformed prior the 90\% confidence intervals for the specific viscosities would be even wider than what is indicated by the solid lines in Fig.~\ref{fig:visc_post_red_prior}. Future improvements of the precision of our knowledge of the QGP viscosities require progress along at least one of the following two directions: (i) theoretical work leading to more objective priors, if not for the parameters of primary interest (i.e. the viscosities) then at least for the ``nuisance parameters"; (ii) inclusion of additional  measurements into the Bayesian analysis that have the potential to provide tighter likelihood functions for both the parameters of primary interest and the ``nuisance parameters". In the absence of theoretical progress towards tighter first-principles constraints on the viscosities, less subjective priors for these parameters of primary interest should be employed, to minimize sensitivity of the posterior to prior knowledge.  

\section{Model Sensitivity}
\label{sec:model_sensitivity}

\begin{figure*}[tbp]
\noindent\makebox[\textwidth]{%
  \centering
  \begin{minipage}{0.5\textwidth}
    \includegraphics[trim=0 0 0 18, clip, width=9cm]{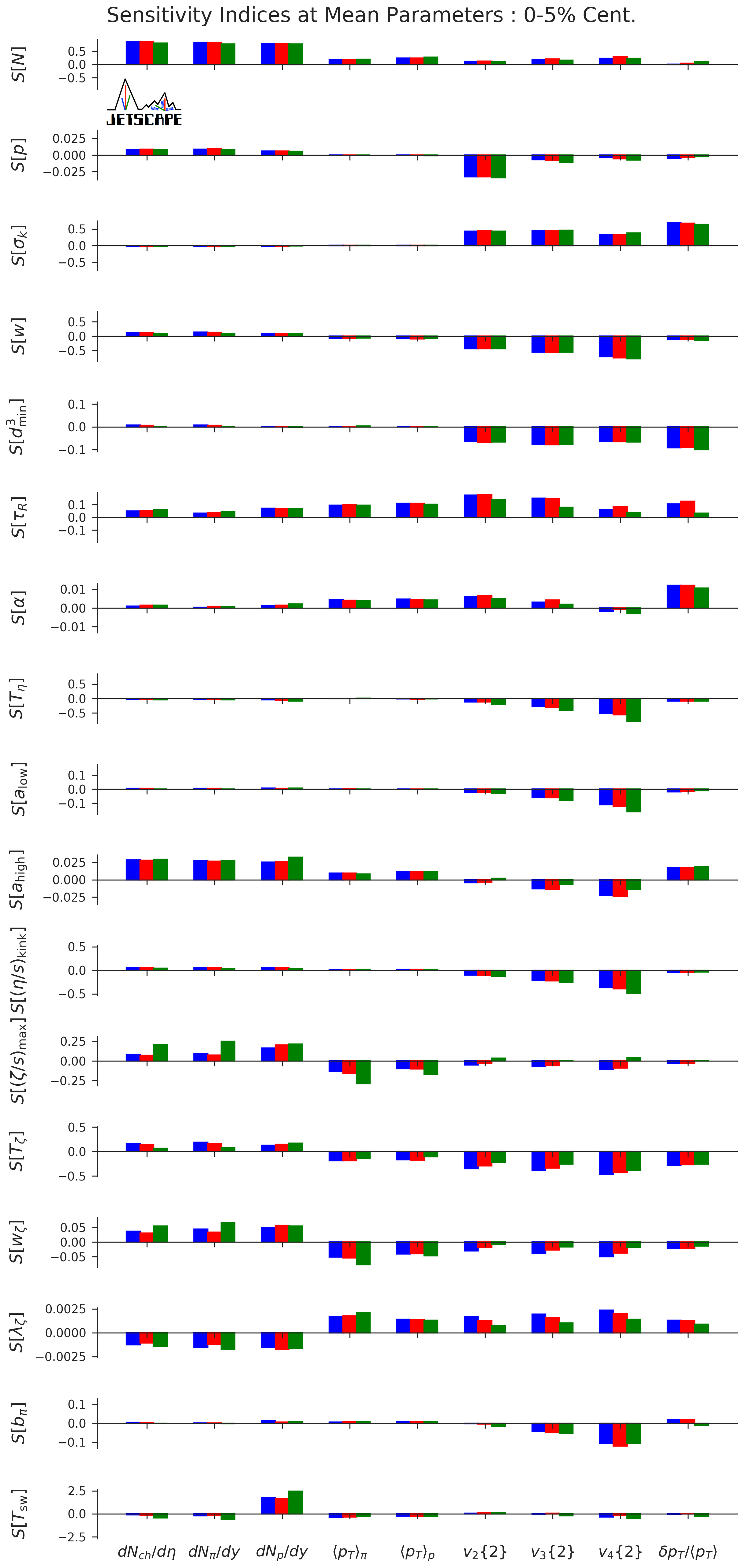}
  \end{minipage}
  \begin{minipage}{0.5\textwidth}
    \includegraphics[trim=0 0 0 18, clip,width=9cm]{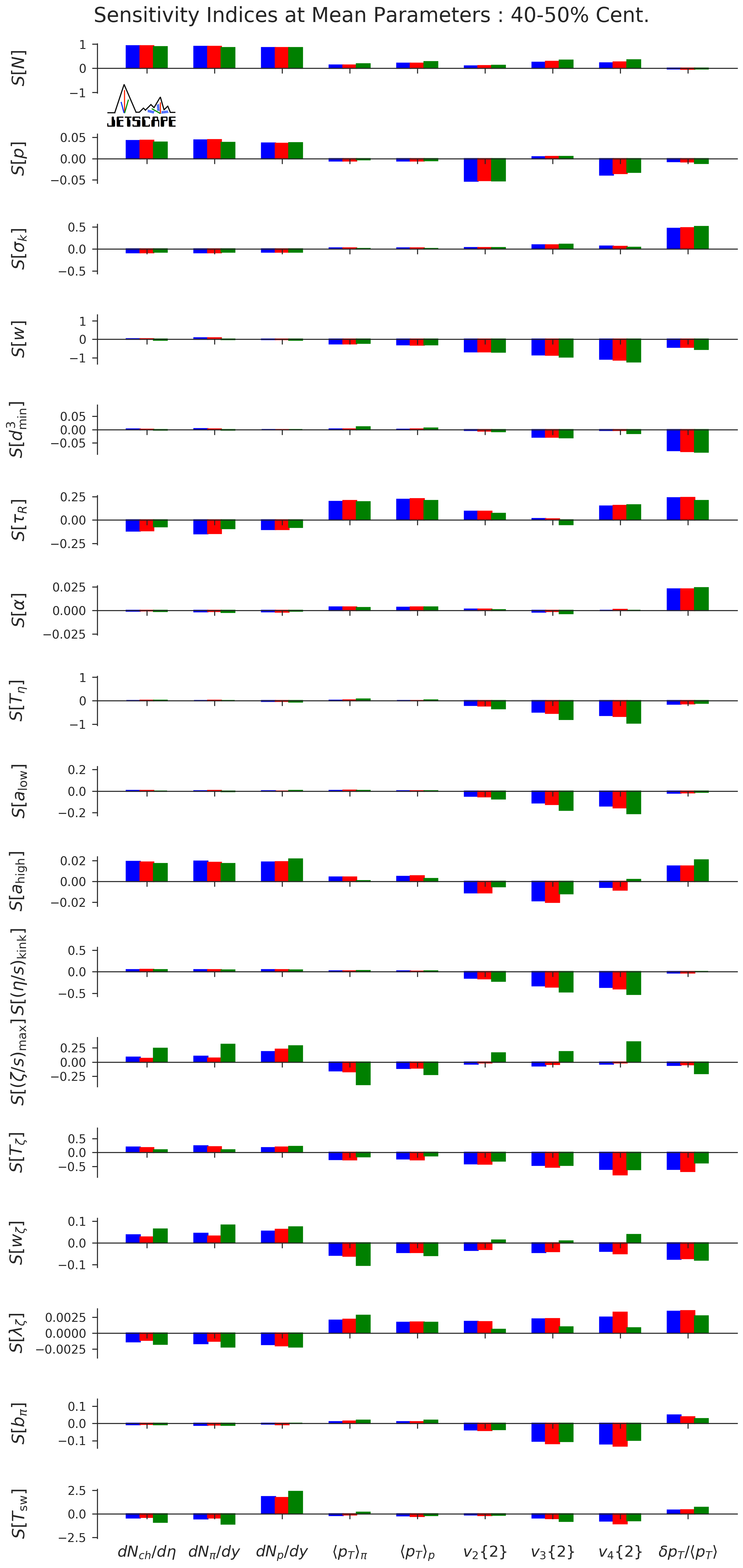}
  \end{minipage}
  }
  \caption{Sensitivity indices for LHC observables measured in the $0-5$\% (left) and $40-50$\% (right) centrality bin (except for the mean $p_T$ event fluctuation $\delta p_T / p_T$ for which the $40-45$\% bin is plotted on the right), as a function of all model parameters. Plotted in blue is the Grad viscous correction model, in red the Chapman-Enskog model, and in green the Pratt-Torrieri-Bernhard model. The bars show the sensitivity to a 10\% change in each parameter ($\delta = 0.1$). }
  \label{obs_sensitivity}
\end{figure*}

To understand the posterior obtained by performing parameter estimation, it is useful to quantify which observables carry information about which model parameters. This can be achieved as follows: 
\begin{itemize}
    \item Select a set of parameter values at which the model sensitivity is explored
    \item Vary a single parameter at a time and quantify how much each observable responds
\end{itemize}
Note that the model is non-linear, and consequently this is a \emph{local} measure of observables sensitivity, at a given point in the multidimensional parameter space.

Following Ref.~\cite{Hamby} we define a local sensitivity index as follows: define two points in parameter space by $\mathbf{x} = (x_1, x_2, ...,x_j, ..., x_p)$ and $\mathbf{x'} = (x_1, x_2, ..., (1 + \delta)x_j , ..., x_p)$ where $\delta$ is a fixed percent difference. We use our emulator to predict all of the observables at these two points in parameter space. Suppose for some particular observable $O$, the emulator predicts $\hat{O} = \hat{O}(\mathbf{x})$. Then, defining the percent difference in the observable by 
\begin{equation}
    \Delta \equiv \frac{ \hat{O}(\mathbf{x'}) - \hat{O}(\mathbf{x}) }{\hat{O}(\mathbf{x})},
\end{equation}
our ``sensitivity index'' $S[x_j]$ for observable $O$ under a change in parameter $x_j$ is given by 
\begin{equation}
    S[x_j] \equiv \Delta / \delta.
\end{equation}

We chose $\mathbf{x}$ to be defined as the average of the three different Maximum A Posteriori (MAP) parameters (see \Section{section:bayes:MAP}) of each three viscous correction models, listed in \Table{table_MAP_grad}.

These sensitivity indices $S[x_j]$ for pairs of observables and parameters are shown in \fig{obs_sensitivity} for select Pb-Pb observables at $\sqrts=2.76$ TeV and a step size $\delta = 0.1$. We verified that we obtain quantitatively similar results with a larger parameter step size $\delta = 0.4$, indicating that the parameter dependence of the model is reasonably close to linear in the region of parameter space studied. Note that the inclusion of emulator uncertainties in the sensitivity analysis is left as future work.

Although a local measure of the response of the model observables to changes in parameters is a strong approximation, it can nonetheless help guide our understanding regarding which observables carry information about each of the parameters. First note that the scale of the sensitivity indices is different for each parameter. Changing the shear relaxation time normalization $b_\pi$ has a very small effect on all observables investigated in this work, with a $10$\% change in $b_\pi$ leading to less than $1$\% change in observables.

On the other hand, very strong dependence on the model parameters have been found. The proton yield shows strong sensitivity to the switching temperature. Increasing the switching temperature by $10\%$ increases the proton yield by about $20\%$. Consequently, most of the constraining power (or information) about the switching temperature is carried by the proton yield among the observables used herein. 

As noted throughout this study, many of the observables show stronger sensitivity to the maximum of the specific bulk viscosity when the Pratt-Torrieri-Benhard distribution was employed, compared to any other viscous corrections used here. Looking again at \fig{compare_visc_posteriors}, the strong sensitivity of the Pratt-Torierri-Bernhard viscous correction causes the $90$\% posterior credible interval for the specific bulk viscosity to be most tightly constrained among the viscous correction models explored here. The narrower posterior of $\zeta/s$ for this viscous correction model is a direct consequence of these larger sensitivity indices.

The parameter $w$ in \trento{} is largely responsible for controlling the eccentricities of the initial state. We find that the elliptic, triangular and quadrangular flows $v_{2}\{2\}$, $v_{3}\{2\}$, and $v_{4}\{2\}$ show strongest sensitivity among the observables plotted in \fig{obs_sensitivity}. This may be expected from hydrodynamic response, in which $v_{2}\{2\} \propto \epsilon_2$, $v_{3}\{2\} \propto \epsilon_3$ and $v_{4}\{2\} \propto v_{2}\{2\}^2$. In addition, the initial geometry is more sensitive to the nucleon width $w$ for peripheral collisions: we see that the harmonic flows for 40--50\% centrality bins show close to twice the sensitivity to the width parameter than
for 0--5\% centrality. 

As discussed in \Section{sec:model_sensitivity}, the triangular flow $v_3\{2\}$ and quadrangular flow  $v_4\{2\}$ show strongest  sensitivity to $b_{\pi}$ in our model. This sensitivity remains small however: a $10$\% change in the shear relaxation time leads to a $1$\% change in $v_4\{2\}$ for example. This explains the challenge of constraining the shear relaxation time.

We highlight that in addition to guiding our understanding of the model, sensitivity measures can also direct where future experimental efforts can be focused \cite{Sangaline:2015isa} in order to best constrain model parameters, which is of interest to the entire community. We hope to expand our efforts in this direction in future studies with global (rather than local) sensitivity measures. 

\section{Bayesian Model Selection}
\label{sec:model_selection}

In addition to the estimation of parameters of a given model, Bayesian inference can also be used to quantify which model, among several competing models, is better supported by the experimental observations. There exist several metrics for comparison; we choose to employ the Bayes factor. We will first illustrate the application of the Bayes factor toward three of the viscous correction models for particlization that were used throughout this work. We then use it to compare our hydrodynamic model with simpler models which are `nested' inside. Finally, the Bayes factor is applied towards answering whether a consistent model, with the same set of parameters describing the system created in RHIC Au-Au and LHC Pb-Pb collisions, or more complicated models where some parameters are allowed to differ, are better justified in light of the experimental data.   

In the context of model selection, a `model' refers to a specific set of parameters together with their prior, and a unique map from the parameters to a set of observables. As an example, a polynomial of second degree is a different model than a third-degree polynomial. In this case, however, the second degree polynomial model is `nested' inside the polynomial of third degree; when the coefficient of the cubic term is fixed to zero, we can recover the second-order polynomial model. The polynomial of third degree has additional model complexity, given by the additional parameter and its prior. On the other hand, we can also compare a model given by a second degree polynomial with for example a model given by a sinusoid, with an uncertain amplitude and phase velocity. These models give altogether different predictions, and their coefficients have different meanings. In both cases above, whether the models share similar features or not, we will refer to them as different models when comparing them with the Bayes factor (which is the standard terminology). If one of the models happens to be nested inside the other, we will make note. 

It may be useful to provide a word of caution at this point, as the use of the Bayes factor for comparing models has been met with some skepticism in the past; this skepticism is not necessarily undue. It is well known that the Bayes factor becomes ill-defined when one or both models have an `improper' (non-normalizable) prior. In addition, the marginal evidence of each model in principle depends on the priors that were chosen. Neither of these potential issues is a roadblock for our current purposes, however. The priors for all model parameters considered are not improper, and have been purposely selected, weighing the applicability of each theoretical model along with reasonable information stemming from other sources of experimental data. A more thorough elucidation of the Bayesian methodology, parameter estimation, model selection, and examples in cosmology can be found in ~\cite{Trotta:2008qt}. 

\subsection{Overview and estimation of Bayes factors}
\label{sec:model_selection_overview}
\subsubsection{Bayes factor definition and interpretation}
\label{sec9a1}
The Bayes factor is a useful measure for evaluating the relative merit of two competing models $A$ and $B$ in light of a given set of experimental data $\mathbf{y}_{\exp}$. It is the ratio of the conditional probabilities, or the `odds',
\begin{equation}
    B_{A/B} \equiv \frac{ \mathcal{P}( A | \mathbf{y}_{\exp} ) }{ \mathcal{P}( B | \mathbf{y}_{\exp} ) } = \frac{ \mathcal{P}( \mathbf{y}_{\exp} | A )  }{  \mathcal{P}( \mathbf{y}_{\exp} | B )   } \frac{ \mathcal{P}(A) }{ \mathcal{P}(B)  } ,
\end{equation}
where Bayes' theorem was applied to both the numerator and denominator. The odds depend on both the ratio of the likelihoods, marginalized over all parameters, multiplied by the ratio of prior beliefs about the validity of these two models. Unless there is a strong reason to believe that one model is much more likely than another, the ratio of priors is taken to be unity. In this case, the odds are simply the ratio of the Bayesian evidences of the two models:
\begin{equation}
  B_{A/B} = \frac{ \mathcal{P}( \mathbf{y}_{\exp} | A )  }{  \mathcal{P}( \mathbf{y}_{\exp} | B )   } .
\end{equation}

Using the marginalization and product rules for probabilities we `integrate in' the model parameters $\mathbf{x}_A$ for model $A$, 
\begin{equation}
\label{marginal_likelihood}
    \mathcal{P}( \mathbf{y}_{\exp} | A ) = \int d\mathbf{x}_A \mathcal{P}( \mathbf{y}_{\exp} | \mathbf{x}_A, A ) \mathcal{P}(\mathbf{x}_A)
\end{equation}
and likewise for model $B$. The integrand appearing in  \eq{marginal_likelihood} is the product of the same likelihood and prior which have been discussed in section \ref{overview_param_est}. Therefore, the Bayesian evidence is simply the average of the likelihood with respect to the prior probability density. 
We note that when using priors which are uniform distributions on a finite domain, this expression can be simplified, yielding
\begin{equation}
\label{eqn:bayes_evidence_vol}
    \mathcal{P}( \mathbf{y}_{\exp} | A ) = \frac{1}{\mathcal{V}_A} \int_{\mathcal{D}_A} d\mathbf{x}_A \mathcal{P}( \mathbf{y}_{\exp} | \mathbf{x}_A, A ),
\end{equation}
where we have defined the total volume of the prior $\mathcal{V}_A$ for model $A$. This is the volume of the hypercube $\mathcal{D}_A$ inside which the prior for model $A$ is nonzero. 

Because all of the models for which we perform model comparisons have uniform priors, the interpretations of our results are more straightforward. The Bayesian evidence of each model is the integral over the likelihood for the model inside the prior bounds, divided by the volume of the prior. Belief in a model is increased by ability to fit the data (larger likelihood), averaged inside of the prior bounds. But belief in the model is decreased by its `complexity' (the `Occam penalty'), which is the volume of its prior which is excluded by the data. In a situation where the likelihood $\mathcal{P}( \mathbf{y}_{\exp} | \mathbf{x}_A, A )$ does not actually depend on a particular model parameter $x_{A, i}$, we see from \eq{eqn:bayes_evidence_vol} that there is no Occam penalty because the size of the prior region cancels in the numerator and denominator. Therefore, the Bayes factor does not penalize a model for having a parameter which is unconstrained by the data. For a more thorough explanation with examples, see Ref.~\cite{Trotta:2008qt}. 

\subsubsection{Numerical methods for estimating the Bayes evidence}
\label{sec9a2}

The integral over all the model parameters is very high-dimensional and does not lend itself to elementary methods. Fortunately, there exist methods for estimating the evidence that are ready to use in the existing Markov Chain Monte Carlo implementation~\cite{ptemcee_code}  used throughout this work. A `parallel-tempered' Markov Chain Monte Carlo routine defines a ladder of inverse `temperatures' $\beta_i$, and then evolves an ensemble of walkers by sampling from distributions defined by 
\begin{equation}
    \left[\mathcal{P}( \mathbf{y}_{\exp} | \mathbf{x}_A, A ) \right]^{\beta_i}  \mathcal{P}(\mathbf{x}_A).
\end{equation}

We see that in the limit $\beta \rightarrow 0$, we recover our prior $\mathcal{P}(\mathbf{x}_A)$. At regular intervals each walker inside of each tempered distribution has the opportunity to swap positions with walkers at adjacent temperatures. Walkers at very high temperatures are not strongly affected by peaks in the likelihood function, while walkers at $\beta = 1$ are sampling from the target posterior. This gives this algorithm the advantage that it can efficiently sample multimodal distributions, which can be more difficult for other algorithms, including the ordinary Metropolis-Hastings, to sample accurately. Besides these advantages, the ladder of tempered distributions also gives an estimation of the Bayes evidence by the following trick. Defining the Bayesian evidence as a function of inverse temperature:
\begin{equation}
    Z(\beta) = \int d\mathbf{x}_A \left[\mathcal{P}( \mathbf{y}_{\exp} | \mathbf{x}_A, A ) \right]^{\beta}  \mathcal{P}(\mathbf{x}_A)
\end{equation}

we note that it satisfies a differential equation
\begin{eqnarray}
\nonumber
    && \frac{d \ln Z}{d\beta}  \\\nonumber
    &=&  \frac{1}{Z(\beta)} \int d\mathbf{x}_A \mathcal{P}(\mathbf{x}_A) \ln[\mathcal{P}( \mathbf{y}_{\exp} | \mathbf{x}_A, A )] [\mathcal{P}( \mathbf{y}_{\exp} | \mathbf{x}_A, A ) ]^{\beta} \\ 
     &\equiv&  \langle \ln\left[ \mathcal{P}\left( \mathbf{y}_{\exp} | \mathbf{x}_A, A \right)\right] \rangle_{\beta}.
\label{eq:thermo_int_evidence}
\end{eqnarray}

Therefore, $\ln Z(\beta = 1)$ can be estimated by integrating by quadrature the average at each temperature. The uncertainty in this estimate $\delta \ln Z$ is primarily from using a finite number of points in the quadrature (finite number of `temperatures'). 

\subsection{Comparing viscous correction models}
\label{sec:model_selection:df}

As a first illustration of Bayesian model selection, we quantify if our experimental data give evidence to prefer one viscous correction model over another. We have estimated the logarithm of the Bayes evidence $\ln Z$ as well as the integration uncertainty $\delta \ln Z$ for three of the four models using the parallel-tempering described above. Their mutual Bayes factors are shown in Table ~\ref{bayes_fac_estimates}. 

\begin{table}[!htb]
\centering
\begin{tabular}{|l|l|l|}
\hline
Model $A$ & Model $B$ & $\ln B_{A/B}$   \\ \hline
Grad & CE & $8.2 \pm 2.3$  \\ \hline
Grad & PTB & $1.4 \pm 2.5$  \\ \hline
PTB & CE & $6.8 \pm 2.4$ \\ \hline
\end{tabular}
\caption{ A table of the logarithm of the Bayes factor $\ln B_{A/B}$ for each pair of viscous correction models and its integration uncertainty for the Grad, Chapman-Enskog (CE) and Pratt-Torrieri-Bernhard (PTB) viscous correction models. }
\label{bayes_fac_estimates}
\end{table}

From the Table we see that the Grad and Pratt-Torrieri-Bernhard models have Bayesian evidences that are compatible within the numerical uncertainty. The odds that the Grad model is better than the Pratt-Torrieri-Bernhard model are about 3:1, given our $0.6 \sigma$ observation.\footnote{%
    The probability is given by the left-tailed $p$-value.}
Therefore, the $p_{T}$-integrated calibration observables cannot distinguish which of these two models is more likely. However, we have moderate evidence to conclude that both of these models work better to describe the hadronic observables studied in this work than the Chapman-Enskog model, with the Grad versus Chapman-Enskog comparison being a $3.6 \sigma$ observation (odds about 5000:1), and Pratt-Torrieri-Bernhard comparison a $2.8 \sigma$ observation (odds about 400:1). 

For the Chapman-Enskog model, we note from \fig{initial_state_posterior} that the marginal posterior of the free-streaming energy dependence $\alpha$ has a local maximum for $\alpha \lesssim -0.3$. It is possible that widening our prior to include smaller values of $\alpha$ would also increase the Bayes evidence for the Chapman-Enskog model. Unfortunately, this would require a new set of model calculations at new design points, which is beyond the scope of the present work. Due to the very large odds of the Grad model compared to the Chapman-Enskog model given by the Bayes factor, we also considered the frequentist odds defined by the maximum likelihood ratio. If $L_A$ is the maximum value of the likelihood function for model $A$, and $L_B$ the same for model $B$, this maximum likelihood ratio is simply defined by $L_A / L_B$. These maximum-likelihood odds were found to be close to 300:1 for the ratio of Grad to Chapman-Enskog models.

\begin{figure}[tb]
  \centering
    \includegraphics[width=8.cm]{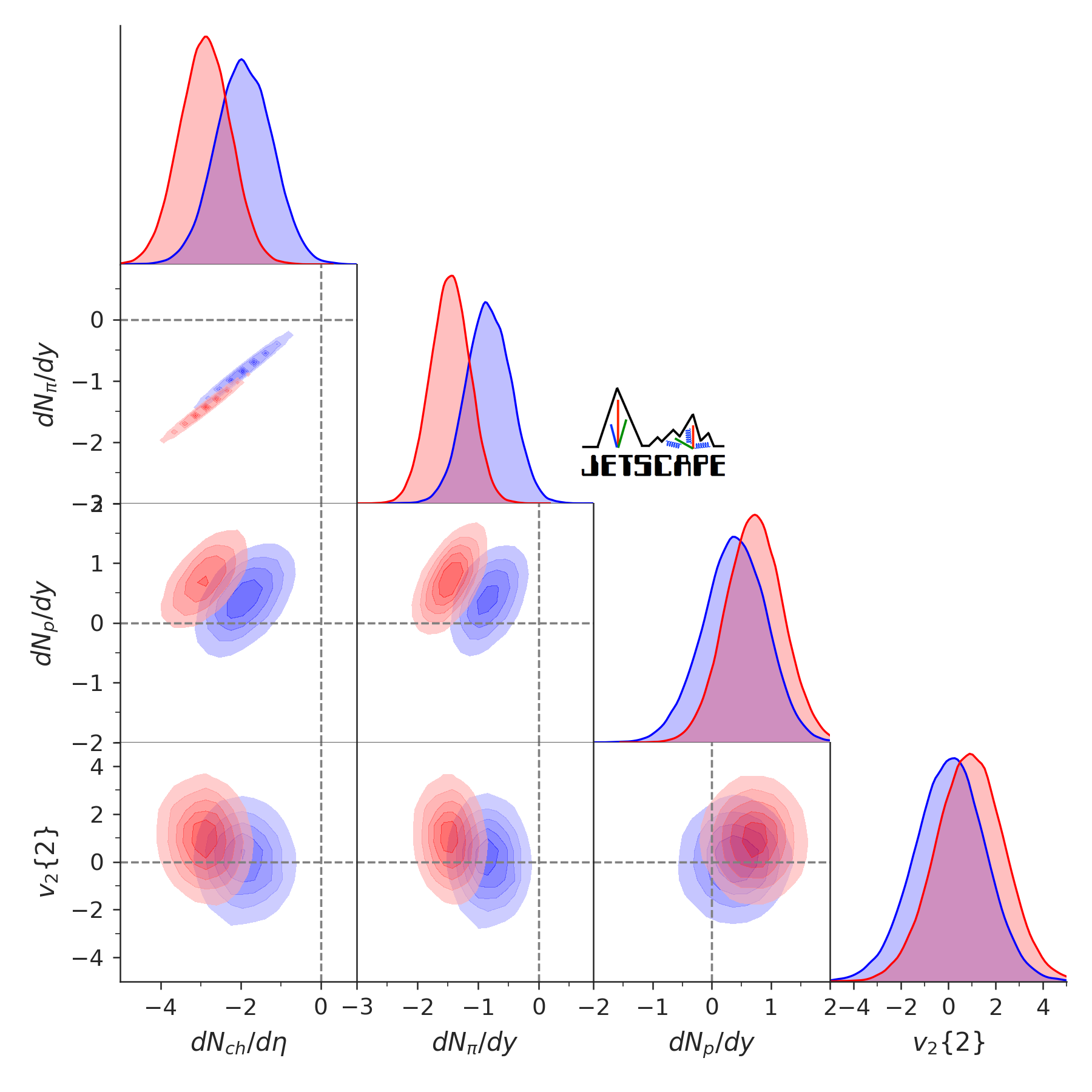}
    \caption{Diagonal and off-diagonal panels show one- and two-dimensional projections of the 
    $n$-dimensional posterior predictive distributions for selected Pb-Pb observables at fixed 
    collision centrality of 0--5\%. Plotted are the discrepancies between prediction and 
    measurements in units of the experimental standard deviation; axes are labeled with 
    shorthand notation $y \equiv (y_{\rm model}{-}y_{\rm exp}) / \sigma_{\rm exp}$ where 
    $y$ stands for the observable whose model discrepancy is shown. The Grad model is 
    shown in blue and Chapman-Enskog in red.
    }
    \label{fig:posterior_predictive_discrepancy}
\end{figure}

The Chapman-Enskog model is not able to simultaneously fit the proton multiplicity together with the other observables, such as the pion multiplicity. This puts the model under tension, and reduces the likelihood (averaged across the parameter space) of the Chapman-Enskog model. This is illustrated by ~\fig{fig:posterior_predictive_discrepancy}, which displays the single and joint posterior predictive distributions of select observables for the most central bin 0--5\% for Pb-Pb collisions at $\sqrts{} = 2.76$\,TeV. For each of the Grad model (blue) and Chapman-Enskog model (red), parameter samples are drawn from the posteriors calibrated to all observables at both LHC and RHIC. Then, the model prediction is calculated using the emulator for all observables, and plotted is the model-experiment discrepancy, i.e., the difference between model prediction and experimental mean normalized by experimental standard deviation. That the chemical abundances disfavor the Chapman-Enskog model was further strengthened by recalculating the posteriors and Bayes factor for the Grad and Chapman-Enskog models excluding the LHC proton multiplicity from the data. In this case, the odds were greatly reduced to only about 5:1 in favor of Grad. For both the linearized Grad and Chapman-Enskog models, it is the bulk viscous correction which changes the chemical abundances from their equilibrium values (the shear viscous correction does not correct the equilibrium yields). Therefore, in light of the chemical abundances being a strong discriminator, it is specifically the bulk viscous correction given by the Chapman-Enskog model which is disfavored by the particle yields. 

In conclusion, the hadronic observables studied in this work favor the Grad and Pratt-Torrieri-Bernhard models of viscous corrections over the Chapman-Enskog model. This is due in large part because the Chapman-Enskog model is worse at simultaneously fitting the chemical abundances. In light of the caveats, we do not believe this finding should be taken as a blanket statement on the validity of the Chapman-Enskog model in studying heavy ion collisions. Future studies will be necessary to clarify if viscous corrections can be systematically constrained from measurements.

Knowing the relative odds between the different particlization models, a model-averaged posterior with improved uncertainties for the inferred parameters can be derived using Bayesian Model Averaging which calculates the posterior as a weighted average of the individual model posteriors, weighted by the Bayes evidence for each model. We reported such a result in Ref.~\cite{Everett:2020xug}. 

\subsection{Comparing hydrodynamic models}
As another application of Bayesian model selection, we quantify whether simpler models, which are nested within the model described in \Section{sec:model_overview}, are favored or disfavored by the data. We will make comparisons with models with simplified assumptions for the shear viscosity. As a reminder, the more complex model will be penalized by the additional parameters (the `Occam penalty') that are constrained by the data, and will only yield a larger evidence than the simpler model if the extra constrained parameters significantly improve the fit to the data. An additional model parameter that is not well-constrained by the data within the range of the prior will have an insignificant Occam penalty.  

\subsubsection{Temperature independent specific shear viscosity}
\label{sec:model_selection:flat_shear}

We consider whether our full model, which includes a temperature-dependent specific shear viscosity, is preferred by the data to a simpler model with a temperature-independent specific shear viscosity. In both cases we use the Grad viscous correction. We denote by model $A$ our usual model with temperature dependent specific shear viscosity. We denote by $B$ the model in which the low-temperature and high-temperature slopes $a_{\rm low}$ and $a_{\rm high}$ are fixed to zero; the temperature of the kink $T_{\eta}$ is irrelevant in this scenario, and is also fixed to an arbitrary value. We find the logarithm of the Bayes factor to be consistent with zero within its uncertainty, $\ln B_{A/B} = -0.2 \pm 2.4$. Hence, the selected experimental data provide no evidence in favor of the common theoretical preference for a temperature-dependent specific shear viscosity of QCD matter. As noted above, the Occam penalty for including the additional parameters, which here are the slopes of the specific shear viscosity and position of its inflection, is minimal; this is because these parameters are not well constrained within the range of the prior. In any case, this ``null'' result suggests we should include more discriminating observables in future studies, beyond the subset we have chosen here. 

\subsubsection{Zero specific shear viscosity}

We also study if our data provide strong evidence for a non-zero specific shear viscosity. This can be quantified in the same way as above, setting the parameters for the specific shear viscosity such that $(\eta/s)(T)\equiv0$. We again use the Grad viscous correction model for this comparison and allow the specific bulk viscosity (as well as all other parameters) within their full prior ranges. We find the logarithm of the Bayes factor $\ln B_{A/B} = 11.7 \pm 2.6$ where model $A$ is the default model with nonzero specific shear viscosity while model $B$ has $\eta/s{\,\equiv\,}0$. We conclude that the data provide strong evidence that the hydrodynamic models with nonzero specific shear viscosity are preferred.

\subsection{Quantifying tension between LHC and RHIC}
\label{sec:bayes_factor_tension}

The Bayes factor is also useful for quantifying if our model is under significant tension when trying to simultaneously fit the data at both collision energies \cite{Marshall_2006}. Throughout this work we have assumed that all model parameters are shared between the two systems except for their initial energy density normalizations. We can however relax these assumptions, and allow parameters to be different for the two different systems.

\subsubsection{No common parameters between collision systems}
\label{sec:model_selection:no_common_params}

Suppose that we allow all of the parameters to be different for the two systems defined by collisions at RHIC and LHC, respectively, including the initial conditions, viscosities, and switching temperature. In this case, our model has a total of 32 parameters. We want to compute the Bayes factor $\ln B_{A/B}$ where model $A$ is the default model while model $B$ assumes independent sets of model parameters for describing the data collected at different collision energies.

As usual, we take the ratio of our prior beliefs about these two models to be unity, $\mathcal{P}(A) / \mathcal{P}(B) = 1$, such that the Bayes factor reduces to the ratio of marginal evidences:
\begin{equation}
  B_{A/B} = \frac{ \mathcal{P}( \mathbf{y}_{\rm LHC}, \mathbf{y}_{\rm RHIC} | A )  }{  \mathcal{P}( \mathbf{y}_{\rm LHC}, \mathbf{y}_{\rm RHIC} | B )}.
\end{equation}
As a consequence of the assumed statistical independence of measurements performed with different detectors at different collision energies, we can estimate the model evidence in the denominator as follows:
\begin{equation}
    \mathcal{P}( \mathbf{y}_{\rm LHC}, \mathbf{y}_{\rm RHIC} | B ) = \mathcal{P}( \mathbf{y}_{\rm LHC} | B ) \mathcal{P}( \mathbf{y}_{\rm RHIC} | B ).
\end{equation}
Integrating over the model parameters for the LHC model yields
\begin{multline}
    \mathcal{P}( \mathbf{y}_{\rm LHC} | B ) 
    \\ = \int d\mathbf{x}_{\rm LHC} \mathcal{P}( \mathbf{y}_{\rm LHC} | B, \mathbf{x}_{\rm LHC}) \mathcal{P}( \mathbf{x}_{\rm LHC} | B )
\end{multline}
which we have estimated using \eq{eq:thermo_int_evidence}. A similar result holds for the model describing the RHIC data. 

In this way we find $\ln B_{A/B} = 24.1 \pm 2.6$. We conclude that our data from the LHC and RHIC give very strong evidence that a model in which all parameters except the initial energy density normalizations are the same is strongly preferred by the data over a model in which all parameters are allowed to be different. The Occam penalty for basically doubling the number of model parameters far outweighs the gain of precision in the data description. We take this as strong evidence that a hybrid viscous hydrodynamic model with a single set of parameters provides a coherent physics picture for the experimental data measured at two collision energies that differ by over an order of magnitude.

Admittedly, allowing all of the parameters to be different leads to a very drastic comparison, adding far more model complexity than perhaps reasonable, thus entailing an outsized Occam penalty. A more meaningful study might have tried to identify tensions between a few specific observations and their predictions from the calibrated model, use the sensitivity analysis to zero in on one or a small number of model parameters to which these data are most sensitive, and introduce a small number of extra parameters to introduce additional collision energy dependence in that sector, with the aim of reducing the tensions. We leave this for future work.    

\subsubsection{Different transverse length scales in the initial conditions}
\label{sec:model_selection:trento_width}
\begin{figure}[tb]
  \centering
    \includegraphics[width=8.cm]{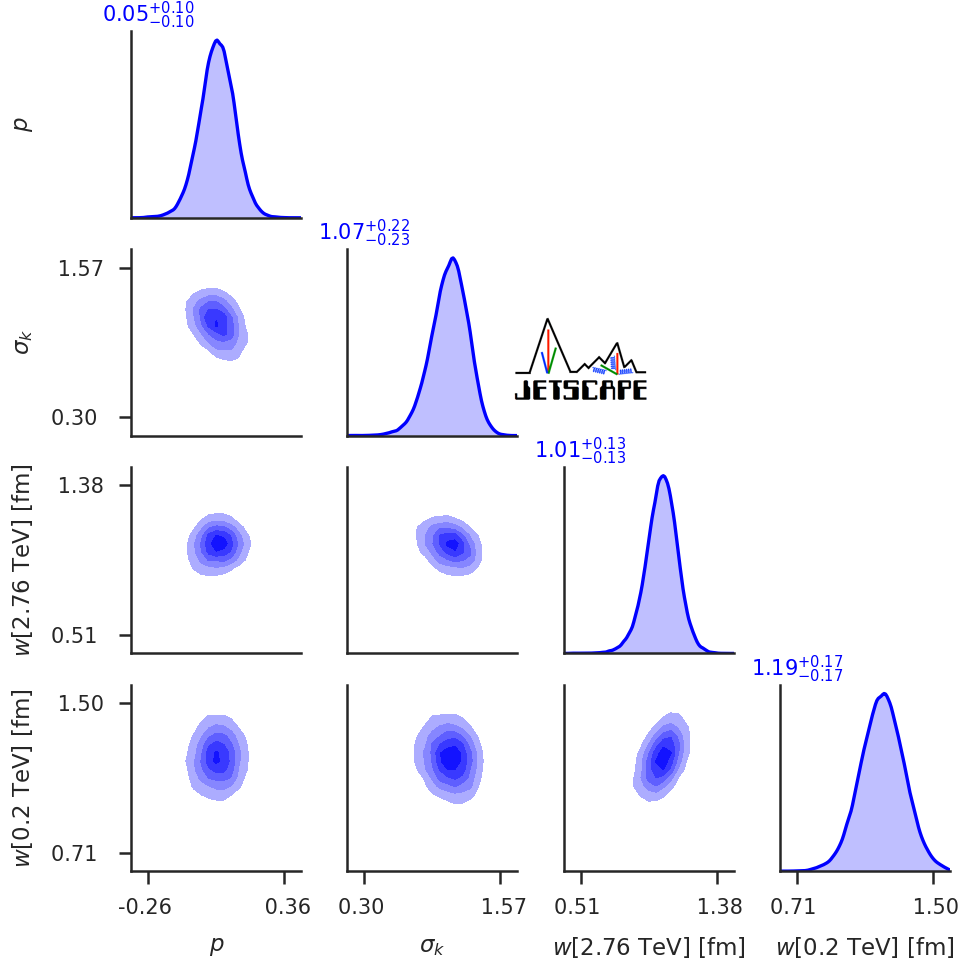}
    \caption{Partial representation of the posterior  for a model that uses the Grad particlization model and allows for different nucleon width parameters $w$ at RHIC and LHC energies. The estimated nucleon widths at the two collision energies inferred from the Bayesian analysis are found to agree within the 90\% confidence limits.}
    \label{fig:posterior_diff_nuc_width}
\end{figure}

%
We mentioned earlier that some theoretical models of the energy deposition in a heavy ion collision feature transverse length scales that depend on the collision energy. In our \trento{} model, it is the ``nucleon width'' $w$ which controls the transverse length scale for fluctuations in the initial conditions, and we have so far assumed that its value is independent of the collision energy. To test this assumption, we calculate the posterior for a model that introduces one additional parameter to allow the nucleon width $w$ to change from one collision energy to the other. The posterior for this model is shown in \fig{fig:posterior_diff_nuc_width}. 

We see that the most probable value for the nucleon width $w[0.2$\,TeV$]$ in Au-Au collisions at RHIC is about $20$\% larger than the width $w[2.76$\,TeV$]$ in Pb-Pb collisions at the LHC, though both agree within uncertainty as shown in Fig.~\ref{fig:posterior_diff_nuc_width}. We note that the Color Glass Condensate model predicts roughly a factor of two difference between the color flux tube diameters at RHIC and LHC energies \cite{Gelis:2014qga}. The measured total inelastic nucleon-nucleon cross section also increases by about a factor two from RHIC to LHC, indicating a possible growth of $w$ by a factor $\sim \sqrt{2}$. If $A$ denotes the default model and $B$ the model where the nucleon widths at the two collision energies are allowed to differ, we find $\ln B_{A/B} = 0.7 \pm 2.5$. Within the uncertainty of the estimate, we can thus not distinguish which model is preferred. The amount of tension that is caused by ignoring energy dependence of the nucleon width is not significant, and if there is indeed a small gain in the model evidence due to a slightly improved description of the data it is erased by Occam's penalty for the increase in model complexity. 

\section{Predicting $p_T$-differential observables}
\label{sec:calib_model_predictions}

\begin{figure*}[tb]
  \centering
    \includegraphics[width=16cm]{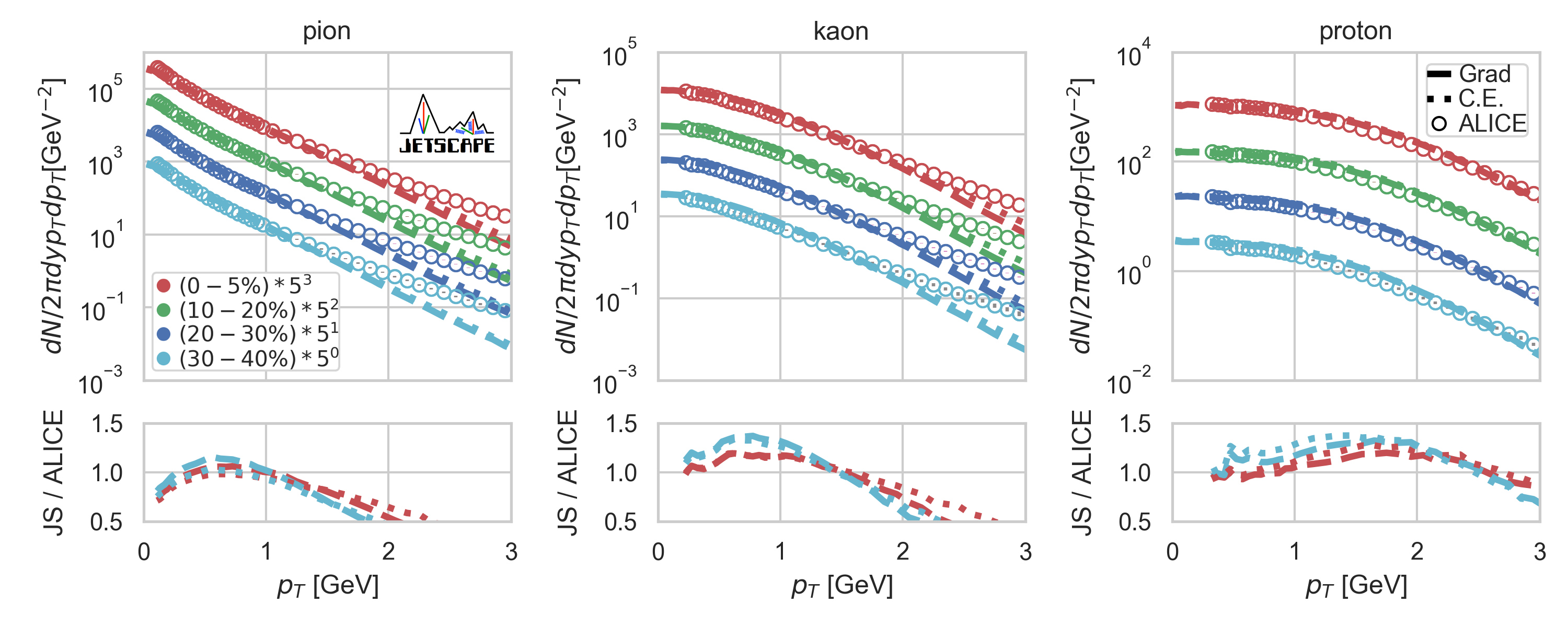}
    \caption{The transverse momentum spectra for pions (left), kaons (center) and protons (right) averaged over five thousand fluctuating events predicted by the Grad (dashed lines) and Chapman-Enskog (dotted lines) models, each run at their respective MAP parameters. Shown are the predictions for the $0-5\%$(red), $10-20\%$(green), $20-30\%$(blue), and $30-40\%$(cyan) centralities, each having been scaled by a power of five for visualization. Also shown are the measurements from ALICE (open circles). The bottom panel shows the ratio of the model prediction `JS' divided by the ALICE data. }
    \label{fig:pT_spectra}
\end{figure*}

\begin{figure}[tb]
    \includegraphics[width=6cm]{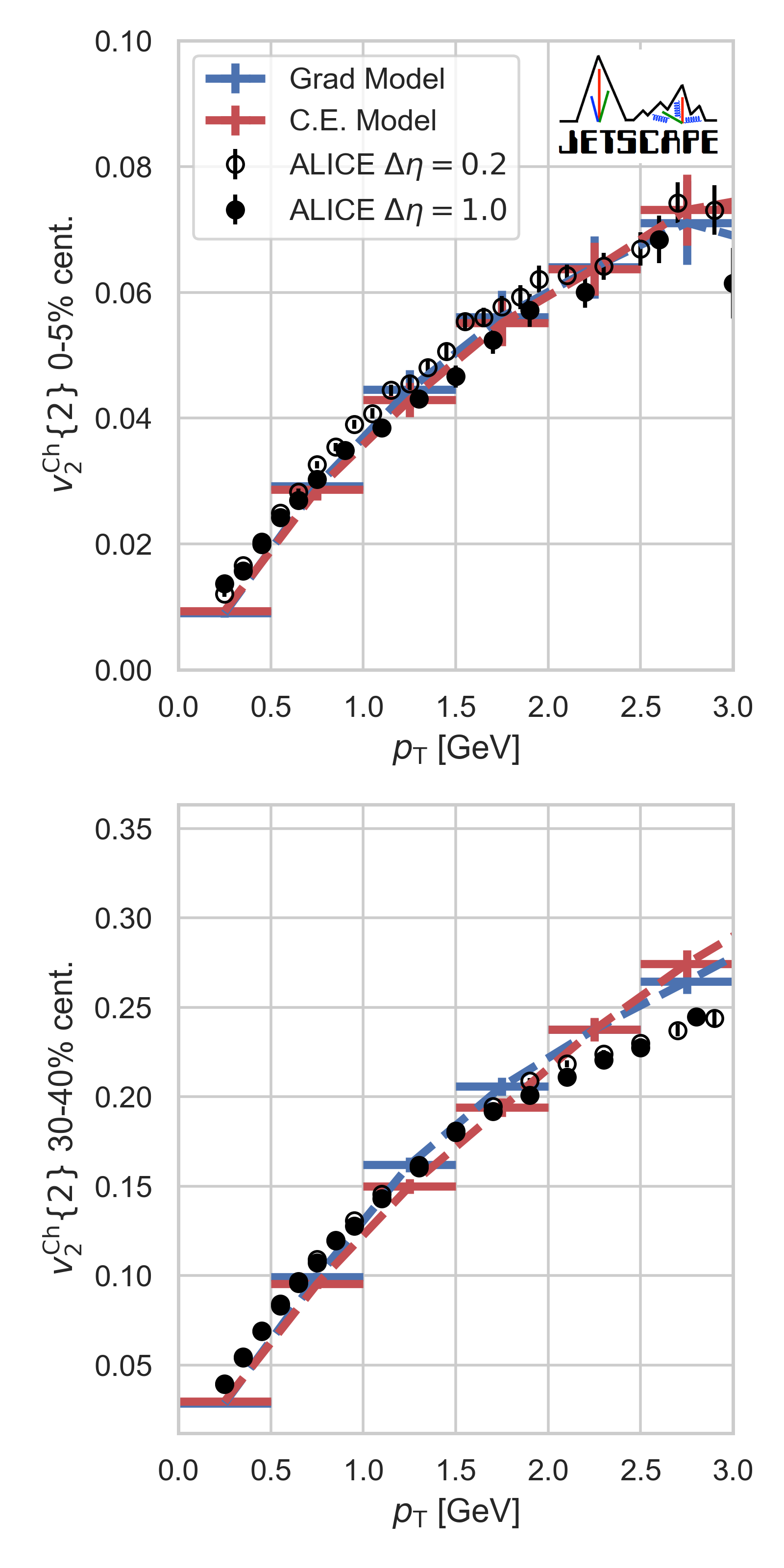}
    \caption{The $p_T$-differential two-particle cumulant elliptic flow of charged particles averaged over five thousand fluctuating events predicted by the Grad (blue) and Chapman-Enskog (red) models, run at their respective MAP parameters. Also shown are measurements from ALICE taking a pseudorapidity gap $\Delta \eta = 0.2$ (open circles) or $\Delta \eta = 1.0$ (filled circles)}
    \label{fig:pT_v2}
\end{figure}

From the Bayesian point of view, a model is more useful if it is capable of making accurate predictions for observables that were not used for its calibration. This fits the physicist's frame of mind in which belief in a model's veracity is increased when the model makes an accurate {\it pre}diction of some observable (and similarly, models that make inaccurate predictions are held in lower esteem). 

We should thus check whether our calibrated model for heavy-ion collision evolution makes accurate predictions. We consider as a prediction any observable calculated from the model using the Maximum A Posteriori (MAP) parameters (see \Section{section:bayes:MAP} and \Table{table_MAP_grad}) that has not been used for the model calibration, neither through a prior nor via the likelihood. As our model is intended to describe the physics of particles with soft momenta $p_T \lesssim 2$\,GeV, accurately predicted soft observables should increase our belief in the model, while soft observables that are inaccurately predicted will decrease it. As an example, in this section we use our model to predict the shapes of the $p_T$-differential identified hadron spectra and charged hadron elliptic flow measured by ALICE at the LHC, shown in Figs.~\ref{fig:pT_spectra} and \ref{fig:pT_v2} for the Grad and Chapman-Enskog particlization models.\footnote{%
    We remind the reader that the posterior of our model parameters was estimated using only $p_T$-integrated observables, e.g. the multiplicities and mean transverse momenta for pions, kaons and protons, the $p_T$-integrated harmonic flows, etc.} 

Because the multiplicities and mean transverse momenta are dominated by particles with typical (flow-boosted) thermal momenta, the model tends to fit the slope of the pion differential spectra better at soft momenta $p_T \lesssim 1.5$\,GeV. The stronger boost from radial flow experienced by heavier hadrons \cite{Schnedermann:1993ws, Heinz:2004qz} extends this agreement with the model to higher $p_T\lesssim 2.5$\,GeV for protons. This finding is consistent with that of Ref.~\cite{Novak:2013bqa} which showed that the shape of the pion and proton spectra could be characterized very well by the mean transverse momenta and yields.

For the differential elliptic flow, the agreement between model prediction and experiment is generally good for both the Grad and Chapman-Enskog models; neither model performs qualitatively better than the other. To what extent each of these models' predictions also agree with additional experimental results that were not used for model calibration will be further explored in future studies. We note that the Chapman-Enskog viscous correction model is not able to fit the experimental multiplicities of pions and protons as well as the Grad model, but in the $p_T$-differential elliptic flow the normalizations of these spectra drop out and only their shapes as a function of $p_T$ matter.

\section{Summary}

Building upon previous studies \cite{Petersen:2010zt, Novak:2013bqa, Sangaline:2015isa, Bernhard:2015hxa,Bernhard:2016tnd,Bernhard:2019bmu},  this work presents a comprehensive framework to perform Bayesian inference in heavy-ion collisions. A major new addition compared to these earlier studies is the use of closure tests (\Section{sec:closure_tests}), which we used to perform an extensive validation of our analysis before comparison to data. We further discussed how closure tests can be used to explore the constraining power of different observables, before measurements are even performed, thus helping prioritize which observables need better measurements.

We highlighted the important role played by the parameter ``priors'' in \Section{sec:priors}. The prior probability distribution encodes how likely we believe the model parameters to take certain values, before comparison with data. Failure to quantify the impact of the priors on the posterior can lead to incorrect or misleading conclusions about the constraining power of heavy-ion measurements. 

Using the above Bayesian framework, duly validated with closure tests, a parameter estimation was performed using RHIC and LHC data, first using measurements from these two colliders separately and then combining them (\Section{sec:post_param}). Our analysis shows that these RHIC and LHC data can set a strong constraint on the viscosity of hot nuclear matter in a temperature window between 150 and 200 MeV, which is probed by a large fraction of the space-time volume filled by the expanding medium created in the collision. However, as hinted by closure tests (\Section{sec:closure_tests}), we observed that the bulk viscosity of QCD is difficult to constrain for temperatures above $\sim$200 MeV, at least with the hadronic observables used in this work. The shear viscosity was also found to be difficult to constrain for $T\gtrsim 250$\,MeV. 

A good simultaneous agreement with measurements from RHIC and the LHC was found, both when comparing with random samples from the parameter posterior (\fig{grad_lhc_rhic_expt_fit}) and when comparing with the Maximum A Posteriori parameters (\fig{grad_lhc_rhic_MAP_observables}). Results from simulations utilizing the Maximum A Posteriori parameters furthermore agreed well with the $p_T$-differential spectra of identified hadrons (\fig{fig:pT_spectra}), and the $p_T$-differential $v_2$ of charged hadrons (\fig{fig:pT_v2}).

All results presented in this work were performed with a new comprehensive simulation framework for heavy-ion collisions, described in \Section{sec:model_overview}, which combines \trento{} initial conditions, free-streaming, second-order relativistic hydrodynamics, a flexible particlization module and the recently developed SMASH hadronic transport as afterburner. The diversity of physics ingredients entails a large set of model parameters. The significance of these parameters was investigated by studying the sensitivity of predicted values for the experimental data to changes in these parameters (\Section{sec:model_sensitivity}). The Maximum A Posteriori (MAP) values for these parameters are listed in \Table{table_MAP_grad}, for three different particlization models that employ different parametrizations of viscous corrections to the momentum distributions and provide the mapping between the energy-momentum tensor of hydrodynamics and the momentum distributions of hadrons. We emphasize that the posterior probability distributions of the parameters, such as those in \fig{compare_visc_posteriors} and \fig{initial_state_posterior}, contain much more information than just the Maximum A Posteriori. Nevertheless, for practical applications, a single set of model parameters must often be used. For such applications, we put forward the Maximum A Posteriori parameters from \Table{table_MAP_grad} as a sensible choice. 

The Maximum A Posteriori parameters from \Table{table_MAP_grad} take somewhat different values for the three different particlization models investigated herein. This is one of several examples of model uncertainty that were investigated in this work. Overall, these studies lend a considerable amount of credence to our constraints on the model parameters, such as the shear and bulk viscosities, and we expect them to provide valuable guidance for analyses that use somewhat different models of heavy-ion collisions.

This work makes clear that constraints on the viscosities of QCD still have substantial theoretical uncertainties from viscous corrections to the hadronic distributions. We showed in \fig{compare_visc_posteriors} that constraints on $\zeta/s$ and $\eta/s$ can shift significantly when different viscous corrections are used. Not all model parameters were found to be equally sensitive to these viscous corrections. For example, the initial condition parameters from the \trento{} ansatz do not generally depend heavily on the viscous corrections (\fig{initial_state_posterior}). The free-streaming time, on the other hand, was found to have a significant sensitivity (\fig{tau_fs_posterior}). This highlights the challenge of phenomenological constraints on properties of QCD: the details of how energy and momentum are distributed across species at the late particlization stage can have a significant effect on other model parameters describing much earlier evolution stages. The dependence of the free-streaming time on viscous corrections is significant. There have been associations made in the past between the free-streaming time $\taufs$ and the ``hydrodynamization time'' in heavy-ion collisions. We would like to express some  wariness about this interpretation of $\taufs$: our difficulty in constraining the centrality and center-of-mass energy dependence of the free-streaming time (\eq{eq:taufs} and \fig{tau_fs_posterior}) suggests that there are still significant model biases in our treatment of the initial stage of the collision.

The effect of the shear relaxation time on the inferred values for shear and bulk viscosity was quantified in \fig{visc_post_shear_relax}. We believe this inclusion of a second-order transport coefficient in the model calibration to be a significant advance, given that significant theoretical uncertainties remain in the treatment of second-order transport coefficients in general (see the end of \Section{sec:hydro}).

The result of our Bayesian parameter estimation is characterized by a non-trivial combination of the sensitivities of each observable to the model parameters. In \fig{obs_sensitivity}, we showed explicitly the manner in which the observables react to changes in the values of the model parameters used in this work. As this dependence is non-linear, the local model sensitivity results are not universal. Nevertheless \fig{obs_sensitivity} provides valuable intuition on the contribution of different observables to constraining model parameters, connecting with previous works on the topic~\cite{Sangaline:2015isa}.

A further important tool explored here is Bayesian model selection (\Section{sec:model_selection}). Bayes factors were used to compare the level of tension with measurements of different viscous correction models (\Section{sec:model_selection:df}). We found that chemical abundance measurements disfavor the Chapman-Enskog particlization model in comparison with the two other viscous correction models studied in this work; this begs for additional studies to more firmly assess the robustness of this conclusion.

Bayes factors reward improved model agreement with measurements and penalize increased model complexity. We used this feature in \Section{sec:model_selection:no_common_params} to verify that a joint description of both RHIC and LHC data by a single dynamical model with a common set of model parameters is favored over individual descriptions with separate sets of model parameter values. In \Section{sec:model_selection:trento_width} we used the same approach to quantify the odds in favor of, or against, a collision energy dependence of the nucleon width parameter $w$ in \trento{}, finding no statistically significant evidence against using the same value at RHIC and LHC. 

The entire model presented in this work, including the Bayesian inference tools employed for its calibration and the uncertainty quantification for the inferred model parameters will soon become publicly available as part of a new release of the JETSCAPE framework \cite{Putschke:2019yrg}. As described in \Appendix{appendix:validation}, we performed a careful and thorough validation of these new numerical implementations. An important outcome of this validation exercise was a systematic comparison of the SMASH and UrQMD afterburners. We found that (at least for the model parameters used here) both afterburners produce very similar results for the hadronic observables studied in this work (\Appendix{app:smash}). Importantly, we found that exact consistency of the particle species included in the equation of state and the hadron lists for the particlization and afterburner modules is essential for this test to succeed: an inconsistency in any of these ingredients could easily be misinterpreted as a difference in the afterburners themselves.

We believe that there is great value in the ability to easily (i.e. at little numerical cost) visualize how observables react to changes in individual or combinations of model parameters --- a piece of information provided by the emulator. We have constructed a user-friendly ``widget'' that offers this feature for some of the LHC data used in this work and make it publicly available online in Ref.~\cite{widget}. We believe this visualization tool can be of use to the heavy-ion community for better understanding the relation between the QGP viscosities and the hadronic observables measured experimentally, for example.

The present analysis builds on pioneering work published in several previous studies, in particular in Jonah Bernhard's recent PhD thesis \cite{Bernhard:2019bmu}. The results presented here differ from this latest analysis in several respects and for a number of reasons that are discussed in detail in \Appendix{appendix:comparison_nature_physics}. We consider identifying the role of the parameter prior as particularly important. The smaller range of values for $\zeta/s$ explored in Ref.~\cite{Bernhard:2019bmu} could give the impression that good constraints were possible on the bulk viscosity at high temperature; however, by allowing $\zeta/s$ to take a wider range of values in this work, we observed that $\zeta/s$ is poorly constrained at temperature above $200$\,MeV.

The work reported here presents the state of the art in heavy-ion collision modeling using model calibration via Bayesian inference. The posterior probability distributions obtained here provide the most realistic and robust constraints available on key properties of the quark-gluon plasma created in relativistic heavy-ion collisions at RHIC and LHC. While the uncertainties quoted by us are quantified to the extent possible with the tools in our hands, they are still large for several parameters of primary interest. In the following Section we offer an outlook on future work that can perhaps help to further improve this situation. 

\section{Outlook}

There are many ways to build upon the results presented here. Within the  model of heavy-ion collisions used in this work, one can proceed to include additional observables in the Bayesian analysis, to improve the still limited constraints on shear and bulk viscosity. Observables that could be especially interesting to include are the HBT radii, which are expected to provide complementary information to the development of radial flow and switching temperature \cite{Wiedemann:1997cn, Sangaline:2015isa}. The inclusion of jet observables and electromagnetic probes in the Bayesian analysis, while a major undertaking, would represent milestones with considerable potential to constrain the viscosities at higher temperature as well as the properties of the early stage in heavy-ion collision \cite{Vujanovic:2017psb, Gale:2018vuh, Hauksson:2020etn, Gale:2020xlg}. The realization of the present study within the JETSCAPE Collaboration, using the JETSCAPE Framework, is meant as a step in this direction.

Evidently, the Bayesian analysis framework presented in this work can be extended to multiple collision systems with different sizes, from the ones at intermediate size such as LHC's Xe-Xe collisions and RHIC's isobar run (Ru-Ru and Zr-Zr), to smaller and asymmetric collision systems, such as RHIC's p-Au, d-Au and He-Au collisions, LHC's p-Pb and high-multiplicity p-p collisions, and O+O collisions at both colliders. Performing a Bayesian analysis with small systems has its challenges but also provides highly valuable insights of a systematic model-to-data comparison of collision systems of varying sizes using a single model \cite{Moreland:2018gsh, Nijs:2020ors, Nijs:2020roc}.

As discussed throughout the manuscript, it is inevitable that the results of a Bayesian analysis are tied to the exact details of the model. For example, it is possible that the inferred values of the shear and bulk viscosities of QCD would change non-trivially if different models of pre-hydrodynamic physics were used. The flexibility of our \trento{}+free-streaming ansatz captures much of the uncertainty from the pre-hydrodynamic phase, and folds this uncertainty into our posteriors for the viscosities of QCD. Nevertheless we do believe that the pre-hydrodynamic stage presents one of the major remaining sources of uncertainty in constraining the viscosities of QCD from heavy ion measurements; it will be important to explore different pre-hydrodynamic models in the future.\footnote{%
    An example of Bayesian study with a different pre-hydrodynamic model that allows for breaking the assumption of conformal symmetry can be found in Refs.~\cite{Nijs:2020ors, Nijs:2020roc}. In that model the expansion velocity is parametrized and can be used to control the sign and magnitude of the bulk pressure. This may improve the description of the pre-hydrodynamic phase.} 

Another important source of uncertainty is expected to be the bulk relaxation time. The current initialization of bulk pressure has known issues, and the bulk relaxation time will have a direct impact on the propagation of this initialization uncertainty onto the final hadronic observables. While we did not find a strong dependence on the bulk relaxation time for our hadronic observables, at least for our Maximum A Posteriori parameters (\Appendix{app_bulk_relax}), we do believe it is important for this second-order transport coefficients to be explored systematically in the future.\footnote{%
    We note that Refs.\cite{Nijs:2020ors, Nijs:2020roc} include a first study of the bulk relaxation time using Bayesian inference.}

Finally, we note that this work includes no theoretical uncertainty from the equation of state. Such uncertainties may be larger than commonly expected \cite{Auvinen:2020mpc} and should be explored in the future, building on works such as Refs~\cite{Sangaline:2015isa, Moreland:2015dvc, Auvinen:2020mpc}.

These and other theoretical uncertainties, such as the mapping between hydrodynamics and the hadronic momentum distribution discussed in this work, currently limit our ability to constrain the QGP viscosities from heavy-ion collision measurements. Statistical methods for quantifying, within the framework of Bayesian inference, these modeling uncertainties are currently being developed; targeted strategies for further reducing them require additional theoretical effort. 

\section*{Acknowledgments}

We thank Jonah Bernhard, Gabriel Denicol, Scott Moreland, Scott Pratt and Derek Teaney for useful discussions, and Richard J. Furnstahl and Xilin Zhang for their insights on Bayesian Inference and Markov Chain Monte Carlo. This work was supported in part by the National Science Foundation (NSF) within the framework of the JETSCAPE collaboration, under grant numbers ACI-1550172 (Y.C. and G.R.), ACI-1550221 (R.J.F., F.G., M.K. and B.K.), ACI-1550223 (D.E., M.M., U.H., L.D., and D.L.), ACI-1550225 (S.A.B., J.C., T.D., W.F., W.K., R.W., S.M., and Y.X.), ACI-1550228 (J.M., B.J., P.J., W.K., X.-N.W.), and ACI-1550300 (S.C., L.C., A.K., A.M., C.N., A.S., J.P., L.S., C.Si., R.A.S. and G.V.). It was also supported in part by the NSF under grant numbers PHY-1516590, 1812431 and PHY-2012922 (R.J.F., B.K., F.G., M.K., and C.S.), and by the US Department of Energy, Office of Science, Office of Nuclear Physics under grant numbers \rm{DE-AC02-05CH11231} (D.O., X.-N.W.), \rm{DE-AC52-07NA27344} (A.A., R.A.S.), \rm{DE-SC0013460} (S.C., A.K., A.M., C.S. and C.Si.), \rm{DE-SC0004286} (L.D., M.M., D.E., U.H. and D.L.), \rm{DE-SC0012704} (B.S. and C.S.), \rm{DE-FG02-92ER40713} (J.P.) and \rm{DE-FG02-05ER41367} (T.D., W.K., J.-F.P., S.A.B. and Y.X.). The work was also supported in part by the National Science Foundation of China (NSFC) under grant numbers 11935007, 11861131009 and 11890714 (Y.H. and X.-N.W.), by the Natural Sciences and Engineering Research Council of Canada (C.G., M.H., S.J., C.P. and G.V.), by the Fonds de recherche du Qu\'{e}bec -- Nature et technologies (FRQNT) (G.V.), by the Office of the Vice President for Research (OVPR) at Wayne State University (Y.T.), by the S\~{a}o Paulo Research Foundation (FAPESP) under projects 2016/24029-6, 2017/05685-2 and 2018/24720-6 (M.L.), and by the University of California, Berkeley - Central China Normal University Collaboration Grant (W.K.). U.H. would like to acknowledge support by the Alexander von Humboldt Foundation through a Humboldt Research Award. Allocation of supercomputing resources (Project: PHY180035) were obtained in part through the Extreme Science and Engineering Discovery Environment (XSEDE), which is supported by National Science Foundation grant number ACI-1548562. Calculation were performed in part on Stampede2 compute nodes, generously funded by the National Science Foundation (NSF) through award ACI-1134872, within the Texas Advanced Computing Center (TACC) at the University of Texas at Austin \cite{TACC}, and in part on the Ohio Supercomputer \cite{OhioSupercomputerCenter1987} (Project PAS0254). Computations were also carried out on the Wayne State Grid funded by the Wayne State OVPR, and on the supercomputer \emph{Guillimin} from McGill University, managed by Calcul Qu\'{e}bec and Compute Canada. The operation of the supercomputer \emph{Guillimin} is funded by the Canada Foundation for Innovation (CFI), NanoQu\'{e}bec, R\'{e}seau de M\'{e}dicine G\'{e}n\'{e}tique Appliqu\'{e}e~(RMGA) and FRQ-NT.  Data storage was provided in part by the OSIRIS project supported by the National Science Foundation under grant number OAC-1541335.

\begin{appendix}
\section{Full posterior of model parameters}
\label{app:full_post}

\begin{figure*}[tb] 
\centering
\includegraphics[width=18cm]{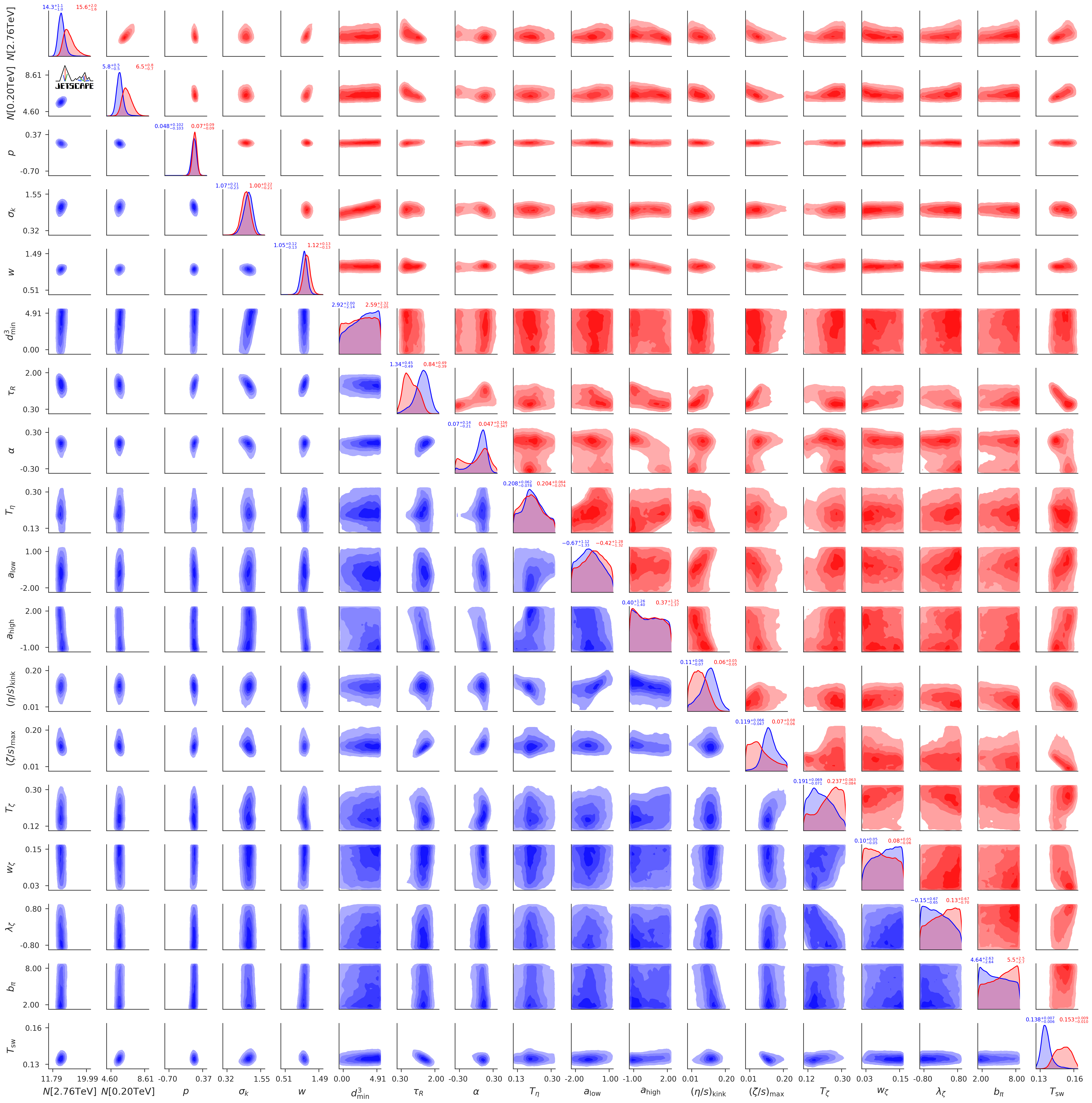}
\caption{
The posterior for Grad (blue) and Chapman-Enskog (red) viscous correction models for all model parameters, combining both RHIC and LHC experimental results. Units for dimensionful quantities are those given in Table~\ref{prior_table}.
}
\label{fig:full_posterior}
\end{figure*}

For completeness, we show in Fig.~\ref{fig:full_posterior} the posterior of all model parameters single and joint-parameter marginal distributions for the Grad (blue) and Chapman-Enskog (red) viscous correction models, combining both RHIC and LHC experimental results. 

\section{Posterior for LHC and RHIC independently}
\label{app:post_LHC_RHIC_separate}

\begin{figure}
\centering
\includegraphics[width=8.5cm]{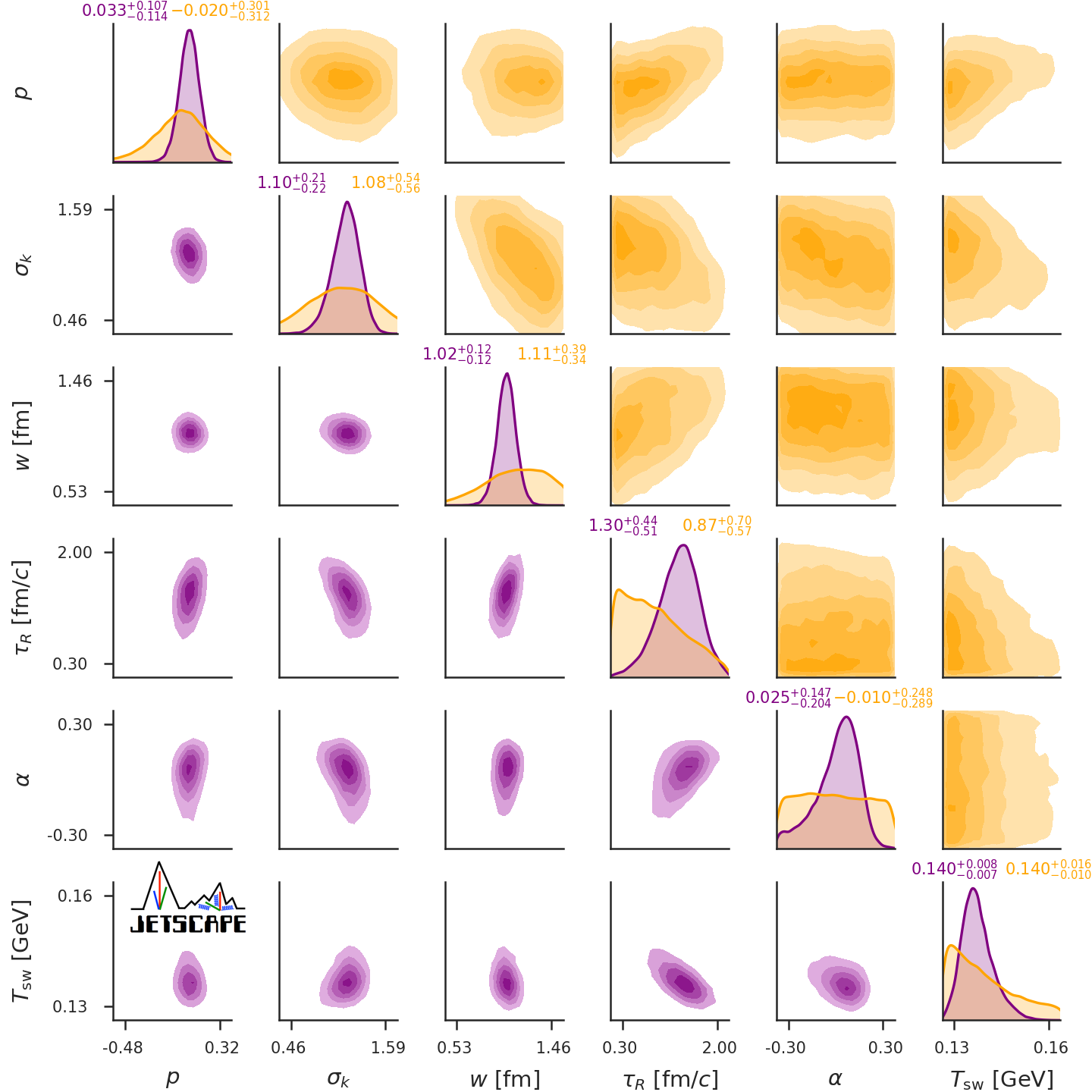}
\caption{The posterior of initial conditions and switching temperature for the Grad viscous correction model using only LHC data (purple) or only RHIC data (orange). }
\label{fig:post_LHC_RHIC_separate}
\end{figure}

In Fig.~\ref{fig:post_LHC_RHIC_separate}
we show the parameter estimates for select \trento{} initial condition parameters, as well as the switching temperature. Each posterior was estimated using only observables from a single system; LHC observables (purple) or RHIC observables (orange).

In this study, we have included more observables for Pb-Pb $\sqrts{} =2.76$ TeV collisions at LHC, so the likelihood functions are more tightly constrained, while those for Au-Au $\sqrts{} =0.2$ TeV collisions at RHIC are broader. In addition, the switching temperature for RHIC posterior is poorly constrained because we have omitted the proton yield, which should be among the most sensitive observables. Overall, we see that there is good agreement in the estimates of these parameters whether one uses Pb-Pb $\sqrts{} = 2.76$ TeV observables or Au-Au $\sqrts{} = 0.2$ TeV observables.

\section{Validation of principal component analysis}\label{pca_valid}

\begin{figure*}
\centering
\includegraphics[width=18cm]{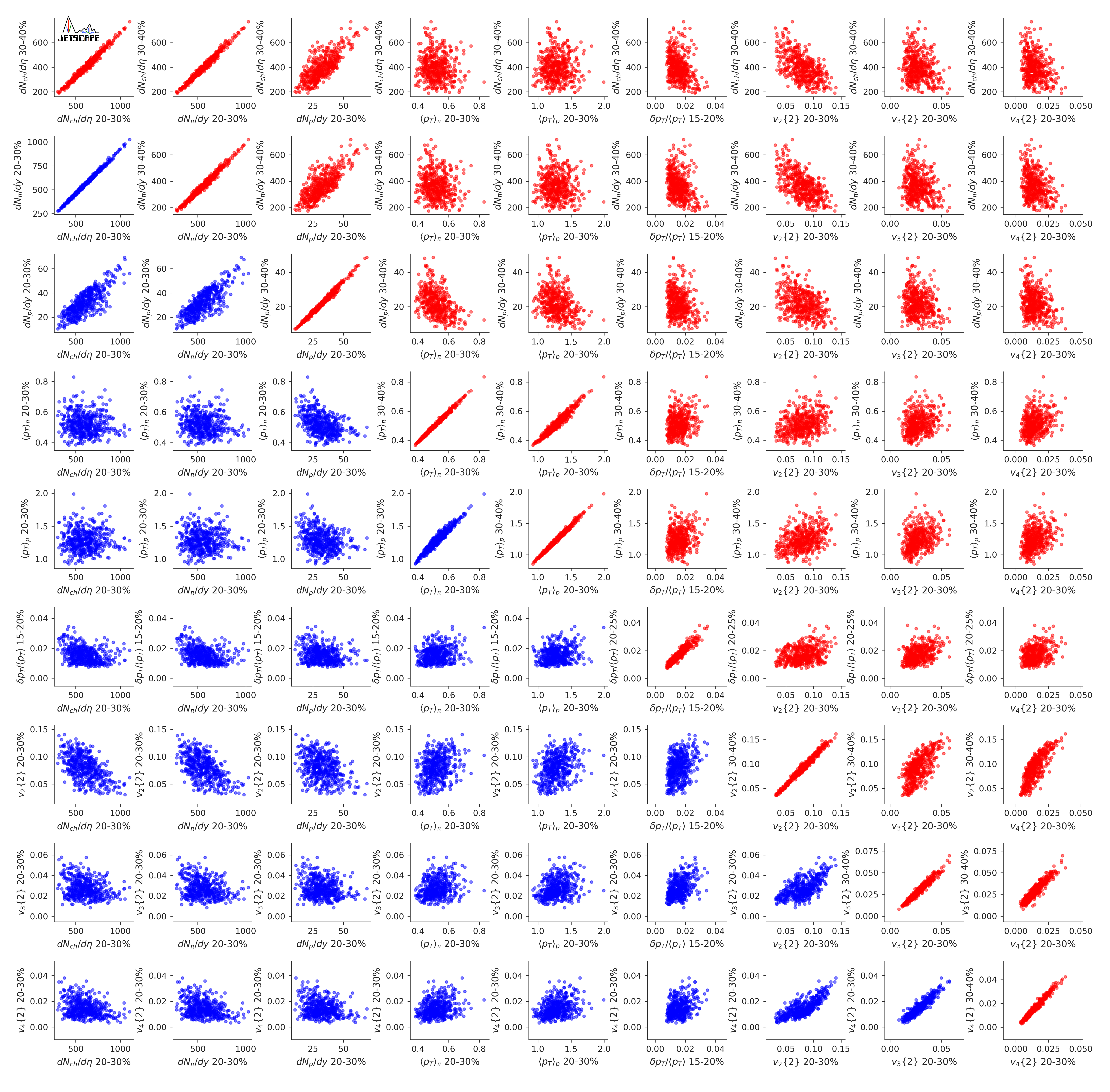}
\caption{Scatter plots of selected pairs of observables predicted by our model for Pb-Pb collisions at $\sqrts{} = 2.76$ TeV for the 500-point parameter design. Some pairs in the same centrality bin are shown in blue, while all pairs of different centrality bins are shown in red. Many pairs of observables have strong linear correlations, in which case they do not contain significant mutual information (knowing one is nearly sufficient). Pairs of observables which do not have strong linear correlations carry independent information about the parameters.}
\label{fig:observables_corr}
\end{figure*}

Principal component analysis acts to identify the linear correlations among pairs of observables. A figure showing the correlations among all possible pairs of observables would be far too large to plot, but we plot a subset of possible pairs in Fig.~\ref{fig:observables_corr} and make some important observations.  

\begin{figure}
\centering
\includegraphics[trim=0 0 0 16, clip, width=8cm]{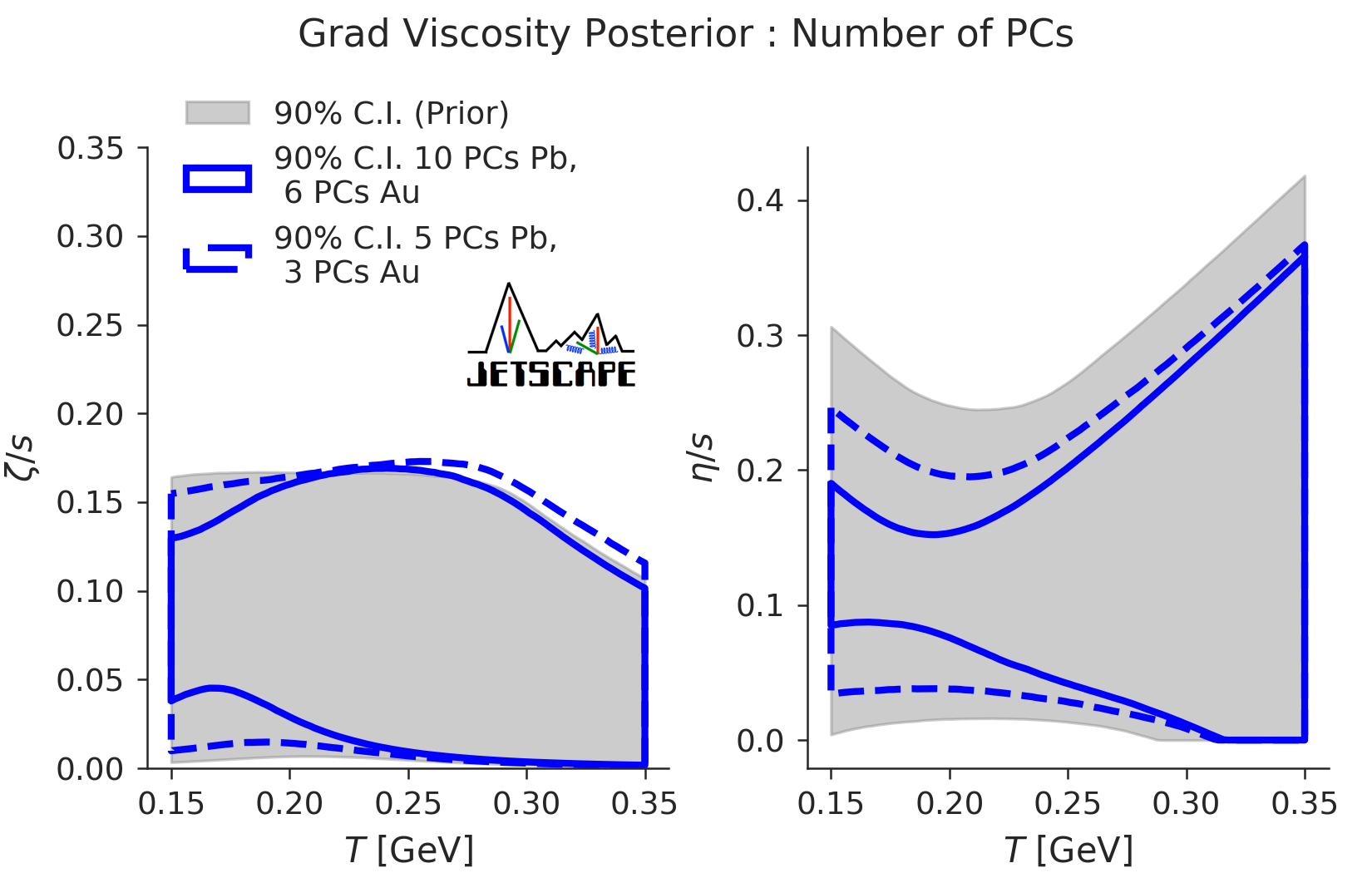}
\caption{Comparing the viscosity posteriors when we perform Bayesian parameter estimation with less principal components. The solid blue results from estimation using 10 principal components for the Pb-Pb $\sqrts{} = 2.76$ TeV emulator and $6$ principal components for the Au-Au $\sqrts{} = 0.2$ TeV emulator. The dotted blue results from $5$ principal components for Pb-Pb $\sqrts{} = 2.76$ TeV and $3$ principal components for Au-Au $\sqrts{} = 0.2$ TeV. }
\label{compare_posterior_npc}
\end{figure}

We see that certain pairs of observables have a strong linear correlation: for instance the yield $dN/dy$ of pions in the $20-30$\% centrality bin and yield of charged particles $dN_{\rm ch}/d\eta$ in the $30-40$\% centrality bin. For such pairs, nearly all the information about the model parameters is contained in just one of the observables. Uncorrelated pairs contain independent information about the model parameters. No pair of observables displays a significant non-linear correlation except for the elliptic flow $v_2\{2\}$ and quadrangular flow $v_4\{2\}$, which shows a correlation $v_4\{2\} \sim (v_2\{2\})^2$. The scarcity of strong non-linear correlations suggests that ordinary principal component analysis is a suitable method for dimensionality reduction. 
To test that our model emulator used for Bayesian inference is not overfit to features from statistical noise in the hybrid model, we have examined the effect on the posterior when we reduce the number of principal components by a factor of two for each system. In this case, five principal components explain about $94$\% of the variance of Pb-Pb collisions at $\sqrts{} = 2.76$ TeV and three principal components about $91$\% of the variance of Au-Au collisions at $\sqrts{} = 200$ GeV data. The posteriors of the specific viscosities in these two cases are compared in \fig{compare_posterior_npc}. 

The uncertainty contributed by the principal components that we omit contributes to the total emulator uncertainty and the posterior of specific shear and bulk viscosities is broadened in the case with fewer principal components included. To be sure that an emulator (and the choice of the number of principal components) is not underfit or overfit, one must perform emulator validation (see \Section{sec:closure_tests}). An emulator which is overfit will fit the training points very well, but will perform poorly in predicting the observables for a novel testing point.

\section{Experimental covariance matrix} \label{app_exp_cov}

\begin{figure}
  \centering
  \includegraphics[trim=0 0 0 16, clip, width=8cm]{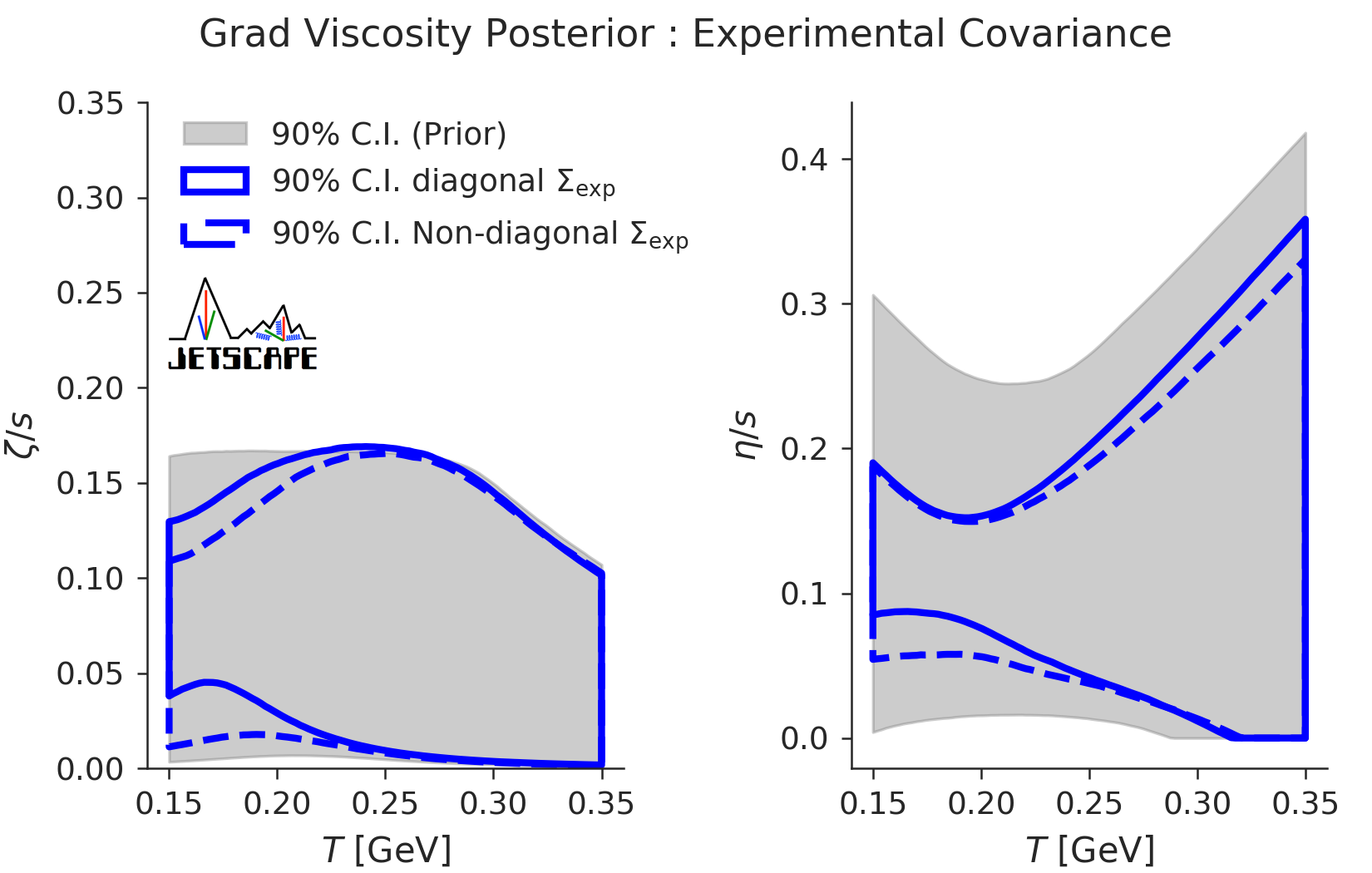}
    \caption{The change in the viscous posterior resulting from assuming a diagonal experimental covariance matrix (solid blue band) or correlated experimental covariance matrix (dashed blue band). Details regarding the magnitude of correlations in text body. }
    \label{fig:posterior_exp_cov}
\end{figure}

Currently, only the diagonal terms in the experimental covariance matrix are reported by the ALICE and STAR experiments. We have assumed a diagonal covariance matrix when performing parameter estimation. However there are undeniably nontrivial correlations in the systematic uncertainties of measured observables and centrality bins. This is important, since systematic uncertainties are generally the dominant source, larger than statistical uncertainties. We test qualitatively how these correlated uncertainties may affect our analysis. The assumed covariance matrix will affect the posterior for all model parameters, but for simplicity we quantify its effect on the posteriors of specific shear and bulk viscosities. This is shown in \fig{fig:posterior_exp_cov} for the Grad viscous correction model. 

In the case of the correlated experimental covariance matrix, given centrality bins $c_i$ and $c_j$ of the same observable, the experimental covariance is assumed to be
\begin{equation}
    \Sigma^{\text{exp}}_{i,j} = \rho \sigma_i \sigma_j
\end{equation}
where
\begin{equation}
    \rho = \exp\left(-(c_i - c_j)^2 / l^2\right)
\end{equation}
and $\sigma_i$ is the standard deviation of the observable in centrality bin $c_i$.
Observables are organized in groups: (i) multiplicities; (ii) mean transverse momenta; (iii) harmonic flows; and (iv) transverse momentum fluctuations. For pairs of different observables within the same `group', we take the same correlation coefficient defined above and multiply by an overall factor of $0.8$. For pairs of different observables in different groups, we assume zero correlation. The ``correlation length'' between centrality bins is assumed $l = 0.5$~\cite{Bernhard:2019bmu}. 

We see that an ansatz for the covariance matrix that includes nonzero correlations has a significant effect of broadening the viscous posterior, increasing the overall uncertainty. In the absence of a reported experimental covariance matrix, a more Bayesian approach would be to treat the correlation length $l$ and magnitude $\rho$ as uncertain nuisance parameters in the Bayesian parameter estimation, with priors guided by the knowledge and study of the experimental collaborations, and marginalize over them. This is an important extension that we leave for future studies.  

\section{Reducing experimental uncertainty} \label{app_exp_uncertainty}

\begin{figure}[!htb] 
\centering
\includegraphics[trim=0 0 0 16, clip, width=8cm]{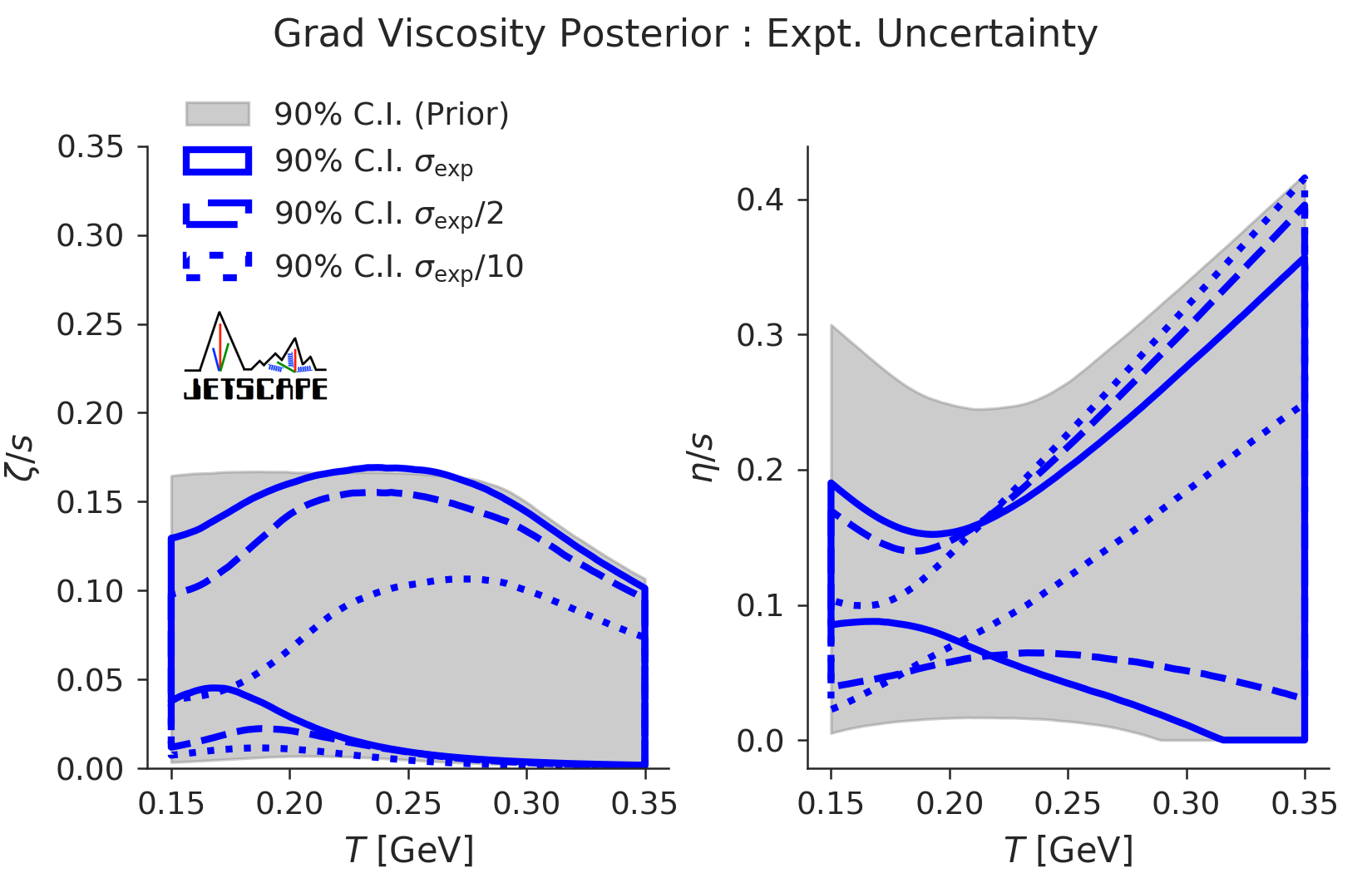}
\caption{The posterior for specific bulk (left) and shear (right) viscosities depending on whether includes the full experimental uncertainties (filled blue) or divides them by a factor of 2 (dashed blue) or 10 (dotted blue). The model emulator always contributes non-zero uncertainty. }
\label{visc_posterior_exp_uncertainty}
\end{figure}

We quantify the extent to which the experimental uncertainty contributes to the total uncertainty in our posterior for the specific shear and bulk viscosities. Besides experimental uncertainty, there is always nonzero uncertainty contributed by the use of a model emulator. We quantify this  by changing artificially the uncertainty on experimentally measured observables during the parameter estimation; the result is shown in \fig{visc_posterior_exp_uncertainty}. 

We see that significantly reducing the experimental error has the potential to qualitatively move our posterior for the specific bulk and shear viscosities. Perhaps more importantly, even if we reduce all of the experimental uncertainty by a factor of two, the credible intervals for the specific bulk and shear viscosities still remain quite large at high temperatures. This hints that in the future we should include additional observables and systems which are more sensitive to the viscosities at high temperatures.

\section{Bulk relaxation time} 
\label{app_bulk_relax}

\begin{figure}[tb]
  \centering
    \includegraphics[trim=0 0 0 25, clip,width=5cm]{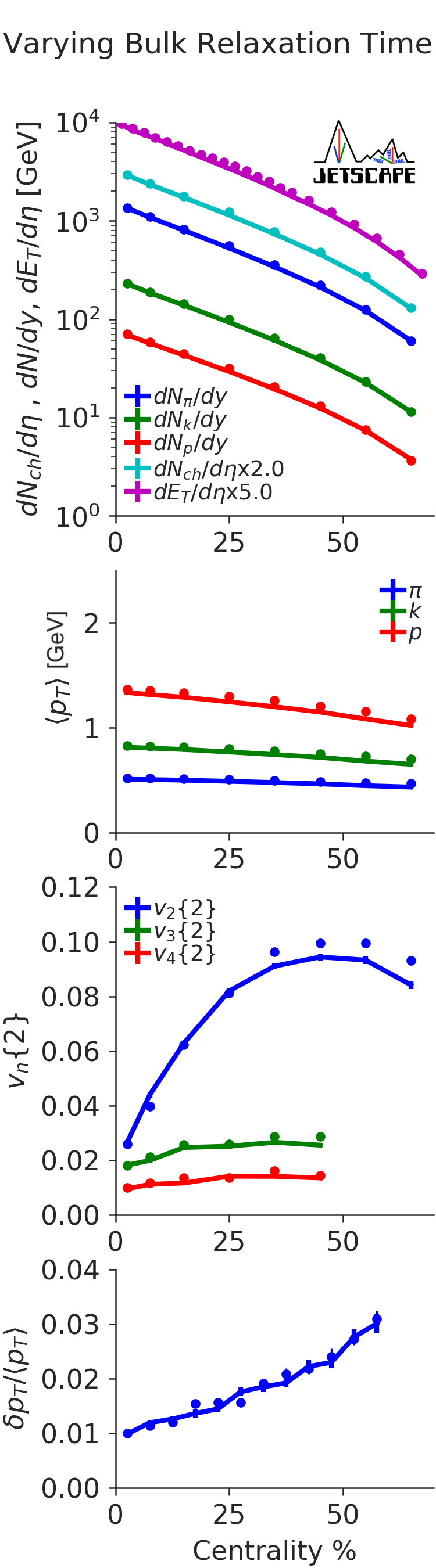}
    \caption{The solid lines are averages over five thousand Pb-Pb $\sqrts{} = 2.76$ TeV events generated with the MAP parameters for the Grad model, and the default bulk relaxation time factor $b_{\Pi} = 1/14.55$. The circles are generated with the same set of parameters except $b_{\Pi} = 2/14.55$.}
    \label{fig:obs_bulk_relax}
\end{figure}

Throughout this study, we have used a parametrization of the specific bulk viscosity given by
\begin{equation}
    \tau_{\Pi} = b_{\Pi} \frac{\zeta}{\left(\frac{1}{3} - c_s^2\right)^2 (\epsilon + p)}
\end{equation}
where $b_{\Pi} = 1/14.55$ \cite{Denicol:2014vaa}. We study how a change in $b_{\Pi}$ translates into a change in our observables using the Maximum A Posteriori parameters for the Grad viscous correction model. This is shown in \fig{fig:obs_bulk_relax}. 

Because our pre-hydrodynamic model is free-streaming and conformal, the resulting bulk pressure at Landau matching is large and positive. The bulk pressure will relax to its Navier--Stokes value $\Pi_{\text{NS}} = -\zeta \theta$ on a time scale given by $\tau_{\Pi}$. Increasing $\tau_{\Pi}$ we find that the bulk pressure stays positive for a longer time. Therefore, comparing the two sets of events, the calculations with the larger bulk relaxation time have larger mean transverse momenta and transverse energy. In future studies, it will be important to study to what extent the bulk relaxation time affects the posterior for the specific bulk and shear viscosities. 

As the bulk relaxation time is further increased, our model also has the feature that the bulk pressure may not have time to relax to its Navier-Stokes value $\Pi_{\rm NS} = -\zeta \theta$ during the lifetime of the hydrodynamic phase. In this case, the evolution of the bulk pressure becomes less sensitive to the value of the specific bulk viscosity and is dominated by its initial conditions. Inferring the likely values of the specific bulk viscosity then becomes more challenging. 

\section{Comparison to previous studies}
\label{appendix:comparison_nature_physics}

In this section we enumerate the largest differences between the parameter estimation presented in this analysis and the analysis found in Ref.~\cite{Bernhard:2019bmu}.

\subsection{Physics models}

\paragraph{Pre-hydrodynamic free-streaming:}
Both Ref.~\cite{Bernhard:2019bmu} and our own used free-streaming as a pre-hydrodynamic expansion model. Different numerical implementations were used, but they were validated against each other and found to be in excellent numerical agreement. However, in this work we have allowed the free-streaming time to be dependent on the energy of each collision. This additional feature is manifest in the parameter $\alpha$; when $\alpha$ is fixed to zero, both studies have the same physics for the pre-hydrodynamic free-streaming. 

\paragraph{Hydrodynamics: equation of state and viscosities:}

The largest differences in the hydrodynamic models include the equation of state and the parametrization of specific shear and bulk viscosities. 

The equation of state used in Ref.~\cite{Bernhard:2019bmu} was given by the HotQCD lattice result at high temperatures matched to the 2017 PDG table of hadronic resonances at low temperatures. In particular, this included a very light $\sigma$ meson with a mass of about $500$ MeV. Our study has matched the same HotQCD lattice equation of state at high temperatures to a table of hadronic resonances entering in the \SMASH{} afterburner.
In particular, we excluded the $\sigma$ meson entirely in the construction of the equation of state.

Besides the list of resonances which compose the hadron resonance gas component, Ref.~\cite{Bernhard:2019bmu} also computed the hadronic equation of state assuming relativistic Breit-Wigner resonances with nonzero width, while this study assumed all resonances on mass-shell in constructing the equation of state. 

In the parametrization of the specific shear viscosity, Ref.~\cite{Bernhard:2019bmu} included a curvature parameter for the specific shear viscosity at high temperatures, which was not included in this work. On the other hand, we varied the slope of the low-temperature specific shear viscosity, as well as the position of the ``kink'' in this work, while both of these were fixed in Bernhard's study. For the specific bulk viscosity, we allowed the parametrization to have a nonzero skewness, which was not present in Bernhard's study. 

\paragraph{Particlization, resonance width and $\sigma$ resonance:}

Ref.~\cite{Bernhard:2019bmu} fixed the particlization model to be what we have referred to as the Pratt-Torrieri-Bernhard viscous correction model (\Section{sec:model:particlization}), while in this work we also investigated other models. 

For all viscous correction models in this work, the particles were sampled on their mass-shell, while particlization in Ref.~\cite{Bernhard:2019bmu}  sampled the particles mass from a relativistic Breit-Wigner function. 

In addition, as already mentioned, Ref.~\cite{Bernhard:2019bmu} sampled unstable $\sigma$ resonances with a mass of about $500$ MeV, which significantly increased the number of pions at low momenta once they decayed. This study excluded the $\sigma$ resonance from sampling during particlization.

\paragraph{Hadronic afterburner:}

Finally, the hadronic afterburner used in Ref.~\cite{Bernhard:2019bmu} was \URQMD{}, while we use \SMASH{}. Although these two models include somewhat different lists of resonances as well as slightly different hadronic cross-sections, we checked in \Appendix{app:smash} that \URQMD{} and \SMASH{} have excellent agreement when used with the model parameters that agree well with data. For that reason, we believe at this time that the difference in hadronic afterburners is negligible in comparison to the other differences listed above. 

\subsection{Prior distributions}

The prior used in Ref.~\cite{Bernhard:2019bmu} is nearly a subspace of the prior used in this study, with the exception of the high-temperature behavior of the specific shear viscosity. Ref.~\cite{Bernhard:2019bmu} allowed the specific shear viscosity to have a nonzero curvature, i.e. quadratic temperature dependence at high temperatures. In this study, we have not allowed such a quadratic temperature dependence in the specific shear viscosity at high temperature. 

\subsection{Experimental data}

Both Ref.~\cite{Bernhard:2019bmu} and this study have included the ALICE $p_T$-integrated, centrality-dependent data for Pb-Pb collisions at $\sqrt{s_{\rm NN}} = 2.76$ TeV. However, Ref.~\cite{Bernhard:2019bmu} additionally included data for Pb-Pb collisions at $\sqrt{s_{\rm NN}} = 5.02$ TeV, which are not included in this work. Instead, we have included STAR data for Au-Au collisions at $\sqrt{s_{\rm NN}} = 200$ GeV, which were not included in Ref.~\cite{Bernhard:2019bmu}.

\section{Multistage model validation}
\label{appendix:validation}

Several of the numerical implementations of models used in this work are used for the first time; other required modifications and expansions.
For this reason we include validations of these codes against counterparts which have been used extensively in previous studies.

\subsection{Validation of second-order viscous hydrodynamics implementation}\label{sec:app_hydro_val}

In this section, we compare two different numerical implementations of the same underlying second-order relativistic hydrodynamics equations~\cite{Denicol:2012es}. The first implementation is the one used throughout this work is \music{}~\cite{Schenke:2010nt,Schenke:2010rr,Paquet:2015lta}. The second implementation is a slightly modified version of the \vishnew{} 2+1D hydrodynamics code~\cite{Song:2009gc, Shen:2014vra}, \texttt{osu-hydro}~\cite{osuhydro}, used in previous studies~\cite{Bernhard:2016tnd, Bernhard:2018hnz, Moreland:2019szz}. Both \music{} and \vishnew{} solve the same hydrodynamic equations of motion~\cite{Denicol:2012es} but with two different numerical schemes: \vishnew{} uses SHASTA~\cite{Boris:1973tjt} while \music{} uses the Kurganov-Tadmor algorithm \cite{2000JCoPh.160..241K}. Despite differences in the numerical algorithms --- amounting to approximations of spatial derivatives --- for sufficiently smooth hydrodynamic fields the two codes should agree well.

\begin{figure*}[!htb]
\noindent\makebox[\textwidth]{%
  \centering
  \begin{minipage}{0.5\textwidth}
    \includegraphics[width=\textwidth]{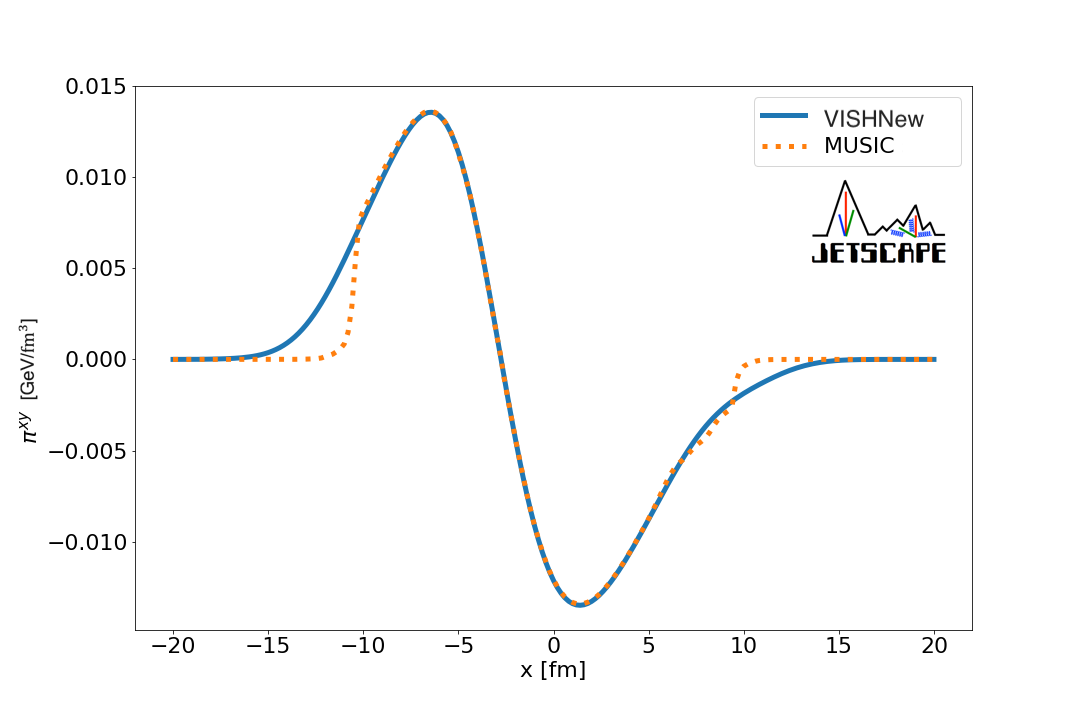}
  \end{minipage}
  \begin{minipage}{0.5\textwidth}
    \includegraphics[trim=0 0 0 16, clip, width=\textwidth]{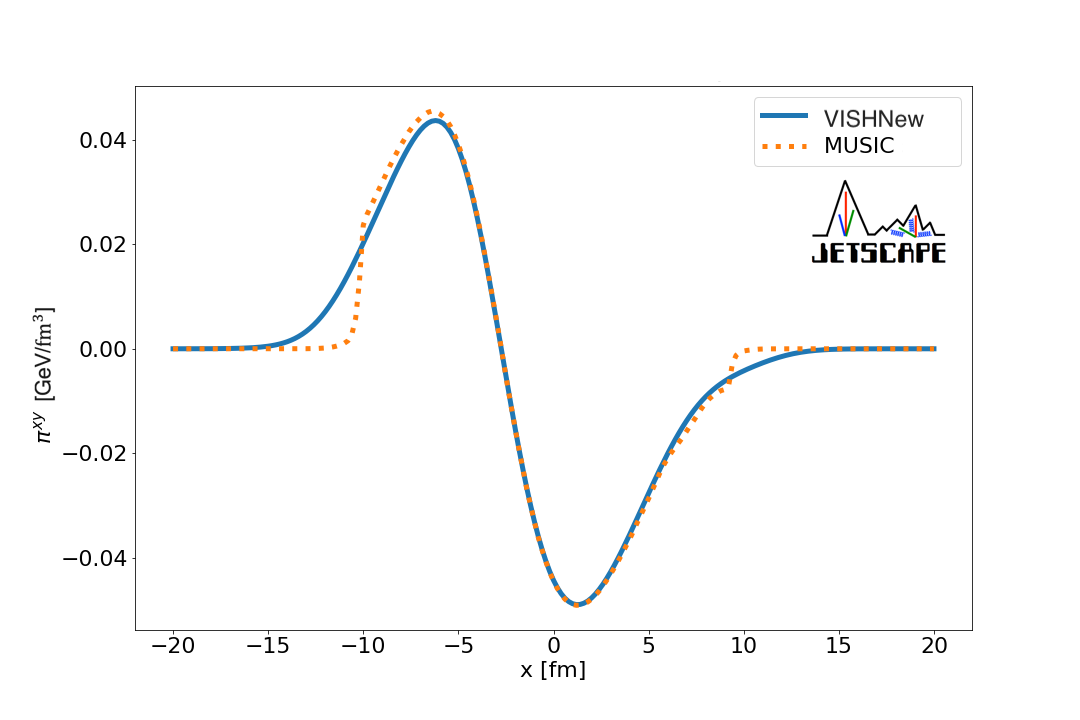}
  \end{minipage}
  }
  \caption{The results of the hydrodynamic evolution of the shear stress for a smooth initial condition, just before freeze-out, for $\eta/s = 0.08$ (left) and $\eta/s = 0.3$ (right). The \music{} regulation scheme allows larger inverse Reynolds numbers inside of the switching surface than the \vishnew{} scheme. }
  \label{shear_reg}
\end{figure*}

Besides the different numerical schemes, \vishnew{} and \music{} have different viscous current regulation schemes. The regulation scheme used in \vishnew{} is described in Ref.~\cite{Shen:2014vra} while that used in \music{} can be found in Ref.~\cite{Denicol:2018wdp, Schenke:2020mbo}. For small to moderate values of $\eta/s$ and $\zeta/s$, neither of these schemes should regulate the viscous currents close to or inside the constant energy density (or temperature) switching hypersurface. Because our hydrodynamic model will explore moderate and large values of $\eta/s$ and $\zeta/s$, it is important to compare the hydrodynamic fields. For a fixed $\eta/s = 0.08$, we run the same smooth initial conditions used for the ideal hydrodynamic comparison through free-streaming and either \music{} or \vishnew{} with zero bulk viscosity and the conformal equation of state $\epsilon = 3p$. These are shown in \fig{shear_reg}.

At late times, there are differences in the shear stress $\pi^{xy}$ near the dilute regions of the grid. These differences do not propagate into the region inside the particlization surface ($\epsilon \gtrsim 0.2$ GeV/fm$^3$).
We have also run the exact same event through viscous hydro with a fixed $\eta/s = 0.3$. The larger specific shear viscosity will incur stronger regulation.  We find that the \music{} scheme, while aggressive in low temperature regions, allows larger values of shear pressure inside the region $\epsilon > 0.2 $ GeV/fm$^3$.

\begin{figure*}[!htb]
\noindent\makebox[\textwidth]{%
  \centering
  \begin{minipage}{0.5\textwidth}
    \includegraphics[width=\textwidth]{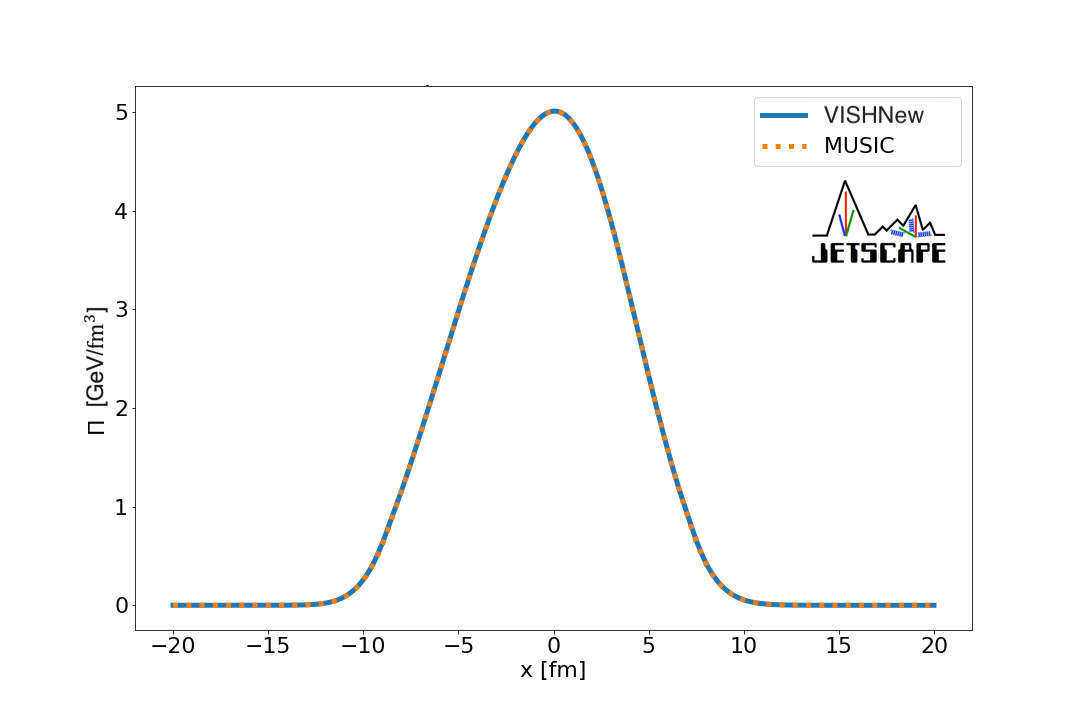}
  \end{minipage}
  \begin{minipage}{0.5\textwidth}
    \includegraphics[width=\textwidth]{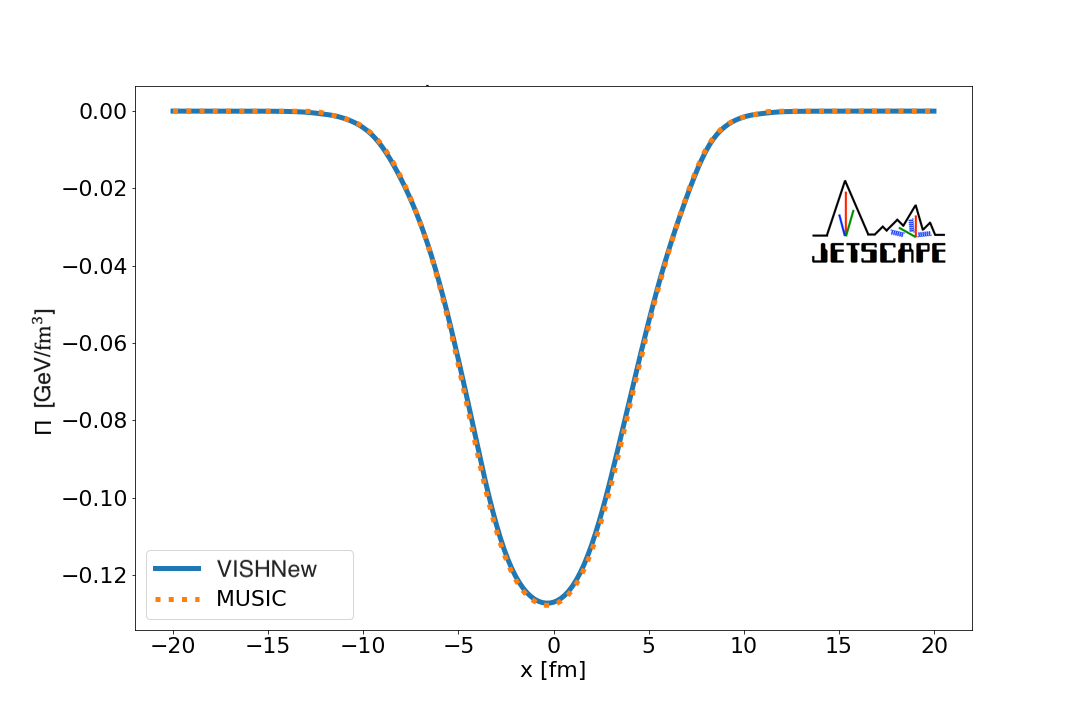}
  \end{minipage}
  }
  \caption{The initial bulk pressure (left) and bulk pressure just before freeze-out (right), resulting from hydrodynamic evolution of a smooth initial condition. The specific shear viscosity was fixed $\eta/s = 0.08$, and specific bulk viscosity $(\zeta/s)(T)$ was given by~\cite{Bernhard:2018hnz} for this test.}
  \label{bulk_final_Pi}
\end{figure*}

\begin{figure*}[!htb]
\noindent\makebox[\textwidth]{%
  \centering
  \begin{minipage}{0.5\textwidth}
    \includegraphics[width=\textwidth]{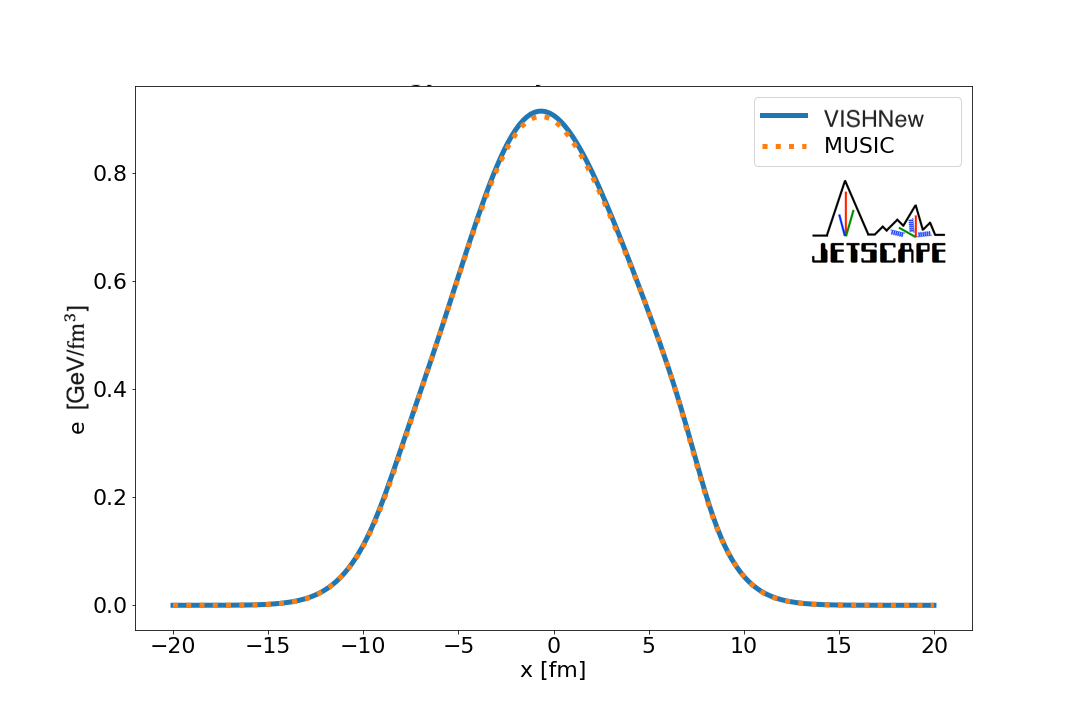}
  \end{minipage}
  \begin{minipage}{0.5\textwidth}
    \includegraphics[width=\textwidth]{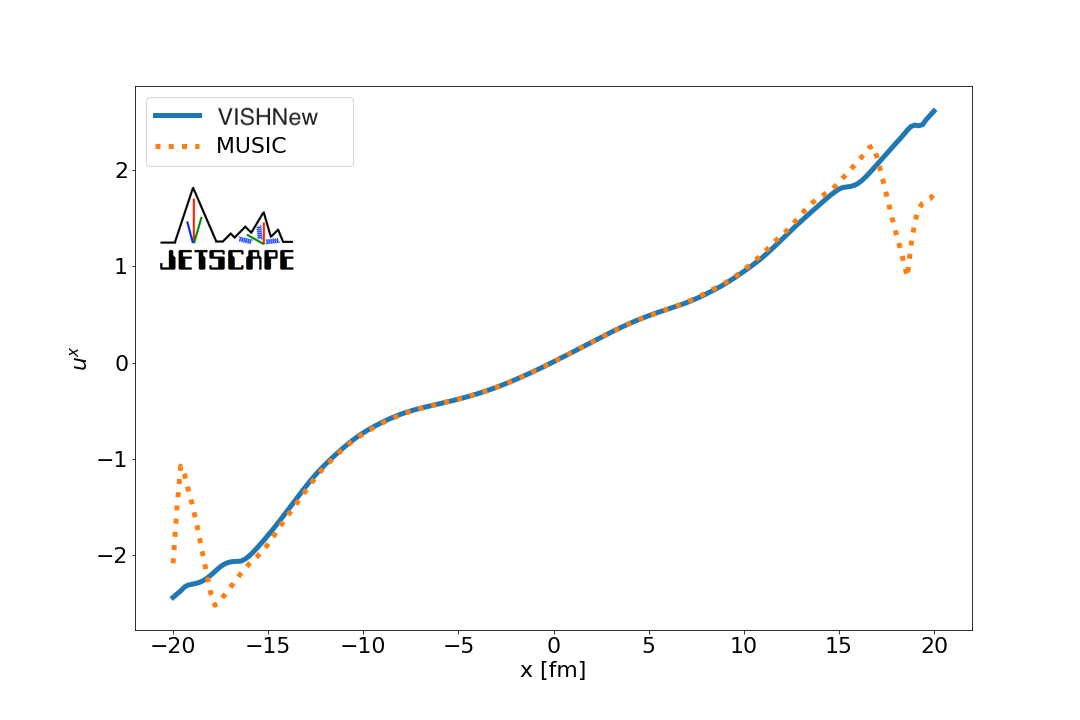}
  \end{minipage}
  }
  \caption{The energy density (left) and flow (right) after hydrodynamic evolution of a smooth initial condition. The specific shear viscosity was fixed $\eta/s = 0.08$, and specific bulk viscosity $(\zeta/s)(T)$ was given by~\cite{Bernhard:2018hnz} for this test.}
  \label{bulk_final_u}
\end{figure*}

As additional validation, we repeated the previous test with a QCD equation of state (as described in \Section{sec:hydro}), again with a fixed specific shear viscosity $\eta/s = 0.08$ but this time with a temperature dependent specific bulk viscosity $(\zeta/s)(T)$ from Ref.~\cite{Bernhard:2018hnz}.
The bulk pressure, energy density and flow are shown in \figs{bulk_final_Pi} and \ref{bulk_final_u}. Good agreement is found between the two codes.

\begin{figure*}[!htb]
\noindent\makebox[\textwidth]{%
  \centering
  \begin{minipage}{0.5\textwidth}
    \includegraphics[width=8cm]{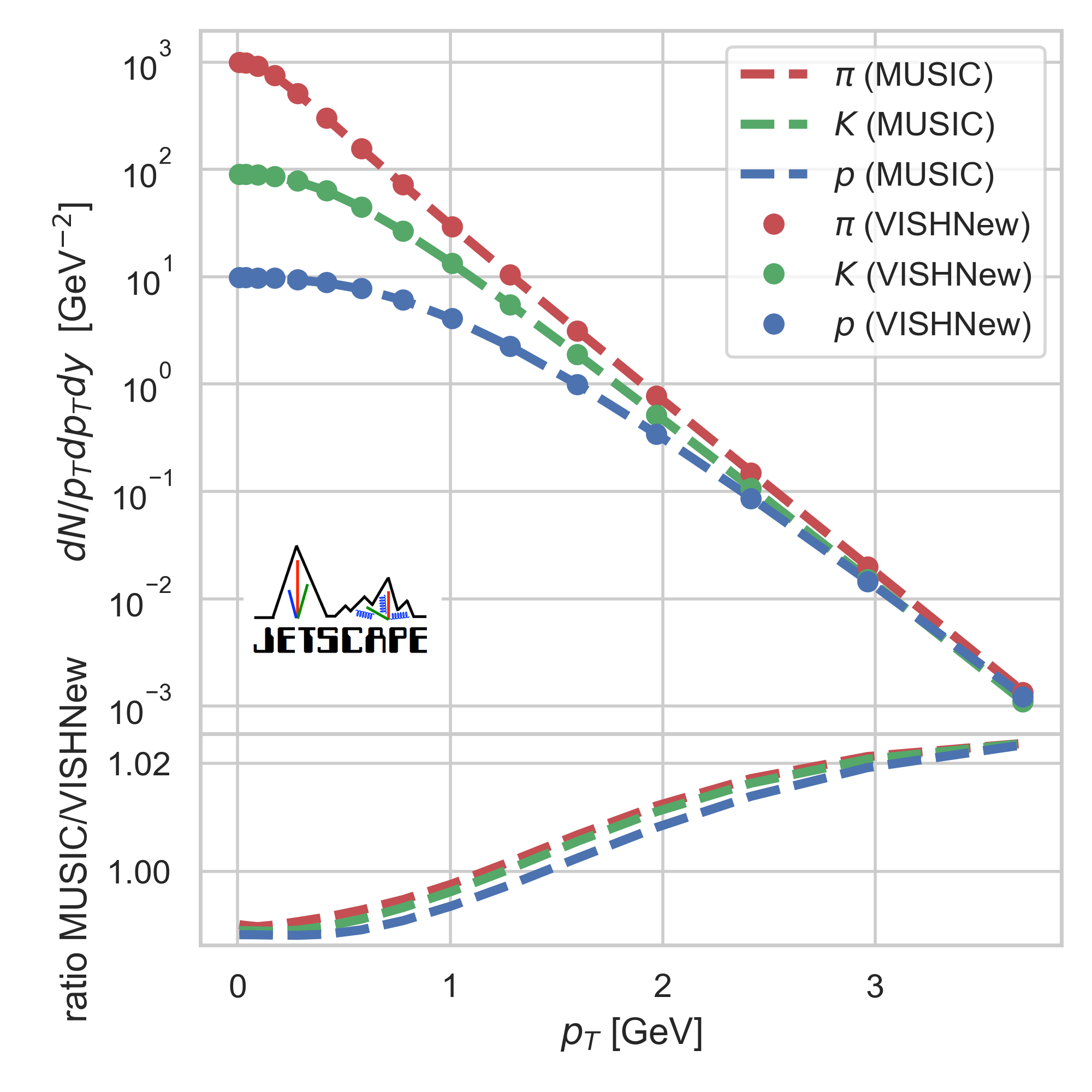}
  \end{minipage}
  \begin{minipage}{0.5\textwidth}
    \includegraphics[width=8cm]{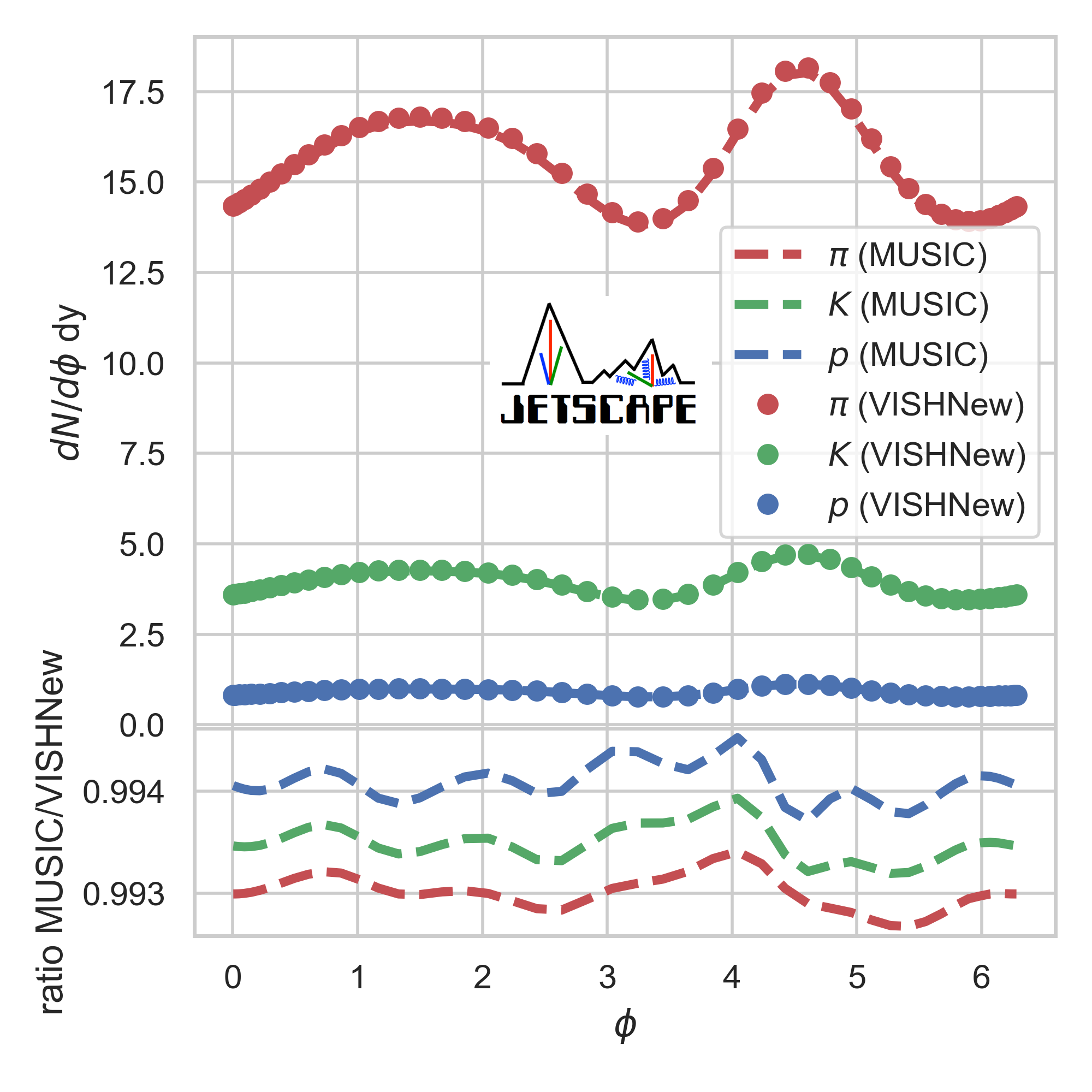}
  \end{minipage}
  }
  \caption{Comparison of the transverse momentum $p_T$ spectra (left) and azimuthal $\phi_p$ spectra (right) generated from the \music{} and \vishnew{} freezeout surfaces. The freezeout surface was generated using the events compared above, with fixed $\eta/s = 0.08$, and specific bulk viscosity $(\zeta/s)(T)$ was given by~\cite{Bernhard:2018hnz}. }
\label{spectra_ratio_pT_phi}
\end{figure*}

In order to quantify the effects of any small differences that the hydrodynamics may have on our hadronic observables, we have evaluated the smooth Cooper-Frye integral over the switching surface generated by each hydrodynamics code. The hydrodynamic event used was the same event with bulk and shear pressures for which the hydrodynamic evolution was compared above. We used \texttt{iS3D} to perform the smooth Cooper-Frye integral over each surface, including bulk and shear Grad viscous corrections, and plotted the comparisons below for pions, kaons and protons. In general, the agreement in the spectra is very good. These are shown in 
\fig{spectra_ratio_pT_phi}. These differences of about 1\% or less in the differential observables yield differences $\lesssim 1\%$ in the $p_T$ and $\phi_p$ integrated observables. 

\subsubsection{Validation against cylindrically symmetric external solution}

\begin{figure*}[tb]
  \centering
  \includegraphics[width=0.45\textwidth]{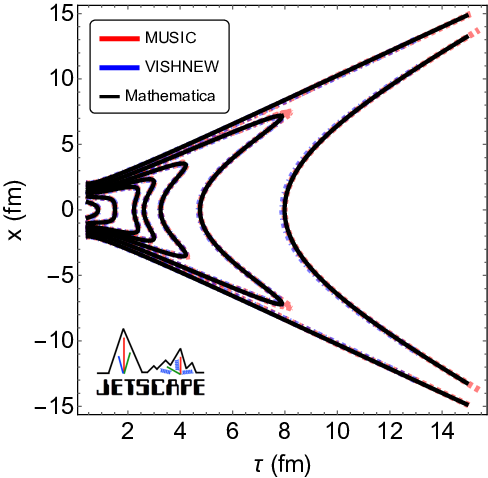}
  \hfill
    \includegraphics[width=0.45\textwidth]{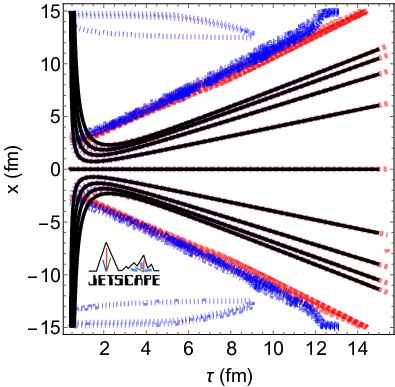}
    \caption{Temperature (left) and $u^x$ profiles (right) for ideal hydrodynamics initialized with $T(\tau_0,r)=T_0 \exp(-r^2/\sigma^2)$ with $\tau_0=0.4$\,fm/$c$, $T_0=600$~MeV and $\sigma=1$~fm. Conformal equation of state. Result from new version of \music{} and from \vishnew{}, compared with a solution obtained from Mathematica. The temperature contours in the left panel are (from left to right) for 10, 25, 50, 75, 100, 200 and 400~MeV, while the flow rapidity contours are (from top to bottom) for rapidities ranging from -4 to 4 by step of 1.}
    \label{fig:mathematica_hydro_sol_sigma1}
\end{figure*}

In this section, we compare the ideal hydrodynamic numerical solution obtained with \music{} and \vishnew{} with a solution obtained with Mathematica \cite{Mathematica_v12_1}. Cylindrically symmetric initial conditions simplify considerably the equations of hydrodynamics \cite{Paquet:2019npk}, making them solvable in Mathematica \cite{mathematica_ideal_hydro_cylindrical}. Since the differential equation solver in Mathematica is adaptive and completely different from that of \music{} and \vishnew{}, it provides complementary validation of their numerical solutions. We focus on a scenario where large gradients develop, to challenge the numerical algorithms of \music{} and \vishnew{}. We use the initial temperature profile $T(\tau_0,r)=T_0 \exp(-r^2/\sigma^2)$ with $\tau_0=0.4$\,fm/$c$, $T_0=600$~MeV and $\sigma=1$~fm. We use a conformal equation of state. The result is shown in \fig{fig:mathematica_hydro_sol_sigma1} for the temperature profile (a) and the flow rapidity $u^x$ (b).

Overall, all three solutions agree well in regions of larger temperatures, which are the physically relevant regions of spacetime. Near the edge of the fireball, in regions of very low temperatures (small $\tau$ and large $x$), neither \music{} and \vishnew{} can reconstruct the flow velocity well. This is a well-know challenge of many numerical solution of relativistic hydrodynamic equation used in heavy ion physics: regions of very low temperatures are difficult to solve accurately and are susceptible to numerical instabilities. These issues at very low temperatures typically do not propagate to the higher-temperature regions and, as such, are not serious in practice.

\subsection{SMASH}\label{app:smash}

The use of \SMASH{} as an afterburner for event-by-event studies of heavy-ion collisions is still fairly new. For this reason, we have made a comparison between \URQMD{} and \SMASH{} as they relate to our transverse-momentum-integrated observables. We generated five thousand fluctuating initial conditions for Au-Au $\sqrts{}=0.2$\,TeV collisions with parameters fixed by the Maximum A Posteriori parameters found in \cite{Bernhard:2018hnz} except for the initial energy density normalization, which was scaled to fit the multiplicities. 

We allowed each initial condition to free-stream for the same time and then used these initial conditions for hydrodynamics in two different models:
\begin{enumerate}[topsep=1pt,itemsep=0ex,partopsep=1ex,parsep=1ex]
    \item \SMASH{}: We matched the HotQCD lattice equation of state to the \SMASH{} list of resonances (excluding the $\sigma$ meson). Each initial condition was propagated through viscous hydrodynamics with this equation of state, followed by particlization using the Pratt-Torrieri-Bernhard viscous correction ansatz, followed by dynamics in \SMASH{}. 
    \item \URQMD{}: We matched the HotQCD lattice equation of state to the list of resonances which can be propagated in \URQMD{}. Each initial condition was propagated through viscous hydrodynamics with this equation of state, followed by particlization using the Pratt-Torrieri-Bernhard viscous correction ansatz, followed by dynamics in \URQMD{}. 
\end{enumerate}

\begin{figure*}
  \centering
    \includegraphics[width=16cm]{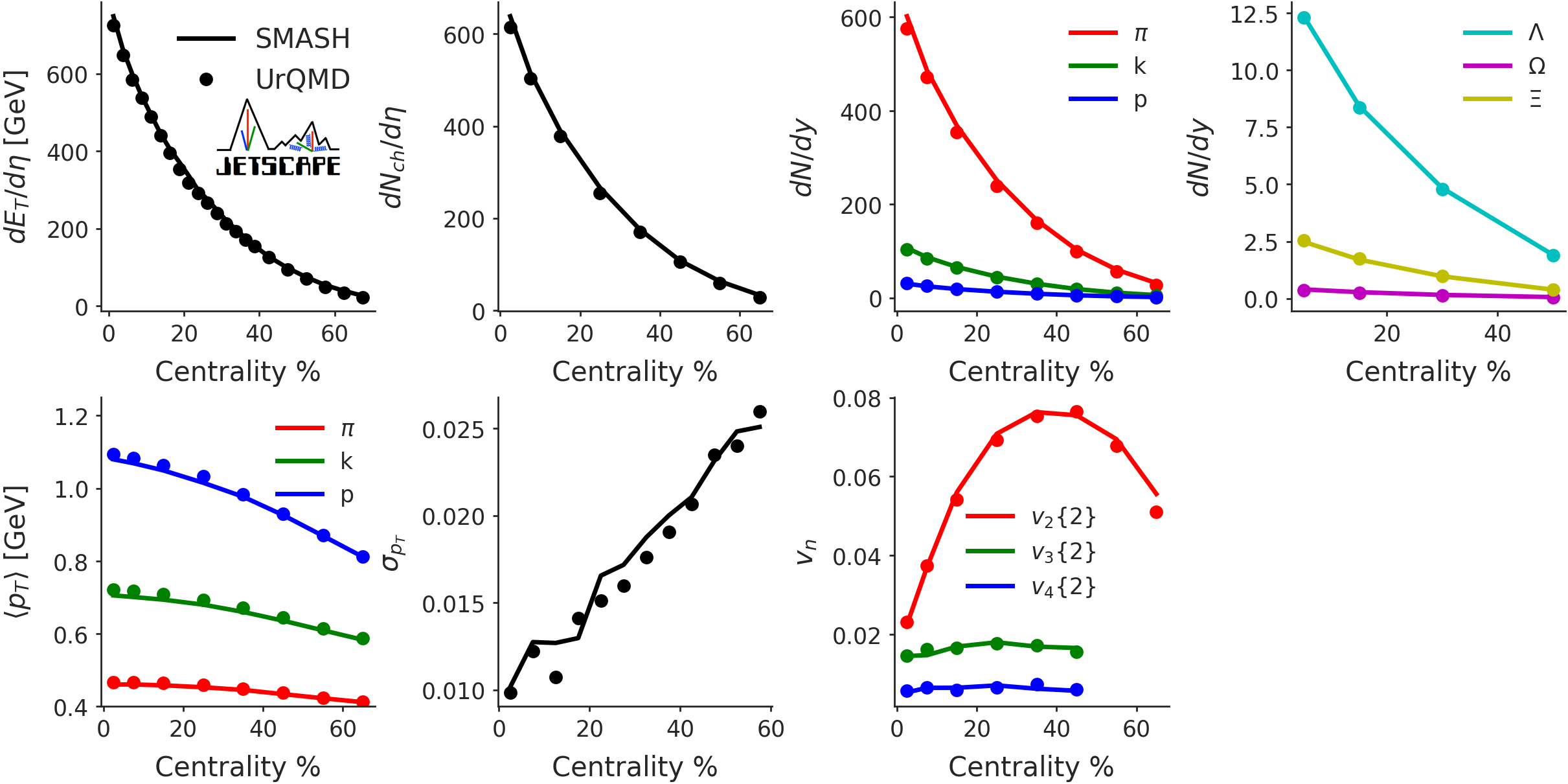}
    \caption{
    Comparison of soft hadronic observables for Au-Au $\sqrts{}=0.2$\,TeV collisions using the \SMASH{} (solid lines) or \URQMD{} (dots) afterburner.}
    \label{compare_urqmd_smash}
\end{figure*}

Finally we compared the observables predicted by the two models, shown in \fig{compare_urqmd_smash}. 
For the observables we considered, we found very good agreement. In particular, heavier resonances have spectra that are more strongly influenced by the hadronic afterburner than lighter resonances and the agreement in the multiplicity and transverse momenta of the proton and $\Lambda$ is strong. 

We also ran hydrodynamics with a fixed equation of state matched to the \SMASH{} hadron resonance gas particle content and then switched at the same temperature to \URQMD{} or \SMASH{}. Because of the mismatch between the equation of state generated with the \SMASH{}  and the \URQMD{} resonance gases, there is a discrepancy at particlization in all of the thermodynamic variables. For example, at the same temperature, the energy density of the \SMASH{} resonance gas and \URQMD{}'s are different. This leads to a disagreement in observables. In particular,  observables sensitive to the normalization of energy density, such as multiplicities and the transverse energy, showed a discrepancy at the level of approximately five percent. It is easy to understand that the energy density of the \URQMD{} resonance gas is a few percent smaller than \SMASH{}'s at the same temperature because of the different species and masses of hadrons. More details can be found in \Appendix{appendix:eos}. 

Given the novelty of using \SMASH{} as an afterburner, we share for completeness the numerical parameters that we used with \SMASH{}. These parameters, shown in \Table{smash_params}, gave sufficient accuracy without unreasonable loss of speed.
\begin{table}
\begin{tabular}{|l|l|}
\hline
Modus            & Afterburner \\ \hline
Time\_Step\_Mode & Fixed       \\ \hline
Delta\_Time      & 1.0         \\ \hline
End\_Time        & 1000.0      \\ \hline
\end{tabular}
\caption{\SMASH{} parameters used event-by-event throughout this study.}
\label{smash_params}
\end{table}

\subsection{Comparison of JETSCAPE with hic-eventgen}

\begin{figure*}
  \centering
    \includegraphics[width=16cm]{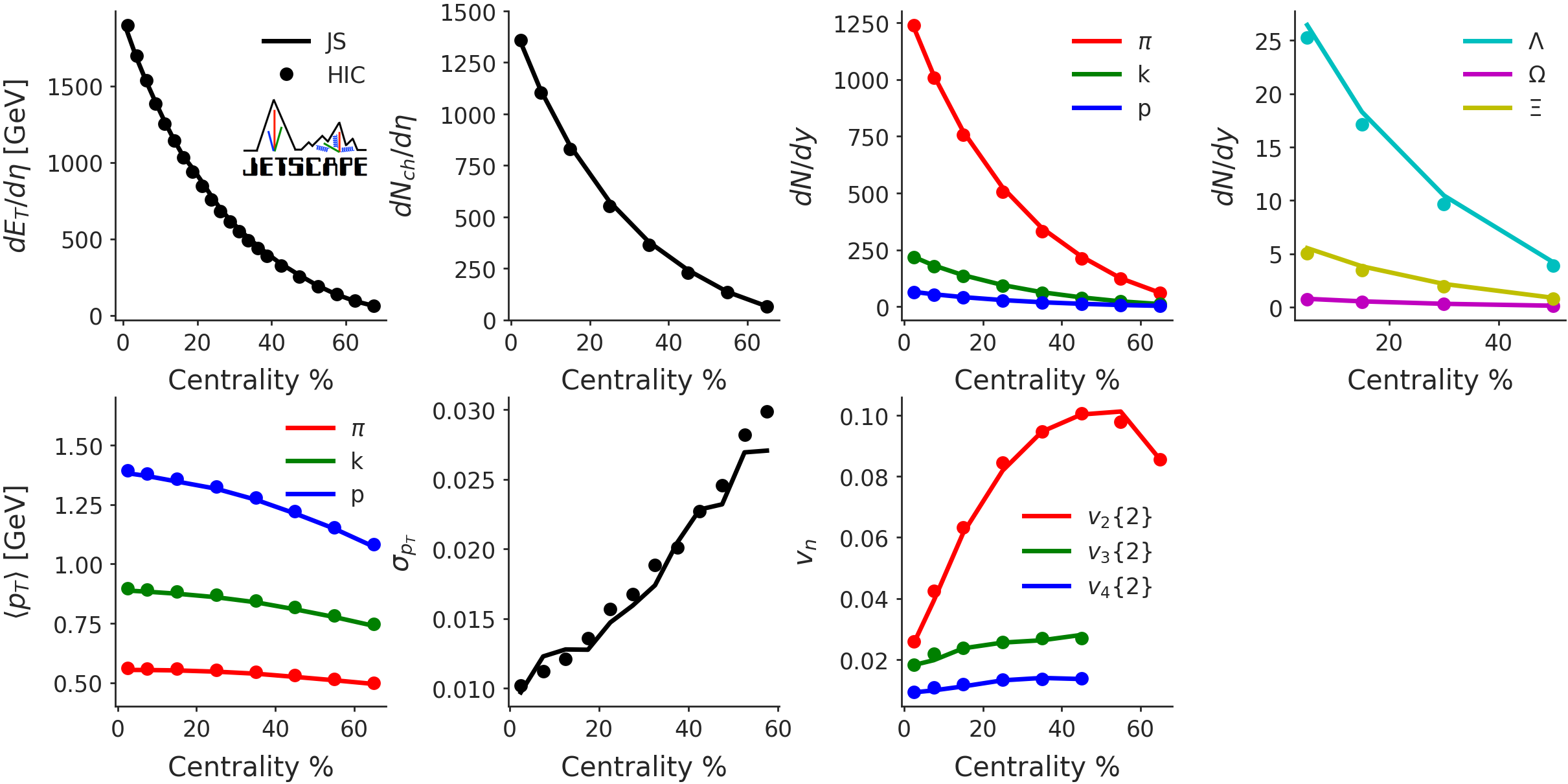}
    \caption{
    Comparison of soft hadronic observables depending on whether one uses the \texttt{JETSCAPE} event generator (solid lines) or \texttt{hic-eventgen} (dots), averaged over five-thousand Pb-Pb $\sqrts{}=2.76$ TeV events.}
    \label{compare_js_hiceventgen}
\end{figure*}

In addition to validating of all the separate model components, we also have checked that the centrality-averaged observables predicted by our \texttt{JETSCAPE} model agree very well with a version of \texttt{hic-eventgen}, the event generator used in Ref.~\cite{Bernhard:2018hnz}. This was performed by restricting our parametrizations to be the same as the Maximum A Posteriori parameters found in that study. The results of this comparison are shown in \fig{compare_js_hiceventgen}, in which we have averaged over five thousand fluctuating Pb-Pb $\sqrts{}=2.76$ TeV collision events.

In general, we find excellent agreement between the two hybrid models. For this level of agreement, the $\sigma$ meson had to be excluded from the \texttt{hic-eventgen} model; all resonances were also sampled on their mass-shell in \texttt{frzout}~\cite{frzout}, the particle sampler in \texttt{hic-eventgen}. The equation of state used during the hydrodynamic evolution was constructed to match the hadron resonance gas used in \texttt{frzout} (excluding the $\sigma$ meson). 

\subsection{The $\sigma$ meson}
\label{app:sigma_effect}

The effects of including a $\sigma$ meson resonance in our hadron resonance gas are studied using the \texttt{frzout} module~\cite{frzout}, which is designed with the option to sample the $\sigma$ resonance as a thermal resonance and perform its decay to pions. In particular, we compare three scenarios:
\begin{itemize}
    \item Excluding the $\sigma$ meson from sampling (labeled by $m\rightarrow \infty$).
    \item Sampling the $\sigma$ meson with the PDG pole mass ($\sim 500$ MeV) \cite{Tanabashi:2018oca}.
    \item Sampling the $\sigma$ meson with the mass used in \SMASH{} ($\sim 800$ MeV).
\end{itemize}

The \texttt{frzout} module was used to sample particles from a hypersurface generated by the \music{} simulation of a mid-central Pb-Pb event. The initial condition, free-streaming, and hydrodynamic transport parameters were set by the Maximum A Posteriori parameters given in \cite{Bernhard:2018hnz}. The switching temperature was 151 MeV. Sampled particles are then propagated to \URQMD{} to perform hadronic rescatterings. Note that \URQMD{} does not have a $\sigma$ meson: the effect of the  $\sigma$ meson is purely being tested at the level of the particlization, not in the afterburner. A total number of $100$ over-samples were generated to increase the statistics. The results on charged-particle multiplicity, transverse energy, and pion multiplicity and mean transverse momentum are shown in \Table{sigma_sensitivity}.

\begin{table}
\begin{tabular}{||c|c|c|c|c||}
\hline
$\sigma$       & $dN_\text{ch}/d\eta$ & $dE_T/d\eta$ {[}GeV{]} & $dN_{\pi}/dy$ & $\langle p_T^{\pi}\rangle$ {[}GeV{]} \\ 
\hline
$m = 475$ MeV & 615       & 777                  & 569        & 0.54                                       \\ 
\hline
$m = 800$ MeV & 583       & 754                  & 534        & 0.55                                       \\ 
\hline
$m\rightarrow \infty$        & 579       & 743                  & 531        & 0.54                                       \\
\hline
\end{tabular}
\caption{Sensitivity of observables to inclusion of the $\sigma$ meson. The observables computed using the same cuts as the ALICE experiment.}
\label{sigma_sensitivity}
\end{table}

Because the $\sigma$ meson decays into pions, we see that the pion yield can differ by 7\% for the lightest $\sigma$ resonance. For the higher mass $\sigma$, the results are close to not sampling a $\sigma$ meson. Additional differences would manifest if we included the effect of varying the $\sigma$ mass in constructing the hadron gas equation of state as this would also have an effect on the hydrodynamic evolution. As has been explained in the main text, we chose to omit the $\sigma$ from the equation of state and particlization, following \cite{Broniowski:2015oha}.

\subsection{Sampling particles on mass-shell}
In general, resonances occupying the hadron resonance gas should be described by spectral functions with a nonzero width. Previous Bayesian analyses have employed the sampling of thermal resonances off-shell \cite{Bernhard:2018hnz, Moreland:2019szz} in their extraction of medium properties. In this work, we chose to sample resonances on their pole-mass, neglecting this effect for now. 

\begin{figure}[tb]
  \centering
    \includegraphics[width=8cm]{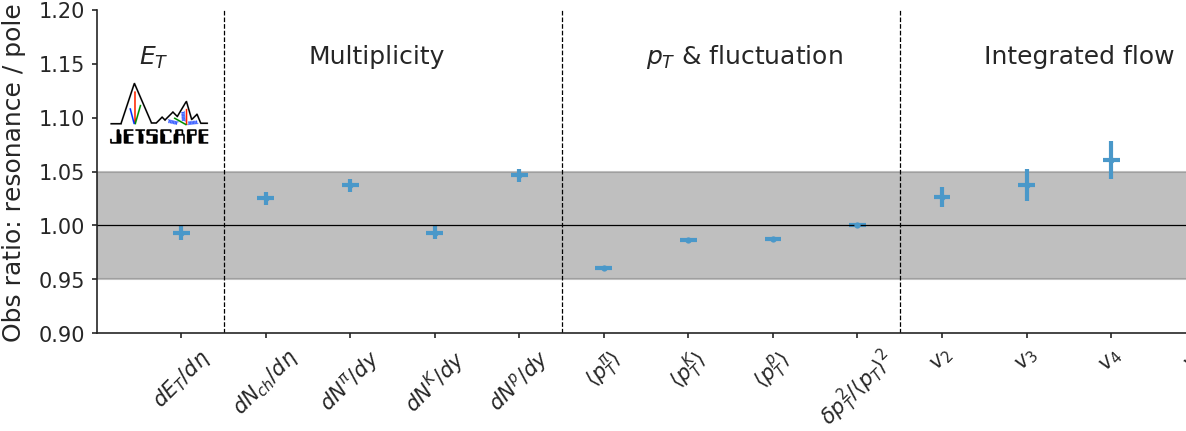}
    \caption{Ratio of observables between sampling resonances with a non-zero width and sampling on the pole-mass. We find that the differences of less than 5\% for the observables used in this study justify neglecting the non-zero width of resonances when performing parameter estimation. }
    \label{res_width_effect}
\end{figure}

In what follows, we perform a simple test to quantify the effect of off-shell sampling of resonance masses for hadronic observables used in this work.\footnote{%
    This test used 4000 Pb-Pb collisions at $\sqrt{s_{\rm NN}} = 5.02$ TeV events within a narrow impact parameter range 9.88 fm < b < 11.02 fm.}
The results are shown in \fig{res_width_effect}. The particles sampled off-shell are put on-shell after one collision in \URQMD{}. All particles occupying the hadron resonance gas are assigned Breit-Wigner spectral functions. The sampling was performed in both cases using the \texttt{frzout} module~\cite{frzout};  details about the hadron resonance gas pole-masses and widths can be found in Ref.~\cite{Bernhard:2018hnz}.

We find a $\sim 5$\% effect from the off-shell sampling in the hadronic observables shown in \fig{res_width_effect}. We believe this effect is larger because of the $\sigma$ meson present in this test: a light $\sigma$ meson with pole mass $m \approx 475$ MeV and a very broad width $\Gamma \approx 500$ MeV, as used in  Ref.~\cite{Bernhard:2018hnz}. As we saw in the previous section, the presence of the $\sigma$ meson (sampled on the pole-mass) contributes significantly to the pion yield. 

\subsection{QCD equations of state with different hadron resonance gases}
\label{appendix:eos}

The QCD equation of state used in hydrodynamic simulations of heavy ion collision matches a lattice calculation at high temperature ($T \gtrsim 120$~MeV) with a hadron resonance gas calculation at low temperature. In this work, we use the lattice calculations from Ref.~\cite{Bazavov:2014pvz}. As explained in Ref.~\cite{Bazavov:2014pvz}, the trace anomaly evaluated from lattice is used to compute the pressure by integration of
\begin{equation}
    \frac{p(T)}{T^4}=\frac{p_0(T)}{T_0^4}+\int_{T_0}^{T} \frac{d T^\prime}{T^\prime} \frac{\Theta^{\mu\mu}(T^\prime)}{T^{\prime 4}}.
\end{equation}
Energy density and entropy density then follow. The integration constant for the pressure is obtained from a hadron resonance gas calculation at $T=130$~MeV.

Reference~\cite{Bernhard:2018hnz} followed a related but different approach: to ensure energy-momentum conservation at particlization, the lattice QCD trace anomaly is matched to a hadron resonance gas in a $[T_a,T_b]$ temperature range. The trace anomaly below $T_a$ is that of the hadron resonance gas; the trace anomaly between $T_a$ and $T_b$ is an interpolation between the resonance gas and the lattice QCD trace anomaly. Above $T_b$, the trace anomaly is that of the lattice QCD. Using this new trace anomaly, which differs from that of the lattice below $T_b$, the pressure is computed by integration using $p_0(T_0=50 \textrm{MeV})$ as reference; energy density and entropy density are then calculated.

We illustrate first the differences between the lattice pressure, and the pressure obtained with the above matching. If the temperature is below the matching point $T_a$, the pressure from the lattice case is given by
\begin{equation}
    \frac{p_{L}(T)}{T^4}=\frac{p_{L,0}}{T_0^4}+\int_{T_0}^{T} \frac{d T^\prime}{T^\prime} \frac{\Theta_L^{\mu\mu}(T^\prime)}{T^{\prime 4}}.
\end{equation}
where the integration constant $p_{L,0}$ is the only input from the hadron resonance gas that enters in the definition of the pressure.

In the ``matched'' equation of state, however, the entire thermodynamics is determined by the hadron resonance gas below the lower matching temperature $T_a$:
\begin{equation}
    \frac{p_{M}(T)}{T^4}=\frac{p_{M,0}}{T_0^4}+\int_{T_0}^{T} \frac{d T^\prime}{T^\prime} \frac{\Theta_{HRG}^{\mu\mu}(T^\prime)}{T^{\prime 4}}.
    \label{eq:pressure_matched}
\end{equation}
There is no information from the lattice calculations entering in \eq{eq:pressure_matched} if $T<T_a$.
This example makes is clear that any mismatch between the trace anomaly of lattice calculations and that of the hadron resonance gas results in a difference in the equation of state. This is of course the case even if the exact same hadron resonance gas are used to fix $p_{L,0}$ --- $p_{L,0} = p_{M,0}$ --- which is arguably never the case. These uncertainties are difficult to eliminate: any mismatch between the hadron resonance gas and the lattice calculation would result in a discontinuity at particlization. 

\begin{figure}
  \centering
    \includegraphics[width=8cm]{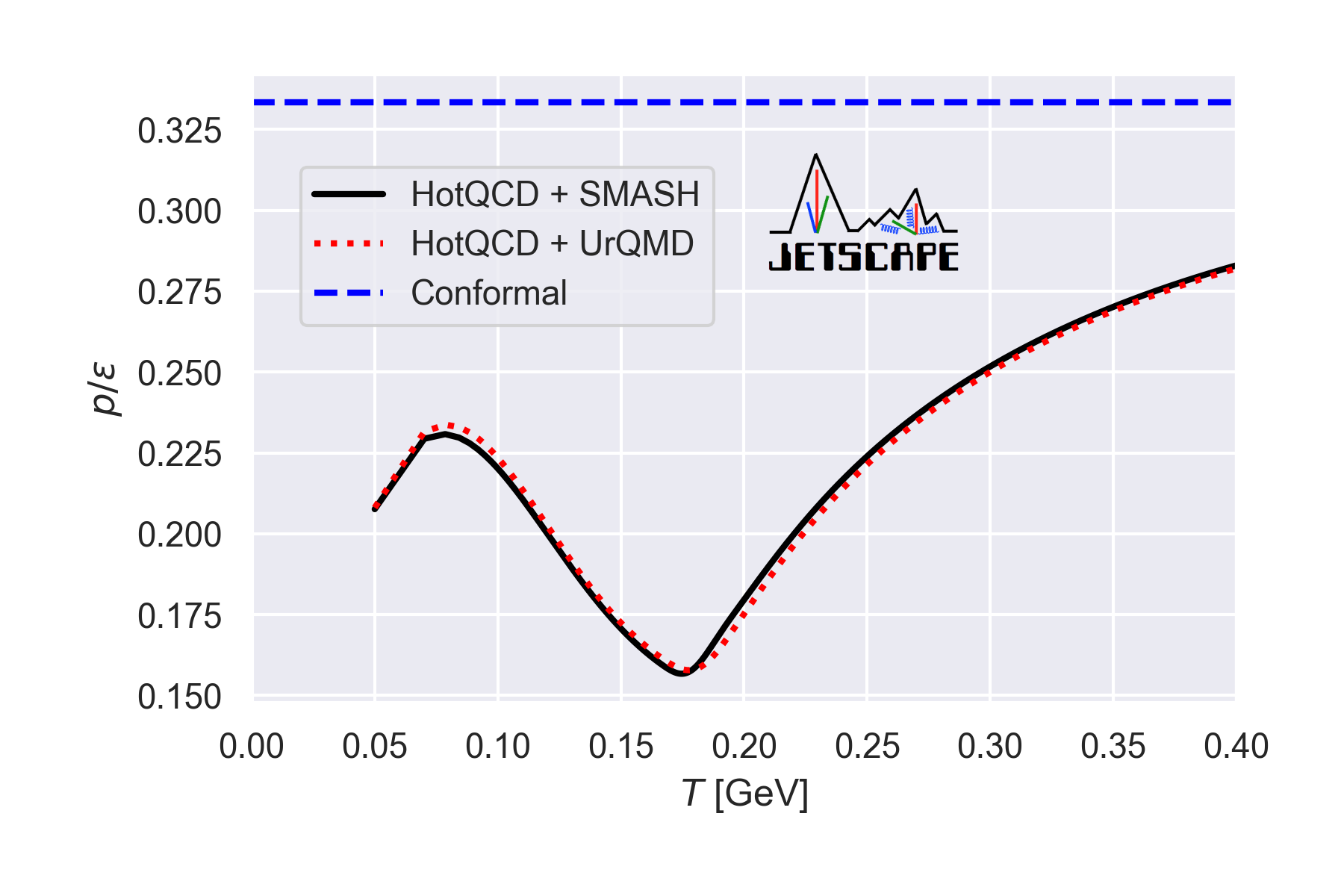}
    \caption{The equation of state used throughout this work for parameter estimation `HotQCD + \SMASH{}' is shown as well as a different equation of state that has been matched to the list of resonances propagated in \URQMD{}. The conformal equation of state is included as a visual reference.}
    \label{fig:eos_comp}
\end{figure}

\begin{figure}
  \centering
    \includegraphics[width=8cm]{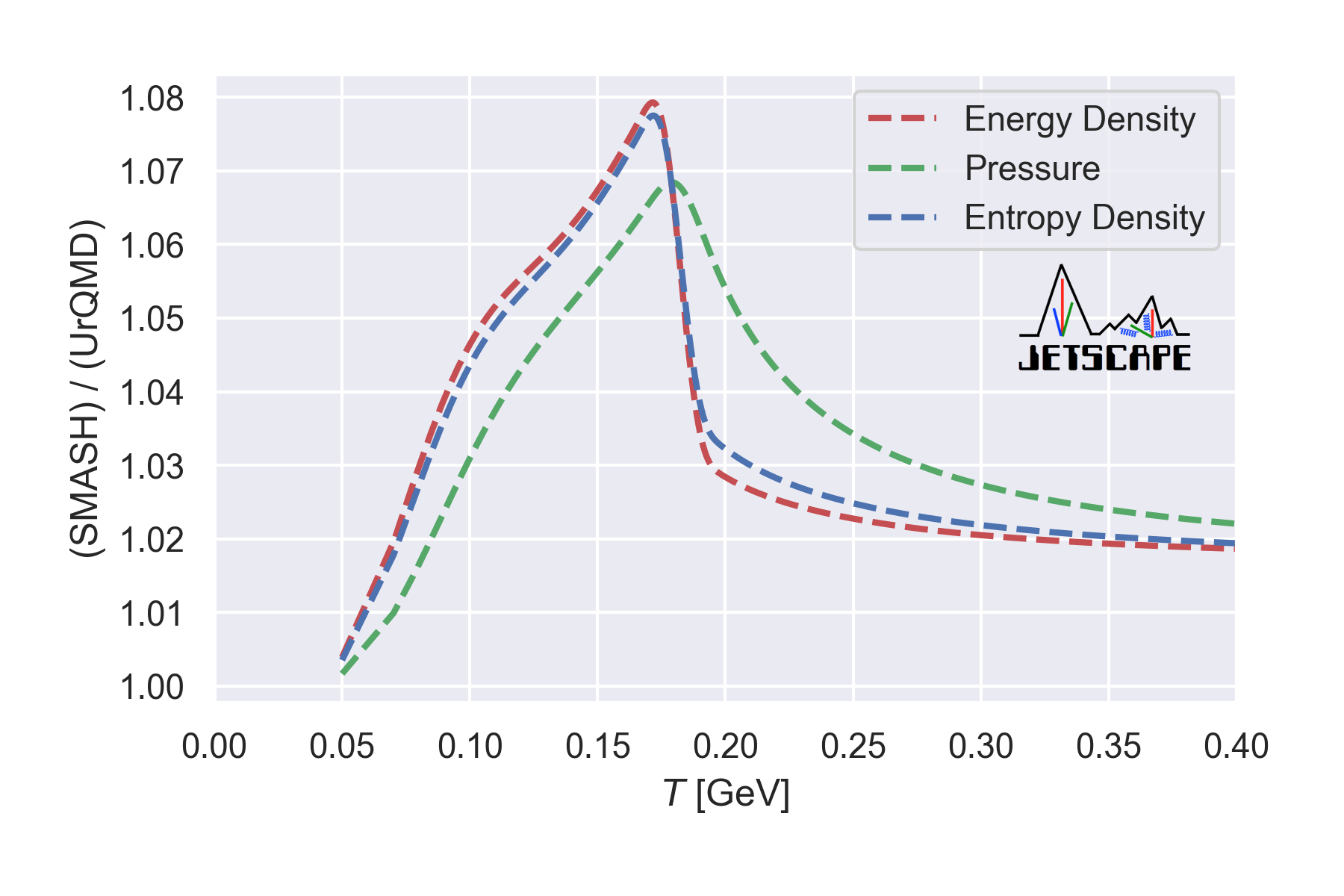}
    \caption{Ratio of \SMASH{} / \URQMD{} at the same temperature for three thermodynamic quantities: energy density (red), equilibrium pressure (green), and entropy density (blue). Each equation of state is constructed by matching the $l$QCD equation of state to a hadron resonance gas matching the list/masses of particles for each code. We see that the disagreement is largest near the region of the switching temperature.}
    \label{fig:eos_ratio}
\end{figure}

Evidently, even with the same matching procedure between the hadron resonance gas and the lattice calculation, the exact content of the hadron resonance gas is important. In the present case, we are interested in two configurations: one used the particle content from \SMASH{}, while the other uses \URQMD{}'s. Both are matched to HotQCD's lattice calculation as described above. The equation of state with the \SMASH{} hadron resonance gas is the one that has been used to perform parameter estimation in this study. We compare the two equations of state in \figs{fig:eos_comp} and~\ref{fig:eos_ratio}. The differences between the two equations of state amount to up to 8\%.

In \Appendix{app:smash}, we compared the predictions of two hybrid models, one model using \SMASH{} as afterburner and the other using \URQMD{}. To obtain such a level of agreement in the observables, it was necessary to use, in the hydrodynamics, equations of state that matched consistently the chosen hadronic transport afterburner. This is consistent with what we see in \fig{fig:eos_comp}: inside the window of particlization temperature, the differences between the equations of state can be larger than $\sim$5\%, and can undeniably produce noticeably different hadronic observables.

We note that this work uses a fixed equation of state which does not parametrize any potential theoretical uncertainties. See Ref.~\cite{Auvinen:2020mpc} for a recent study which includes uncertainty in the lattice-matched equation of state. 

\vfill

\end{appendix}

\bibliography{biblio}

\begin{thebibliography}{177}%
\makeatletter
\providecommand \@ifxundefined [1]{%
 \@ifx{#1\undefined}
}%
\providecommand \@ifnum [1]{%
 \ifnum #1\expandafter \@firstoftwo
 \else \expandafter \@secondoftwo
 \fi
}%
\providecommand \@ifx [1]{%
 \ifx #1\expandafter \@firstoftwo
 \else \expandafter \@secondoftwo
 \fi
}%
\providecommand \natexlab [1]{#1}%
\providecommand \enquote  [1]{``#1''}%
\providecommand \bibnamefont  [1]{#1}%
\providecommand \bibfnamefont [1]{#1}%
\providecommand \citenamefont [1]{#1}%
\providecommand \href@noop [0]{\@secondoftwo}%
\providecommand \href [0]{\begingroup \@sanitize@url \@href}%
\providecommand \@href[1]{\@@startlink{#1}\@@href}%
\providecommand \@@href[1]{\endgroup#1\@@endlink}%
\providecommand \@sanitize@url [0]{\catcode `\\12\catcode `\$12\catcode
  `\&12\catcode `\#12\catcode `\^12\catcode `\_12\catcode `\%12\relax}%
\providecommand \@@startlink[1]{}%
\providecommand \@@endlink[0]{}%
\providecommand \url  [0]{\begingroup\@sanitize@url \@url }%
\providecommand \@url [1]{\endgroup\@href {#1}{\urlprefix }}%
\providecommand \urlprefix  [0]{URL }%
\providecommand \Eprint [0]{\href }%
\providecommand \doibase [0]{http://dx.doi.org/}%
\providecommand \selectlanguage [0]{\@gobble}%
\providecommand \bibinfo  [0]{\@secondoftwo}%
\providecommand \bibfield  [0]{\@secondoftwo}%
\providecommand \translation [1]{[#1]}%
\providecommand \BibitemOpen [0]{}%
\providecommand \bibitemStop [0]{}%
\providecommand \bibitemNoStop [0]{.\EOS\space}%
\providecommand \EOS [0]{\spacefactor3000\relax}%
\providecommand \BibitemShut  [1]{\csname bibitem#1\endcsname}%
\let\auto@bib@innerbib\@empty
\bibitem [{\citenamefont {Borsanyi}\ \emph {et~al.}(2014)\citenamefont
  {Borsanyi}, \citenamefont {Fodor}, \citenamefont {Hoelbling}, \citenamefont
  {Katz}, \citenamefont {Krieg},\ and\ \citenamefont
  {Szabo}}]{Borsanyi:2013bia}%
  \BibitemOpen
  \bibfield  {author} {\bibinfo {author} {\bibfnamefont {S.}~\bibnamefont
  {Borsanyi}}, \bibinfo {author} {\bibfnamefont {Z.}~\bibnamefont {Fodor}},
  \bibinfo {author} {\bibfnamefont {C.}~\bibnamefont {Hoelbling}}, \bibinfo
  {author} {\bibfnamefont {S.~D.}\ \bibnamefont {Katz}}, \bibinfo {author}
  {\bibfnamefont {S.}~\bibnamefont {Krieg}}, \ and\ \bibinfo {author}
  {\bibfnamefont {K.~K.}\ \bibnamefont {Szabo}},\ }\href {\doibase
  10.1016/j.physletb.2014.01.007} {\bibfield  {journal} {\bibinfo  {journal}
  {Phys. Lett.}\ }\textbf {\bibinfo {volume} {B730}},\ \bibinfo {pages} {99}
  (\bibinfo {year} {2014})},\ \Eprint {http://arxiv.org/abs/1309.5258}
  {arXiv:1309.5258 [hep-lat]} \BibitemShut {NoStop}%
\bibitem [{\citenamefont {Bazavov}\ \emph {et~al.}(2014)\citenamefont {Bazavov}
  \emph {et~al.}}]{Bazavov:2014pvz}%
  \BibitemOpen
  \bibfield  {author} {\bibinfo {author} {\bibfnamefont {A.}~\bibnamefont
  {Bazavov}} \emph {et~al.} (\bibinfo {collaboration} {HotQCD}),\ }\href
  {\doibase 10.1103/PhysRevD.90.094503} {\bibfield  {journal} {\bibinfo
  {journal} {Phys. Rev.}\ }\textbf {\bibinfo {volume} {D90}},\ \bibinfo {pages}
  {094503} (\bibinfo {year} {2014})},\ \Eprint {http://arxiv.org/abs/1407.6387}
  {arXiv:1407.6387 [hep-lat]} \BibitemShut {NoStop}%
\bibitem [{\citenamefont {Shen}(2020)}]{Shen:2020gef}%
  \BibitemOpen
  \bibfield  {author} {\bibinfo {author} {\bibfnamefont {C.}~\bibnamefont
  {Shen}},\ }in\ \href@noop {} {\emph {\bibinfo {booktitle} {{28th
  International Conference on Ultrarelativistic Nucleus-Nucleus Collisions}}}}\
  (\bibinfo {year} {2020})\ \Eprint {http://arxiv.org/abs/2001.11858}
  {arXiv:2001.11858 [nucl-th]} \BibitemShut {NoStop}%
\bibitem [{\citenamefont {Bazavov}\ \emph {et~al.}(2019)\citenamefont
  {Bazavov}, \citenamefont {Karsch}, \citenamefont {Mukherjee},\ and\
  \citenamefont {Petreczky}}]{Bazavov:2019lgz}%
  \BibitemOpen
  \bibfield  {author} {\bibinfo {author} {\bibfnamefont {A.}~\bibnamefont
  {Bazavov}}, \bibinfo {author} {\bibfnamefont {F.}~\bibnamefont {Karsch}},
  \bibinfo {author} {\bibfnamefont {S.}~\bibnamefont {Mukherjee}}, \ and\
  \bibinfo {author} {\bibfnamefont {P.}~\bibnamefont {Petreczky}} (\bibinfo
  {collaboration} {USQCD}),\ }\href {\doibase 10.1140/epja/i2019-12922-0}
  {\bibfield  {journal} {\bibinfo  {journal} {Eur. Phys. J. A}\ }\textbf
  {\bibinfo {volume} {55}},\ \bibinfo {pages} {194} (\bibinfo {year} {2019})},\
  \Eprint {http://arxiv.org/abs/1904.09951} {arXiv:1904.09951 [hep-lat]}
  \BibitemShut {NoStop}%
\bibitem [{\citenamefont {Lu}\ and\ \citenamefont {Moore}(2011)}]{Lu:2011df}%
  \BibitemOpen
  \bibfield  {author} {\bibinfo {author} {\bibfnamefont {E.}~\bibnamefont
  {Lu}}\ and\ \bibinfo {author} {\bibfnamefont {G.~D.}\ \bibnamefont {Moore}},\
  }\href {\doibase 10.1103/PhysRevC.83.044901} {\bibfield  {journal} {\bibinfo
  {journal} {Phys. Rev. C}\ }\textbf {\bibinfo {volume} {83}},\ \bibinfo
  {pages} {044901} (\bibinfo {year} {2011})},\ \Eprint
  {http://arxiv.org/abs/1102.0017} {arXiv:1102.0017 [hep-ph]} \BibitemShut
  {NoStop}%
\bibitem [{\citenamefont {Rose}\ \emph {et~al.}(2018)\citenamefont {Rose},
  \citenamefont {Torres-Rincon}, \citenamefont {Schäfer}, \citenamefont
  {Oliinychenko},\ and\ \citenamefont {Petersen}}]{Rose:2017bjz}%
  \BibitemOpen
  \bibfield  {author} {\bibinfo {author} {\bibfnamefont {J.~B.}\ \bibnamefont
  {Rose}}, \bibinfo {author} {\bibfnamefont {J.}~\bibnamefont {Torres-Rincon}},
  \bibinfo {author} {\bibfnamefont {A.}~\bibnamefont {Schäfer}}, \bibinfo
  {author} {\bibfnamefont {D.}~\bibnamefont {Oliinychenko}}, \ and\ \bibinfo
  {author} {\bibfnamefont {H.}~\bibnamefont {Petersen}},\ }\href {\doibase
  10.1103/PhysRevC.97.055204} {\bibfield  {journal} {\bibinfo  {journal} {Phys.
  Rev. C}\ }\textbf {\bibinfo {volume} {97}},\ \bibinfo {pages} {055204}
  (\bibinfo {year} {2018})},\ \Eprint {http://arxiv.org/abs/1709.03826}
  {arXiv:1709.03826 [nucl-th]} \BibitemShut {NoStop}%
\bibitem [{\citenamefont {Rose}\ \emph {et~al.}(2020)\citenamefont {Rose},
  \citenamefont {Torres-Rincon},\ and\ \citenamefont {Elfner}}]{Rose:2020lfc}%
  \BibitemOpen
  \bibfield  {author} {\bibinfo {author} {\bibfnamefont {J.-B.}\ \bibnamefont
  {Rose}}, \bibinfo {author} {\bibfnamefont {J.}~\bibnamefont {Torres-Rincon}},
  \ and\ \bibinfo {author} {\bibfnamefont {H.}~\bibnamefont {Elfner}},\
  }\href@noop {} {\  (\bibinfo {year} {2020})},\ \Eprint
  {http://arxiv.org/abs/2005.03647} {arXiv:2005.03647 [hep-ph]} \BibitemShut
  {NoStop}%
\bibitem [{\citenamefont {Karsch}\ \emph {et~al.}(2008)\citenamefont {Karsch},
  \citenamefont {Kharzeev},\ and\ \citenamefont {Tuchin}}]{Karsch:2007jc}%
  \BibitemOpen
  \bibfield  {author} {\bibinfo {author} {\bibfnamefont {F.}~\bibnamefont
  {Karsch}}, \bibinfo {author} {\bibfnamefont {D.}~\bibnamefont {Kharzeev}}, \
  and\ \bibinfo {author} {\bibfnamefont {K.}~\bibnamefont {Tuchin}},\ }\href
  {\doibase 10.1016/j.physletb.2008.01.080} {\bibfield  {journal} {\bibinfo
  {journal} {Phys. Lett. B}\ }\textbf {\bibinfo {volume} {663}},\ \bibinfo
  {pages} {217} (\bibinfo {year} {2008})},\ \Eprint
  {http://arxiv.org/abs/0711.0914} {arXiv:0711.0914 [hep-ph]} \BibitemShut
  {NoStop}%
\bibitem [{\citenamefont {Noronha-Hostler}\ \emph {et~al.}(2009)\citenamefont
  {Noronha-Hostler}, \citenamefont {Noronha},\ and\ \citenamefont
  {Greiner}}]{NoronhaHostler:2008ju}%
  \BibitemOpen
  \bibfield  {author} {\bibinfo {author} {\bibfnamefont {J.}~\bibnamefont
  {Noronha-Hostler}}, \bibinfo {author} {\bibfnamefont {J.}~\bibnamefont
  {Noronha}}, \ and\ \bibinfo {author} {\bibfnamefont {C.}~\bibnamefont
  {Greiner}},\ }\href {\doibase 10.1103/PhysRevLett.103.172302} {\bibfield
  {journal} {\bibinfo  {journal} {Phys. Rev. Lett.}\ }\textbf {\bibinfo
  {volume} {103}},\ \bibinfo {pages} {172302} (\bibinfo {year} {2009})},\
  \Eprint {http://arxiv.org/abs/0811.1571} {arXiv:0811.1571 [nucl-th]}
  \BibitemShut {NoStop}%
\bibitem [{\citenamefont {Arnold}\ \emph {et~al.}(2006)\citenamefont {Arnold},
  \citenamefont {Dogan},\ and\ \citenamefont {Moore}}]{Arnold:2006fz}%
  \BibitemOpen
  \bibfield  {author} {\bibinfo {author} {\bibfnamefont {P.~B.}\ \bibnamefont
  {Arnold}}, \bibinfo {author} {\bibfnamefont {C.}~\bibnamefont {Dogan}}, \
  and\ \bibinfo {author} {\bibfnamefont {G.~D.}\ \bibnamefont {Moore}},\ }\href
  {\doibase 10.1103/PhysRevD.74.085021} {\bibfield  {journal} {\bibinfo
  {journal} {Phys. Rev. D}\ }\textbf {\bibinfo {volume} {74}},\ \bibinfo
  {pages} {085021} (\bibinfo {year} {2006})},\ \Eprint
  {http://arxiv.org/abs/hep-ph/0608012} {arXiv:hep-ph/0608012} \BibitemShut
  {NoStop}%
\bibitem [{\citenamefont {Ghiglieri}\ \emph
  {et~al.}(2018{\natexlab{a}})\citenamefont {Ghiglieri}, \citenamefont
  {Moore},\ and\ \citenamefont {Teaney}}]{Ghiglieri:2018dib}%
  \BibitemOpen
  \bibfield  {author} {\bibinfo {author} {\bibfnamefont {J.}~\bibnamefont
  {Ghiglieri}}, \bibinfo {author} {\bibfnamefont {G.~D.}\ \bibnamefont
  {Moore}}, \ and\ \bibinfo {author} {\bibfnamefont {D.}~\bibnamefont
  {Teaney}},\ }\href {\doibase 10.1007/JHEP03(2018)179} {\bibfield  {journal}
  {\bibinfo  {journal} {JHEP}\ }\textbf {\bibinfo {volume} {03}},\ \bibinfo
  {pages} {179} (\bibinfo {year} {2018}{\natexlab{a}})},\ \Eprint
  {http://arxiv.org/abs/1802.09535} {arXiv:1802.09535 [hep-ph]} \BibitemShut
  {NoStop}%
\bibitem [{\citenamefont {Heinz}\ and\ \citenamefont
  {Snellings}(2013)}]{Heinz:2013th}%
  \BibitemOpen
  \bibfield  {author} {\bibinfo {author} {\bibfnamefont {U.}~\bibnamefont
  {Heinz}}\ and\ \bibinfo {author} {\bibfnamefont {R.}~\bibnamefont
  {Snellings}},\ }\href {\doibase 10.1146/annurev-nucl-102212-170540}
  {\bibfield  {journal} {\bibinfo  {journal} {Ann. Rev. Nucl. Part. Sci.}\
  }\textbf {\bibinfo {volume} {63}},\ \bibinfo {pages} {123} (\bibinfo {year}
  {2013})},\ \Eprint {http://arxiv.org/abs/1301.2826} {arXiv:1301.2826
  [nucl-th]} \BibitemShut {NoStop}%
\bibitem [{\citenamefont {Gale}\ \emph
  {et~al.}(2013{\natexlab{a}})\citenamefont {Gale}, \citenamefont {Jeon},\ and\
  \citenamefont {Schenke}}]{Gale:2013da}%
  \BibitemOpen
  \bibfield  {author} {\bibinfo {author} {\bibfnamefont {C.}~\bibnamefont
  {Gale}}, \bibinfo {author} {\bibfnamefont {S.}~\bibnamefont {Jeon}}, \ and\
  \bibinfo {author} {\bibfnamefont {B.}~\bibnamefont {Schenke}},\ }\href
  {\doibase 10.1142/S0217751X13400113} {\bibfield  {journal} {\bibinfo
  {journal} {Int. J. Mod. Phys.}\ }\textbf {\bibinfo {volume} {A28}},\ \bibinfo
  {pages} {1340011} (\bibinfo {year} {2013}{\natexlab{a}})},\ \Eprint
  {http://arxiv.org/abs/1301.5893} {arXiv:1301.5893 [nucl-th]} \BibitemShut
  {NoStop}%
\bibitem [{\citenamefont {Derradi~de Souza}\ \emph {et~al.}(2016)\citenamefont
  {Derradi~de Souza}, \citenamefont {Koide},\ and\ \citenamefont
  {Kodama}}]{deSouza:2015ena}%
  \BibitemOpen
  \bibfield  {author} {\bibinfo {author} {\bibfnamefont {R.}~\bibnamefont
  {Derradi~de Souza}}, \bibinfo {author} {\bibfnamefont {T.}~\bibnamefont
  {Koide}}, \ and\ \bibinfo {author} {\bibfnamefont {T.}~\bibnamefont
  {Kodama}},\ }\href {\doibase 10.1016/j.ppnp.2015.09.002} {\bibfield
  {journal} {\bibinfo  {journal} {Prog. Part. Nucl. Phys.}\ }\textbf {\bibinfo
  {volume} {86}},\ \bibinfo {pages} {35} (\bibinfo {year} {2016})},\ \Eprint
  {http://arxiv.org/abs/1506.03863} {arXiv:1506.03863 [nucl-th]} \BibitemShut
  {NoStop}%
\bibitem [{\citenamefont {Schenke}\ \emph {et~al.}(2020)\citenamefont
  {Schenke}, \citenamefont {Shen},\ and\ \citenamefont
  {Tribedy}}]{Schenke:2020mbo}%
  \BibitemOpen
  \bibfield  {author} {\bibinfo {author} {\bibfnamefont {B.}~\bibnamefont
  {Schenke}}, \bibinfo {author} {\bibfnamefont {C.}~\bibnamefont {Shen}}, \
  and\ \bibinfo {author} {\bibfnamefont {P.}~\bibnamefont {Tribedy}},\
  }\href@noop {} {\  (\bibinfo {year} {2020})},\ \Eprint
  {http://arxiv.org/abs/2005.14682} {arXiv:2005.14682 [nucl-th]} \BibitemShut
  {NoStop}%
\bibitem [{\citenamefont {Romatschke}\ and\ \citenamefont
  {Romatschke}(2007)}]{Romatschke:2007mq}%
  \BibitemOpen
  \bibfield  {author} {\bibinfo {author} {\bibfnamefont {P.}~\bibnamefont
  {Romatschke}}\ and\ \bibinfo {author} {\bibfnamefont {U.}~\bibnamefont
  {Romatschke}},\ }\href {\doibase 10.1103/PhysRevLett.99.172301} {\bibfield
  {journal} {\bibinfo  {journal} {Phys. Rev. Lett.}\ }\textbf {\bibinfo
  {volume} {99}},\ \bibinfo {pages} {172301} (\bibinfo {year} {2007})},\
  \Eprint {http://arxiv.org/abs/0706.1522} {arXiv:0706.1522 [nucl-th]}
  \BibitemShut {NoStop}%
\bibitem [{\citenamefont {Song}\ and\ \citenamefont
  {Heinz}(2008{\natexlab{a}})}]{Song:2007fn}%
  \BibitemOpen
  \bibfield  {author} {\bibinfo {author} {\bibfnamefont {H.}~\bibnamefont
  {Song}}\ and\ \bibinfo {author} {\bibfnamefont {U.}~\bibnamefont {Heinz}},\
  }\href {\doibase 10.1016/j.physletb.2007.11.019} {\bibfield  {journal}
  {\bibinfo  {journal} {Phys. Lett. B}\ }\textbf {\bibinfo {volume} {658}},\
  \bibinfo {pages} {279} (\bibinfo {year} {2008}{\natexlab{a}})},\ \Eprint
  {http://arxiv.org/abs/0709.0742} {arXiv:0709.0742 [nucl-th]} \BibitemShut
  {NoStop}%
\bibitem [{\citenamefont {Song}\ and\ \citenamefont
  {Heinz}(2009)}]{Song:2008hj}%
  \BibitemOpen
  \bibfield  {author} {\bibinfo {author} {\bibfnamefont {H.}~\bibnamefont
  {Song}}\ and\ \bibinfo {author} {\bibfnamefont {U.}~\bibnamefont {Heinz}},\
  }in\ \href {\doibase 10.1088/0954-3899/36/6/064033} {\emph {\bibinfo
  {booktitle} {{Strangeness in Quark Matter 2008. Proceedings, International
  Conference, SQM 2008, Beijing, P.R. China, October 5-10, 2008}}}},\ Vol.\
  \bibinfo {volume} {G36}\ (\bibinfo {year} {2009})\ p.\ \bibinfo {pages}
  {064033},\ \Eprint {http://arxiv.org/abs/0812.4274} {arXiv:0812.4274
  [nucl-th]} \BibitemShut {NoStop}%
\bibitem [{\citenamefont {Denicol}\ \emph {et~al.}(2009)\citenamefont
  {Denicol}, \citenamefont {Kodama}, \citenamefont {Koide},\ and\ \citenamefont
  {Mota}}]{Denicol:2009am}%
  \BibitemOpen
  \bibfield  {author} {\bibinfo {author} {\bibfnamefont {G.~S.}\ \bibnamefont
  {Denicol}}, \bibinfo {author} {\bibfnamefont {T.}~\bibnamefont {Kodama}},
  \bibinfo {author} {\bibfnamefont {T.}~\bibnamefont {Koide}}, \ and\ \bibinfo
  {author} {\bibfnamefont {P.}~\bibnamefont {Mota}},\ }\href {\doibase
  10.1103/PhysRevC.80.064901} {\bibfield  {journal} {\bibinfo  {journal} {Phys.
  Rev.}\ }\textbf {\bibinfo {volume} {C80}},\ \bibinfo {pages} {064901}
  (\bibinfo {year} {2009})},\ \Eprint {http://arxiv.org/abs/0903.3595}
  {arXiv:0903.3595 [hep-ph]} \BibitemShut {NoStop}%
\bibitem [{hot(2012)}]{hotQCDwhitepaper}%
  \BibitemOpen
  \href@noop {} {\enquote {\bibinfo {title} {Hot and {D}ense {QCD} {W}hite
  {P}aper},}\ }\bibinfo {howpublished}
  {\url{http://www.bnl.gov/npp/docs/Bass_RHI_WP_final.pdf}} (\bibinfo {year}
  {2012})\BibitemShut {NoStop}%
\bibitem [{\citenamefont {Akiba}\ \emph {et~al.}(2015)\citenamefont {Akiba}
  \emph {et~al.}}]{Akiba:2015jwa}%
  \BibitemOpen
  \bibfield  {author} {\bibinfo {author} {\bibfnamefont {Y.}~\bibnamefont
  {Akiba}} \emph {et~al.},\ }\href@noop {} {\  (\bibinfo {year} {2015})},\
  \Eprint {http://arxiv.org/abs/1502.02730} {arXiv:1502.02730 [nucl-ex]}
  \BibitemShut {NoStop}%
\bibitem [{\citenamefont {Petersen}\ \emph {et~al.}(2011)\citenamefont
  {Petersen}, \citenamefont {Coleman-Smith}, \citenamefont {Bass},\ and\
  \citenamefont {Wolpert}}]{Petersen:2010zt}%
  \BibitemOpen
  \bibfield  {author} {\bibinfo {author} {\bibfnamefont {H.}~\bibnamefont
  {Petersen}}, \bibinfo {author} {\bibfnamefont {C.}~\bibnamefont
  {Coleman-Smith}}, \bibinfo {author} {\bibfnamefont {S.~A.}\ \bibnamefont
  {Bass}}, \ and\ \bibinfo {author} {\bibfnamefont {R.}~\bibnamefont
  {Wolpert}},\ }\href {\doibase 10.1088/0954-3899/38/4/045102} {\bibfield
  {journal} {\bibinfo  {journal} {J. Phys. G}\ }\textbf {\bibinfo {volume}
  {38}},\ \bibinfo {pages} {045102} (\bibinfo {year} {2011})},\ \Eprint
  {http://arxiv.org/abs/1012.4629} {arXiv:1012.4629 [nucl-th]} \BibitemShut
  {NoStop}%
\bibitem [{\citenamefont {Novak}\ \emph {et~al.}(2014)\citenamefont {Novak},
  \citenamefont {Novak}, \citenamefont {Pratt}, \citenamefont {Vredevoogd},
  \citenamefont {Coleman-Smith},\ and\ \citenamefont
  {Wolpert}}]{Novak:2013bqa}%
  \BibitemOpen
  \bibfield  {author} {\bibinfo {author} {\bibfnamefont {J.}~\bibnamefont
  {Novak}}, \bibinfo {author} {\bibfnamefont {K.}~\bibnamefont {Novak}},
  \bibinfo {author} {\bibfnamefont {S.}~\bibnamefont {Pratt}}, \bibinfo
  {author} {\bibfnamefont {J.}~\bibnamefont {Vredevoogd}}, \bibinfo {author}
  {\bibfnamefont {C.}~\bibnamefont {Coleman-Smith}}, \ and\ \bibinfo {author}
  {\bibfnamefont {R.}~\bibnamefont {Wolpert}},\ }\href {\doibase
  10.1103/PhysRevC.89.034917} {\bibfield  {journal} {\bibinfo  {journal} {Phys.
  Rev.}\ }\textbf {\bibinfo {volume} {C89}},\ \bibinfo {pages} {034917}
  (\bibinfo {year} {2014})},\ \Eprint {http://arxiv.org/abs/1303.5769}
  {arXiv:1303.5769 [nucl-th]} \BibitemShut {NoStop}%
\bibitem [{\citenamefont {Sangaline}\ and\ \citenamefont
  {Pratt}(2016)}]{Sangaline:2015isa}%
  \BibitemOpen
  \bibfield  {author} {\bibinfo {author} {\bibfnamefont {E.}~\bibnamefont
  {Sangaline}}\ and\ \bibinfo {author} {\bibfnamefont {S.}~\bibnamefont
  {Pratt}},\ }\href {\doibase 10.1103/PhysRevC.93.024908} {\bibfield  {journal}
  {\bibinfo  {journal} {Phys. Rev.}\ }\textbf {\bibinfo {volume} {C93}},\
  \bibinfo {pages} {024908} (\bibinfo {year} {2016})},\ \Eprint
  {http://arxiv.org/abs/1508.07017} {arXiv:1508.07017 [nucl-th]} \BibitemShut
  {NoStop}%
\bibitem [{\citenamefont {Bernhard}\ \emph {et~al.}(2015)\citenamefont
  {Bernhard}, \citenamefont {Marcy}, \citenamefont {Coleman-Smith},
  \citenamefont {Huzurbazar}, \citenamefont {Wolpert},\ and\ \citenamefont
  {Bass}}]{Bernhard:2015hxa}%
  \BibitemOpen
  \bibfield  {author} {\bibinfo {author} {\bibfnamefont {J.~E.}\ \bibnamefont
  {Bernhard}}, \bibinfo {author} {\bibfnamefont {P.~W.}\ \bibnamefont {Marcy}},
  \bibinfo {author} {\bibfnamefont {C.~E.}\ \bibnamefont {Coleman-Smith}},
  \bibinfo {author} {\bibfnamefont {S.}~\bibnamefont {Huzurbazar}}, \bibinfo
  {author} {\bibfnamefont {R.~L.}\ \bibnamefont {Wolpert}}, \ and\ \bibinfo
  {author} {\bibfnamefont {S.~A.}\ \bibnamefont {Bass}},\ }\href {\doibase
  10.1103/PhysRevC.91.054910} {\bibfield  {journal} {\bibinfo  {journal} {Phys.
  Rev. C}\ }\textbf {\bibinfo {volume} {91}},\ \bibinfo {pages} {054910}
  (\bibinfo {year} {2015})},\ \Eprint {http://arxiv.org/abs/1502.00339}
  {arXiv:1502.00339 [nucl-th]} \BibitemShut {NoStop}%
\bibitem [{\citenamefont {Bernhard}\ \emph {et~al.}(2016)\citenamefont
  {Bernhard}, \citenamefont {Moreland}, \citenamefont {Bass}, \citenamefont
  {Liu},\ and\ \citenamefont {Heinz}}]{Bernhard:2016tnd}%
  \BibitemOpen
  \bibfield  {author} {\bibinfo {author} {\bibfnamefont {J.~E.}\ \bibnamefont
  {Bernhard}}, \bibinfo {author} {\bibfnamefont {J.~S.}\ \bibnamefont
  {Moreland}}, \bibinfo {author} {\bibfnamefont {S.~A.}\ \bibnamefont {Bass}},
  \bibinfo {author} {\bibfnamefont {J.}~\bibnamefont {Liu}}, \ and\ \bibinfo
  {author} {\bibfnamefont {U.}~\bibnamefont {Heinz}},\ }\href {\doibase
  10.1103/PhysRevC.94.024907} {\bibfield  {journal} {\bibinfo  {journal} {Phys.
  Rev.}\ }\textbf {\bibinfo {volume} {C94}},\ \bibinfo {pages} {024907}
  (\bibinfo {year} {2016})},\ \Eprint {http://arxiv.org/abs/1605.03954}
  {arXiv:1605.03954 [nucl-th]} \BibitemShut {NoStop}%
\bibitem [{\citenamefont {Bernhard}\ \emph {et~al.}(2019)\citenamefont
  {Bernhard}, \citenamefont {Moreland},\ and\ \citenamefont
  {Bass}}]{Bernhard:2019bmu}%
  \BibitemOpen
  \bibfield  {author} {\bibinfo {author} {\bibfnamefont {J.~E.}\ \bibnamefont
  {Bernhard}}, \bibinfo {author} {\bibfnamefont {J.~S.}\ \bibnamefont
  {Moreland}}, \ and\ \bibinfo {author} {\bibfnamefont {S.~A.}\ \bibnamefont
  {Bass}},\ }\href {\doibase 10.1038/s41567-019-0611-8} {\bibfield  {journal}
  {\bibinfo  {journal} {Nature Phys.}\ }\textbf {\bibinfo {volume} {15}},\
  \bibinfo {pages} {1113} (\bibinfo {year} {2019})}\BibitemShut {NoStop}%
\bibitem [{\citenamefont {Everett}\ \emph {et~al.}(2020)\citenamefont {Everett}
  \emph {et~al.}}]{Everett:2020xug}%
  \BibitemOpen
  \bibfield  {author} {\bibinfo {author} {\bibfnamefont {D.}~\bibnamefont
  {Everett}} \emph {et~al.} (\bibinfo {collaboration} {JETSCAPE}),\ }\href@noop
  {} {\  (\bibinfo {year} {2020})},\ \Eprint {http://arxiv.org/abs/2011.01430}
  {arXiv:2011.01430 [hep-ph]} \BibitemShut {NoStop}%
\bibitem [{\citenamefont {Song}\ \emph
  {et~al.}(2011{\natexlab{a}})\citenamefont {Song}, \citenamefont {Bass},
  \citenamefont {Heinz}, \citenamefont {Hirano},\ and\ \citenamefont
  {Shen}}]{Song:2010mg}%
  \BibitemOpen
  \bibfield  {author} {\bibinfo {author} {\bibfnamefont {H.}~\bibnamefont
  {Song}}, \bibinfo {author} {\bibfnamefont {S.~A.}\ \bibnamefont {Bass}},
  \bibinfo {author} {\bibfnamefont {U.}~\bibnamefont {Heinz}}, \bibinfo
  {author} {\bibfnamefont {T.}~\bibnamefont {Hirano}}, \ and\ \bibinfo {author}
  {\bibfnamefont {C.}~\bibnamefont {Shen}},\ }\href {\doibase
  10.1103/PhysRevLett.106.192301} {\bibfield  {journal} {\bibinfo  {journal}
  {Phys. Rev. Lett.}\ }\textbf {\bibinfo {volume} {106}},\ \bibinfo {pages}
  {192301} (\bibinfo {year} {2011}{\natexlab{a}})},\ \bibinfo {note} {[Erratum:
  Phys.Rev.Lett. 109, 139904 (2012)]},\ \Eprint
  {http://arxiv.org/abs/1011.2783} {arXiv:1011.2783 [nucl-th]} \BibitemShut
  {NoStop}%
\bibitem [{\citenamefont {Nijs}\ \emph
  {et~al.}(2020{\natexlab{a}})\citenamefont {Nijs}, \citenamefont {van~der
  Schee}, \citenamefont {G\"ursoy},\ and\ \citenamefont
  {Snellings}}]{Nijs:2020ors}%
  \BibitemOpen
  \bibfield  {author} {\bibinfo {author} {\bibfnamefont {G.}~\bibnamefont
  {Nijs}}, \bibinfo {author} {\bibfnamefont {W.}~\bibnamefont {van~der Schee}},
  \bibinfo {author} {\bibfnamefont {U.}~\bibnamefont {G\"ursoy}}, \ and\
  \bibinfo {author} {\bibfnamefont {R.}~\bibnamefont {Snellings}},\ }\href@noop
  {} {\  (\bibinfo {year} {2020}{\natexlab{a}})},\ \Eprint
  {http://arxiv.org/abs/2010.15130} {arXiv:2010.15130 [nucl-th]} \BibitemShut
  {NoStop}%
\bibitem [{\citenamefont {Nijs}\ \emph
  {et~al.}(2020{\natexlab{b}})\citenamefont {Nijs}, \citenamefont {van~der
  Schee}, \citenamefont {G\"ursoy},\ and\ \citenamefont
  {Snellings}}]{Nijs:2020roc}%
  \BibitemOpen
  \bibfield  {author} {\bibinfo {author} {\bibfnamefont {G.}~\bibnamefont
  {Nijs}}, \bibinfo {author} {\bibfnamefont {W.}~\bibnamefont {van~der Schee}},
  \bibinfo {author} {\bibfnamefont {U.}~\bibnamefont {G\"ursoy}}, \ and\
  \bibinfo {author} {\bibfnamefont {R.}~\bibnamefont {Snellings}},\ }\href@noop
  {} {\  (\bibinfo {year} {2020}{\natexlab{b}})},\ \Eprint
  {http://arxiv.org/abs/2010.15134} {arXiv:2010.15134 [nucl-th]} \BibitemShut
  {NoStop}%
\bibitem [{\citenamefont {Putschke}\ \emph {et~al.}(2019)\citenamefont
  {Putschke} \emph {et~al.}}]{Putschke:2019yrg}%
  \BibitemOpen
  \bibfield  {author} {\bibinfo {author} {\bibfnamefont {J.~H.}\ \bibnamefont
  {Putschke}} \emph {et~al.} (\bibinfo {collaboration} {JETSCAPE}),\
  }\href@noop {} {\  (\bibinfo {year} {2019})},\ \Eprint
  {http://arxiv.org/abs/1903.07706} {arXiv:1903.07706 [nucl-th]} \BibitemShut
  {NoStop}%
\bibitem [{\citenamefont {Sivia}\ and\ \citenamefont
  {Skilling}(2006)}]{Sivia2006}%
  \BibitemOpen
  \bibfield  {author} {\bibinfo {author} {\bibfnamefont {D.~S.}\ \bibnamefont
  {Sivia}}\ and\ \bibinfo {author} {\bibfnamefont {J.}~\bibnamefont
  {Skilling}},\ }\href@noop {} {\emph {\bibinfo {title} {{Data Analysis - A
  Bayesian Tutorial}}}},\ \bibinfo {edition} {2nd}\ ed.,\ Oxford Science
  Publications\ (\bibinfo  {publisher} {Oxford University Press},\ \bibinfo
  {year} {2006})\BibitemShut {NoStop}%
\bibitem [{\citenamefont {Bass}\ and\ \citenamefont
  {Dumitru}(2000)}]{Bass:2000ib}%
  \BibitemOpen
  \bibfield  {author} {\bibinfo {author} {\bibfnamefont {S.}~\bibnamefont
  {Bass}}\ and\ \bibinfo {author} {\bibfnamefont {A.}~\bibnamefont {Dumitru}},\
  }\href {\doibase 10.1103/PhysRevC.61.064909} {\bibfield  {journal} {\bibinfo
  {journal} {Phys. Rev. C}\ }\textbf {\bibinfo {volume} {61}},\ \bibinfo
  {pages} {064909} (\bibinfo {year} {2000})},\ \Eprint
  {http://arxiv.org/abs/nucl-th/0001033} {arXiv:nucl-th/0001033} \BibitemShut
  {NoStop}%
\bibitem [{\citenamefont {Nonaka}\ and\ \citenamefont
  {Bass}(2007)}]{Nonaka:2006yn}%
  \BibitemOpen
  \bibfield  {author} {\bibinfo {author} {\bibfnamefont {C.}~\bibnamefont
  {Nonaka}}\ and\ \bibinfo {author} {\bibfnamefont {S.~A.}\ \bibnamefont
  {Bass}},\ }\href {\doibase 10.1103/PhysRevC.75.014902} {\bibfield  {journal}
  {\bibinfo  {journal} {Phys. Rev.}\ }\textbf {\bibinfo {volume} {C75}},\
  \bibinfo {pages} {014902} (\bibinfo {year} {2007})},\ \Eprint
  {http://arxiv.org/abs/nucl-th/0607018} {arXiv:nucl-th/0607018 [nucl-th]}
  \BibitemShut {NoStop}%
\bibitem [{\citenamefont {Hirano}\ \emph {et~al.}(2008)\citenamefont {Hirano},
  \citenamefont {Heinz}, \citenamefont {Kharzeev}, \citenamefont {Lacey},\ and\
  \citenamefont {Nara}}]{Hirano:2007ei}%
  \BibitemOpen
  \bibfield  {author} {\bibinfo {author} {\bibfnamefont {T.}~\bibnamefont
  {Hirano}}, \bibinfo {author} {\bibfnamefont {U.}~\bibnamefont {Heinz}},
  \bibinfo {author} {\bibfnamefont {D.}~\bibnamefont {Kharzeev}}, \bibinfo
  {author} {\bibfnamefont {R.}~\bibnamefont {Lacey}}, \ and\ \bibinfo {author}
  {\bibfnamefont {Y.}~\bibnamefont {Nara}},\ }\href {\doibase
  10.1103/PhysRevC.77.044909} {\bibfield  {journal} {\bibinfo  {journal} {Phys.
  Rev.}\ }\textbf {\bibinfo {volume} {C77}},\ \bibinfo {pages} {044909}
  (\bibinfo {year} {2008})},\ \Eprint {http://arxiv.org/abs/0710.5795}
  {arXiv:0710.5795 [nucl-th]} \BibitemShut {NoStop}%
\bibitem [{\citenamefont {Petersen}\ \emph {et~al.}(2008)\citenamefont
  {Petersen}, \citenamefont {Steinheimer}, \citenamefont {Burau}, \citenamefont
  {Bleicher},\ and\ \citenamefont {Stocker}}]{Petersen:2008dd}%
  \BibitemOpen
  \bibfield  {author} {\bibinfo {author} {\bibfnamefont {H.}~\bibnamefont
  {Petersen}}, \bibinfo {author} {\bibfnamefont {J.}~\bibnamefont
  {Steinheimer}}, \bibinfo {author} {\bibfnamefont {G.}~\bibnamefont {Burau}},
  \bibinfo {author} {\bibfnamefont {M.}~\bibnamefont {Bleicher}}, \ and\
  \bibinfo {author} {\bibfnamefont {H.}~\bibnamefont {Stocker}},\ }\href
  {\doibase 10.1103/PhysRevC.78.044901} {\bibfield  {journal} {\bibinfo
  {journal} {Phys. Rev.}\ }\textbf {\bibinfo {volume} {C78}},\ \bibinfo {pages}
  {044901} (\bibinfo {year} {2008})},\ \Eprint {http://arxiv.org/abs/0806.1695}
  {arXiv:0806.1695 [nucl-th]} \BibitemShut {NoStop}%
\bibitem [{\citenamefont {Song}\ \emph
  {et~al.}(2011{\natexlab{b}})\citenamefont {Song}, \citenamefont {Bass},\ and\
  \citenamefont {Heinz}}]{Song:2010aq}%
  \BibitemOpen
  \bibfield  {author} {\bibinfo {author} {\bibfnamefont {H.}~\bibnamefont
  {Song}}, \bibinfo {author} {\bibfnamefont {S.~A.}\ \bibnamefont {Bass}}, \
  and\ \bibinfo {author} {\bibfnamefont {U.}~\bibnamefont {Heinz}},\ }\href
  {\doibase 10.1103/PhysRevC.83.024912} {\bibfield  {journal} {\bibinfo
  {journal} {Phys. Rev.}\ }\textbf {\bibinfo {volume} {C83}},\ \bibinfo {pages}
  {024912} (\bibinfo {year} {2011}{\natexlab{b}})},\ \Eprint
  {http://arxiv.org/abs/1012.0555} {arXiv:1012.0555 [nucl-th]} \BibitemShut
  {NoStop}%
\bibitem [{\citenamefont {Heinz}\ \emph {et~al.}(2012)\citenamefont {Heinz},
  \citenamefont {Shen},\ and\ \citenamefont {Song}}]{Heinz:2011kt}%
  \BibitemOpen
  \bibfield  {author} {\bibinfo {author} {\bibfnamefont {U.}~\bibnamefont
  {Heinz}}, \bibinfo {author} {\bibfnamefont {C.}~\bibnamefont {Shen}}, \ and\
  \bibinfo {author} {\bibfnamefont {H.}~\bibnamefont {Song}},\ }\href {\doibase
  10.1063/1.3700674} {\bibfield  {journal} {\bibinfo  {journal} {AIP Conf.
  Proc.}\ }\textbf {\bibinfo {volume} {1441}},\ \bibinfo {pages} {766}
  (\bibinfo {year} {2012})},\ \Eprint {http://arxiv.org/abs/1108.5323}
  {arXiv:1108.5323 [nucl-th]} \BibitemShut {NoStop}%
\bibitem [{\citenamefont {Song}\ \emph {et~al.}(2014)\citenamefont {Song},
  \citenamefont {Bass},\ and\ \citenamefont {Heinz}}]{Song:2013qma}%
  \BibitemOpen
  \bibfield  {author} {\bibinfo {author} {\bibfnamefont {H.}~\bibnamefont
  {Song}}, \bibinfo {author} {\bibfnamefont {S.}~\bibnamefont {Bass}}, \ and\
  \bibinfo {author} {\bibfnamefont {U.}~\bibnamefont {Heinz}},\ }\href
  {\doibase 10.1103/PhysRevC.89.034919} {\bibfield  {journal} {\bibinfo
  {journal} {Phys. Rev.}\ }\textbf {\bibinfo {volume} {C89}},\ \bibinfo {pages}
  {034919} (\bibinfo {year} {2014})},\ \Eprint {http://arxiv.org/abs/1311.0157}
  {arXiv:1311.0157 [nucl-th]} \BibitemShut {NoStop}%
\bibitem [{\citenamefont {Zhu}\ \emph {et~al.}(2015)\citenamefont {Zhu},
  \citenamefont {Meng}, \citenamefont {Song},\ and\ \citenamefont
  {Liu}}]{Zhu:2015dfa}%
  \BibitemOpen
  \bibfield  {author} {\bibinfo {author} {\bibfnamefont {X.}~\bibnamefont
  {Zhu}}, \bibinfo {author} {\bibfnamefont {F.}~\bibnamefont {Meng}}, \bibinfo
  {author} {\bibfnamefont {H.}~\bibnamefont {Song}}, \ and\ \bibinfo {author}
  {\bibfnamefont {Y.-X.}\ \bibnamefont {Liu}},\ }\href {\doibase
  10.1103/PhysRevC.91.034904} {\bibfield  {journal} {\bibinfo  {journal} {Phys.
  Rev.}\ }\textbf {\bibinfo {volume} {C91}},\ \bibinfo {pages} {034904}
  (\bibinfo {year} {2015})},\ \Eprint {http://arxiv.org/abs/1501.03286}
  {arXiv:1501.03286 [nucl-th]} \BibitemShut {NoStop}%
\bibitem [{\citenamefont {Ryu}\ \emph {et~al.}(2018)\citenamefont {Ryu},
  \citenamefont {Paquet}, \citenamefont {Shen}, \citenamefont {Denicol},
  \citenamefont {Schenke}, \citenamefont {Jeon},\ and\ \citenamefont
  {Gale}}]{Ryu:2017qzn}%
  \BibitemOpen
  \bibfield  {author} {\bibinfo {author} {\bibfnamefont {S.}~\bibnamefont
  {Ryu}}, \bibinfo {author} {\bibfnamefont {J.-F.}\ \bibnamefont {Paquet}},
  \bibinfo {author} {\bibfnamefont {C.}~\bibnamefont {Shen}}, \bibinfo {author}
  {\bibfnamefont {G.}~\bibnamefont {Denicol}}, \bibinfo {author} {\bibfnamefont
  {B.}~\bibnamefont {Schenke}}, \bibinfo {author} {\bibfnamefont
  {S.}~\bibnamefont {Jeon}}, \ and\ \bibinfo {author} {\bibfnamefont
  {C.}~\bibnamefont {Gale}},\ }\href {\doibase 10.1103/PhysRevC.97.034910}
  {\bibfield  {journal} {\bibinfo  {journal} {Phys. Rev.}\ }\textbf {\bibinfo
  {volume} {C97}},\ \bibinfo {pages} {034910} (\bibinfo {year} {2018})},\
  \Eprint {http://arxiv.org/abs/1704.04216} {arXiv:1704.04216 [nucl-th]}
  \BibitemShut {NoStop}%
\bibitem [{\citenamefont {Vardanyan}\ \emph {et~al.}(2011)\citenamefont
  {Vardanyan}, \citenamefont {Trotta},\ and\ \citenamefont
  {Silk}}]{Vardanyan_2011}%
  \BibitemOpen
  \bibfield  {author} {\bibinfo {author} {\bibfnamefont {M.}~\bibnamefont
  {Vardanyan}}, \bibinfo {author} {\bibfnamefont {R.}~\bibnamefont {Trotta}}, \
  and\ \bibinfo {author} {\bibfnamefont {J.}~\bibnamefont {Silk}},\ }\href
  {\doibase 10.1111/j.1745-3933.2011.01040.x} {\bibfield  {journal} {\bibinfo
  {journal} {Monthly Notices of the Royal Astronomical Society: Letters}\
  }\textbf {\bibinfo {volume} {413}},\ \bibinfo {pages} {L91–L95} (\bibinfo
  {year} {2011})}\BibitemShut {NoStop}%
\bibitem [{\citenamefont {Moreland}\ \emph {et~al.}(2015)\citenamefont
  {Moreland}, \citenamefont {Bernhard},\ and\ \citenamefont
  {Bass}}]{Moreland:2014oya}%
  \BibitemOpen
  \bibfield  {author} {\bibinfo {author} {\bibfnamefont {J.~S.}\ \bibnamefont
  {Moreland}}, \bibinfo {author} {\bibfnamefont {J.~E.}\ \bibnamefont
  {Bernhard}}, \ and\ \bibinfo {author} {\bibfnamefont {S.~A.}\ \bibnamefont
  {Bass}},\ }\href {\doibase 10.1103/PhysRevC.92.011901} {\bibfield  {journal}
  {\bibinfo  {journal} {Phys. Rev.}\ }\textbf {\bibinfo {volume} {C92}},\
  \bibinfo {pages} {011901} (\bibinfo {year} {2015})},\ \Eprint
  {http://arxiv.org/abs/1412.4708} {arXiv:1412.4708 [nucl-th]} \BibitemShut
  {NoStop}%
\bibitem [{tre()}]{trento_code}%
  \BibitemOpen
  \href@noop {} {}\bibinfo {howpublished}
  {\url{https://github.com/Duke-QCD/trento.git}}\BibitemShut {NoStop}%
\bibitem [{\citenamefont {Liu}\ \emph {et~al.}(2015)\citenamefont {Liu},
  \citenamefont {Shen},\ and\ \citenamefont {Heinz}}]{Liu:2015nwa}%
  \BibitemOpen
  \bibfield  {author} {\bibinfo {author} {\bibfnamefont {J.}~\bibnamefont
  {Liu}}, \bibinfo {author} {\bibfnamefont {C.}~\bibnamefont {Shen}}, \ and\
  \bibinfo {author} {\bibfnamefont {U.}~\bibnamefont {Heinz}},\ }\href
  {\doibase 10.1103/PhysRevC.92.049904, 10.1103/PhysRevC.91.064906} {\bibfield
  {journal} {\bibinfo  {journal} {Phys. Rev.}\ }\textbf {\bibinfo {volume}
  {C91}},\ \bibinfo {pages} {064906} (\bibinfo {year} {2015})},\ \bibinfo
  {note} {[Erratum: Phys. Rev.C92,no.4,049904(2015)]},\ \Eprint
  {http://arxiv.org/abs/1504.02160} {arXiv:1504.02160 [nucl-th]} \BibitemShut
  {NoStop}%
\bibitem [{\citenamefont {Broniowski}\ \emph {et~al.}(2009)\citenamefont
  {Broniowski}, \citenamefont {Florkowski}, \citenamefont {Chojnacki},\ and\
  \citenamefont {Kisiel}}]{Broniowski:2008qk}%
  \BibitemOpen
  \bibfield  {author} {\bibinfo {author} {\bibfnamefont {W.}~\bibnamefont
  {Broniowski}}, \bibinfo {author} {\bibfnamefont {W.}~\bibnamefont
  {Florkowski}}, \bibinfo {author} {\bibfnamefont {M.}~\bibnamefont
  {Chojnacki}}, \ and\ \bibinfo {author} {\bibfnamefont {A.}~\bibnamefont
  {Kisiel}},\ }\href {\doibase 10.1103/PhysRevC.80.034902} {\bibfield
  {journal} {\bibinfo  {journal} {Phys. Rev.}\ }\textbf {\bibinfo {volume}
  {C80}},\ \bibinfo {pages} {034902} (\bibinfo {year} {2009})},\ \Eprint
  {http://arxiv.org/abs/0812.3393} {arXiv:0812.3393 [nucl-th]} \BibitemShut
  {NoStop}%
\bibitem [{fs_()}]{fs_code}%
  \BibitemOpen
  \href@noop {} {}\bibinfo {howpublished}
  {\url{https://github.com/derekeverett/freestream-milne}}\BibitemShut
  {NoStop}%
\bibitem [{\citenamefont {Schenke}\ \emph {et~al.}(2010)\citenamefont
  {Schenke}, \citenamefont {Jeon},\ and\ \citenamefont
  {Gale}}]{Schenke:2010nt}%
  \BibitemOpen
  \bibfield  {author} {\bibinfo {author} {\bibfnamefont {B.}~\bibnamefont
  {Schenke}}, \bibinfo {author} {\bibfnamefont {S.}~\bibnamefont {Jeon}}, \
  and\ \bibinfo {author} {\bibfnamefont {C.}~\bibnamefont {Gale}},\ }\href
  {\doibase 10.1103/PhysRevC.82.014903} {\bibfield  {journal} {\bibinfo
  {journal} {Phys. Rev.}\ }\textbf {\bibinfo {volume} {C82}},\ \bibinfo {pages}
  {014903} (\bibinfo {year} {2010})},\ \Eprint {http://arxiv.org/abs/1004.1408}
  {arXiv:1004.1408 [hep-ph]} \BibitemShut {NoStop}%
\bibitem [{\citenamefont {Schenke}\ \emph {et~al.}(2011)\citenamefont
  {Schenke}, \citenamefont {Jeon},\ and\ \citenamefont
  {Gale}}]{Schenke:2010rr}%
  \BibitemOpen
  \bibfield  {author} {\bibinfo {author} {\bibfnamefont {B.}~\bibnamefont
  {Schenke}}, \bibinfo {author} {\bibfnamefont {S.}~\bibnamefont {Jeon}}, \
  and\ \bibinfo {author} {\bibfnamefont {C.}~\bibnamefont {Gale}},\ }\href
  {\doibase 10.1103/PhysRevLett.106.042301} {\bibfield  {journal} {\bibinfo
  {journal} {Phys. Rev. Lett.}\ }\textbf {\bibinfo {volume} {106}},\ \bibinfo
  {pages} {042301} (\bibinfo {year} {2011})},\ \Eprint
  {http://arxiv.org/abs/1009.3244} {arXiv:1009.3244 [hep-ph]} \BibitemShut
  {NoStop}%
\bibitem [{\citenamefont {Paquet}\ \emph {et~al.}(2016)\citenamefont {Paquet},
  \citenamefont {Shen}, \citenamefont {Denicol}, \citenamefont {Luzum},
  \citenamefont {Schenke}, \citenamefont {Jeon},\ and\ \citenamefont
  {Gale}}]{Paquet:2015lta}%
  \BibitemOpen
  \bibfield  {author} {\bibinfo {author} {\bibfnamefont {J.-F.}\ \bibnamefont
  {Paquet}}, \bibinfo {author} {\bibfnamefont {C.}~\bibnamefont {Shen}},
  \bibinfo {author} {\bibfnamefont {G.~S.}\ \bibnamefont {Denicol}}, \bibinfo
  {author} {\bibfnamefont {M.}~\bibnamefont {Luzum}}, \bibinfo {author}
  {\bibfnamefont {B.}~\bibnamefont {Schenke}}, \bibinfo {author} {\bibfnamefont
  {S.}~\bibnamefont {Jeon}}, \ and\ \bibinfo {author} {\bibfnamefont
  {C.}~\bibnamefont {Gale}},\ }\href {\doibase 10.1103/PhysRevC.93.044906}
  {\bibfield  {journal} {\bibinfo  {journal} {Phys. Rev.}\ }\textbf {\bibinfo
  {volume} {C93}},\ \bibinfo {pages} {044906} (\bibinfo {year} {2016})},\
  \Eprint {http://arxiv.org/abs/1509.06738} {arXiv:1509.06738 [hep-ph]}
  \BibitemShut {NoStop}%
\bibitem [{\citenamefont {{Kurganov}}\ and\ \citenamefont
  {{Tadmor}}(2000)}]{2000JCoPh.160..241K}%
  \BibitemOpen
  \bibfield  {author} {\bibinfo {author} {\bibfnamefont {A.}~\bibnamefont
  {{Kurganov}}}\ and\ \bibinfo {author} {\bibfnamefont {E.}~\bibnamefont
  {{Tadmor}}},\ }\href {\doibase 10.1006/jcph.2000.6459} {\bibfield  {journal}
  {\bibinfo  {journal} {Journal of Computational Physics}\ }\textbf {\bibinfo
  {volume} {160}},\ \bibinfo {pages} {241} (\bibinfo {year}
  {2000})}\BibitemShut {NoStop}%
\bibitem [{\citenamefont {Bernhard}(4 19)}]{Bernhard:2018hnz}%
  \BibitemOpen
  \bibfield  {author} {\bibinfo {author} {\bibfnamefont {J.~E.}\ \bibnamefont
  {Bernhard}},\ }\emph {\bibinfo {title} {{Bayesian parameter estimation for
  relativistic heavy-ion collisions}}},\ \href@noop {} {Ph.D. thesis},\
  \bibinfo  {school} {Duke U.} (\bibinfo {year} {2018-04-19}),\ \Eprint
  {http://arxiv.org/abs/1804.06469} {arXiv:1804.06469 [nucl-th]} \BibitemShut
  {NoStop}%
\bibitem [{eos()}]{eos_code}%
  \BibitemOpen
  \href@noop {} {}\bibinfo {howpublished}
  {\url{https://github.com/j-f-paquet/eos_maker}}\BibitemShut {NoStop}%
\bibitem [{\citenamefont {McNelis}\ \emph {et~al.}(2021)\citenamefont
  {McNelis}, \citenamefont {Everett},\ and\ \citenamefont
  {Heinz}}]{McNelis:2019auj}%
  \BibitemOpen
  \bibfield  {author} {\bibinfo {author} {\bibfnamefont {M.}~\bibnamefont
  {McNelis}}, \bibinfo {author} {\bibfnamefont {D.}~\bibnamefont {Everett}}, \
  and\ \bibinfo {author} {\bibfnamefont {U.}~\bibnamefont {Heinz}},\ }\href
  {\doibase 10.1016/j.cpc.2020.107604} {\bibfield  {journal} {\bibinfo
  {journal} {Comput. Phys. Commun.}\ }\textbf {\bibinfo {volume} {258}},\
  \bibinfo {pages} {107604} (\bibinfo {year} {2021})},\ \Eprint
  {http://arxiv.org/abs/1912.08271} {arXiv:1912.08271 [nucl-th]} \BibitemShut
  {NoStop}%
\bibitem [{is3()}]{is3d_code}%
  \BibitemOpen
  \href@noop {} {}\bibinfo {howpublished}
  {\url{https://github.com/derekeverett/iS3D}}\BibitemShut {NoStop}%
\bibitem [{\citenamefont {Weil}\ \emph {et~al.}(2016)\citenamefont {Weil} \emph
  {et~al.}}]{Weil:2016zrk}%
  \BibitemOpen
  \bibfield  {author} {\bibinfo {author} {\bibfnamefont {J.}~\bibnamefont
  {Weil}} \emph {et~al.},\ }\href {\doibase 10.1103/PhysRevC.94.054905}
  {\bibfield  {journal} {\bibinfo  {journal} {Phys. Rev.}\ }\textbf {\bibinfo
  {volume} {C94}},\ \bibinfo {pages} {054905} (\bibinfo {year} {2016})},\
  \Eprint {http://arxiv.org/abs/1606.06642} {arXiv:1606.06642 [nucl-th]}
  \BibitemShut {NoStop}%
\bibitem [{sma()}]{smash_code}%
  \BibitemOpen
  \href@noop {} {}\bibinfo {howpublished}
  {\url{https://github.com/smash-transport/smash}}\BibitemShut {NoStop}%
\bibitem [{\citenamefont {Florkowski}\ \emph {et~al.}(2018)\citenamefont
  {Florkowski}, \citenamefont {Heller},\ and\ \citenamefont
  {Spalinski}}]{Florkowski:2017olj}%
  \BibitemOpen
  \bibfield  {author} {\bibinfo {author} {\bibfnamefont {W.}~\bibnamefont
  {Florkowski}}, \bibinfo {author} {\bibfnamefont {M.~P.}\ \bibnamefont
  {Heller}}, \ and\ \bibinfo {author} {\bibfnamefont {M.}~\bibnamefont
  {Spalinski}},\ }\href {\doibase 10.1088/1361-6633/aaa091} {\bibfield
  {journal} {\bibinfo  {journal} {Rept. Prog. Phys.}\ }\textbf {\bibinfo
  {volume} {81}},\ \bibinfo {pages} {046001} (\bibinfo {year} {2018})},\
  \Eprint {http://arxiv.org/abs/1707.02282} {arXiv:1707.02282 [hep-ph]}
  \BibitemShut {NoStop}%
\bibitem [{\citenamefont {Vredevoogd}\ and\ \citenamefont
  {Pratt}(2009)}]{Vredevoogd:2008id}%
  \BibitemOpen
  \bibfield  {author} {\bibinfo {author} {\bibfnamefont {J.}~\bibnamefont
  {Vredevoogd}}\ and\ \bibinfo {author} {\bibfnamefont {S.}~\bibnamefont
  {Pratt}},\ }\href {\doibase 10.1103/PhysRevC.79.044915} {\bibfield  {journal}
  {\bibinfo  {journal} {Phys. Rev.}\ }\textbf {\bibinfo {volume} {C79}},\
  \bibinfo {pages} {044915} (\bibinfo {year} {2009})},\ \Eprint
  {http://arxiv.org/abs/0810.4325} {arXiv:0810.4325 [nucl-th]} \BibitemShut
  {NoStop}%
\bibitem [{\citenamefont {Keegan}\ \emph {et~al.}(2016)\citenamefont {Keegan},
  \citenamefont {Kurkela}, \citenamefont {Mazeliauskas},\ and\ \citenamefont
  {Teaney}}]{Keegan:2016cpi}%
  \BibitemOpen
  \bibfield  {author} {\bibinfo {author} {\bibfnamefont {L.}~\bibnamefont
  {Keegan}}, \bibinfo {author} {\bibfnamefont {A.}~\bibnamefont {Kurkela}},
  \bibinfo {author} {\bibfnamefont {A.}~\bibnamefont {Mazeliauskas}}, \ and\
  \bibinfo {author} {\bibfnamefont {D.}~\bibnamefont {Teaney}},\ }\href
  {\doibase 10.1007/JHEP08(2016)171} {\bibfield  {journal} {\bibinfo  {journal}
  {JHEP}\ }\textbf {\bibinfo {volume} {08}},\ \bibinfo {pages} {171} (\bibinfo
  {year} {2016})},\ \Eprint {http://arxiv.org/abs/1605.04287} {arXiv:1605.04287
  [hep-ph]} \BibitemShut {NoStop}%
\bibitem [{\citenamefont {Kurkela}\ \emph
  {et~al.}(2019{\natexlab{a}})\citenamefont {Kurkela}, \citenamefont
  {Mazeliauskas}, \citenamefont {Paquet}, \citenamefont {Schlichting},\ and\
  \citenamefont {Teaney}}]{Kurkela:2018vqr}%
  \BibitemOpen
  \bibfield  {author} {\bibinfo {author} {\bibfnamefont {A.}~\bibnamefont
  {Kurkela}}, \bibinfo {author} {\bibfnamefont {A.}~\bibnamefont
  {Mazeliauskas}}, \bibinfo {author} {\bibfnamefont {J.-F.}\ \bibnamefont
  {Paquet}}, \bibinfo {author} {\bibfnamefont {S.}~\bibnamefont {Schlichting}},
  \ and\ \bibinfo {author} {\bibfnamefont {D.}~\bibnamefont {Teaney}},\ }\href
  {\doibase 10.1103/PhysRevC.99.034910} {\bibfield  {journal} {\bibinfo
  {journal} {Phys. Rev.}\ }\textbf {\bibinfo {volume} {C99}},\ \bibinfo {pages}
  {034910} (\bibinfo {year} {2019}{\natexlab{a}})},\ \Eprint
  {http://arxiv.org/abs/1805.00961} {arXiv:1805.00961 [hep-ph]} \BibitemShut
  {NoStop}%
\bibitem [{\citenamefont {Kurkela}\ \emph
  {et~al.}(2019{\natexlab{b}})\citenamefont {Kurkela}, \citenamefont
  {Mazeliauskas}, \citenamefont {Paquet}, \citenamefont {Schlichting},\ and\
  \citenamefont {Teaney}}]{Kurkela:2018wud}%
  \BibitemOpen
  \bibfield  {author} {\bibinfo {author} {\bibfnamefont {A.}~\bibnamefont
  {Kurkela}}, \bibinfo {author} {\bibfnamefont {A.}~\bibnamefont
  {Mazeliauskas}}, \bibinfo {author} {\bibfnamefont {J.-F.}\ \bibnamefont
  {Paquet}}, \bibinfo {author} {\bibfnamefont {S.}~\bibnamefont {Schlichting}},
  \ and\ \bibinfo {author} {\bibfnamefont {D.}~\bibnamefont {Teaney}},\ }\href
  {\doibase 10.1103/PhysRevLett.122.122302} {\bibfield  {journal} {\bibinfo
  {journal} {Phys. Rev. Lett.}\ }\textbf {\bibinfo {volume} {122}},\ \bibinfo
  {pages} {122302} (\bibinfo {year} {2019}{\natexlab{b}})},\ \Eprint
  {http://arxiv.org/abs/1805.01604} {arXiv:1805.01604 [hep-ph]} \BibitemShut
  {NoStop}%
\bibitem [{\citenamefont {Schlichting}\ and\ \citenamefont
  {Teaney}(2019)}]{Schlichting:2019abc}%
  \BibitemOpen
  \bibfield  {author} {\bibinfo {author} {\bibfnamefont {S.}~\bibnamefont
  {Schlichting}}\ and\ \bibinfo {author} {\bibfnamefont {D.}~\bibnamefont
  {Teaney}},\ }\href {\doibase 10.1146/annurev-nucl-101918-023825} {\bibfield
  {journal} {\bibinfo  {journal} {Ann. Rev. Nucl. Part. Sci.}\ }\textbf
  {\bibinfo {volume} {69}},\ \bibinfo {pages} {447} (\bibinfo {year} {2019})},\
  \Eprint {http://arxiv.org/abs/1908.02113} {arXiv:1908.02113 [nucl-th]}
  \BibitemShut {NoStop}%
\bibitem [{\citenamefont {Kurkela}\ \emph {et~al.}(2020)\citenamefont
  {Kurkela}, \citenamefont {van~der Schee}, \citenamefont {Wiedemann},\ and\
  \citenamefont {Wu}}]{Kurkela:2019set}%
  \BibitemOpen
  \bibfield  {author} {\bibinfo {author} {\bibfnamefont {A.}~\bibnamefont
  {Kurkela}}, \bibinfo {author} {\bibfnamefont {W.}~\bibnamefont {van~der
  Schee}}, \bibinfo {author} {\bibfnamefont {U.~A.}\ \bibnamefont {Wiedemann}},
  \ and\ \bibinfo {author} {\bibfnamefont {B.}~\bibnamefont {Wu}},\ }\href
  {\doibase 10.1103/PhysRevLett.124.102301} {\bibfield  {journal} {\bibinfo
  {journal} {Phys. Rev. Lett.}\ }\textbf {\bibinfo {volume} {124}},\ \bibinfo
  {pages} {102301} (\bibinfo {year} {2020})},\ \Eprint
  {http://arxiv.org/abs/1907.08101} {arXiv:1907.08101 [hep-ph]} \BibitemShut
  {NoStop}%
\bibitem [{\citenamefont {Ke}\ \emph {et~al.}(2017)\citenamefont {Ke},
  \citenamefont {Moreland}, \citenamefont {Bernhard},\ and\ \citenamefont
  {Bass}}]{Ke:2016jrd}%
  \BibitemOpen
  \bibfield  {author} {\bibinfo {author} {\bibfnamefont {W.}~\bibnamefont
  {Ke}}, \bibinfo {author} {\bibfnamefont {J.~S.}\ \bibnamefont {Moreland}},
  \bibinfo {author} {\bibfnamefont {J.~E.}\ \bibnamefont {Bernhard}}, \ and\
  \bibinfo {author} {\bibfnamefont {S.~A.}\ \bibnamefont {Bass}},\ }\href
  {\doibase 10.1103/PhysRevC.96.044912} {\bibfield  {journal} {\bibinfo
  {journal} {Phys. Rev.}\ }\textbf {\bibinfo {volume} {C96}},\ \bibinfo {pages}
  {044912} (\bibinfo {year} {2017})},\ \Eprint
  {http://arxiv.org/abs/1610.08490} {arXiv:1610.08490 [nucl-th]} \BibitemShut
  {NoStop}%
\bibitem [{\citenamefont {Moreland}\ \emph {et~al.}(2020)\citenamefont
  {Moreland}, \citenamefont {Bernhard},\ and\ \citenamefont
  {Bass}}]{Moreland:2018gsh}%
  \BibitemOpen
  \bibfield  {author} {\bibinfo {author} {\bibfnamefont {J.~S.}\ \bibnamefont
  {Moreland}}, \bibinfo {author} {\bibfnamefont {J.~E.}\ \bibnamefont
  {Bernhard}}, \ and\ \bibinfo {author} {\bibfnamefont {S.~A.}\ \bibnamefont
  {Bass}},\ }\href {\doibase 10.1103/PhysRevC.101.024911} {\bibfield  {journal}
  {\bibinfo  {journal} {Phys. Rev. C}\ }\textbf {\bibinfo {volume} {101}},\
  \bibinfo {pages} {024911} (\bibinfo {year} {2020})},\ \Eprint
  {http://arxiv.org/abs/1808.02106} {arXiv:1808.02106 [nucl-th]} \BibitemShut
  {NoStop}%
\bibitem [{\citenamefont {Schenke}\ \emph {et~al.}(2012)\citenamefont
  {Schenke}, \citenamefont {Tribedy},\ and\ \citenamefont
  {Venugopalan}}]{Schenke:2012wb}%
  \BibitemOpen
  \bibfield  {author} {\bibinfo {author} {\bibfnamefont {B.}~\bibnamefont
  {Schenke}}, \bibinfo {author} {\bibfnamefont {P.}~\bibnamefont {Tribedy}}, \
  and\ \bibinfo {author} {\bibfnamefont {R.}~\bibnamefont {Venugopalan}},\
  }\href {\doibase 10.1103/PhysRevLett.108.252301} {\bibfield  {journal}
  {\bibinfo  {journal} {Phys. Rev. Lett.}\ }\textbf {\bibinfo {volume} {108}},\
  \bibinfo {pages} {252301} (\bibinfo {year} {2012})},\ \Eprint
  {http://arxiv.org/abs/1202.6646} {arXiv:1202.6646 [nucl-th]} \BibitemShut
  {NoStop}%
\bibitem [{\citenamefont {Shen}\ and\ \citenamefont
  {Alzhrani}(2020)}]{Shen:2020jwv}%
  \BibitemOpen
  \bibfield  {author} {\bibinfo {author} {\bibfnamefont {C.}~\bibnamefont
  {Shen}}\ and\ \bibinfo {author} {\bibfnamefont {S.}~\bibnamefont
  {Alzhrani}},\ }\href@noop {} {\  (\bibinfo {year} {2020})},\ \Eprint
  {http://arxiv.org/abs/2003.05852} {arXiv:2003.05852 [nucl-th]} \BibitemShut
  {NoStop}%
\bibitem [{\citenamefont {Miller}\ \emph {et~al.}(2007)\citenamefont {Miller},
  \citenamefont {Reygers}, \citenamefont {Sanders},\ and\ \citenamefont
  {Steinberg}}]{Miller:2007ri}%
  \BibitemOpen
  \bibfield  {author} {\bibinfo {author} {\bibfnamefont {M.~L.}\ \bibnamefont
  {Miller}}, \bibinfo {author} {\bibfnamefont {K.}~\bibnamefont {Reygers}},
  \bibinfo {author} {\bibfnamefont {S.~J.}\ \bibnamefont {Sanders}}, \ and\
  \bibinfo {author} {\bibfnamefont {P.}~\bibnamefont {Steinberg}},\ }\href
  {\doibase 10.1146/annurev.nucl.57.090506.123020} {\bibfield  {journal}
  {\bibinfo  {journal} {Ann. Rev. Nucl. Part. Sci.}\ }\textbf {\bibinfo
  {volume} {57}},\ \bibinfo {pages} {205} (\bibinfo {year} {2007})},\ \Eprint
  {http://arxiv.org/abs/nucl-ex/0701025} {arXiv:nucl-ex/0701025 [nucl-ex]}
  \BibitemShut {NoStop}%
\bibitem [{\citenamefont {Oliinychenko}\ and\ \citenamefont
  {Petersen}(2016)}]{Oliinychenko:2015lva}%
  \BibitemOpen
  \bibfield  {author} {\bibinfo {author} {\bibfnamefont {D.}~\bibnamefont
  {Oliinychenko}}\ and\ \bibinfo {author} {\bibfnamefont {H.}~\bibnamefont
  {Petersen}},\ }\href {\doibase 10.1103/PhysRevC.93.034905} {\bibfield
  {journal} {\bibinfo  {journal} {Phys. Rev. C}\ }\textbf {\bibinfo {volume}
  {93}},\ \bibinfo {pages} {034905} (\bibinfo {year} {2016})},\ \Eprint
  {http://arxiv.org/abs/1508.04378} {arXiv:1508.04378 [nucl-th]} \BibitemShut
  {NoStop}%
\bibitem [{\citenamefont {Ba\c{s}ar}\ and\ \citenamefont
  {Teaney}(2014)}]{Basar:2013hea}%
  \BibitemOpen
  \bibfield  {author} {\bibinfo {author} {\bibfnamefont {G.}~\bibnamefont
  {Ba\c{s}ar}}\ and\ \bibinfo {author} {\bibfnamefont {D.}~\bibnamefont
  {Teaney}},\ }\href {\doibase 10.1103/PhysRevC.90.054903} {\bibfield
  {journal} {\bibinfo  {journal} {Phys. Rev. C}\ }\textbf {\bibinfo {volume}
  {90}},\ \bibinfo {pages} {054903} (\bibinfo {year} {2014})},\ \Eprint
  {http://arxiv.org/abs/1312.6770} {arXiv:1312.6770 [nucl-th]} \BibitemShut
  {NoStop}%
\bibitem [{\citenamefont {Denicol}\ \emph
  {et~al.}(2012{\natexlab{a}})\citenamefont {Denicol}, \citenamefont {Molnár},
  \citenamefont {Niemi},\ and\ \citenamefont {Rischke}}]{Denicol:2012es}%
  \BibitemOpen
  \bibfield  {author} {\bibinfo {author} {\bibfnamefont {G.~S.}\ \bibnamefont
  {Denicol}}, \bibinfo {author} {\bibfnamefont {E.}~\bibnamefont {Molnár}},
  \bibinfo {author} {\bibfnamefont {H.}~\bibnamefont {Niemi}}, \ and\ \bibinfo
  {author} {\bibfnamefont {D.~H.}\ \bibnamefont {Rischke}},\ }\href {\doibase
  10.1140/epja/i2012-12170-x} {\bibfield  {journal} {\bibinfo  {journal} {Eur.
  Phys. J.}\ }\textbf {\bibinfo {volume} {A48}},\ \bibinfo {pages} {170}
  (\bibinfo {year} {2012}{\natexlab{a}})},\ \Eprint
  {http://arxiv.org/abs/1206.1554} {arXiv:1206.1554 [nucl-th]} \BibitemShut
  {NoStop}%
\bibitem [{\citenamefont {Shen}\ \emph {et~al.}(2016)\citenamefont {Shen},
  \citenamefont {Qiu}, \citenamefont {Song}, \citenamefont {Bernhard},
  \citenamefont {Bass},\ and\ \citenamefont {Heinz}}]{Shen:2014vra}%
  \BibitemOpen
  \bibfield  {author} {\bibinfo {author} {\bibfnamefont {C.}~\bibnamefont
  {Shen}}, \bibinfo {author} {\bibfnamefont {Z.}~\bibnamefont {Qiu}}, \bibinfo
  {author} {\bibfnamefont {H.}~\bibnamefont {Song}}, \bibinfo {author}
  {\bibfnamefont {J.}~\bibnamefont {Bernhard}}, \bibinfo {author}
  {\bibfnamefont {S.}~\bibnamefont {Bass}}, \ and\ \bibinfo {author}
  {\bibfnamefont {U.}~\bibnamefont {Heinz}},\ }\href {\doibase
  10.1016/j.cpc.2015.08.039} {\bibfield  {journal} {\bibinfo  {journal}
  {Comput. Phys. Commun.}\ }\textbf {\bibinfo {volume} {199}},\ \bibinfo
  {pages} {61} (\bibinfo {year} {2016})},\ \Eprint
  {http://arxiv.org/abs/1409.8164} {arXiv:1409.8164 [nucl-th]} \BibitemShut
  {NoStop}%
\bibitem [{hyd()}]{hydro_code}%
  \BibitemOpen
  \href@noop {} {}\bibinfo {howpublished}
  {\url{http://www.physics.mcgill.ca/music/}}\BibitemShut {NoStop}%
\bibitem [{\citenamefont {Auvinen}\ \emph {et~al.}(2020)\citenamefont
  {Auvinen}, \citenamefont {Eskola}, \citenamefont {Huovinen}, \citenamefont
  {Niemi}, \citenamefont {Paatelainen},\ and\ \citenamefont
  {Petreczky}}]{Auvinen:2020mpc}%
  \BibitemOpen
  \bibfield  {author} {\bibinfo {author} {\bibfnamefont {J.}~\bibnamefont
  {Auvinen}}, \bibinfo {author} {\bibfnamefont {K.~J.}\ \bibnamefont {Eskola}},
  \bibinfo {author} {\bibfnamefont {P.}~\bibnamefont {Huovinen}}, \bibinfo
  {author} {\bibfnamefont {H.}~\bibnamefont {Niemi}}, \bibinfo {author}
  {\bibfnamefont {R.}~\bibnamefont {Paatelainen}}, \ and\ \bibinfo {author}
  {\bibfnamefont {P.}~\bibnamefont {Petreczky}},\ }\href@noop {} {\  (\bibinfo
  {year} {2020})},\ \Eprint {http://arxiv.org/abs/2006.12499} {arXiv:2006.12499
  [nucl-th]} \BibitemShut {NoStop}%
\bibitem [{\citenamefont {Liu}(2015)}]{Liu-thesis}%
  \BibitemOpen
  \bibfield  {author} {\bibinfo {author} {\bibfnamefont {J.}~\bibnamefont
  {Liu}},\ }\emph {\bibinfo {title} {{Pre-equilibrium evolution effects on
  relativistic heavy-ion collision observables}}},\ \href
  {http://rave.ohiolink.edu/etdc/view?acc_num=osu1449185522} {Ph.D. thesis},\
  \bibinfo  {school} {Ohio State University} (\bibinfo {year}
  {2015})\BibitemShut {NoStop}%
\bibitem [{\citenamefont {Schenke}\ \emph {et~al.}(2019)\citenamefont
  {Schenke}, \citenamefont {Shen},\ and\ \citenamefont
  {Tribedy}}]{Schenke:2019pmk}%
  \BibitemOpen
  \bibfield  {author} {\bibinfo {author} {\bibfnamefont {B.}~\bibnamefont
  {Schenke}}, \bibinfo {author} {\bibfnamefont {C.}~\bibnamefont {Shen}}, \
  and\ \bibinfo {author} {\bibfnamefont {P.}~\bibnamefont {Tribedy}},\
  }\href@noop {} {\  (\bibinfo {year} {2019})},\ \Eprint
  {http://arxiv.org/abs/1908.06212} {arXiv:1908.06212 [nucl-th]} \BibitemShut
  {NoStop}%
\bibitem [{\citenamefont {Denicol}(2014)}]{Denicol:2014loa}%
  \BibitemOpen
  \bibfield  {author} {\bibinfo {author} {\bibfnamefont {G.}~\bibnamefont
  {Denicol}},\ }\href {\doibase 10.1088/0954-3899/41/12/124004} {\bibfield
  {journal} {\bibinfo  {journal} {J. Phys. G}\ }\textbf {\bibinfo {volume}
  {41}},\ \bibinfo {pages} {124004} (\bibinfo {year} {2014})}\BibitemShut
  {NoStop}%
\bibitem [{\citenamefont {Denicol}\ \emph {et~al.}(2014)\citenamefont
  {Denicol}, \citenamefont {Jeon},\ and\ \citenamefont
  {Gale}}]{Denicol:2014vaa}%
  \BibitemOpen
  \bibfield  {author} {\bibinfo {author} {\bibfnamefont {G.~S.}\ \bibnamefont
  {Denicol}}, \bibinfo {author} {\bibfnamefont {S.}~\bibnamefont {Jeon}}, \
  and\ \bibinfo {author} {\bibfnamefont {C.}~\bibnamefont {Gale}},\ }\href
  {\doibase 10.1103/PhysRevC.90.024912} {\bibfield  {journal} {\bibinfo
  {journal} {Phys. Rev.}\ }\textbf {\bibinfo {volume} {C90}},\ \bibinfo {pages}
  {024912} (\bibinfo {year} {2014})},\ \Eprint {http://arxiv.org/abs/1403.0962}
  {arXiv:1403.0962 [nucl-th]} \BibitemShut {NoStop}%
\bibitem [{\citenamefont {Niemi}\ \emph {et~al.}(2016)\citenamefont {Niemi},
  \citenamefont {Eskola},\ and\ \citenamefont {Paatelainen}}]{Niemi:2015qia}%
  \BibitemOpen
  \bibfield  {author} {\bibinfo {author} {\bibfnamefont {H.}~\bibnamefont
  {Niemi}}, \bibinfo {author} {\bibfnamefont {K.}~\bibnamefont {Eskola}}, \
  and\ \bibinfo {author} {\bibfnamefont {R.}~\bibnamefont {Paatelainen}},\
  }\href {\doibase 10.1103/PhysRevC.93.024907} {\bibfield  {journal} {\bibinfo
  {journal} {Phys. Rev. C}\ }\textbf {\bibinfo {volume} {93}},\ \bibinfo
  {pages} {024907} (\bibinfo {year} {2016})},\ \Eprint
  {http://arxiv.org/abs/1505.02677} {arXiv:1505.02677 [hep-ph]} \BibitemShut
  {NoStop}%
\bibitem [{\citenamefont {Csernai}\ \emph {et~al.}(2006)\citenamefont
  {Csernai}, \citenamefont {Kapusta},\ and\ \citenamefont
  {McLerran}}]{Csernai:2006zz}%
  \BibitemOpen
  \bibfield  {author} {\bibinfo {author} {\bibfnamefont {L.~P.}\ \bibnamefont
  {Csernai}}, \bibinfo {author} {\bibfnamefont {J.}~\bibnamefont {Kapusta}}, \
  and\ \bibinfo {author} {\bibfnamefont {L.~D.}\ \bibnamefont {McLerran}},\
  }\href {\doibase 10.1103/PhysRevLett.97.152303} {\bibfield  {journal}
  {\bibinfo  {journal} {Phys.\ Rev.\ Lett.}\ }\textbf {\bibinfo {volume}
  {97}},\ \bibinfo {pages} {152303} (\bibinfo {year} {2006})},\ \Eprint
  {http://arxiv.org/abs/nucl-th/0604032} {arXiv:nucl-th/0604032} \BibitemShut
  {NoStop}%
\bibitem [{\citenamefont {Kharzeev}\ and\ \citenamefont
  {Tuchin}(2008)}]{Kharzeev:2007wb}%
  \BibitemOpen
  \bibfield  {author} {\bibinfo {author} {\bibfnamefont {D.}~\bibnamefont
  {Kharzeev}}\ and\ \bibinfo {author} {\bibfnamefont {K.}~\bibnamefont
  {Tuchin}},\ }\href {\doibase 10.1088/1126-6708/2008/09/093} {\bibfield
  {journal} {\bibinfo  {journal} {JHEP}\ }\textbf {\bibinfo {volume} {09}},\
  \bibinfo {pages} {093} (\bibinfo {year} {2008})},\ \Eprint
  {http://arxiv.org/abs/0705.4280} {arXiv:0705.4280 [hep-ph]} \BibitemShut
  {NoStop}%
\bibitem [{\citenamefont {Baier}\ \emph {et~al.}(2008)\citenamefont {Baier},
  \citenamefont {Romatschke}, \citenamefont {Son}, \citenamefont {Starinets},\
  and\ \citenamefont {Stephanov}}]{Baier:2007ix}%
  \BibitemOpen
  \bibfield  {author} {\bibinfo {author} {\bibfnamefont {R.}~\bibnamefont
  {Baier}}, \bibinfo {author} {\bibfnamefont {P.}~\bibnamefont {Romatschke}},
  \bibinfo {author} {\bibfnamefont {D.~T.}\ \bibnamefont {Son}}, \bibinfo
  {author} {\bibfnamefont {A.~O.}\ \bibnamefont {Starinets}}, \ and\ \bibinfo
  {author} {\bibfnamefont {M.~A.}\ \bibnamefont {Stephanov}},\ }\href {\doibase
  10.1088/1126-6708/2008/04/100} {\bibfield  {journal} {\bibinfo  {journal}
  {JHEP}\ }\textbf {\bibinfo {volume} {04}},\ \bibinfo {pages} {100} (\bibinfo
  {year} {2008})},\ \Eprint {http://arxiv.org/abs/0712.2451} {arXiv:0712.2451
  [hep-th]} \BibitemShut {NoStop}%
\bibitem [{\citenamefont {Bhattacharyya}\ \emph {et~al.}(2008)\citenamefont
  {Bhattacharyya}, \citenamefont {Hubeny}, \citenamefont {Minwalla},\ and\
  \citenamefont {Rangamani}}]{Bhattacharyya:2008jc}%
  \BibitemOpen
  \bibfield  {author} {\bibinfo {author} {\bibfnamefont {S.}~\bibnamefont
  {Bhattacharyya}}, \bibinfo {author} {\bibfnamefont {V.~E.}\ \bibnamefont
  {Hubeny}}, \bibinfo {author} {\bibfnamefont {S.}~\bibnamefont {Minwalla}}, \
  and\ \bibinfo {author} {\bibfnamefont {M.}~\bibnamefont {Rangamani}},\ }\href
  {\doibase 10.1088/1126-6708/2008/02/045} {\bibfield  {journal} {\bibinfo
  {journal} {JHEP}\ }\textbf {\bibinfo {volume} {02}},\ \bibinfo {pages} {045}
  (\bibinfo {year} {2008})},\ \Eprint {http://arxiv.org/abs/0712.2456}
  {arXiv:0712.2456 [hep-th]} \BibitemShut {NoStop}%
\bibitem [{\citenamefont {Florkowski}\ \emph {et~al.}(2015)\citenamefont
  {Florkowski}, \citenamefont {Jaiswal}, \citenamefont {Maksymiuk},
  \citenamefont {Ryblewski},\ and\ \citenamefont
  {Strickland}}]{Florkowski:2015lra}%
  \BibitemOpen
  \bibfield  {author} {\bibinfo {author} {\bibfnamefont {W.}~\bibnamefont
  {Florkowski}}, \bibinfo {author} {\bibfnamefont {A.}~\bibnamefont {Jaiswal}},
  \bibinfo {author} {\bibfnamefont {E.}~\bibnamefont {Maksymiuk}}, \bibinfo
  {author} {\bibfnamefont {R.}~\bibnamefont {Ryblewski}}, \ and\ \bibinfo
  {author} {\bibfnamefont {M.}~\bibnamefont {Strickland}},\ }\href {\doibase
  10.1103/PhysRevC.91.054907} {\bibfield  {journal} {\bibinfo  {journal} {Phys.
  Rev.}\ }\textbf {\bibinfo {volume} {C91}},\ \bibinfo {pages} {054907}
  (\bibinfo {year} {2015})},\ \Eprint {http://arxiv.org/abs/1503.03226}
  {arXiv:1503.03226 [nucl-th]} \BibitemShut {NoStop}%
\bibitem [{\citenamefont {Czajka}\ \emph {et~al.}(2018)\citenamefont {Czajka},
  \citenamefont {Hauksson}, \citenamefont {Shen}, \citenamefont {Jeon},\ and\
  \citenamefont {Gale}}]{Czajka:2017wdo}%
  \BibitemOpen
  \bibfield  {author} {\bibinfo {author} {\bibfnamefont {A.}~\bibnamefont
  {Czajka}}, \bibinfo {author} {\bibfnamefont {S.}~\bibnamefont {Hauksson}},
  \bibinfo {author} {\bibfnamefont {C.}~\bibnamefont {Shen}}, \bibinfo {author}
  {\bibfnamefont {S.}~\bibnamefont {Jeon}}, \ and\ \bibinfo {author}
  {\bibfnamefont {C.}~\bibnamefont {Gale}},\ }\href {\doibase
  10.1103/PhysRevC.97.044914} {\bibfield  {journal} {\bibinfo  {journal} {Phys.
  Rev.}\ }\textbf {\bibinfo {volume} {C97}},\ \bibinfo {pages} {044914}
  (\bibinfo {year} {2018})}\BibitemShut {NoStop}%
\bibitem [{\citenamefont {Ghiglieri}\ \emph
  {et~al.}(2018{\natexlab{b}})\citenamefont {Ghiglieri}, \citenamefont
  {Moore},\ and\ \citenamefont {Teaney}}]{Ghiglieri:2018dgf}%
  \BibitemOpen
  \bibfield  {author} {\bibinfo {author} {\bibfnamefont {J.}~\bibnamefont
  {Ghiglieri}}, \bibinfo {author} {\bibfnamefont {G.~D.}\ \bibnamefont
  {Moore}}, \ and\ \bibinfo {author} {\bibfnamefont {D.}~\bibnamefont
  {Teaney}},\ }\href {\doibase 10.1103/PhysRevLett.121.052302} {\bibfield
  {journal} {\bibinfo  {journal} {Phys. Rev. Lett.}\ }\textbf {\bibinfo
  {volume} {121}},\ \bibinfo {pages} {052302} (\bibinfo {year}
  {2018}{\natexlab{b}})},\ \Eprint {http://arxiv.org/abs/1805.02663}
  {arXiv:1805.02663 [hep-ph]} \BibitemShut {NoStop}%
\bibitem [{\citenamefont {Pu}\ \emph {et~al.}(2010)\citenamefont {Pu},
  \citenamefont {Koide},\ and\ \citenamefont {Rischke}}]{Pu:2009fj}%
  \BibitemOpen
  \bibfield  {author} {\bibinfo {author} {\bibfnamefont {S.}~\bibnamefont
  {Pu}}, \bibinfo {author} {\bibfnamefont {T.}~\bibnamefont {Koide}}, \ and\
  \bibinfo {author} {\bibfnamefont {D.~H.}\ \bibnamefont {Rischke}},\ }\href
  {\doibase 10.1103/PhysRevD.81.114039} {\bibfield  {journal} {\bibinfo
  {journal} {Phys. Rev.}\ }\textbf {\bibinfo {volume} {D81}},\ \bibinfo {pages}
  {114039} (\bibinfo {year} {2010})},\ \Eprint {http://arxiv.org/abs/0907.3906}
  {arXiv:0907.3906 [hep-ph]} \BibitemShut {NoStop}%
\bibitem [{\citenamefont {Song}\ and\ \citenamefont
  {Heinz}(2008{\natexlab{b}})}]{Song:2008si}%
  \BibitemOpen
  \bibfield  {author} {\bibinfo {author} {\bibfnamefont {H.}~\bibnamefont
  {Song}}\ and\ \bibinfo {author} {\bibfnamefont {U.}~\bibnamefont {Heinz}},\
  }\href {\doibase 10.1103/PhysRevC.78.024902} {\bibfield  {journal} {\bibinfo
  {journal} {Phys. Rev. C}\ }\textbf {\bibinfo {volume} {78}},\ \bibinfo
  {pages} {024902} (\bibinfo {year} {2008}{\natexlab{b}})},\ \Eprint
  {http://arxiv.org/abs/0805.1756} {arXiv:0805.1756 [nucl-th]} \BibitemShut
  {NoStop}%
\bibitem [{\citenamefont {Schaefer}(2014)}]{Schaefer:2014awa}%
  \BibitemOpen
  \bibfield  {author} {\bibinfo {author} {\bibfnamefont {T.}~\bibnamefont
  {Schaefer}},\ }\href {\doibase 10.1146/annurev-nucl-102313-025439} {\bibfield
   {journal} {\bibinfo  {journal} {Ann. Rev. Nucl. Part. Sci.}\ }\textbf
  {\bibinfo {volume} {64}},\ \bibinfo {pages} {125} (\bibinfo {year} {2014})},\
  \Eprint {http://arxiv.org/abs/1403.0653} {arXiv:1403.0653 [hep-ph]}
  \BibitemShut {NoStop}%
\bibitem [{\citenamefont {Cooper}\ and\ \citenamefont
  {Frye}(1974)}]{Cooper:1974mv}%
  \BibitemOpen
  \bibfield  {author} {\bibinfo {author} {\bibfnamefont {F.}~\bibnamefont
  {Cooper}}\ and\ \bibinfo {author} {\bibfnamefont {G.}~\bibnamefont {Frye}},\
  }\href {\doibase 10.1103/PhysRevD.10.186} {\bibfield  {journal} {\bibinfo
  {journal} {Phys. Rev. D}\ }\textbf {\bibinfo {volume} {10}},\ \bibinfo
  {pages} {186} (\bibinfo {year} {1974})}\BibitemShut {NoStop}%
\bibitem [{\citenamefont {Cooper}\ \emph {et~al.}(1975)\citenamefont {Cooper},
  \citenamefont {Frye},\ and\ \citenamefont {Schonberg}}]{Cooper:1974qi}%
  \BibitemOpen
  \bibfield  {author} {\bibinfo {author} {\bibfnamefont {F.}~\bibnamefont
  {Cooper}}, \bibinfo {author} {\bibfnamefont {G.}~\bibnamefont {Frye}}, \ and\
  \bibinfo {author} {\bibfnamefont {E.}~\bibnamefont {Schonberg}},\ }\href
  {\doibase 10.1103/PhysRevD.11.192} {\bibfield  {journal} {\bibinfo  {journal}
  {Phys. Rev.}\ }\textbf {\bibinfo {volume} {D11}},\ \bibinfo {pages} {192}
  (\bibinfo {year} {1975})}\BibitemShut {NoStop}%
\bibitem [{\citenamefont {Huovinen}\ and\ \citenamefont
  {Petersen}(2012)}]{Huovinen:2012is}%
  \BibitemOpen
  \bibfield  {author} {\bibinfo {author} {\bibfnamefont {P.}~\bibnamefont
  {Huovinen}}\ and\ \bibinfo {author} {\bibfnamefont {H.}~\bibnamefont
  {Petersen}},\ }\href {\doibase 10.1140/epja/i2012-12171-9} {\bibfield
  {journal} {\bibinfo  {journal} {Eur. Phys. J.}\ }\textbf {\bibinfo {volume}
  {A48}},\ \bibinfo {pages} {171} (\bibinfo {year} {2012})},\ \Eprint
  {http://arxiv.org/abs/1206.3371} {arXiv:1206.3371 [nucl-th]} \BibitemShut
  {NoStop}%
\bibitem [{\citenamefont {Dusling}\ \emph {et~al.}(2010)\citenamefont
  {Dusling}, \citenamefont {Moore},\ and\ \citenamefont
  {Teaney}}]{Dusling:2009df}%
  \BibitemOpen
  \bibfield  {author} {\bibinfo {author} {\bibfnamefont {K.}~\bibnamefont
  {Dusling}}, \bibinfo {author} {\bibfnamefont {G.~D.}\ \bibnamefont {Moore}},
  \ and\ \bibinfo {author} {\bibfnamefont {D.}~\bibnamefont {Teaney}},\ }\href
  {\doibase 10.1103/PhysRevC.81.034907} {\bibfield  {journal} {\bibinfo
  {journal} {Phys. Rev.}\ }\textbf {\bibinfo {volume} {C81}},\ \bibinfo {pages}
  {034907} (\bibinfo {year} {2010})},\ \Eprint {http://arxiv.org/abs/0909.0754}
  {arXiv:0909.0754 [nucl-th]} \BibitemShut {NoStop}%
\bibitem [{\citenamefont {Molnar}\ and\ \citenamefont
  {Wolff}(2017)}]{Molnar:2014fva}%
  \BibitemOpen
  \bibfield  {author} {\bibinfo {author} {\bibfnamefont {D.}~\bibnamefont
  {Molnar}}\ and\ \bibinfo {author} {\bibfnamefont {Z.}~\bibnamefont {Wolff}},\
  }\href {\doibase 10.1103/PhysRevC.95.024903} {\bibfield  {journal} {\bibinfo
  {journal} {Phys.\ Rev.\ C}\ }\textbf {\bibinfo {volume} {95}},\ \bibinfo
  {pages} {024903} (\bibinfo {year} {2017})},\ \Eprint
  {http://arxiv.org/abs/1404.7850} {arXiv:1404.7850 [nucl-th]} \BibitemShut
  {NoStop}%
\bibitem [{\citenamefont {Damodaran}\ \emph {et~al.}(2017)\citenamefont
  {Damodaran}, \citenamefont {Molnar}, \citenamefont {Barnaföldi},
  \citenamefont {Berényi},\ and\ \citenamefont {Ferenc
  Nagy-Egri}}]{Damodaran:2017ior}%
  \BibitemOpen
  \bibfield  {author} {\bibinfo {author} {\bibfnamefont {M.}~\bibnamefont
  {Damodaran}}, \bibinfo {author} {\bibfnamefont {D.}~\bibnamefont {Molnar}},
  \bibinfo {author} {\bibfnamefont {G.~G.}\ \bibnamefont {Barnaföldi}},
  \bibinfo {author} {\bibfnamefont {D.}~\bibnamefont {Berényi}}, \ and\
  \bibinfo {author} {\bibfnamefont {M.}~\bibnamefont {Ferenc Nagy-Egri}},\
  }\href@noop {} {\  (\bibinfo {year} {2017})},\ \Eprint
  {http://arxiv.org/abs/1707.00793} {arXiv:1707.00793 [nucl-th]} \BibitemShut
  {NoStop}%
\bibitem [{\citenamefont {Damodaran}\ \emph {et~al.}(2020)\citenamefont
  {Damodaran}, \citenamefont {Molnar}, \citenamefont {Barnaf\"oldi},
  \citenamefont {Ber\'enyi},\ and\ \citenamefont
  {Nagy-Egri}}]{Damodaran:2020qxx}%
  \BibitemOpen
  \bibfield  {author} {\bibinfo {author} {\bibfnamefont {M.}~\bibnamefont
  {Damodaran}}, \bibinfo {author} {\bibfnamefont {D.}~\bibnamefont {Molnar}},
  \bibinfo {author} {\bibfnamefont {G.~G.}\ \bibnamefont {Barnaf\"oldi}},
  \bibinfo {author} {\bibfnamefont {D.}~\bibnamefont {Ber\'enyi}}, \ and\
  \bibinfo {author} {\bibfnamefont {M.~F.}\ \bibnamefont {Nagy-Egri}},\ }\href
  {\doibase 10.1103/PhysRevC.102.014907} {\bibfield  {journal} {\bibinfo
  {journal} {Phys. Rev. C}\ }\textbf {\bibinfo {volume} {102}},\ \bibinfo
  {pages} {014907} (\bibinfo {year} {2020})}\BibitemShut {NoStop}%
\bibitem [{\citenamefont {Heinz}(1999)}]{Heinz:1999kb}%
  \BibitemOpen
  \bibfield  {author} {\bibinfo {author} {\bibfnamefont {U.}~\bibnamefont
  {Heinz}},\ }\href {\doibase 10.1016/S0375-9474(99)85016-7} {\bibfield
  {journal} {\bibinfo  {journal} {Nucl. Phys. A}\ }\textbf {\bibinfo {volume}
  {661}},\ \bibinfo {pages} {140} (\bibinfo {year} {1999})},\ \Eprint
  {http://arxiv.org/abs/nucl-th/9907060} {arXiv:nucl-th/9907060} \BibitemShut
  {NoStop}%
\bibitem [{\citenamefont {Andronic}\ \emph {et~al.}(2006)\citenamefont
  {Andronic}, \citenamefont {Braun-Munzinger},\ and\ \citenamefont
  {Stachel}}]{Andronic:2005yp}%
  \BibitemOpen
  \bibfield  {author} {\bibinfo {author} {\bibfnamefont {A.}~\bibnamefont
  {Andronic}}, \bibinfo {author} {\bibfnamefont {P.}~\bibnamefont
  {Braun-Munzinger}}, \ and\ \bibinfo {author} {\bibfnamefont {J.}~\bibnamefont
  {Stachel}},\ }\href {\doibase 10.1016/j.nuclphysa.2006.03.012} {\bibfield
  {journal} {\bibinfo  {journal} {Nucl. Phys. A}\ }\textbf {\bibinfo {volume}
  {772}},\ \bibinfo {pages} {167} (\bibinfo {year} {2006})},\ \Eprint
  {http://arxiv.org/abs/nucl-th/0511071} {arXiv:nucl-th/0511071} \BibitemShut
  {NoStop}%
\bibitem [{\citenamefont {Andronic}\ \emph {et~al.}(2018)\citenamefont
  {Andronic}, \citenamefont {Braun-Munzinger}, \citenamefont {Redlich},\ and\
  \citenamefont {Stachel}}]{Andronic:2017pug}%
  \BibitemOpen
  \bibfield  {author} {\bibinfo {author} {\bibfnamefont {A.}~\bibnamefont
  {Andronic}}, \bibinfo {author} {\bibfnamefont {P.}~\bibnamefont
  {Braun-Munzinger}}, \bibinfo {author} {\bibfnamefont {K.}~\bibnamefont
  {Redlich}}, \ and\ \bibinfo {author} {\bibfnamefont {J.}~\bibnamefont
  {Stachel}},\ }\href {\doibase 10.1038/s41586-018-0491-6} {\bibfield
  {journal} {\bibinfo  {journal} {Nature}\ }\textbf {\bibinfo {volume} {561}},\
  \bibinfo {pages} {321} (\bibinfo {year} {2018})},\ \Eprint
  {http://arxiv.org/abs/1710.09425} {arXiv:1710.09425 [nucl-th]} \BibitemShut
  {NoStop}%
\bibitem [{\citenamefont {Grad}(1949)}]{Grad}%
  \BibitemOpen
  \bibfield  {author} {\bibinfo {author} {\bibfnamefont {H.}~\bibnamefont
  {Grad}},\ }\href {\doibase 10.1002/cpa.3160020403} {\bibfield  {journal}
  {\bibinfo  {journal} {Commun. Pure Appl. Math}\ }\textbf {\bibinfo {volume}
  {2}},\ \bibinfo {pages} {331} (\bibinfo {year} {1949})}\BibitemShut {NoStop}%
\bibitem [{\citenamefont {Israel}(1976)}]{Israel:1976tn}%
  \BibitemOpen
  \bibfield  {author} {\bibinfo {author} {\bibfnamefont {W.}~\bibnamefont
  {Israel}},\ }\href {\doibase 10.1016/0003-4916(76)90064-6} {\bibfield
  {journal} {\bibinfo  {journal} {Annals Phys.}\ }\textbf {\bibinfo {volume}
  {100}},\ \bibinfo {pages} {310} (\bibinfo {year} {1976})}\BibitemShut
  {NoStop}%
\bibitem [{\citenamefont {Israel}\ and\ \citenamefont
  {Stewart}(1979)}]{Israel:1979wp}%
  \BibitemOpen
  \bibfield  {author} {\bibinfo {author} {\bibfnamefont {W.}~\bibnamefont
  {Israel}}\ and\ \bibinfo {author} {\bibfnamefont {J.~M.}\ \bibnamefont
  {Stewart}},\ }\href {\doibase 10.1016/0003-4916(79)90130-1} {\bibfield
  {journal} {\bibinfo  {journal} {Annals Phys.}\ }\textbf {\bibinfo {volume}
  {118}},\ \bibinfo {pages} {341} (\bibinfo {year} {1979})}\BibitemShut
  {NoStop}%
\bibitem [{\citenamefont {Teaney}(2003)}]{Teaney:2003kp}%
  \BibitemOpen
  \bibfield  {author} {\bibinfo {author} {\bibfnamefont {D.}~\bibnamefont
  {Teaney}},\ }\href {\doibase 10.1103/PhysRevC.68.034913} {\bibfield
  {journal} {\bibinfo  {journal} {Phys. Rev.}\ }\textbf {\bibinfo {volume}
  {C68}},\ \bibinfo {pages} {034913} (\bibinfo {year} {2003})},\ \Eprint
  {http://arxiv.org/abs/nucl-th/0301099} {arXiv:nucl-th/0301099 [nucl-th]}
  \BibitemShut {NoStop}%
\bibitem [{\citenamefont {Monnai}\ and\ \citenamefont
  {Hirano}(2009)}]{Monnai:2009ad}%
  \BibitemOpen
  \bibfield  {author} {\bibinfo {author} {\bibfnamefont {A.}~\bibnamefont
  {Monnai}}\ and\ \bibinfo {author} {\bibfnamefont {T.}~\bibnamefont
  {Hirano}},\ }\href {\doibase 10.1103/PhysRevC.80.054906} {\bibfield
  {journal} {\bibinfo  {journal} {Phys. Rev.}\ }\textbf {\bibinfo {volume}
  {C80}},\ \bibinfo {pages} {054906} (\bibinfo {year} {2009})},\ \Eprint
  {http://arxiv.org/abs/0903.4436} {arXiv:0903.4436 [nucl-th]} \BibitemShut
  {NoStop}%
\bibitem [{\citenamefont {Dusling}\ and\ \citenamefont
  {Schäfer}(2012)}]{Dusling:2011fd}%
  \BibitemOpen
  \bibfield  {author} {\bibinfo {author} {\bibfnamefont {K.}~\bibnamefont
  {Dusling}}\ and\ \bibinfo {author} {\bibfnamefont {T.}~\bibnamefont
  {Schäfer}},\ }\href {\doibase 10.1103/PhysRevC.85.044909} {\bibfield
  {journal} {\bibinfo  {journal} {Phys. Rev.}\ }\textbf {\bibinfo {volume}
  {C85}},\ \bibinfo {pages} {044909} (\bibinfo {year} {2012})},\ \Eprint
  {http://arxiv.org/abs/1109.5181} {arXiv:1109.5181 [hep-ph]} \BibitemShut
  {NoStop}%
\bibitem [{\citenamefont {Denicol}\ \emph
  {et~al.}(2012{\natexlab{b}})\citenamefont {Denicol}, \citenamefont {Niemi},
  \citenamefont {Molnar},\ and\ \citenamefont {Rischke}}]{Denicol:2012cn}%
  \BibitemOpen
  \bibfield  {author} {\bibinfo {author} {\bibfnamefont {G.~S.}\ \bibnamefont
  {Denicol}}, \bibinfo {author} {\bibfnamefont {H.}~\bibnamefont {Niemi}},
  \bibinfo {author} {\bibfnamefont {E.}~\bibnamefont {Molnar}}, \ and\ \bibinfo
  {author} {\bibfnamefont {D.~H.}\ \bibnamefont {Rischke}},\ }\href {\doibase
  10.1103/PhysRevD.85.114047, 10.1103/PhysRevD.91.039902} {\bibfield  {journal}
  {\bibinfo  {journal} {Phys. Rev.}\ }\textbf {\bibinfo {volume} {D85}},\
  \bibinfo {pages} {114047} (\bibinfo {year} {2012}{\natexlab{b}})},\ \bibinfo
  {note} {[Erratum: Phys. Rev.D91,no.3,039902(2015)]},\ \Eprint
  {http://arxiv.org/abs/1202.4551} {arXiv:1202.4551 [nucl-th]} \BibitemShut
  {NoStop}%
\bibitem [{\citenamefont {Chapman}\ \emph {et~al.}(1990)\citenamefont
  {Chapman}, \citenamefont {Cowling},\ and\ \citenamefont
  {Burnett}}]{chapman1990mathematical}%
  \BibitemOpen
  \bibfield  {author} {\bibinfo {author} {\bibfnamefont {S.}~\bibnamefont
  {Chapman}}, \bibinfo {author} {\bibfnamefont {T.~G.}\ \bibnamefont
  {Cowling}}, \ and\ \bibinfo {author} {\bibfnamefont {D.}~\bibnamefont
  {Burnett}},\ }\href@noop {} {\emph {\bibinfo {title} {The mathematical theory
  of non-uniform gases: an account of the kinetic theory of viscosity, thermal
  conduction and diffusion in gases}}}\ (\bibinfo  {publisher} {Cambridge
  university press},\ \bibinfo {year} {1990})\BibitemShut {NoStop}%
\bibitem [{\citenamefont {Anderson}\ and\ \citenamefont
  {Witting}(1974)}]{ANDERSON1974466}%
  \BibitemOpen
  \bibfield  {author} {\bibinfo {author} {\bibfnamefont {J.}~\bibnamefont
  {Anderson}}\ and\ \bibinfo {author} {\bibfnamefont {H.}~\bibnamefont
  {Witting}},\ }\href {\doibase https://doi.org/10.1016/0031-8914(74)90355-3}
  {\bibfield  {journal} {\bibinfo  {journal} {Physica}\ }\textbf {\bibinfo
  {volume} {74}},\ \bibinfo {pages} {466 } (\bibinfo {year}
  {1974})}\BibitemShut {NoStop}%
\bibitem [{\citenamefont {Jaiswal}\ \emph {et~al.}(2014)\citenamefont
  {Jaiswal}, \citenamefont {Ryblewski},\ and\ \citenamefont
  {Strickland}}]{Jaiswal:2014isa}%
  \BibitemOpen
  \bibfield  {author} {\bibinfo {author} {\bibfnamefont {A.}~\bibnamefont
  {Jaiswal}}, \bibinfo {author} {\bibfnamefont {R.}~\bibnamefont {Ryblewski}},
  \ and\ \bibinfo {author} {\bibfnamefont {M.}~\bibnamefont {Strickland}},\
  }\href {\doibase 10.1103/PhysRevC.90.044908} {\bibfield  {journal} {\bibinfo
  {journal} {Phys. Rev.}\ }\textbf {\bibinfo {volume} {C90}},\ \bibinfo {pages}
  {044908} (\bibinfo {year} {2014})},\ \Eprint {http://arxiv.org/abs/1407.7231}
  {arXiv:1407.7231 [hep-ph]} \BibitemShut {NoStop}%
\bibitem [{\citenamefont {Pratt}\ and\ \citenamefont
  {Torrieri}(2010)}]{Pratt:2010jt}%
  \BibitemOpen
  \bibfield  {author} {\bibinfo {author} {\bibfnamefont {S.}~\bibnamefont
  {Pratt}}\ and\ \bibinfo {author} {\bibfnamefont {G.}~\bibnamefont
  {Torrieri}},\ }\href {\doibase 10.1103/PhysRevC.82.044901} {\bibfield
  {journal} {\bibinfo  {journal} {Phys. Rev.}\ }\textbf {\bibinfo {volume}
  {C82}},\ \bibinfo {pages} {044901} (\bibinfo {year} {2010})},\ \Eprint
  {http://arxiv.org/abs/1003.0413} {arXiv:1003.0413 [nucl-th]} \BibitemShut
  {NoStop}%
\bibitem [{\citenamefont {Bhatnagar}\ \emph {et~al.}(1954)\citenamefont
  {Bhatnagar}, \citenamefont {Gross},\ and\ \citenamefont
  {Krook}}]{Bhatnagar:1954zz}%
  \BibitemOpen
  \bibfield  {author} {\bibinfo {author} {\bibfnamefont {P.~L.}\ \bibnamefont
  {Bhatnagar}}, \bibinfo {author} {\bibfnamefont {E.~P.}\ \bibnamefont
  {Gross}}, \ and\ \bibinfo {author} {\bibfnamefont {M.}~\bibnamefont
  {Krook}},\ }\href {\doibase 10.1103/PhysRev.94.511} {\bibfield  {journal}
  {\bibinfo  {journal} {Phys. Rev.}\ }\textbf {\bibinfo {volume} {94}},\
  \bibinfo {pages} {511} (\bibinfo {year} {1954})}\BibitemShut {NoStop}%
\bibitem [{\citenamefont {Romatschke}\ and\ \citenamefont
  {Strickland}(2003)}]{Romatschke:2003ms}%
  \BibitemOpen
  \bibfield  {author} {\bibinfo {author} {\bibfnamefont {P.}~\bibnamefont
  {Romatschke}}\ and\ \bibinfo {author} {\bibfnamefont {M.}~\bibnamefont
  {Strickland}},\ }\href {\doibase 10.1103/PhysRevD.68.036004} {\bibfield
  {journal} {\bibinfo  {journal} {Phys. Rev.}\ }\textbf {\bibinfo {volume}
  {D68}},\ \bibinfo {pages} {036004} (\bibinfo {year} {2003})},\ \Eprint
  {http://arxiv.org/abs/hep-ph/0304092} {arXiv:hep-ph/0304092 [hep-ph]}
  \BibitemShut {NoStop}%
\bibitem [{\citenamefont {Romatschke}\ and\ \citenamefont
  {Strickland}(2004)}]{Romatschke:2004jh}%
  \BibitemOpen
  \bibfield  {author} {\bibinfo {author} {\bibfnamefont {P.}~\bibnamefont
  {Romatschke}}\ and\ \bibinfo {author} {\bibfnamefont {M.}~\bibnamefont
  {Strickland}},\ }\href {\doibase 10.1103/PhysRevD.70.116006} {\bibfield
  {journal} {\bibinfo  {journal} {Phys. Rev.}\ }\textbf {\bibinfo {volume}
  {D70}},\ \bibinfo {pages} {116006} (\bibinfo {year} {2004})},\ \Eprint
  {http://arxiv.org/abs/hep-ph/0406188} {arXiv:hep-ph/0406188 [hep-ph]}
  \BibitemShut {NoStop}%
\bibitem [{\citenamefont {Martinez}\ \emph {et~al.}(2012)\citenamefont
  {Martinez}, \citenamefont {Ryblewski},\ and\ \citenamefont
  {Strickland}}]{Martinez:2012tu}%
  \BibitemOpen
  \bibfield  {author} {\bibinfo {author} {\bibfnamefont {M.}~\bibnamefont
  {Martinez}}, \bibinfo {author} {\bibfnamefont {R.}~\bibnamefont {Ryblewski}},
  \ and\ \bibinfo {author} {\bibfnamefont {M.}~\bibnamefont {Strickland}},\
  }\href {\doibase 10.1103/PhysRevC.85.064913} {\bibfield  {journal} {\bibinfo
  {journal} {Phys. Rev.}\ }\textbf {\bibinfo {volume} {C85}},\ \bibinfo {pages}
  {064913} (\bibinfo {year} {2012})},\ \Eprint {http://arxiv.org/abs/1204.1473}
  {arXiv:1204.1473 [nucl-th]} \BibitemShut {NoStop}%
\bibitem [{\citenamefont {Florkowski}\ \emph {et~al.}(2013)\citenamefont
  {Florkowski}, \citenamefont {Ryblewski},\ and\ \citenamefont
  {Strickland}}]{Florkowski:2013lya}%
  \BibitemOpen
  \bibfield  {author} {\bibinfo {author} {\bibfnamefont {W.}~\bibnamefont
  {Florkowski}}, \bibinfo {author} {\bibfnamefont {R.}~\bibnamefont
  {Ryblewski}}, \ and\ \bibinfo {author} {\bibfnamefont {M.}~\bibnamefont
  {Strickland}},\ }\href {\doibase 10.1103/PhysRevC.88.024903} {\bibfield
  {journal} {\bibinfo  {journal} {Phys. Rev.}\ }\textbf {\bibinfo {volume}
  {C88}},\ \bibinfo {pages} {024903} (\bibinfo {year} {2013})},\ \Eprint
  {http://arxiv.org/abs/1305.7234} {arXiv:1305.7234 [nucl-th]} \BibitemShut
  {NoStop}%
\bibitem [{\citenamefont {Florkowski}\ \emph {et~al.}(2014)\citenamefont
  {Florkowski}, \citenamefont {Ryblewski}, \citenamefont {Strickland},\ and\
  \citenamefont {Tinti}}]{Florkowski:2014bba}%
  \BibitemOpen
  \bibfield  {author} {\bibinfo {author} {\bibfnamefont {W.}~\bibnamefont
  {Florkowski}}, \bibinfo {author} {\bibfnamefont {R.}~\bibnamefont
  {Ryblewski}}, \bibinfo {author} {\bibfnamefont {M.}~\bibnamefont
  {Strickland}}, \ and\ \bibinfo {author} {\bibfnamefont {L.}~\bibnamefont
  {Tinti}},\ }\href {\doibase 10.1103/PhysRevC.89.054909} {\bibfield  {journal}
  {\bibinfo  {journal} {Phys. Rev.}\ }\textbf {\bibinfo {volume} {C89}},\
  \bibinfo {pages} {054909} (\bibinfo {year} {2014})},\ \Eprint
  {http://arxiv.org/abs/1403.1223} {arXiv:1403.1223 [hep-ph]} \BibitemShut
  {NoStop}%
\bibitem [{\citenamefont {Tinti}(2016)}]{Tinti:2015xwa}%
  \BibitemOpen
  \bibfield  {author} {\bibinfo {author} {\bibfnamefont {L.}~\bibnamefont
  {Tinti}},\ }\href {\doibase 10.1103/PhysRevC.94.044902} {\bibfield  {journal}
  {\bibinfo  {journal} {Phys. Rev.}\ }\textbf {\bibinfo {volume} {C94}},\
  \bibinfo {pages} {044902} (\bibinfo {year} {2016})},\ \Eprint
  {http://arxiv.org/abs/1506.07164} {arXiv:1506.07164 [hep-ph]} \BibitemShut
  {NoStop}%
\bibitem [{\citenamefont {Molnar}\ \emph {et~al.}(2016)\citenamefont {Molnar},
  \citenamefont {Niemi},\ and\ \citenamefont {Rischke}}]{Molnar:2016vvu}%
  \BibitemOpen
  \bibfield  {author} {\bibinfo {author} {\bibfnamefont {E.}~\bibnamefont
  {Molnar}}, \bibinfo {author} {\bibfnamefont {H.}~\bibnamefont {Niemi}}, \
  and\ \bibinfo {author} {\bibfnamefont {D.~H.}\ \bibnamefont {Rischke}},\
  }\href {\doibase 10.1103/PhysRevD.93.114025} {\bibfield  {journal} {\bibinfo
  {journal} {Phys. Rev.}\ }\textbf {\bibinfo {volume} {D93}},\ \bibinfo {pages}
  {114025} (\bibinfo {year} {2016})},\ \Eprint
  {http://arxiv.org/abs/1602.00573} {arXiv:1602.00573 [nucl-th]} \BibitemShut
  {NoStop}%
\bibitem [{\citenamefont {Tinti}\ \emph
  {et~al.}(2019{\natexlab{a}})\citenamefont {Tinti}, \citenamefont {Vujanovic},
  \citenamefont {Noronha},\ and\ \citenamefont {Heinz}}]{Tinti:2018nrp}%
  \BibitemOpen
  \bibfield  {author} {\bibinfo {author} {\bibfnamefont {L.}~\bibnamefont
  {Tinti}}, \bibinfo {author} {\bibfnamefont {G.}~\bibnamefont {Vujanovic}},
  \bibinfo {author} {\bibfnamefont {J.}~\bibnamefont {Noronha}}, \ and\
  \bibinfo {author} {\bibfnamefont {U.}~\bibnamefont {Heinz}},\ }\bibfield
  {booktitle} {\emph {\bibinfo {booktitle} {{Proceedings, 27th International
  Conference on Ultrarelativistic Nucleus-Nucleus Collisions (Quark Matter
  2018): Venice, Italy, May 14-19, 2018}}},\ }\href {\doibase
  10.1016/j.nuclphysa.2018.10.038} {\bibfield  {journal} {\bibinfo  {journal}
  {Nucl. Phys.}\ }\textbf {\bibinfo {volume} {A982}},\ \bibinfo {pages} {919}
  (\bibinfo {year} {2019}{\natexlab{a}})},\ \Eprint
  {http://arxiv.org/abs/1808.06212} {arXiv:1808.06212 [nucl-th]} \BibitemShut
  {NoStop}%
\bibitem [{\citenamefont {Tinti}\ \emph
  {et~al.}(2019{\natexlab{b}})\citenamefont {Tinti}, \citenamefont {Vujanovic},
  \citenamefont {Noronha},\ and\ \citenamefont {Heinz}}]{Tinti:2018qfb}%
  \BibitemOpen
  \bibfield  {author} {\bibinfo {author} {\bibfnamefont {L.}~\bibnamefont
  {Tinti}}, \bibinfo {author} {\bibfnamefont {G.}~\bibnamefont {Vujanovic}},
  \bibinfo {author} {\bibfnamefont {J.}~\bibnamefont {Noronha}}, \ and\
  \bibinfo {author} {\bibfnamefont {U.}~\bibnamefont {Heinz}},\ }\href
  {\doibase 10.1103/PhysRevD.99.016009} {\bibfield  {journal} {\bibinfo
  {journal} {Phys. Rev.}\ }\textbf {\bibinfo {volume} {D99}},\ \bibinfo {pages}
  {016009} (\bibinfo {year} {2019}{\natexlab{b}})},\ \Eprint
  {http://arxiv.org/abs/1808.06436} {arXiv:1808.06436 [nucl-th]} \BibitemShut
  {NoStop}%
\bibitem [{\citenamefont {Hirano}\ and\ \citenamefont
  {Tsuda}(2002)}]{Hirano:2002ds}%
  \BibitemOpen
  \bibfield  {author} {\bibinfo {author} {\bibfnamefont {T.}~\bibnamefont
  {Hirano}}\ and\ \bibinfo {author} {\bibfnamefont {K.}~\bibnamefont {Tsuda}},\
  }\href {\doibase 10.1103/PhysRevC.66.054905} {\bibfield  {journal} {\bibinfo
  {journal} {Phys. Rev. C}\ }\textbf {\bibinfo {volume} {66}},\ \bibinfo
  {pages} {054905} (\bibinfo {year} {2002})},\ \Eprint
  {http://arxiv.org/abs/nucl-th/0205043} {arXiv:nucl-th/0205043} \BibitemShut
  {NoStop}%
\bibitem [{\citenamefont {Steinheimer}\ \emph {et~al.}(2013)\citenamefont
  {Steinheimer}, \citenamefont {Aichelin},\ and\ \citenamefont
  {Bleicher}}]{Steinheimer:2012rd}%
  \BibitemOpen
  \bibfield  {author} {\bibinfo {author} {\bibfnamefont {J.}~\bibnamefont
  {Steinheimer}}, \bibinfo {author} {\bibfnamefont {J.}~\bibnamefont
  {Aichelin}}, \ and\ \bibinfo {author} {\bibfnamefont {M.}~\bibnamefont
  {Bleicher}},\ }\href {\doibase 10.1103/PhysRevLett.110.042501} {\bibfield
  {journal} {\bibinfo  {journal} {Phys. Rev. Lett.}\ }\textbf {\bibinfo
  {volume} {110}},\ \bibinfo {pages} {042501} (\bibinfo {year} {2013})},\
  \Eprint {http://arxiv.org/abs/1203.5302} {arXiv:1203.5302 [nucl-th]}
  \BibitemShut {NoStop}%
\bibitem [{\citenamefont {Broniowski}\ \emph {et~al.}(2015)\citenamefont
  {Broniowski}, \citenamefont {Giacosa},\ and\ \citenamefont
  {Begun}}]{Broniowski:2015oha}%
  \BibitemOpen
  \bibfield  {author} {\bibinfo {author} {\bibfnamefont {W.}~\bibnamefont
  {Broniowski}}, \bibinfo {author} {\bibfnamefont {F.}~\bibnamefont {Giacosa}},
  \ and\ \bibinfo {author} {\bibfnamefont {V.}~\bibnamefont {Begun}},\ }\href
  {\doibase 10.1103/PhysRevC.92.034905} {\bibfield  {journal} {\bibinfo
  {journal} {Phys. Rev.}\ }\textbf {\bibinfo {volume} {C92}},\ \bibinfo {pages}
  {034905} (\bibinfo {year} {2015})},\ \Eprint
  {http://arxiv.org/abs/1506.01260} {arXiv:1506.01260 [nucl-th]} \BibitemShut
  {NoStop}%
\bibitem [{\citenamefont {Jaynes}(1957)}]{Jaynes:1957zza}%
  \BibitemOpen
  \bibfield  {author} {\bibinfo {author} {\bibfnamefont {E.~T.}\ \bibnamefont
  {Jaynes}},\ }\href {\doibase 10.1103/PhysRev.106.620} {\bibfield  {journal}
  {\bibinfo  {journal} {Phys. Rev.}\ }\textbf {\bibinfo {volume} {106}},\
  \bibinfo {pages} {620} (\bibinfo {year} {1957})}\BibitemShut {NoStop}%
\bibitem [{\citenamefont {Gelis}(2014)}]{Gelis:2014qga}%
  \BibitemOpen
  \bibfield  {author} {\bibinfo {author} {\bibfnamefont {F.}~\bibnamefont
  {Gelis}},\ }\href {\doibase 10.5506/APhysPolB.45.2257} {\bibfield  {journal}
  {\bibinfo  {journal} {Acta Phys. Polon. B}\ }\textbf {\bibinfo {volume}
  {45}},\ \bibinfo {pages} {2257} (\bibinfo {year} {2014})}\BibitemShut
  {NoStop}%
\bibitem [{\citenamefont {Bass}\ \emph {et~al.}(1999)\citenamefont {Bass},
  \citenamefont {Dumitru}, \citenamefont {Bleicher}, \citenamefont {Bravina},
  \citenamefont {Zabrodin}, \citenamefont {Stoecker},\ and\ \citenamefont
  {Greiner}}]{Bass:1999tu}%
  \BibitemOpen
  \bibfield  {author} {\bibinfo {author} {\bibfnamefont {S.}~\bibnamefont
  {Bass}}, \bibinfo {author} {\bibfnamefont {A.}~\bibnamefont {Dumitru}},
  \bibinfo {author} {\bibfnamefont {M.}~\bibnamefont {Bleicher}}, \bibinfo
  {author} {\bibfnamefont {L.}~\bibnamefont {Bravina}}, \bibinfo {author}
  {\bibfnamefont {E.}~\bibnamefont {Zabrodin}}, \bibinfo {author}
  {\bibfnamefont {H.}~\bibnamefont {Stoecker}}, \ and\ \bibinfo {author}
  {\bibfnamefont {W.}~\bibnamefont {Greiner}},\ }\href {\doibase
  10.1103/PhysRevC.60.021902} {\bibfield  {journal} {\bibinfo  {journal} {Phys.
  Rev. C}\ }\textbf {\bibinfo {volume} {60}},\ \bibinfo {pages} {021902}
  (\bibinfo {year} {1999})},\ \Eprint {http://arxiv.org/abs/nucl-th/9902062}
  {arXiv:nucl-th/9902062} \BibitemShut {NoStop}%
\bibitem [{\citenamefont {Loeppky}\ \emph {et~al.}(2009)\citenamefont
  {Loeppky}, \citenamefont {Sacks},\ and\ \citenamefont {Welch}}]{Loeppky}%
  \BibitemOpen
  \bibfield  {author} {\bibinfo {author} {\bibfnamefont {J.}~\bibnamefont
  {Loeppky}}, \bibinfo {author} {\bibfnamefont {J.}~\bibnamefont {Sacks}}, \
  and\ \bibinfo {author} {\bibfnamefont {W.}~\bibnamefont {Welch}},\ }\href
  {\doibase 10.1198/TECH.2009.08040} {\bibfield  {journal} {\bibinfo  {journal}
  {Technometrics}\ }\textbf {\bibinfo {volume} {51}},\ \bibinfo {pages} {366}
  (\bibinfo {year} {2009})}\BibitemShut {NoStop}%
\bibitem [{\citenamefont {Pedregosa}\ \emph {et~al.}(2011)\citenamefont
  {Pedregosa}, \citenamefont {Varoquaux}, \citenamefont {Gramfort},
  \citenamefont {Michel}, \citenamefont {Thirion}, \citenamefont {Grisel},
  \citenamefont {Blondel}, \citenamefont {Prettenhofer}, \citenamefont {Weiss},
  \citenamefont {Dubourg}, \citenamefont {Vanderplas}, \citenamefont {Passos},
  \citenamefont {Cournapeau}, \citenamefont {Brucher}, \citenamefont {Perrot},\
  and\ \citenamefont {Duchesnay}}]{scikit-learn}%
  \BibitemOpen
  \bibfield  {author} {\bibinfo {author} {\bibfnamefont {F.}~\bibnamefont
  {Pedregosa}}, \bibinfo {author} {\bibfnamefont {G.}~\bibnamefont
  {Varoquaux}}, \bibinfo {author} {\bibfnamefont {A.}~\bibnamefont {Gramfort}},
  \bibinfo {author} {\bibfnamefont {V.}~\bibnamefont {Michel}}, \bibinfo
  {author} {\bibfnamefont {B.}~\bibnamefont {Thirion}}, \bibinfo {author}
  {\bibfnamefont {O.}~\bibnamefont {Grisel}}, \bibinfo {author} {\bibfnamefont
  {M.}~\bibnamefont {Blondel}}, \bibinfo {author} {\bibfnamefont
  {P.}~\bibnamefont {Prettenhofer}}, \bibinfo {author} {\bibfnamefont
  {R.}~\bibnamefont {Weiss}}, \bibinfo {author} {\bibfnamefont
  {V.}~\bibnamefont {Dubourg}}, \bibinfo {author} {\bibfnamefont
  {J.}~\bibnamefont {Vanderplas}}, \bibinfo {author} {\bibfnamefont
  {A.}~\bibnamefont {Passos}}, \bibinfo {author} {\bibfnamefont
  {D.}~\bibnamefont {Cournapeau}}, \bibinfo {author} {\bibfnamefont
  {M.}~\bibnamefont {Brucher}}, \bibinfo {author} {\bibfnamefont
  {M.}~\bibnamefont {Perrot}}, \ and\ \bibinfo {author} {\bibfnamefont
  {E.}~\bibnamefont {Duchesnay}},\ }\href@noop {} {\bibfield  {journal}
  {\bibinfo  {journal} {Journal of Machine Learning Research}\ }\textbf
  {\bibinfo {volume} {12}},\ \bibinfo {pages} {2825} (\bibinfo {year}
  {2011})}\BibitemShut {NoStop}%
\bibitem [{\citenamefont {Brynjarsdóttir}\ and\ \citenamefont
  {O'Hagan}(2014)}]{Brynjarsdottir_2014}%
  \BibitemOpen
  \bibfield  {author} {\bibinfo {author} {\bibfnamefont {J.}~\bibnamefont
  {Brynjarsdóttir}}\ and\ \bibinfo {author} {\bibfnamefont {A.}~\bibnamefont
  {O'Hagan}},\ }\href {\doibase 10.1088/0266-5611/30/11/114007} {\bibfield
  {journal} {\bibinfo  {journal} {Inverse Problems}\ }\textbf {\bibinfo
  {volume} {30}} (\bibinfo {year} {2014}),\
  10.1088/0266-5611/30/11/114007}\BibitemShut {NoStop}%
\bibitem [{\citenamefont {Hogg}\ and\ \citenamefont
  {Foreman-Mackey}(2018)}]{Hogg:2017akh}%
  \BibitemOpen
  \bibfield  {author} {\bibinfo {author} {\bibfnamefont {D.~W.}\ \bibnamefont
  {Hogg}}\ and\ \bibinfo {author} {\bibfnamefont {D.}~\bibnamefont
  {Foreman-Mackey}},\ }\href {\doibase 10.3847/1538-4365/aab76e} {\bibfield
  {journal} {\bibinfo  {journal} {The Astrophysical Journal Supplement Series}\
  }\textbf {\bibinfo {volume} {236}},\ \bibinfo {pages} {11} (\bibinfo {year}
  {2018})}\BibitemShut {NoStop}%
\bibitem [{\citenamefont {Brooks}\ \emph {et~al.}(2011)\citenamefont {Brooks},
  \citenamefont {Gelman}, \citenamefont {Jones},\ and\ \citenamefont
  {Meng}}]{brooks2011handbook}%
  \BibitemOpen
  \bibfield  {author} {\bibinfo {author} {\bibfnamefont {S.}~\bibnamefont
  {Brooks}}, \bibinfo {author} {\bibfnamefont {A.}~\bibnamefont {Gelman}},
  \bibinfo {author} {\bibfnamefont {G.}~\bibnamefont {Jones}}, \ and\ \bibinfo
  {author} {\bibfnamefont {X.}~\bibnamefont {Meng}},\ }\href
  {https://books.google.com/books?id=qfRsAIKZ4rIC} {\emph {\bibinfo {title}
  {Handbook of Markov Chain Monte Carlo}}},\ Chapman \& Hall/CRC Handbooks of
  Modern Statistical Methods\ (\bibinfo  {publisher} {CRC Press},\ \bibinfo
  {year} {2011})\BibitemShut {NoStop}%
\bibitem [{\citenamefont {Vousden}\ \emph {et~al.}(2015)\citenamefont
  {Vousden}, \citenamefont {Farr},\ and\ \citenamefont
  {Mandel}}]{Vousden_2015}%
  \BibitemOpen
  \bibfield  {author} {\bibinfo {author} {\bibfnamefont {W.~D.}\ \bibnamefont
  {Vousden}}, \bibinfo {author} {\bibfnamefont {W.~M.}\ \bibnamefont {Farr}}, \
  and\ \bibinfo {author} {\bibfnamefont {I.}~\bibnamefont {Mandel}},\ }\href
  {\doibase 10.1093/mnras/stv2422} {\bibfield  {journal} {\bibinfo  {journal}
  {Monthly Notices of the Royal Astronomical Society}\ }\textbf {\bibinfo
  {volume} {455}},\ \bibinfo {pages} {1919} (\bibinfo {year}
  {2015})}\BibitemShut {NoStop}%
\bibitem [{\citenamefont {Shen}\ \emph {et~al.}(2014)\citenamefont {Shen},
  \citenamefont {Heinz}, \citenamefont {Paquet},\ and\ \citenamefont
  {Gale}}]{Shen:2013vja}%
  \BibitemOpen
  \bibfield  {author} {\bibinfo {author} {\bibfnamefont {C.}~\bibnamefont
  {Shen}}, \bibinfo {author} {\bibfnamefont {U.}~\bibnamefont {Heinz}},
  \bibinfo {author} {\bibfnamefont {J.-F.}\ \bibnamefont {Paquet}}, \ and\
  \bibinfo {author} {\bibfnamefont {C.}~\bibnamefont {Gale}},\ }\href {\doibase
  10.1103/PhysRevC.89.044910} {\bibfield  {journal} {\bibinfo  {journal} {Phys.
  Rev. C}\ }\textbf {\bibinfo {volume} {89}},\ \bibinfo {pages} {044910}
  (\bibinfo {year} {2014})},\ \Eprint {http://arxiv.org/abs/1308.2440}
  {arXiv:1308.2440 [nucl-th]} \BibitemShut {NoStop}%
\bibitem [{\citenamefont {Chaloner}\ and\ \citenamefont
  {Verdinelli}(1995)}]{chaloner1995}%
  \BibitemOpen
  \bibfield  {author} {\bibinfo {author} {\bibfnamefont {K.}~\bibnamefont
  {Chaloner}}\ and\ \bibinfo {author} {\bibfnamefont {I.}~\bibnamefont
  {Verdinelli}},\ }\href {\doibase 10.1214/ss/1177009939} {\bibfield  {journal}
  {\bibinfo  {journal} {Statist. Sci.}\ }\textbf {\bibinfo {volume} {10}},\
  \bibinfo {pages} {273} (\bibinfo {year} {1995})}\BibitemShut {NoStop}%
\bibitem [{\citenamefont {Aamodt}\ \emph
  {et~al.}(2011{\natexlab{a}})\citenamefont {Aamodt} \emph
  {et~al.}}]{Aamodt:2010cz}%
  \BibitemOpen
  \bibfield  {author} {\bibinfo {author} {\bibfnamefont {K.}~\bibnamefont
  {Aamodt}} \emph {et~al.} (\bibinfo {collaboration} {ALICE}),\ }\href
  {\doibase 10.1103/PhysRevLett.106.032301} {\bibfield  {journal} {\bibinfo
  {journal} {Phys. Rev. Lett.}\ }\textbf {\bibinfo {volume} {106}},\ \bibinfo
  {pages} {032301} (\bibinfo {year} {2011}{\natexlab{a}})},\ \Eprint
  {http://arxiv.org/abs/1012.1657} {arXiv:1012.1657 [nucl-ex]} \BibitemShut
  {NoStop}%
\bibitem [{\citenamefont {Adam}\ \emph {et~al.}(2016)\citenamefont {Adam} \emph
  {et~al.}}]{Adam:2016thv}%
  \BibitemOpen
  \bibfield  {author} {\bibinfo {author} {\bibfnamefont {J.}~\bibnamefont
  {Adam}} \emph {et~al.} (\bibinfo {collaboration} {ALICE}),\ }\href {\doibase
  10.1103/PhysRevC.94.034903} {\bibfield  {journal} {\bibinfo  {journal} {Phys.
  Rev.}\ }\textbf {\bibinfo {volume} {C94}},\ \bibinfo {pages} {034903}
  (\bibinfo {year} {2016})},\ \Eprint {http://arxiv.org/abs/1603.04775}
  {arXiv:1603.04775 [nucl-ex]} \BibitemShut {NoStop}%
\bibitem [{\citenamefont {Abelev}\ \emph {et~al.}(2013)\citenamefont {Abelev}
  \emph {et~al.}}]{Abelev:2013vea}%
  \BibitemOpen
  \bibfield  {author} {\bibinfo {author} {\bibfnamefont {B.}~\bibnamefont
  {Abelev}} \emph {et~al.} (\bibinfo {collaboration} {ALICE}),\ }\href
  {\doibase 10.1103/PhysRevC.88.044910} {\bibfield  {journal} {\bibinfo
  {journal} {Phys. Rev.}\ }\textbf {\bibinfo {volume} {C88}},\ \bibinfo {pages}
  {044910} (\bibinfo {year} {2013})},\ \Eprint {http://arxiv.org/abs/1303.0737}
  {arXiv:1303.0737 [hep-ex]} \BibitemShut {NoStop}%
\bibitem [{\citenamefont {Aamodt}\ \emph
  {et~al.}(2011{\natexlab{b}})\citenamefont {Aamodt} \emph
  {et~al.}}]{ALICE:2011ab}%
  \BibitemOpen
  \bibfield  {author} {\bibinfo {author} {\bibfnamefont {K.}~\bibnamefont
  {Aamodt}} \emph {et~al.} (\bibinfo {collaboration} {ALICE}),\ }\href
  {\doibase 10.1103/PhysRevLett.107.032301} {\bibfield  {journal} {\bibinfo
  {journal} {Phys. Rev. Lett.}\ }\textbf {\bibinfo {volume} {107}},\ \bibinfo
  {pages} {032301} (\bibinfo {year} {2011}{\natexlab{b}})},\ \Eprint
  {http://arxiv.org/abs/1105.3865} {arXiv:1105.3865 [nucl-ex]} \BibitemShut
  {NoStop}%
\bibitem [{\citenamefont {Abelev}\ \emph {et~al.}(2014)\citenamefont {Abelev}
  \emph {et~al.}}]{Abelev:2014ckr}%
  \BibitemOpen
  \bibfield  {author} {\bibinfo {author} {\bibfnamefont {B.~B.}\ \bibnamefont
  {Abelev}} \emph {et~al.} (\bibinfo {collaboration} {ALICE}),\ }\href
  {\doibase 10.1140/epjc/s10052-014-3077-y} {\bibfield  {journal} {\bibinfo
  {journal} {Eur. Phys. J.}\ }\textbf {\bibinfo {volume} {C74}},\ \bibinfo
  {pages} {3077} (\bibinfo {year} {2014})},\ \Eprint
  {http://arxiv.org/abs/1407.5530} {arXiv:1407.5530 [nucl-ex]} \BibitemShut
  {NoStop}%
\bibitem [{\citenamefont {Abelev}\ \emph {et~al.}(2009)\citenamefont {Abelev}
  \emph {et~al.}}]{Abelev:2008ab}%
  \BibitemOpen
  \bibfield  {author} {\bibinfo {author} {\bibfnamefont {B.~I.}\ \bibnamefont
  {Abelev}} \emph {et~al.} (\bibinfo {collaboration} {STAR}),\ }\href {\doibase
  10.1103/PhysRevC.79.034909} {\bibfield  {journal} {\bibinfo  {journal} {Phys.
  Rev.}\ }\textbf {\bibinfo {volume} {C79}},\ \bibinfo {pages} {034909}
  (\bibinfo {year} {2009})},\ \Eprint {http://arxiv.org/abs/0808.2041}
  {arXiv:0808.2041 [nucl-ex]} \BibitemShut {NoStop}%
\bibitem [{\citenamefont {Adams}\ \emph {et~al.}(2005)\citenamefont {Adams}
  \emph {et~al.}}]{Adams:2004bi}%
  \BibitemOpen
  \bibfield  {author} {\bibinfo {author} {\bibfnamefont {J.}~\bibnamefont
  {Adams}} \emph {et~al.} (\bibinfo {collaboration} {STAR}),\ }\href {\doibase
  10.1103/PhysRevC.72.014904} {\bibfield  {journal} {\bibinfo  {journal} {Phys.
  Rev.}\ }\textbf {\bibinfo {volume} {C72}},\ \bibinfo {pages} {014904}
  (\bibinfo {year} {2005})},\ \Eprint {http://arxiv.org/abs/nucl-ex/0409033}
  {arXiv:nucl-ex/0409033 [nucl-ex]} \BibitemShut {NoStop}%
\bibitem [{\citenamefont {Adamczyk}\ \emph {et~al.}(2013)\citenamefont
  {Adamczyk} \emph {et~al.}}]{Adamczyk:2013waa}%
  \BibitemOpen
  \bibfield  {author} {\bibinfo {author} {\bibfnamefont {L.}~\bibnamefont
  {Adamczyk}} \emph {et~al.} (\bibinfo {collaboration} {STAR}),\ }\href
  {\doibase 10.1103/PhysRevC.88.014904} {\bibfield  {journal} {\bibinfo
  {journal} {Phys. Rev.}\ }\textbf {\bibinfo {volume} {C88}},\ \bibinfo {pages}
  {014904} (\bibinfo {year} {2013})},\ \Eprint {http://arxiv.org/abs/1301.2187}
  {arXiv:1301.2187 [nucl-ex]} \BibitemShut {NoStop}%
\bibitem [{\citenamefont {Adler}\ \emph {et~al.}(2004)\citenamefont {Adler}
  \emph {et~al.}}]{Adler:2003cb}%
  \BibitemOpen
  \bibfield  {author} {\bibinfo {author} {\bibfnamefont {S.}~\bibnamefont
  {Adler}} \emph {et~al.} (\bibinfo {collaboration} {PHENIX}),\ }\href
  {\doibase 10.1103/PhysRevC.69.034909} {\bibfield  {journal} {\bibinfo
  {journal} {Phys. Rev. C}\ }\textbf {\bibinfo {volume} {69}},\ \bibinfo
  {pages} {034909} (\bibinfo {year} {2004})},\ \Eprint
  {http://arxiv.org/abs/nucl-ex/0307022} {arXiv:nucl-ex/0307022} \BibitemShut
  {NoStop}%
\bibitem [{\citenamefont {Pratt}\ \emph {et~al.}(2015)\citenamefont {Pratt},
  \citenamefont {Sangaline}, \citenamefont {Sorensen},\ and\ \citenamefont
  {Wang}}]{Pratt:2015zsa}%
  \BibitemOpen
  \bibfield  {author} {\bibinfo {author} {\bibfnamefont {S.}~\bibnamefont
  {Pratt}}, \bibinfo {author} {\bibfnamefont {E.}~\bibnamefont {Sangaline}},
  \bibinfo {author} {\bibfnamefont {P.}~\bibnamefont {Sorensen}}, \ and\
  \bibinfo {author} {\bibfnamefont {H.}~\bibnamefont {Wang}},\ }\href {\doibase
  10.1103/PhysRevLett.114.202301} {\bibfield  {journal} {\bibinfo  {journal}
  {Phys. Rev. Lett.}\ }\textbf {\bibinfo {volume} {114}},\ \bibinfo {pages}
  {202301} (\bibinfo {year} {2015})},\ \Eprint
  {http://arxiv.org/abs/1501.04042} {arXiv:1501.04042 [nucl-th]} \BibitemShut
  {NoStop}%
\bibitem [{\citenamefont {Wolff}\ and\ \citenamefont
  {Molnar}(2017)}]{Wolff:2016vcm}%
  \BibitemOpen
  \bibfield  {author} {\bibinfo {author} {\bibfnamefont {Z.}~\bibnamefont
  {Wolff}}\ and\ \bibinfo {author} {\bibfnamefont {D.}~\bibnamefont {Molnar}},\
  }\href {\doibase 10.1103/PhysRevC.96.044909} {\bibfield  {journal} {\bibinfo
  {journal} {Phys.\ Rev.\ C}\ }\textbf {\bibinfo {volume} {96}},\ \bibinfo
  {pages} {044909} (\bibinfo {year} {2017})},\ \Eprint
  {http://arxiv.org/abs/1611.09185} {arXiv:1611.09185 [nucl-th]} \BibitemShut
  {NoStop}%
\bibitem [{\citenamefont {Lappi}(2006)}]{Lappi:2006hq}%
  \BibitemOpen
  \bibfield  {author} {\bibinfo {author} {\bibfnamefont {T.}~\bibnamefont
  {Lappi}},\ }\href {\doibase 10.1016/j.physletb.2006.10.017} {\bibfield
  {journal} {\bibinfo  {journal} {Phys. Lett. B}\ }\textbf {\bibinfo {volume}
  {643}},\ \bibinfo {pages} {11} (\bibinfo {year} {2006})},\ \Eprint
  {http://arxiv.org/abs/hep-ph/0606207} {arXiv:hep-ph/0606207} \BibitemShut
  {NoStop}%
\bibitem [{\citenamefont {Gale}\ \emph
  {et~al.}(2013{\natexlab{b}})\citenamefont {Gale}, \citenamefont {Jeon},
  \citenamefont {Schenke}, \citenamefont {Tribedy},\ and\ \citenamefont
  {Venugopalan}}]{Gale:2012rq}%
  \BibitemOpen
  \bibfield  {author} {\bibinfo {author} {\bibfnamefont {C.}~\bibnamefont
  {Gale}}, \bibinfo {author} {\bibfnamefont {S.}~\bibnamefont {Jeon}}, \bibinfo
  {author} {\bibfnamefont {B.}~\bibnamefont {Schenke}}, \bibinfo {author}
  {\bibfnamefont {P.}~\bibnamefont {Tribedy}}, \ and\ \bibinfo {author}
  {\bibfnamefont {R.}~\bibnamefont {Venugopalan}},\ }\href {\doibase
  10.1103/PhysRevLett.110.012302} {\bibfield  {journal} {\bibinfo  {journal}
  {Phys. Rev. Lett.}\ }\textbf {\bibinfo {volume} {110}},\ \bibinfo {pages}
  {012302} (\bibinfo {year} {2013}{\natexlab{b}})},\ \Eprint
  {http://arxiv.org/abs/1209.6330} {arXiv:1209.6330 [nucl-th]} \BibitemShut
  {NoStop}%
\bibitem [{\citenamefont {McDonald}\ \emph {et~al.}(2017)\citenamefont
  {McDonald}, \citenamefont {Shen}, \citenamefont {Fillion-Gourdeau},
  \citenamefont {Jeon},\ and\ \citenamefont {Gale}}]{McDonald:2016vlt}%
  \BibitemOpen
  \bibfield  {author} {\bibinfo {author} {\bibfnamefont {S.}~\bibnamefont
  {McDonald}}, \bibinfo {author} {\bibfnamefont {C.}~\bibnamefont {Shen}},
  \bibinfo {author} {\bibfnamefont {F.}~\bibnamefont {Fillion-Gourdeau}},
  \bibinfo {author} {\bibfnamefont {S.}~\bibnamefont {Jeon}}, \ and\ \bibinfo
  {author} {\bibfnamefont {C.}~\bibnamefont {Gale}},\ }\href {\doibase
  10.1103/PhysRevC.95.064913} {\bibfield  {journal} {\bibinfo  {journal} {Phys.
  Rev. C}\ }\textbf {\bibinfo {volume} {95}},\ \bibinfo {pages} {064913}
  (\bibinfo {year} {2017})},\ \Eprint {http://arxiv.org/abs/1609.02958}
  {arXiv:1609.02958 [hep-ph]} \BibitemShut {NoStop}%
\bibitem [{\citenamefont {Nunes~da Silva}\ \emph {et~al.}(2020)\citenamefont
  {Nunes~da Silva}, \citenamefont {Chinellato}, \citenamefont {Hippert},
  \citenamefont {Serenone}, \citenamefont {Takahashi}, \citenamefont {Denicol},
  \citenamefont {Luzum},\ and\ \citenamefont {Noronha}}]{NunesdaSilva:2020bfs}%
  \BibitemOpen
  \bibfield  {author} {\bibinfo {author} {\bibfnamefont {T.}~\bibnamefont
  {Nunes~da Silva}}, \bibinfo {author} {\bibfnamefont {D.}~\bibnamefont
  {Chinellato}}, \bibinfo {author} {\bibfnamefont {M.}~\bibnamefont {Hippert}},
  \bibinfo {author} {\bibfnamefont {W.}~\bibnamefont {Serenone}}, \bibinfo
  {author} {\bibfnamefont {J.}~\bibnamefont {Takahashi}}, \bibinfo {author}
  {\bibfnamefont {G.~S.}\ \bibnamefont {Denicol}}, \bibinfo {author}
  {\bibfnamefont {M.}~\bibnamefont {Luzum}}, \ and\ \bibinfo {author}
  {\bibfnamefont {J.}~\bibnamefont {Noronha}},\ }\href@noop {} {\  (\bibinfo
  {year} {2020})},\ \Eprint {http://arxiv.org/abs/2006.02324} {arXiv:2006.02324
  [nucl-th]} \BibitemShut {NoStop}%
\bibitem [{\citenamefont {Luzum}\ and\ \citenamefont
  {Romatschke}(2008)}]{Luzum:2008cw}%
  \BibitemOpen
  \bibfield  {author} {\bibinfo {author} {\bibfnamefont {M.}~\bibnamefont
  {Luzum}}\ and\ \bibinfo {author} {\bibfnamefont {P.}~\bibnamefont
  {Romatschke}},\ }\href {\doibase 10.1103/PhysRevC.78.034915,
  10.1103/PhysRevC.79.039903} {\bibfield  {journal} {\bibinfo  {journal} {Phys.
  Rev.}\ }\textbf {\bibinfo {volume} {C78}},\ \bibinfo {pages} {034915}
  (\bibinfo {year} {2008})},\ \bibinfo {note} {[Erratum: Phys.
  Rev.C79,039903(2009)]},\ \Eprint {http://arxiv.org/abs/0804.4015}
  {arXiv:0804.4015 [nucl-th]} \BibitemShut {NoStop}%
\bibitem [{\citenamefont {Song}(2009)}]{Song:2009gc}%
  \BibitemOpen
  \bibfield  {author} {\bibinfo {author} {\bibfnamefont {H.}~\bibnamefont
  {Song}},\ }\emph {\bibinfo {title} {{Causal Viscous Hydrodynamics for
  Relativistic Heavy Ion Collisions}}},\ \href@noop {} {Ph.D. thesis},\
  \bibinfo  {school} {Ohio State U.} (\bibinfo {year} {2009}),\ \Eprint
  {http://arxiv.org/abs/0908.3656} {arXiv:0908.3656 [nucl-th]} \BibitemShut
  {NoStop}%
\bibitem [{\citenamefont {Hamby}(1994)}]{Hamby}%
  \BibitemOpen
  \bibfield  {author} {\bibinfo {author} {\bibfnamefont {D.}~\bibnamefont
  {Hamby}},\ }\href {\doibase 10.1007/BF00547132} {\bibfield  {journal}
  {\bibinfo  {journal} {Environ Monit Assess}\ }\textbf {\bibinfo {volume}
  {32}},\ \bibinfo {pages} {135} (\bibinfo {year} {1994})}\BibitemShut
  {NoStop}%
\bibitem [{\citenamefont {Trotta}(2008)}]{Trotta:2008qt}%
  \BibitemOpen
  \bibfield  {author} {\bibinfo {author} {\bibfnamefont {R.}~\bibnamefont
  {Trotta}},\ }\href {\doibase 10.1080/00107510802066753} {\bibfield  {journal}
  {\bibinfo  {journal} {Contemp. Phys.}\ }\textbf {\bibinfo {volume} {49}},\
  \bibinfo {pages} {71} (\bibinfo {year} {2008})},\ \Eprint
  {http://arxiv.org/abs/0803.4089} {arXiv:0803.4089 [astro-ph]} \BibitemShut
  {NoStop}%
\bibitem [{pte()}]{ptemcee_code}%
  \BibitemOpen
  \href@noop {} {}\bibinfo {howpublished}
  {\url{https://github.com/willvousden/ptemcee}}\BibitemShut {NoStop}%
\bibitem [{\citenamefont {Marshall}\ \emph {et~al.}(2006)\citenamefont
  {Marshall}, \citenamefont {Rajguru},\ and\ \citenamefont
  {Slosar}}]{Marshall_2006}%
  \BibitemOpen
  \bibfield  {author} {\bibinfo {author} {\bibfnamefont {P.}~\bibnamefont
  {Marshall}}, \bibinfo {author} {\bibfnamefont {N.}~\bibnamefont {Rajguru}}, \
  and\ \bibinfo {author} {\bibfnamefont {A.}~\bibnamefont {Slosar}},\ }\href
  {\doibase 10.1103/physrevd.73.067302} {\bibfield  {journal} {\bibinfo
  {journal} {Physical Review D}\ }\textbf {\bibinfo {volume} {73}} (\bibinfo
  {year} {2006}),\ 10.1103/physrevd.73.067302}\BibitemShut {NoStop}%
\bibitem [{\citenamefont {Schnedermann}\ \emph {et~al.}(1993)\citenamefont
  {Schnedermann}, \citenamefont {Sollfrank},\ and\ \citenamefont
  {Heinz}}]{Schnedermann:1993ws}%
  \BibitemOpen
  \bibfield  {author} {\bibinfo {author} {\bibfnamefont {E.}~\bibnamefont
  {Schnedermann}}, \bibinfo {author} {\bibfnamefont {J.}~\bibnamefont
  {Sollfrank}}, \ and\ \bibinfo {author} {\bibfnamefont {U.}~\bibnamefont
  {Heinz}},\ }\href {\doibase 10.1103/PhysRevC.48.2462} {\bibfield  {journal}
  {\bibinfo  {journal} {Phys. Rev. C}\ }\textbf {\bibinfo {volume} {48}},\
  \bibinfo {pages} {2462} (\bibinfo {year} {1993})},\ \Eprint
  {http://arxiv.org/abs/nucl-th/9307020} {arXiv:nucl-th/9307020} \BibitemShut
  {NoStop}%
\bibitem [{\citenamefont {Heinz}(2004)}]{Heinz:2004qz}%
  \BibitemOpen
  \bibfield  {author} {\bibinfo {author} {\bibfnamefont {U.}~\bibnamefont
  {Heinz}},\ }in\ \href@noop {} {\emph {\bibinfo {booktitle} {{2nd CERN-CLAF
  School of High Energy Physics}}}}\ (\bibinfo {year} {2004})\ pp.\ \bibinfo
  {pages} {165--238},\ \Eprint {http://arxiv.org/abs/hep-ph/0407360}
  {arXiv:hep-ph/0407360} \BibitemShut {NoStop}%
\bibitem [{wid()}]{widget}%
  \BibitemOpen
  \href@noop {} {}\bibinfo {howpublished}
  {\url{http://jetscape.org/sims-widget}}\BibitemShut {NoStop}%
\bibitem [{\citenamefont {Wiedemann}\ \emph {et~al.}(1998)\citenamefont
  {Wiedemann}, \citenamefont {Tomasik},\ and\ \citenamefont
  {Heinz}}]{Wiedemann:1997cn}%
  \BibitemOpen
  \bibfield  {author} {\bibinfo {author} {\bibfnamefont {U.~A.}\ \bibnamefont
  {Wiedemann}}, \bibinfo {author} {\bibfnamefont {B.}~\bibnamefont {Tomasik}},
  \ and\ \bibinfo {author} {\bibfnamefont {U.}~\bibnamefont {Heinz}},\ }\href
  {\doibase 10.1016/S0375-9474(98)00391-1} {\bibfield  {journal} {\bibinfo
  {journal} {Nucl. Phys. A}\ }\textbf {\bibinfo {volume} {638}},\ \bibinfo
  {pages} {475C} (\bibinfo {year} {1998})},\ \Eprint
  {http://arxiv.org/abs/nucl-th/9801017} {arXiv:nucl-th/9801017} \BibitemShut
  {NoStop}%
\bibitem [{\citenamefont {Vujanovic}\ \emph {et~al.}(2018)\citenamefont
  {Vujanovic}, \citenamefont {Denicol}, \citenamefont {Luzum}, \citenamefont
  {Jeon},\ and\ \citenamefont {Gale}}]{Vujanovic:2017psb}%
  \BibitemOpen
  \bibfield  {author} {\bibinfo {author} {\bibfnamefont {G.}~\bibnamefont
  {Vujanovic}}, \bibinfo {author} {\bibfnamefont {G.~S.}\ \bibnamefont
  {Denicol}}, \bibinfo {author} {\bibfnamefont {M.}~\bibnamefont {Luzum}},
  \bibinfo {author} {\bibfnamefont {S.}~\bibnamefont {Jeon}}, \ and\ \bibinfo
  {author} {\bibfnamefont {C.}~\bibnamefont {Gale}},\ }\href {\doibase
  10.1103/PhysRevC.98.014902} {\bibfield  {journal} {\bibinfo  {journal} {Phys.
  Rev.}\ }\textbf {\bibinfo {volume} {C98}},\ \bibinfo {pages} {014902}
  (\bibinfo {year} {2018})},\ \Eprint {http://arxiv.org/abs/1702.02941}
  {arXiv:1702.02941 [nucl-th]} \BibitemShut {NoStop}%
\bibitem [{\citenamefont {Gale}\ \emph {et~al.}(2019)\citenamefont {Gale},
  \citenamefont {Jeon}, \citenamefont {McDonald}, \citenamefont {Paquet},\ and\
  \citenamefont {Shen}}]{Gale:2018vuh}%
  \BibitemOpen
  \bibfield  {author} {\bibinfo {author} {\bibfnamefont {C.}~\bibnamefont
  {Gale}}, \bibinfo {author} {\bibfnamefont {S.}~\bibnamefont {Jeon}}, \bibinfo
  {author} {\bibfnamefont {S.}~\bibnamefont {McDonald}}, \bibinfo {author}
  {\bibfnamefont {J.-F.}\ \bibnamefont {Paquet}}, \ and\ \bibinfo {author}
  {\bibfnamefont {C.}~\bibnamefont {Shen}},\ }\href {\doibase
  10.1016/j.nuclphysa.2018.08.005} {\bibfield  {journal} {\bibinfo  {journal}
  {Nucl. Phys. A}\ }\textbf {\bibinfo {volume} {982}},\ \bibinfo {pages} {767}
  (\bibinfo {year} {2019})},\ \Eprint {http://arxiv.org/abs/1807.09326}
  {arXiv:1807.09326 [nucl-th]} \BibitemShut {NoStop}%
\bibitem [{\citenamefont {Hauksson}\ \emph {et~al.}(2020)\citenamefont
  {Hauksson}, \citenamefont {Jeon},\ and\ \citenamefont
  {Gale}}]{Hauksson:2020etn}%
  \BibitemOpen
  \bibfield  {author} {\bibinfo {author} {\bibfnamefont {S.}~\bibnamefont
  {Hauksson}}, \bibinfo {author} {\bibfnamefont {S.}~\bibnamefont {Jeon}}, \
  and\ \bibinfo {author} {\bibfnamefont {C.}~\bibnamefont {Gale}},\ }in\
  \href@noop {} {\emph {\bibinfo {booktitle} {{28th International Conference on
  Ultrarelativistic Nucleus-Nucleus Collisions}}}}\ (\bibinfo {year} {2020})\
  \Eprint {http://arxiv.org/abs/2001.10046} {arXiv:2001.10046 [hep-ph]}
  \BibitemShut {NoStop}%
\bibitem [{\citenamefont {Gale}\ \emph {et~al.}(2020)\citenamefont {Gale},
  \citenamefont {Paquet}, \citenamefont {Schenke},\ and\ \citenamefont
  {Shen}}]{Gale:2020xlg}%
  \BibitemOpen
  \bibfield  {author} {\bibinfo {author} {\bibfnamefont {C.}~\bibnamefont
  {Gale}}, \bibinfo {author} {\bibfnamefont {J.-F.}\ \bibnamefont {Paquet}},
  \bibinfo {author} {\bibfnamefont {B.}~\bibnamefont {Schenke}}, \ and\
  \bibinfo {author} {\bibfnamefont {C.}~\bibnamefont {Shen}},\ }in\ \href@noop
  {} {\emph {\bibinfo {booktitle} {{28th International Conference on
  Ultrarelativistic Nucleus-Nucleus Collisions}}}}\ (\bibinfo {year} {2020})\
  \Eprint {http://arxiv.org/abs/2002.05191} {arXiv:2002.05191 [hep-ph]}
  \BibitemShut {NoStop}%
\bibitem [{\citenamefont {Moreland}\ and\ \citenamefont
  {Soltz}(2016)}]{Moreland:2015dvc}%
  \BibitemOpen
  \bibfield  {author} {\bibinfo {author} {\bibfnamefont {J.~S.}\ \bibnamefont
  {Moreland}}\ and\ \bibinfo {author} {\bibfnamefont {R.~A.}\ \bibnamefont
  {Soltz}},\ }\href {\doibase 10.1103/PhysRevC.93.044913} {\bibfield  {journal}
  {\bibinfo  {journal} {Phys. Rev. C}\ }\textbf {\bibinfo {volume} {93}},\
  \bibinfo {pages} {044913} (\bibinfo {year} {2016})},\ \Eprint
  {http://arxiv.org/abs/1512.02189} {arXiv:1512.02189 [nucl-th]} \BibitemShut
  {NoStop}%
\bibitem [{TAC()}]{TACC}%
  \BibitemOpen
  \href@noop {} {}\bibinfo {howpublished}
  {\url{http://www.tacc.utexas.edu}}\BibitemShut {NoStop}%
\bibitem [{Ohi(1987)}]{OhioSupercomputerCenter1987}%
  \BibitemOpen
  \href@noop {} {\enquote {\bibinfo {title} {{O}hio {S}upercomputer
  {C}enter},}\ }\bibinfo {howpublished}
  {\url{http://osc.edu/ark:/19495/f5s1ph73}} (\bibinfo {year}
  {1987})\BibitemShut {NoStop}%
\bibitem [{osu()}]{osuhydro}%
  \BibitemOpen
  \href@noop {} {}\bibinfo {howpublished}
  {\url{https://github.com/jbernhard/osu-hydro}}\BibitemShut {NoStop}%
\bibitem [{\citenamefont {Moreland}(2019)}]{Moreland:2019szz}%
  \BibitemOpen
  \bibfield  {author} {\bibinfo {author} {\bibfnamefont {J.~S.}\ \bibnamefont
  {Moreland}},\ }\emph {\bibinfo {title} {{Initial conditions of bulk matter in
  ultrarelativistic nuclear collisions}}},\ \href@noop {} {Ph.D. thesis},\
  \bibinfo  {school} {Duke U.} (\bibinfo {year} {2019}),\ \Eprint
  {http://arxiv.org/abs/1904.08290} {arXiv:1904.08290 [nucl-th]} \BibitemShut
  {NoStop}%
\bibitem [{\citenamefont {Boris}\ and\ \citenamefont
  {Book}(1973)}]{Boris:1973tjt}%
  \BibitemOpen
  \bibfield  {author} {\bibinfo {author} {\bibfnamefont {J.~P.}\ \bibnamefont
  {Boris}}\ and\ \bibinfo {author} {\bibfnamefont {D.~L.}\ \bibnamefont
  {Book}},\ }\href {\doibase 10.1016/0021-9991(73)90147-2} {\bibfield
  {journal} {\bibinfo  {journal} {J. Comput. Phys.}\ }\textbf {\bibinfo
  {volume} {11}},\ \bibinfo {pages} {38} (\bibinfo {year} {1973})}\BibitemShut
  {NoStop}%
\bibitem [{\citenamefont {Denicol}\ \emph {et~al.}(2018)\citenamefont
  {Denicol}, \citenamefont {Gale}, \citenamefont {Jeon}, \citenamefont
  {Monnai}, \citenamefont {Schenke},\ and\ \citenamefont
  {Shen}}]{Denicol:2018wdp}%
  \BibitemOpen
  \bibfield  {author} {\bibinfo {author} {\bibfnamefont {G.~S.}\ \bibnamefont
  {Denicol}}, \bibinfo {author} {\bibfnamefont {C.}~\bibnamefont {Gale}},
  \bibinfo {author} {\bibfnamefont {S.}~\bibnamefont {Jeon}}, \bibinfo {author}
  {\bibfnamefont {A.}~\bibnamefont {Monnai}}, \bibinfo {author} {\bibfnamefont
  {B.}~\bibnamefont {Schenke}}, \ and\ \bibinfo {author} {\bibfnamefont
  {C.}~\bibnamefont {Shen}},\ }\href {\doibase 10.1103/PhysRevC.98.034916}
  {\bibfield  {journal} {\bibinfo  {journal} {Phys. Rev.}\ }\textbf {\bibinfo
  {volume} {C98}},\ \bibinfo {pages} {034916} (\bibinfo {year} {2018})},\
  \Eprint {http://arxiv.org/abs/1804.10557} {arXiv:1804.10557 [nucl-th]}
  \BibitemShut {NoStop}%
\bibitem [{\citenamefont {{Wolfram Research{,} Inc.}}()}]{Mathematica_v12_1}%
  \BibitemOpen
  \bibfield  {author} {\bibinfo {author} {\bibnamefont {{Wolfram Research{,}
  Inc.}}},\ }\href {https://www.wolfram.com/mathematica} {\enquote {\bibinfo
  {title} {Mathematica, {V}ersion 12.1},}\ }\bibinfo {note} {Champaign, IL,
  2020}\BibitemShut {NoStop}%
\bibitem [{\citenamefont {Paquet}\ and\ \citenamefont
  {Bass}(2020)}]{Paquet:2019npk}%
  \BibitemOpen
  \bibfield  {author} {\bibinfo {author} {\bibfnamefont {J.-F.}\ \bibnamefont
  {Paquet}}\ and\ \bibinfo {author} {\bibfnamefont {S.~A.}\ \bibnamefont
  {Bass}},\ }\href {\doibase 10.1103/PhysRevC.102.014903} {\bibfield  {journal}
  {\bibinfo  {journal} {Phys. Rev. C}\ }\textbf {\bibinfo {volume} {102}},\
  \bibinfo {pages} {014903} (\bibinfo {year} {2020})},\ \Eprint
  {http://arxiv.org/abs/1912.06287} {arXiv:1912.06287 [nucl-th]} \BibitemShut
  {NoStop}%
\bibitem [{mat()}]{mathematica_ideal_hydro_cylindrical}%
  \BibitemOpen
  \href@noop {} {}\bibinfo {howpublished}
  {\url{https://github.com/j-f-paquet/ideal_hydro_cylindrical}}\BibitemShut
  {NoStop}%
\bibitem [{frz()}]{frzout}%
  \BibitemOpen
  \href@noop {} {}\bibinfo {howpublished}
  {\url{https://github.com/Duke-QCD/frzout/tree/61b50f627174d8696f33413ab1ff873c4f0c952b}}\BibitemShut
  {NoStop}%
\bibitem [{\citenamefont {Tanabashi}\ \emph {et~al.}(2018)\citenamefont
  {Tanabashi} \emph {et~al.}}]{Tanabashi:2018oca}%
  \BibitemOpen
  \bibfield  {author} {\bibinfo {author} {\bibfnamefont {M.}~\bibnamefont
  {Tanabashi}} \emph {et~al.} (\bibinfo {collaboration} {Particle Data
  Group}),\ }\href {\doibase 10.1103/PhysRevD.98.030001} {\bibfield  {journal}
  {\bibinfo  {journal} {Phys.\ Rev.\ D}\ }\textbf {\bibinfo {volume} {98}},\
  \bibinfo {pages} {030001} (\bibinfo {year} {2018})}\BibitemShut {NoStop}%
\end{thebibliography}%

\end{document}